\documentclass[twoside,fleqn]{ActaStyle}
\usepackage[colorlinks=true, pdfstartview=FitV, linkcolor=black,
            citecolor=black, urlcolor=black]{hyperref}
\usepackage[dvips]{graphicx}
\usepackage{times,cite}
\usepackage{wrapfig}
\usepackage{amsmath,amssymb}
\usepackage{theorem}
\usepackage[usenames]{color}

\usepackage{fancyhdr}
\renewcommand{\subsectionmark}[1]{}
\renewcommand{\thesection}{\arabic{section}}
\renewcommand{\thesubsection}{\thesection.\arabic{subsection}}

\renewcommand{\thefigure}{\thesection.\arabic{figure}}
\renewcommand{\thetable}{\thesection.\arabic{table}}

\def\authorheading{D. Reitzner et al.}
\def\shorttitle{Quantum Walks}

\newcommand{\half}{\frac{1}{2}}
\newcommand{\tr}{\mathrm{tr}}											

\newcommand{\<}{\langle}		
\renewcommand{\>}{\rangle}

\newtheorem{lemma}{Lemma}
\newtheorem{theorem}{Theorem}
{\theorembodyfont{\rmfamily}
 \newtheorem{example}{Example}}
{\theorembodyfont{\rmfamily}
 \newtheorem{exercise}{Exercise}}

\newcommand{\C}{\mathbb{C}}
\newcommand{\cA}{\mathcal{A}}
\newcommand{\cB}{\mathcal{B}}

\newcommand{\E}{\mathrm{\mathbf{E}}}
\newcommand{\Var}{\mathrm{\mathbf{Var}}}
\newcommand{\EE}[1]{\E\left(#1\right)}
\newcommand{\VV}[1]{\Var\left(#1\right)}

\newcommand{\R}{\mathbb{R}}
\newcommand{\ket}[1]{\left| #1\right\rangle}      
\newcommand{\bra}[1]{\left\langle #1\right|}      
\newcommand{\kets}[1]{| #1 \rangle}    				    
\newcommand{\bras}[1]{\langle #1 |}        				
\newcommand{\braket}[2]{\langle #1 | #2 \rangle} 	
\newcommand{\iii}{\mathbb{I}}											
\newcommand{\norm}[1]{\left\| #1\right\|}        	
\newcommand{\ep}{\epsilon}        								

\newcommand{\deriv}[2]{\frac{\textrm{d}#1}{\textrm{d}#2}} 	

\newcommand{\e}{\mathrm{e}}
\newcommand{\dd}{\mathrm{d}}

\newcommand{\sgn}{{\rm sgn\;}}

\newcommand{\spann}{\ell^2}

\newcommand{\ii}{\mathrm{i}}
\newcommand{\op}[1]{{#1}}

\newcommand{\psinit}{\psi_0} 
\newcommand{\order}{O} 
\newcommand{\opt}{\tilde m} 
\newcommand{\ke}[1]{|{#1}\rangle}
\newcommand{\br}[1]{\langle{#1}|}

\newcommand{\Ctrl}{\mathcal C}
\newcommand{\CtrlU}{\Ctrl\op U}
\newcommand{\CtrlWone}{\Ctrl\op W_1}
\newcommand{\CtrlWtwo}{\Ctrl\op W_2}
\newcommand{\id}{\iii}
\newcommand{\spec}{{\mathcal K}}
\newcommand{\all}{{\mathcal N}}
\newcommand{\subspace}{{\mathcal S}}

\renewcommand{\mod}{\mathrm{mod}\,}

\newcommand{\st}{:\ }

\newcommand{\no}[1]{\left|#1\right|}
\newcommand{\diag}{\mathrm{diag}\,}

\newenvironment{algorithm}
	{\begin{center}\rule{10cm}{1pt}
	         \begin{center}\begin{minipage}{8cm}\footnotesize\sf}
	{\end{minipage}\end{center}\rule[0.5cm]{10cm}{1pt}\end{center}\vspace{-1cm}}

\begin{document}
	\pagerange{603}{725}   

\renewcommand{\thefootnote}{\fnsymbol{footnote}}
\setcounter{footnote}{0}

	\title{QUANTUM WALKS\footnote{Small post-processing corrections were made at pages 653, 700, 707, and 711.}}

	\author{Daniel Reitzner}{Department of Mathematics, Technische Universit\"at M\"unchen, 85748 Garching, Germany}

	\author{Daniel Nagaj, Vladim\'{\i}r Bu\v{z}ek}
         {Research Center for Quantum Information, Institute of Physics, Slovak Academy of Sciences, D\'{u}bravsk\'{a} cesta 9, 845 11 Bratislava}

        \datumy{16 July 2012}{20 July 2012}

	\abstract{This tutorial article showcases the many varieties and uses of {\em quantum walks}. Discrete time quantum walks are introduced as counterparts of classical random walks. The emphasis is on the connections and differences between the two types of processes (with rather different underlying dynamics) for producing random distributions. We discuss algorithmic applications for graph-searching and compare the two approaches. Next, we look at quantization of Markov chains and show how it can lead to speedups for sampling schemes. Finally, we turn to continuous time quantum walks and their applications, which provide interesting (even exponential) speedups over classical approaches.}

\begin{flushright}
{\em \dots I walk the line.\\}
Johnny Cash
\end{flushright}

\quad{\small {\sf DOI:}}\ 10.2478/v10155-011-0006-6

\vspace{0.3cm}

\begin{minipage}{2.5cm}
\quad{\small {\sf KEYWORDS:}}
\end{minipage}
\begin{minipage}{10cm}
Quantum Walks, Random Walks, Quantum Algorithms, Markov Chains
\end{minipage}



\tableofcontents

\renewcommand{\thefootnote}{\arabic{footnote}}
\setcounter{footnote}{0}
\setcounter{equation}{0} \setcounter{figure}{0} \setcounter{table}{0}\newpage

\section{Introduction}

For a physicist, a {\em quantum walk} means the dynamics of an excitation in a quantum-mechanical spin system described by the tight-binding (or other similar model), combined with a measurement in a position basis. This procedure produces a random location, with a prescription for computing the probability of finding the excitation at a given place given by the rules of quantum mechanics. 

For a computer scientist or mathematician, a {\em quantum walk} is an analogy of a classical random walk, where instead of transforming probabilities by a stochastic matrix, we transform probability amplitudes by unitary transformations, which brings interesting interference effects into play. 

However, there's more to quantum walks, and the physics and computer science worlds have been both enriched by them. Motivated by classical random walks and algorithms based on them, we are compelled to look for quantum algorithms based not on classical Markov Chains but on quantum walks instead. Sometimes, this ``quantization'' is straightforward, but more often we can utilize the additional features of unitary transformations of amplitudes compared to the transfer-matrix-like evolution of probabilities for algorithmic purposes. This approach is amazingly fruitful, as it continues to produce successful quantum algorithms since its invention. On the other hand, quantum walks have motivated some interesting results in physics and brought forth several interesting experiments involving the dynamics of excitations or the behavior of photons in waveguides.

Our article starts with a review of classical random-walk based methods. The second step is the search for an analogy in the quantum world, with unitary steps transforming amplitudes, arriving at a new model of discrete-time quantum walks. These return back to classical random processes when noise is added to the quantum dynamics, resulting in a loss of coherence. On the other hand, when superpositions come into play, discrete-time quantum walks have strong applications in graph searching and other problems. Next, we will look at how general Markov chains can be quantized, and utilized in physically motivated sampling algorithms. Finally, we will investigate our original motivation -- analyzing the dynamics of an excitation in a spin system, and realize that this is also a quantum walk, in continuous-time. We will investigate its basic properties, natural and powerful algorithms (e.g. graph search and traversal), and computational power (universality for quantum computation). Finally, we will conclude with a view of the analogy and relationship to discrete-time quantum walks.

The paper is writen as a tutorial with the aim of clarifying basic notions and methods for newcomers to the topic. We also include a rich variety of references suitable for more enthusiastic readers and experts in the field. We know that such a reference list can never be truly complete, just as this work can not contain all the things we had on our minds. We had to choose to stop writing, or the work would have remained unpublished forever. As it stands, we believe it contains all the basic information for the start of your journey with quantum walks.

Let's walk!


\setcounter{equation}{0} \setcounter{figure}{0} \setcounter{table}{0}\newpage
\section{Classical Random Walks}

To amuse students, or to catch their attention, it is quite usual to start statistical mechanics university courses with the \emph{drunkard's walk} example. We are told the story of a drunken sailor who gets so drunk that the probability for him to move forward on his way home is equal to the probability to move backward to the tavern where he likes to spend most of his time when he is not at sea. Looking at him at some point between the tavern and his home and letting him make a number of drunken steps, we ask where will we find him with the highest probability?

\begin{figure}[!bh]
\begin{center}
\includegraphics{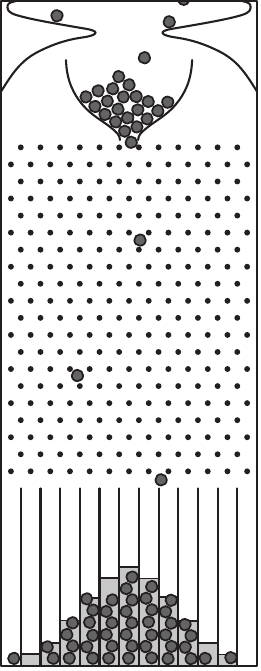}
\end{center}
\caption{A possible realization of the drunkard's walk -- the quincunx (Galton's board). The number of marbles in a bin at the bottom correspond to the probability of the sailor to be at that position after taking 19 random steps (the number of pins a falling marble encounters on its way down).}\label{fig-17}
\end{figure}

This seemingly trivial problem is very important. In fact, it can be retold in many various ways. For example the \emph{Galton's board} \cite{Galton89} device, also known as the \emph{quincunx} (see Fig.~\ref{fig-17}), has the same properties as the drunkard's walk. Balls are dropped from the top orifice and as they move downward, they bounce and scatter off the pins. Finally, they gather and rest in the boxes on the bottom. The number of balls in a box some distance from the horizontal position of the orifice can be mapped (is proportional) to the probability for the drunken sailor to be found at a particular distance from the tavern (after a specific number of steps corresponding to the vertical dimension of the quincunx). This is due to the fact, that the ball on its way downward scatters from the pins with approximately the same probability to the left as to the right --- it moves in the same manner as the sailor on his way from the tavern. Another retelling of the sailor's story is the magnetization of an ideal paramagnetic gas\footnote{It is an ensamble of magnetic particles (in this case without any external field), that do not interact --- they do not feel each other.} of spin-$1/2$ particles. Tossing a coin is yet another example of drunkard's walk.

This simple {\em walk on a line} is not just an example of some simple problems. It serves as a toy model and starting point for more involved variations. These modifications of the drunkard's walk spread across many branches of science and their position is justified by the results they provide. The roots of random walks, however, lie in the areas of mathematics, economy and physics. Probably the earliest (1900) attempt to describe random walks originates in the doctoral thesis of Bachelier \cite{Bachelier00}, who considered the evolution of market prices as a succession of random changes. His thesis, however, was not very rigorous and random walks had to wait to be defined properly from a mathematical viewpoint. This was done by Wiener \cite{Wiener23} in 1923. In physics it was Einstein \cite{Einstein05}, who in his theoretical paper on Brownian motion from 1905 put the first theoretical description of a random walk in natural sciences with explanation based on the kinetic theory, where gases (and liquids) are considered to be composed of many small colliding particles --- it was aimed to explain thermodynamical phenomena as statistical consequences of such movement of particles. Einstein speculated, that if (as he believed) the kinetic theory was right, then the seemingly random Brownian motion \cite{Brown} of large particles would be the result of the action of myriads of small solvent particles. Based on kinetic theory, he was able to describe the properties of the motion of the solute particles --- i.e.~connect the osmotic pressure of these particles with their density --- and show the connection with the diffusive process. Assuming the densities of solvent and solution are known, an experiment based on the barometric formula stating how the pressure changes with altitude provides us with the kinetic energy of solute particles at a given temperature. Hence, the Avogadro number can be estimated. This, indeed, has been done \cite{Perrin09} and strengthened the position of the kinetic theory at that time.

\subsection{Markov Chains}
\label{sec:markov_chains}

Random walks have been formulated in many different ways. Generally, we say that a {\em random walk} is a succession of moves, generated in a stochastic (random) manner, inside a given state space $\Omega$ that generally has a more complex structure than a previously mentioned liear chain (drunkard's walk). If the stochastic moves are correlated in time, we talk about non-Markovian walks (walks with memory), however, for our purposes we will always assume the walks to be Markovian -- the random moves of the walker are uncorrelated in time. Note, though, that the stochastic generator may be position dependent.
The sequences of such moves result in the so-called \emph{Markov Chains}. 

Taking only one instance (path) of a random walk is, naturally, not enough to devise the general properties of the phenomenon of random walks. That is why we consider the evolution of distributions on state space $\Omega$. These distributions are the averages over many instances of a random walk of a single walker (or an ensemble of walkers). In such case, for a Markovian walk in a countable space\footnote{The topic of uncountable spaces is not important for the scope of this work.} $\Omega$, we define a distribution
\[
	p=\begin{bmatrix}
	p_1\\
	p_2\\
	p_3\\
	\vdots\end{bmatrix}
\]
over indexed states, with $p_k$ corresponding to the probability of finding the walker at the position with index $k$. For $p$ to define a probability, it is necessary that
\[
\sum_{k}p_k=1.
\]
In this framework, the stochastic generator that describes the single-step evolution of a distribution function is given as a stochastic matrix\footnote{A \emph{stochastic matrix} is a real matrix with columns summing to 1, which preserves the probability measure, i.e.~all its elements are positive and smaller than one.} $M$, giving the one-step evolution of a distribution $p$ into $p'$ as
\[
	p' = M p.
\]
Having the matrix $M$ and an initial state $p(0)$, the distribution after $m$ steps is described by the formula
\begin{equation}
	p(m)=M^m p(0).
	\label{eq:cevolution}
\end{equation}
The allowed transitions (steps) of the random walker from position $i$ to position $j$ are reflected within $M$ by the condition $M_{ji}\neq 0$. For now, we also assume (although it is not necessary) a weak form of symmetry in the transitions by requiring that $M_{ij}$ is also not zero. Forbidden transitions correspond to the condition $M_{ji}=M_{ij}=0$. The allowed transitions reflected in the non-zero elements of $M$ induce a graph structure $\mathcal G=(V,E)$ on $\Omega$. Here $V=\Omega$ is a set of vertices (corresponding to states) and $E=\{(j,i)\st j,i\in\Omega,M_{ji}\neq0\}$ is a set of ordered pairs of vertices (oriented edges, corresponding to transitions) -- the allowed connections between vertices. Conversely, we may say that the graph structure defines allowed transitions. For simplicity, from now on we will use the notation $ij$ instead of $(i,j)$ for the oriented edges.

The graph structure is not the only thing reflected in $M$. The coefficients of $M$ also define the transitional probabilities to different states and in this manner reproduce the behaviour of the stochastic generator, usually called a \emph{coin}. For an \emph{unbiased coin,} which treats all directions equally, the coefficients of $M$ are defined as
\begin{equation}
\label{eq:unbiasedRW}
M_{ji}=\begin{cases}
\frac{1}{d(i)} & \text{for $(i,j)\in E$,}\\
0 & \text{otherwise,}\end{cases}
\end{equation}
with $d(i)$ being the degree of the vertex $i$. If this is not true, we say that the coin is \emph{biased}, preferring some directions above others.

\subsection{Properties of Random Walks}
\label{sec:RWproperties}

When studying properties of random walks we search for specific properties that (potentially) help us solve various posed problems. Sometimes we want to know where the walker is after some time, in other cases we may want to know how much time does the walker take to arrive at some position or what is the probability of the walker to get there in given time. Another question asks how much time we need to give the walker so that finding him in any position has approximately the same probability. All these questions are not only interesting, but also useful for the construction (and understanding) of randomized algorithms.

If we are concerned with an unbiased random walk in one dimension (the so called drunkard's walk introduced in the previous Chapter), according to the central limit theorem, the distribution very quickly converges to a Gaussian one. The standard deviation of the position of the walker becomes proportional to $\sqrt{m}$, where $m$ is the number of steps taken. Such an evolution describes a diffusive process.

All the routes to get from position $0$ to position $x$ (positive or negative) after performing $m$ steps are composed of $n_+=(m+x)/2$ steps upward and $n_-=(m-x)/2$ steps downward. This is under supposition of what we shall call \emph{modularity} -- after an even (odd) number of steps, the probability to find the walker on odd (even) positions is always zero, constraining $m+x$ to be even (this is consequence of bipartite character of the graph). The number of these paths characterized by specific $n_\pm$ is
\[
N(x,m)={m\choose n_+}=\frac{m!}{n_+!n_-!}.
\]
The number of all possible paths is $N(m)=2^m$ and, hence, the probability to find the walker at position $x$ is $P_\mathrm{rw}(x,m)=N(x,m)/N(m)$. Using Stirling's approximation
\[
n!\sim\sqrt{2\pi n}\left(\frac{n}{\e}\right)^n,
\]
we can express the probability of finding the walker at $x$ after $m$ steps as
\begin{eqnarray*}
	P_\mathrm{rw}(x,m) &=& \frac{m!}{2^m\,n_+!n_-!}\\
	 & \simeq & \sqrt{\frac{2}{m\pi}}\left[\left(1+\frac{x}{m}\right)\left(1-\frac{x}{m}\right)\right]^{-\frac{m}{2}}
	 		\left(1+\frac{x}{m}\right)^{-\frac{x-1}{2}}\left(1-\frac{x}{m}\right)^{\frac{x-1}{2}} \\
	& \simeq & \sqrt{\frac{2}{m\pi}}\left[1-\frac{x^2}{m^2}\right]^{-\frac{m}{2}},
\end{eqnarray*}
where we assumed $m\gg x$, making the last two terms in the second line approximately one. In the limit of large $m$ and small $x$, this approximation can be  further transformed into the formula\footnote{Employ $\lim_{x\to\infty}\left(1-\frac{1}{x}\right)^x=\e$.}
\[
P_\mathrm{rw}(x,m)\simeq\frac{2}{\sqrt{2\pi m}}\exp\left(-\frac{x^2}{2m}\right).
\]
The function $P_\mathrm{rw}$ is not yet a probability distribution as it is normalized to $2$, but we have to recall the modularity of the walk telling us, that after even number of steps, the walker cannot be on odd position and after odd number of steps, the walker cannot be on even position; this leads us to a probability distribution having form
\begin{equation}
	\label{eq:pdf_rw}
	p_\mathrm{rw}(x,m)=\frac{1+(-1)^{m-x}}{\sqrt{2\pi m}}\exp\left(-\frac{x^2}{2m}\right).
\end{equation}
Finally, we observe
\begin{equation}
	\label{eq:rw_slimak}
	\langle x^2\rangle\equiv\sum_{x=-\infty}^\infty x^2 p_\mathrm{rw}(x,m)
	\simeq\frac{1}{2}\int_{-\infty}^\infty x^2 P_\mathrm{rw}(x,m) \dd x=m,
	\qquad
	\sqrt{\langle x^2\rangle}=\sqrt{m},
\end{equation}
where we approximated the modular probability distribution by a smooth one, $P_\mathrm{rw}/2$. 

The standard deviation of a walker is thus proportional to $\sqrt{m}$, and this fact may be deduced even without the knowledge of the limiting distribution (see e.g.~Ref.~\cite{Hughes}). The position of the walker $x$ is in fact a sum of $m$ independent variables (steps up and down) $y_1, \ldots, y_m$:
\[
x=\sum_{j=1}^m y_j.
\]
The square of the standard deviation then reads
\[
	\sigma^2(m) = \langle |x-\langle x\rangle|^2\rangle = \langle x^2\rangle-\langle x\rangle^2
	= \biggl\langle \sum_{j=1}^m y_j\sum_{l=1}^m y_l\biggr\rangle-\bigl\langle \sum_{j=1}^m y_j\bigr\rangle^2.
\]
As the $y_j$'s are independent, the sums and averages can be exchanged, giving us
\[
	\sigma^2(m)=\sum_{j=1}^m\sum_{l=1}^m\bigl(\langle y_jy_l\rangle-\langle y_j\rangle\langle y_l\rangle\bigr).
\]
For $j\neq l$, the independence of the random variables $y_j$ is reflected in $\langle y_jy_l\rangle=\langle y_j\rangle\langle y_l\rangle$, so the square of the standard deviation simplifies to
\[
	\sigma^2(m)=\sum_{j=1}^m\bigl(\langle y_j^2\rangle-\langle y_j\rangle^2\bigr)=\sum_{j=1}^m\sigma^2_j=m\sigma^2,
\]
where $\sigma_j$ are standard deviations of the random variables $y_j$. In this case, they are are all the same and equal to $\sigma_j\equiv\sigma=1$. Thus we finally arrive at $\sigma(m)=\sqrt{m}$ by a different route.

There are also other simple observations we can make. Let us list them here, with the aim of later comparing them to the properties of the distributions arising from quantum walks.

\paragraph{Reaching an absorbing boundary.}\label{sec:classicalabsorbingboundary}
Let us look at the probability with which the walker returns into its starting position. After leaving this position, the walker makes a move forward. Now there is probability $p_{10}$ for him to get back to the original position by taking all the possible routes into consideration. Under closer inspection, we find that this probability consists of the probability for him to make one step backward, which is $1/2$ and the probability for him to get back to the initial position only after moving further away from it first. This latter probability equals $\frac{1}{2}p_{10}^2$, since he moves forward with probability $1/2$ and then has to move two steps back with equal probabilities $p_{10}$ (we suppose, that no matter how far from the initial position the walker is, he always has equal probabilities to move forward and backward). To sum up, we see that
\[
p_{10}=\frac{1}{2}+\frac{1}{2}p_{10}^2.
\]
The only solution to this quadratic equation is $p_{10}=1$. Let us interpret this result: the probability for the walker to return to his initial position is equal to one, i.e.~he always gets back. 
In other words, if we employ an \emph{absorbing boundary} at position 0 and start at position 1, the walker will eventually hit the boundary.

This result can be extended to the statement that the probability of reaching vertex $i$ from any other vertex $j$ is one. As the walk is translationally symmetric, the probability to get from any $x$ to $x\pm 1$ at any time is always the same, $p_{10}=1$. Thus, the probability to get from $j$ to $i$ is
\begin{eqnarray}
	p_{ji}=p_{10}^{|j-i|}=1. \label{classicalabsorb}
\end{eqnarray}
This is quite different from quantum walks, where the probability to hit some boundary is not one, as shown in Sec.~\ref{sec:discreteboundary}.

\begin{exercise}
What is the probability $p_{10}$ in general case, when the probability to move right is $p$ and the probability to move left is $1-p$ in each step?
\end{exercise}


\paragraph{Limiting distribution.}
Now\label{page:convergence} let us step away from the example of a walk on a linear chain and look at the larger picture, considering general graphs. For these, two quantities describing the properties of random walks are widely used in the literature --- the \emph{hitting time} and the \emph{mixing time}. But first, let us have a closer look at walks with unbiased coins on finite graphs. It is interesting to find, that all such graphs (if connected and non-bipartite) converge to its stationary distribution $\pi=(d(1),d(2),\ldots,d(N))/2|E|$, where $d(j)$ stands for the degree of vertex $j$. It is easy to see that this is a stationary distribution. For any vertex $j$ we have
\[
(M\pi)(j)=\sum_{i\st ji\in E}M_{ji}\pi(i)=\sum_{i\st ji\in E}\frac{1}{d(i)}\frac{d(i)}{2|E|}=\frac{d(j)}{2m}=\pi(j).
\]
This distribution is clearly uniform for regular graphs. In Appendix \ref{sec:classicallimit} we also provide a proof that for connected and non-bipartite graphs, this distribution is unique and also is the limiting distribution. With this notion we are ready to proceed to define the quantities of hitting and mixing time.

\paragraph{Hitting time.}
It is the average time (number of steps) that the walker needs to reach a given position $j$, when it starts from a particular vertex ($c$ stands for \emph{classical})
\begin{equation}
	\label{eq:classical_hitting}
	h^\mathrm{c}(j)=\sum_{m=0}^\infty m\, p(j,m).
\end{equation}
This quantity also makes sense for infinite graphs. We will later see (in Sec.~\ref{sec:discretehitting}) that for quantum walks this quantity is defined differently, due to the fact that measurement destroys the quantum characteristics of the walk.

\begin{example}
We have seen, that when the walker starts at position $1$ she will eventually reach position $0$. Let us evaluate now the hitting time between these two positions. As the number of paths of length $2k$ which do not get to position 0 is determined by the Catalan number $C_k$ (see Appendix \ref{sec:catalan_numbers}), the probability for the walker to hit vertex $0$ after $2k+1$ steps is
\[
p_{2k+1}\geq\frac{C_k}{2^{2k+1}},
\]
where the inequality sign comes from the fact that we counted all $2^{2k}$ paths, even those crossing and/or hitting 0. When we employ Eq.~(\ref{eq:classical_hitting}) and Eq.~(\ref{eq:catalan_ineq}) we find, that
\[
h^\mathrm{c}(1\to 0)\geq\sum_{k=0}^\infty \frac{(2k+1)C_k}{2^{2k+1}} \geq \frac{\sqrt{\pi}}{\e^2}\sum_{k=0}^\infty\frac{1}{\sqrt{2k+1}},
\]
which diverges and so the hitting time
is infinite, although the probability of eventually
reaching vertex $0$ is
\[
p_{10}=\frac{1}{2}\sum_{n=0}^\infty \frac{C_n}{2^{2n}}=\frac{1}{2} c(1/4)=1,
\]
where $c(x)$ is the generating function for Catalan numbers.

\end{example}

\begin{exercise}
Consider now, that the probability to jump from position $1$ to $0$ is $p>0$. Starting at position $1$ we wait for one step and look whether the walker hit the boundary. If not we again set the walker to position $1$ and repeat procedure again and again until we hit the boundary. Show that in such experiment the hitting time is
\begin{equation}
\label{eq:hitting_analogue}
h^\mathrm{c}(1\to 0)=\frac{1}{p}.
\end{equation}
In this case the hitting time is finite. We could as well let the walker go for some time $T$ in which case the probability of hitting the boundary within this time would be some other $p$ but the hitting time would be expressed in the same way as in Eq.~(\ref{eq:hitting_analogue}). This definition leads to the emergence of tuples $(T,p)$ with the hitting time given by Eq.~(\ref{eq:classical_hitting}) being just a special case basically when $p<1$ for $T\to\infty$. Such definition seems to coincide more with the notion of absorbing boundary --- the probability of absorption is the smallest $p$ for which $(\infty,p)$ describes correct analouge of the hitting time given by Eq.~(\ref{eq:hitting_analogue}). See also definition of hitting time in the quantum case in Sec.~\ref{sec:discretehitting}.
\end{exercise}

\paragraph{Mixing time.}
The second important quantity is the classical \emph{mixing time} $\mathcal M_\epsilon^\mathrm{c}$. 
As we have seen each unbiased random walk (on connected non-bipartite graphs) reaches a stationary distribution, which we denote $\pi$.
The mixing time is then the time (number of steps) after which the probability distribution remains $\epsilon$-close to the stationary distribution $\pi$, i.e.
\begin{equation}
	\label{eq:classical mixing}
	\mathcal M_\epsilon^\mathrm{c}=\min\{T\st\forall t\geq T, \no{p(\cdot,t)-\pi(\cdot)}_{tvd}\leq\epsilon\},
\end{equation}
where $p(\cdot,t)$ is the distribution in time $t$ and $\no{\ \cdot\ }_{tvd}$ is (total variational) distance of distributions,
\begin{equation}
	\label{eq:TVD}
	\no{p(\cdot,t)-\pi(\cdot)}_{tvd}\equiv\sum_{j}\no{p(j,t)-\pi(j)}.
\end{equation}
Again, we will later see in Sec.~\ref{sec:discretemixing} that the mixing time is defined differently for quantum walks, as they are unitary and converge to a stationary distribution only in a time-averaged sense.

In classical case mixing time depends on the gap between the first two eigenvalues $\lambda_1=1$ (for the stationary distribution $\pi$) and $\lambda_2$. The use is illustrated by the next theorem.
\begin{theorem}[Spectral gap and mixing time]
\[
\frac{\lambda_2}{(1-\lambda_2)\log 2\epsilon}\leq \mathcal M_\epsilon^\mathrm{c}.
\]
\end{theorem}
For closer details see e.g.~Ref.~\cite{Sinclair93}.

\subsection{Classical Random Walk Algorithms}

Random walks are are a powerful computational tool, used in solving optimization problems (e.g. finding spanning trees, shortest paths or minimal-cuts through graphs), geometrical tasks (e.g. finding convex hulls of a set of points) or approximate counting (e.g. sampling-based volume estimation) \cite{MotwaniRaghavanBook}.
In the previous Sections we had an opportunity to get acquainted with several interesting properties of random walks. These are often exploited when constructing new algorithms. For example, small mixing times can lead to more efficient and accurate sampling, while short hitting times can lead to fast search algorithms.

At present, there is a wide range of algorithms that use these properties of random walks to their advantage. These random-walk based algorithms range from searches on graphs, through solving specific mathematical problems such as satisfiability, to physically motivated simulated annealing that searches for ``optimal'' states (ground states, or states that represent minima of complex fitness functions). A large group of algorithms, that (not historically, though) can be considered to be based on random walks are Markov Chain Monte Carlo methods for sampling from low-temperature probability distributions and using them for approximate counting or optimization. 

A huge research effort is devoted today to this vast area of expertise. We will briefly introduce a few of these interesting algorithms in the following pages. Even though the connection to random walks is often not emphasized in the literature, all these algorithms can be viewed from the vantage point of random walks. 


\subsubsection{Graph Searching}
\label{sec:graphsearching}

\begin{figure}[!t]
\begin{center}
\includegraphics[scale=0.76]{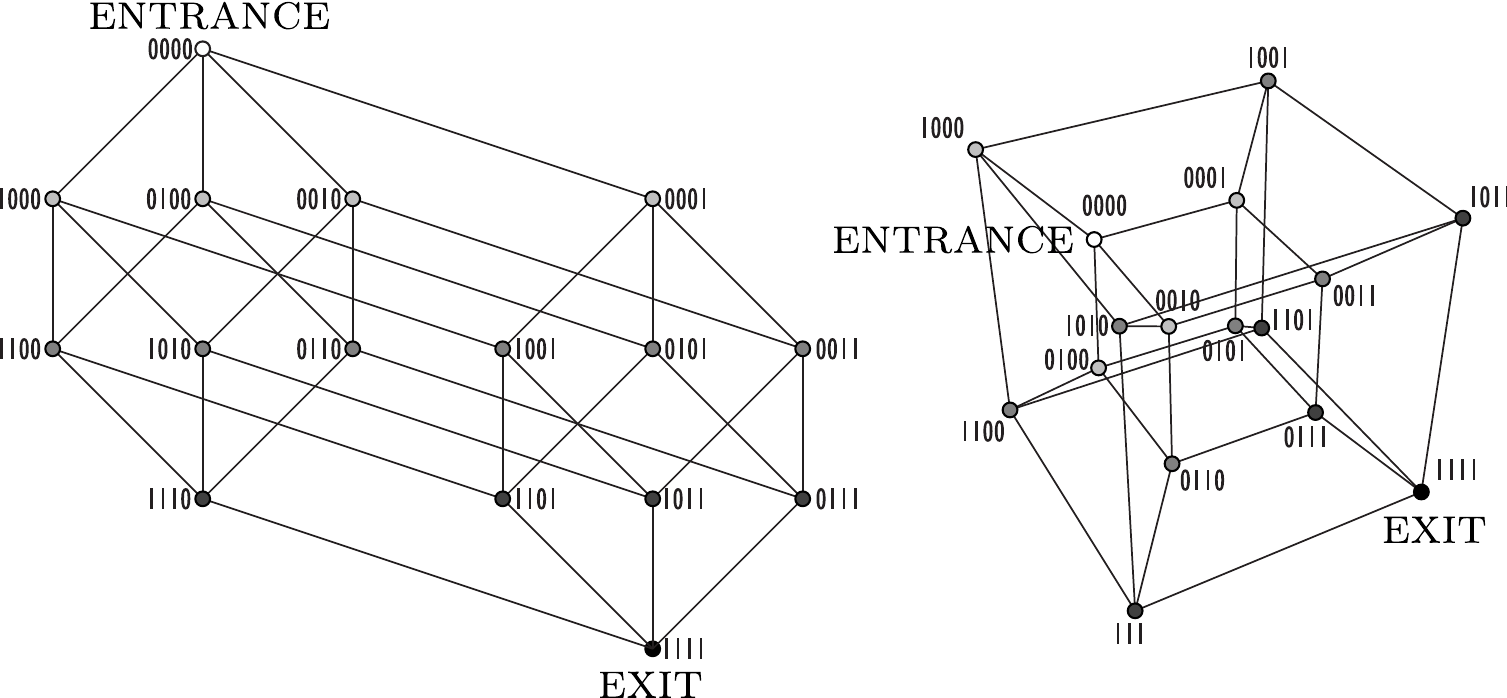}
\end{center}
\vspace{-0.3cm}
\caption{\label{fig:hypercubefigure}
Two graphical representations of a 4-dimensional hypercube with vertices labeled by 4-bit strings, connected when these strings. Connections exist only between vertices whose binary labels differ by a single digit. Vertices with the same number of zeros in their label create a layer of the hypercube. There are $n+1=5$ layers in the 4-cube --- layers 0 and 4 are denoted by black circles, layer 1 by red circles, layer 2 by green circles and layer 3 by blue circles. A quantum walker can traverse from the \textsc{entrance} vertex to the \textsc{exit} vertex in $O(n)$ steps (scaling linearly with the number of layers). A classical random walk is exponentially slower. Note though, that a different efficient classical algorithm (using memory to help the traversal) exists for this problem.}
\end{figure}

Random walks can be used to search for a marked item (or $M$ items) on a graph of size $N$.
Knowing the structure of the graph can sometimes allow us to find efficient deterministic solutions. However, sometimes a random walk approach can become useful. One such example is a search on a complete graph with loops. This task is just a rephrasing of a \emph{blind search} without structure when any element might be the targeted one. One can easily show that on average (even if one remembers vertices that are not targets) one needs $O(N/M)$ queries to the oracle\footnote{In this case the oracle just gives the answer whether the vertex we picked is marked. We will deal with oracles a bit later in Sec.~\ref{sec:oracles_and_searches}.}. Later we will see that in the quantum case, a clever utilization of an oracle in quantum walks can produce a quadratic speedup due to faster mixing.

\begin{figure}[!t]
\begin{center}
\includegraphics[width=3.5in]{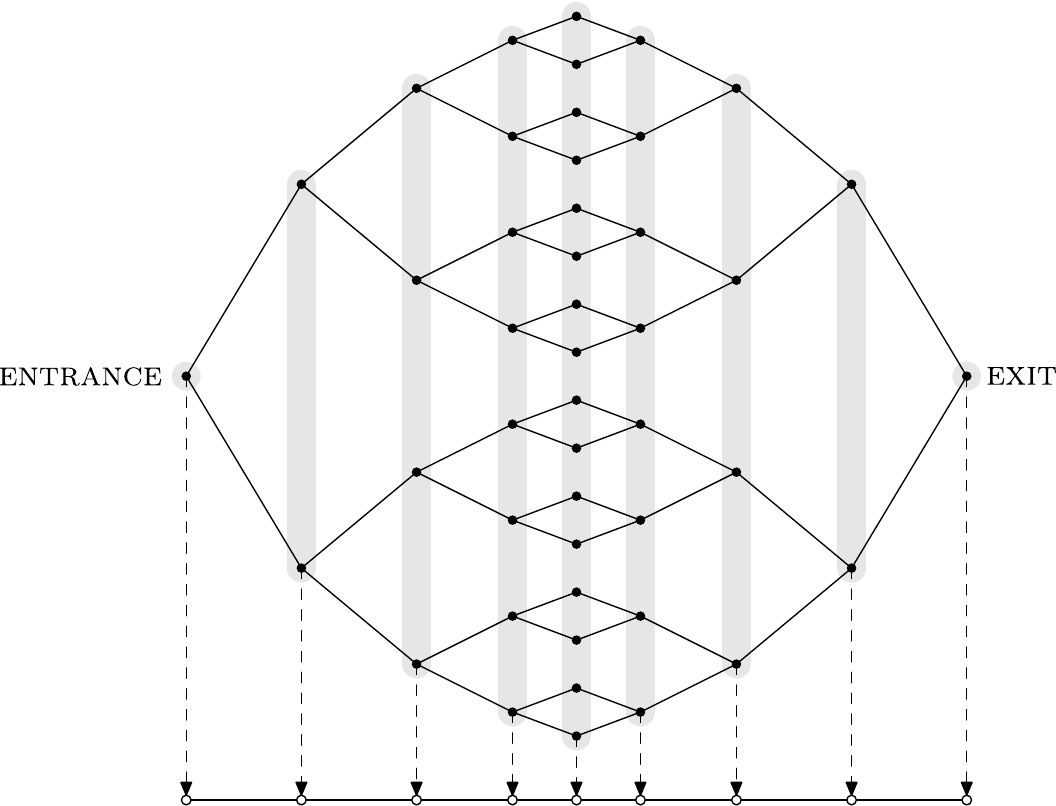}
\end{center}
\vspace{-0.3cm}
\caption{\label{fig:glued_trees_dr} A graph made from two glued trees graph can be efficiently traversed by a quantum walk (or by a clever classical algorithm). However, this graph is not (efficiently) penetrable by a classical random walk. The efficiency of the quantum walk algorithm comes from the graph symmetry -- the walk on this graph is equivalent to a walk on a line of ``column'' states.}
\end{figure}

Two interesting examples of graph-search are hypercube traversal (Fig.~\ref{fig:hypercubefigure}) and glued trees traversal (Fig.~\ref{fig:glued_trees_dr}), in both of which the goal is to find a vertex directly ``opposite'' to the starting one. We are given a description of the graph (with the promised structure) as an oracle that tells us the ``names'' of the neighbors of a given vertex. The goal is to find the ``name'' of the desired opposite vertex. In these examples, employing a quantum walk results in tantalizing exponentially faster hitting times than when using classical random walks. Yet, as we will see in the following examples, this does not mean that there are no deterministic algorithms that can do it as fast as the quantum walk one.
To see an actual exponential speedup by a quantum walk over any possible classical algorithm, we will have to wait until Section \ref{sec:gluedtrees}.

\begin{figure}[!b]
\begin{algorithm}
Set $k=1$, the initial position $a_1$ to a random neighbor of the \textsc{entrance} vertex and $S_0=\{a_0\}$ then for $n-1$ times repeat:
\begin{enumerate}
\item set $k:=k+1$
\item randomly choose $a_k$ to be neighbor of $a_{k-1}$ such that $a_k\not\in S_{k-2}$
\item set $S_{k-1}$ to be the set of all neighbors of $a_k$ that have a neighbor also in $S_{k-2}$
\end{enumerate}
The resulting vertex is \textsc{exit}.
\end{algorithm}
\caption{\label{fig:hypercube_alg} Algorithm for traversing the $n$-hypercube. Starting at position \textsc{entrance} the walker uses structural oracle to navigate through the graph. She has to remember only $O(n)$ vertices in each step which help her propagate thorugh the graph increasing the layer in each step.}
\end{figure}

\begin{example}[Traversing a hypercube]
\label{ex:traversing_hypercube}
An $n$-dimensional {\em hypercube} is a graph with vertices that are binary strings of length $n$ (see Fig.~\ref{fig:hypercubefigure}). For every pair of vertices $a$ and $b$ we can define \emph{Hamming distance,} which is the number of positions in which the two binary representations of $a$ and $b$ differ. Clearly this is the smallest number of steps one has to make in order to get from the vertex $a$ of the hypercube to the vertex $b$ as two vertices are connected by an edge if and only if the two strings corresponding to the vertices differ only by a single bit (thus have Hamming distance 1). In the hypercube, we can further distinguish a layered structure. Let us label two opposite vertices of the hypercube \textsc{entrance} and \textsc{exit}. We say, that vertex $a$ is in the layer $k$, if its Hamming distance from the \textsc{entrance} vertex is $k$. We can see, that any $k$-th level vertex has neighbours only in levels $k-1$ or $k+1$. Also note, that \textsc{entrance} vertex has Hamming distance $0$ and \textsc{exit} vertex has Hamming distance $n$.

Starting at the \textsc{entrance} vertex, our aim is to get to the \textsc{exit} as fast as possible. As between \textsc{entrance} and \textsc{exit} vertices there is rougly $2^n$ other vertices, usual random walk approach would find \textsc{exit} only in time exponential in $n$ --- the hitting time is $O(2^n)$ and as a result mixing time is even longer.

We can however radically shorten the time needed when we employ the walker with memory helping the walker to advance through the hypercube and increase her Hamming distance from the \textsc{entrance} in each step --- see algorithm in Fig.~\ref{fig:hypercube_alg}. Let us label vertices the walker visits as $a_k$, with index $k$ denoting the step (and layer as well). From the \textsc{entrance} vertex, denoted as $a_0$, the walker moves to random neighbour $a_1$, remembering the vertex it came from in memory $S_0=\{a_0\}$. The memory will contain all the vertices that are from lower layer, than actual vertex. Clearly, for $S_0$ this is true.

On each step $k+1$ the walker chooses new neighbour $a_{k+1}$ of actual vertex $a_k$ from the next $(k+1)$-th layer by excluding the vertices from memory $S_{k-1}$ from the selection process, as these are from layer $k-1$. New memory $S_k$ is constructed as a set of all neighbouring vertices of $a_{k+1}$ that are also neighbours of some of the vertices in $S_{k-1}$.

To see that it works as intended, let us say, that $a_{k+1}$ was obtained from $a_k$ by flipping bit at position $q$. From definition of $S_{k-1}$ we know, that $a_k\not\in S_{k-1}$ and so all vertices contained in this set differ from $a_{k+1}$ in two bits, $q$ and some $r\neq q$. Now, to get from $a_{k+1}$ to lower level, we would have to go either to $a_k$, which is a neighbour of $S_{k-1}$, or we would flip the bit on position $r$ taking it to one-bit-flip ($q$) from some vertex from $S_{k-1}$.

It is easy to see now, that $S_k$, as a memory, is a set of all neighbours of vertex $a_{k+1}$ that are at level $k$. In this manner the walker in each step increases its distance from the \textsc{entrance} vertex and decreases its distance from the \textsc{exit} vertex, thus needing $n$ steps to traverse the graph with $\order(n)$ searches in neighbourhood every step. Totally the efficiency of the algorithm is polynomial in $n$ with memory of $O(n)$.
\end{example}

\begin{figure}[!b]
\begin{algorithm}
Set $a_1=$\textsc{entrance} and for $n-1$ times repeat: 
\begin{enumerate}
\item set $k:=k+1$
\item randomly choose $a_k$ to be neighbor of $a_{k-1}$ such that $a_k\neq a_{k-2}$
\end{enumerate}
Set $k=0$ and repeat until $a_{k+n}=$\textsc{exit}:
\begin{enumerate}
\item set $k:=k+1$
\item randomly choose $a_{k+n}$ to be neighbor of $a_{k+n-1}$ such that $a_{k+n}\neq a_{k+n-2}$
\item if vertex $a_k$ has only two neighbors return to $a_{k/2+n}$ (by setting $a_{k+n}:=a_{k/2+n}$ and $k:=k/2$)
\end{enumerate}
\end{algorithm}
\caption{\label{fig:gluedtrees_alg} Algorithm for traversing glued trees. The walker can easily navigate with the help of structural oracle from the \textsc{entrance} vertex to the glued vertices. There she knows for certain her position as these vertices differ from the rest in the number of neighbors being only two. After she passes glued vertices, she can navigate in $O(n^2)$ steps to the \textsc{exit} vertex, as everytime she encounters glued vertex she knows, that halfway between previous encounter of the glued vertex she did a wrong choice of direction. This requires memory of size $O(n)$.}
\end{figure}
\begin{example}[Traversing glued trees]
Another example, where a usual random walker without me\-mo\-ry fails is in traversing a graph made from two glued binary $n$-leveled trees, depicted in Fig.~\ref{fig:glued_trees_dr}. As in the previous example, the random walker gets stuck in exponentially many vertices, that comprise the ``body'' of the graph and she will not be able to reach \textsc{exit} vertex efficiently. However, with memory the walker can exploit the difference between the central vertices and the rest and proceed through the graph as follows (see also Fig.~\ref{fig:gluedtrees_alg}). Starting from the \textsc{entrance} vertex, by simply randomly choosing neighbours other than the previous one, it reaches one of central ``glued'' vertices in time $n$. These vertices are easily recognised since they have only two neighbours. Traversing further through the graph, the walker either reaches the \textsc{exit} vertex or finds itself back in the central region. This region is reached after $2k$ steps after being there for the first time and in this case it is easy to backtrack the last $k$ steps and choose the remaining neighbour. Performing this walk with memory (polynomial in $n$) for $\order(n^2)$ steps will certainly give us the walker ending in the \textsc{exit} vertex whereas in the quantum case we need to perform $O(\mathrm{poly}(n))$ steps. Thus, there is no great speedup coming from a quantum walk. However, this example is just the first step to an actual exponential speedup, discussed in Section~\ref{sec:gluedtrees}.
\end{example}



\subsubsection{Solving Satisfiability Problems}
\label{sec:SATsolving}
A prime example of locally constrained problems is Satisfiability. Its easiest variant, 2-SAT is to determine whether one can find an $n$-bit string $x_1 x_2 \cdots x_n$ that satisfies a boolean formula\footnote{in conjunctive normal form} with exactly 2 literals (bits or their negations) per clause.
A 2-SAT instance with $m$ clauses on $n$ bits has the following general form of the boolean function:
\be
	\phi = (a_1 \vee b_1) \wedge (a_2 \vee b_2) \wedge \dots \wedge (a_m \vee b_m),
\ee
where $a_i, b_i \in \{ x_1,x_2,\dots,x_n, \neg x_1, \neg x_2,\dots \neg x_n\}$.

There exists a deterministic algorithm for solving 2-SAT, but we also know a beautiful random-walk approach to the same problem. The algorithm (first analyzed by Papadimitriou) is this:
\begin{enumerate}
	\item start with $x_1 x_2 \cdots x_n = 11\cdots 1$.
	\item {\bf while} $\phi$ contains an unsatisfied clause 
				(and $\#$steps $<2n^2$) {\bf do}\\
		pick an unsatisfied clause $C$ arbitrarily\\
		pick one of the two literals in $C$ uniformly at random and flip its value\\
		{\bf if} $\phi$ is now satisfied {\bf then} output ``yes'' and stop
	\item output ``no''
\end{enumerate}

This algorithm performs a random walk on the space of strings. One can move from string $s_1$ to string $s_2$ if they differ by a single bit value $x_t$, and bit $t$ belongs to a clause that is unsatisfied for string $s_1$. Why does this algorithm work? Assume there is a solution to the 2-SAT instance. Call $d(s)$ the Hamming distance of string $s$ from the correct solution. In each step of the algorithm, $d(s_{k+1})=d(s_{k})\pm 1$, because we flip a single bit. However, we claim that the probability of the Hamming distance to the correct solution decreases in each step is $\geq \half$. Here's the reason for this. We know that when we choose the clause $C$, it is unsatisfied, so both its bits cannot have the correct value. Imagine both of the bits in the chosen clause $C$ were wrong. We then surely decrease $d$. If one of the bits in $C$ was wrong, we have a 50\% chance of choosing correctly. Together, this means the chance of decreasing $d$ in each step are $\geq \half$. Our random walk algorithm then performs no worse than a random walk on a line graph with $n+1$ vertices $\{0,1,\dots,n\}$, corresponding to Hamming distances of strings from the solution. When we hit $0$, we have the solution! The expected number of steps it takes to hit the endpoint on a line of length $n+1$ is $n^2$. Therefore, using Markov's inequality we can show that running the algorithm for $2n^2$ steps results in finding the satisfying assignment (if it exists) with probability $\geq \half$.

Go to 3-SAT. Basic algorithm and simple analysis gives probability of success $3^{-n/2}$,
which translates to runtime $(1.78)^n$. If you think about taking $3n$ steps, the probability of success increases a lot -- see Luca Trevisan's notes, Sch\"oning's algorithm (1999). Taking $3k$ steps (when starting $k$-away from the solution), we can calculate the probability of success to be at least $\frac{1}{2^k \sqrt{k}}$. When we sum over all starting points and take $3n$ steps, we arrive at an algorithm with runtime $(1.33)^n$.

This can be further improved, small changes -- GSAT, WALKSAT -- restarts, greedy choices of which clause to flip, etc.



\subsubsection{Markov Chain Monte Carlo Algorithms}
\label{sec:metropolisandsuch}

Previously mentioned algorithms are based on a Markov chains. These algorithms are designed for finding a solution within many instances. On the other hand there is a large class of physical problems that do not require the knowledge of one particular instance of the state of the system. In these problems we are interested in statistical properties\footnote{As we will see in the following section, such statistical properties can be, on the other hand, useful in finding optima of various functions.} of various, often physical, systems, such as knowing averages of various quantities describing the system under consideration. For such problems an efficient method of \emph{Monte Carlo} has been devised. We can view random walks as a special approach within this method, that employs Markovianity in the search within the phase space of the problem. The system in these simulations undergoes Markov evolution with specific properties creating a chain of states that are sampled from desired probability distribution.

If we want to estimate a time average of some quantity of a system in equilibrium we either have the option to study the actual time evolution, or we can use \emph{ergodic hypothesis} and estimate this time-average by estimating an average over an ensemble of such systems, where the actual distribution is known but, due to computational restrictions it is difficult to use it directly in (analytical) computation of averages. In simulations, when the sampling process has to be discretized, to find average of some state $\mathbf{x}$ quantity $Q$, we can use formula
\[
\langle Q\rangle\simeq\sum_{j=1}^N Q(\mathbf{x}_j)p[\mathcal H(\mathbf x)],
\]
where $p[\mathcal H(\mathbf x)]$ is the probability (depending only on the energy) of being in the near vicinity of state $\mathbf{x}_j$. These $N$ states are sampled uniformly from the phase space. This is, however, not useful, when only states from a small part of phase space make a major contribution to the average. This makes the sum to converge very slowly.

In order to overcome this drawback, a technique, called \emph{importance sampling} is employed, when the states are not sampled from uniform distribution, but from a distribution that is close to the desired $p[\mathcal H(\mathbf x)]$. Quite surprisingly, this is not so difficult to achieve, as we will see soon, and when we are finally sampling states $\mathbf{x}_j$ from the desired distribution, we can estimate the average as
\[
\langle Q\rangle\simeq\frac{1}{N}\sum_{j=1}^N\left.Q(\mathbf{x}_j)\right|_{\mathrm{pdf}(\mathbf{x}_j=\mathbf{x})=p[\mathcal H(\mathbf x)]}
\]
with $N$ being the normalization for the average.

As said before, random walks come into the play in the form of Markov processes where a walk in the phase space according to preferred distribution is performed. Employing the notation defined in Sec.~\ref{sec:markov_chains} we will show, how to construct such walk in discrete space and time. In general, the change of a distribution $p$ in a step is described by ``continuity'' equation
\begin{equation}
\label{eq:probability_continuity}
\Delta p_j(m)=-\sum_{j'}M_{j',j}p_j(m)+\sum_{j'}M_{j,j'}p_{j'},
\end{equation}
where $M$ is the stochastic transition matrix. This equation just states, that the part of probability that leaves state $j$ (first term) and the part of probability that ``flows'' into the state $j$ correspond to the change of the probability in the state at time $m$, $p_j(m)$. This condition in the case of a system in equilibrium, when $p_j(m)\equiv p_j$ reads
\[
\sum_{j'}M_{j',j}p_j=\sum_{j'}M_{j,j'}p_{j'}
\]
stating that whatever probability flows out of the state $j$ must be replenished. This condition is called \emph{global balance} and is still quite complex for simple utilisation. However, the necessary condition called \emph{detailed balance}
\[
\frac{M_{j',j}}{M_{j,j'}}=\frac{p_{j'}}{p_j}
\]
is suitable for computational purposes.

Indeed, if we take, for example, the choice of Metropolis (and Hastings) \cite{MeRoRoTeTe53,Hastings70},
\begin{equation}
\label{eq:transition_rate}
M_{j',j}=\min\left\{1,\frac{p_{j'}}{p_j}\right\},
\end{equation}
we have a way how to effectively sample the phase space under given distribution --- being in state $j$ we randomly choose state $j'$ and conditioned on transition rate from Eq.~(\ref{eq:transition_rate}), which is easily computable if $p_j$ is known, we replace the state $j$ by $j'$. See also Fig.~\ref{fig:mcmc}.

\begin{figure}
\begin{algorithm}
Initialize $\mathbf x_0=\mathbf x$, $k=0$, $\langle Q\rangle=0$ and for $N$ steps repeat:
\begin{enumerate}
\item choose random state $\mathbf x_{k'}$
\item evaluate $M_{k',k}$ according to Eq.~(\ref{eq:gibbssample})
\item generate random number $q$
\item if $q\leq M_{k',k}$ set $\mathbf x_{k+1}:=\mathbf x_{k'}$ otherwise $\mathbf x_{k+1}:=\mathbf x_{k}$
\item set $k:=k+1$ and if $k> N-N'$ evaluate $\langle Q\rangle:=\langle Q\rangle+Q(\mathbf x_k)/N'$
\end{enumerate}
\end{algorithm}
\caption{\label{fig:mcmc} Monte Carlo algorithm with Metropolis acceptance criterion using Gibbs distribution for estimating the average of $Q$ - this is done only in last $N'$ steps to allow relaxation of the system. Inputs of the algorithm are the initial state $\mathbf x$ and temperature. The choice of $\mathbf x'$ in every step is usually according to some distribution that places $\mathbf x'$ close to $\mathbf x$.}
\end{figure}

\begin{example}[Sampling from the Gibbs canonical distribution] This is a special case of the Me\-tro\-po\-lis-Hastings algorithm. The Gibbs distribution is given as
\[
p_j=\frac{1}{\mathcal Z}\exp[-\mathcal H(j)/k_BT],
\]
where $\mathcal Z$ is partition function, $\mathcal H$ is energy of the system in state $j$, $k_B$ is Boltzmann constant and $T$ is temperature. In this case
\begin{equation}
\label{eq:gibbssample}
M_{j',j}=\min\left\{1,\exp\left[{-\displaystyle{\frac{\mathcal H(j')-\mathcal H(j)}{k_BT}}}\right]\right\}.
\end{equation}
This tells us, that if the energy in the test state $j'$ is lower than actual energy, then accept the state $j'$ as new state, otherwise accept it only with probability
\[
\exp\left[-\frac{\mathcal H(j')-\mathcal H(j)}{k_BT}\right].
\]
\end{example}

Note that these results are in principle applicable also to a continuous phase space, with proper modifications to the above equations. Also Eq.~(\ref{eq:probability_continuity}) can be rewritten to a continuous time form by exchanging the difference in probabilities by a time derivative, and instead of the stochastic matrix $M$, using a transition matrix containing the rates of changes (see Sec.~\ref{sec:ctqw}). 









Preparing and sampling from the Gibbs canonical distribution is interesting, as it is the basis of MCMC methods (based on telescoping sums) for estimating partition functions, allowing one in turn to approximately count the number of ground states of a system. When applied to the Potts model on a particular graph, it becomes a tool for combinatorial problems whose goal is counting (e.g. estimating the number of perfect matchings in a graph). In Section \ref{sec:MCMC}, we will investigate and look at MCMC algorithms and their quantum counterparts in detail.


\subsubsection{Simulated Annealing}

Simulated annealing \cite{KiGeVe83,Cerny85} is a physically motivated method that allows us to search for ground states of simulated systems in the same way as in experiments slowly cooled material tends to get to its ground state (see also Fig.~\ref{fig:sa}). In a more abstract way we can say, that we perform a search for (global) minima\footnote{Note that search for maxima of function $\mathcal F$ is equivalent to the search for minima of function$-\mathcal F$.} of some fitness function $\mathcal F$ that represents energy in these models.

If we look at the annealing as a succession of system states that relaxes under slowly decreased temperature, we can easily devise its computational analog. The succession of system states is the crucial part, where on one hand we tend to choose a state that is close to the previous one. This, in fact, defines us a graph the system (walker) walks on, even more, when the system is described by discrete variables such as spins --- a step in such case may be described as a spin-flip on some position. On the other hand for the walk we need to have the transitional probabilities defined. The sampling of these states is then simply governed by the Metropolis algorithm and its utilization for sampling from Gibbs distribution, which is suitable for the task, as in the limit of zero temperature it becomes a distribution on the ground states of $\mathcal F$ only. So, in the end of the day Eq.~(\ref{eq:gibbssample}) is used with fitness function $\mathcal F$ in the role of the energy $\mathcal H$.

The complete algorithm of simulated annealing starts with the initialization of the system (on random) with the temperature being high\footnote{What ``high'' means is out of the scope of this paper and a lot of attention is devoted to setting all the parameters of the annealing right.}. Then the system is let to evolve by the above-mentioned procedure. When we are sure that the system is in equilibrium in this cycle we decrease the temperature a little. Then, again we let the system evolve for another cycle and decrease temperature further, e.g.~by following rule $(kT)^{-1} = \sqrt{\# cycles}$ or just by setting $T:=\mu T$ for $0\ll\mu<1$. This process is repeated until acceptably small temperature $T_{min}$ is achieved, when the system should be close to its ground state. All the parameters of the annealing have to be chosen carefully so that the cooling would be slow enough to allow the system to get to the state with lowest energy, yet not as slow as not to allow reasonable runnign time of the simulation.

\begin{figure}
\begin{algorithm}
Initialize $\mathbf x_0$ randomly and set $T$ to high enough value. Then while $T>T_min$ repeat:
\begin{enumerate}
\item do MC algorithm (with $N$ interations) from Fig.~\ref{fig:mcmc} with input parameters of $\mathbf x_0$ and given $T$
\item set $T:=\mu T$ and $\mathbf x_0:=\mathbf x_N$
\end{enumerate}
Output $\mathbf x_0$ as (sub)minimum
\end{algorithm}
\caption{\label{fig:sa} Simulated annealing algorithm for finding (sub)optimal value of $\mathbf x$. Value of $0\ll\mu<1$ should be chosen so that the decrease of the temperature would not be fast and the temperature $T$ should be initially high enough; $T_{min}$ should be chosen close enough to zero to prevent jumps to other local minima of the energy function. Also during the runnig of the algorithm one can look at averages of interesting observables.}
\end{figure}

\subsection{Summary}

In this section we have briefly described random walks, their properties and some applications. In the discrete case the evolution is governed by stochastic matrices while the system is described by a probability distribution on given state space. Theoretically studied properties of random walks were successfully applied in various scientific fields. Most prominent but not exclusive are their algorithmic application for probabilistic computations --- Markov Chain Monte Carlo. We have also studied few basic properties of the drukard's walk that will be later used to show the difference between classical random and quantum walks. Such quantities (mixing time, limiting distribution, etc.) serve on the other hand as a merit of efficiency with which the walks can be used in more practical applications.

\setcounter{equation}{0} \setcounter{figure}{0} \setcounter{table}{0}\newpage
\section{Quantum Walks: Using Coins}
\label{sec:DTQW}

Correctly employed, the non-classical features of quantum mechanics can offer us advantages over the classical world in the areas of cryptography, algorithms and quantum simulations. In the classical world, random walks have often been found very practical. It is then quite natural to ask whether there are quantum counterparts to random walks exist and whether it would be possible to utilize them in any way to our profit. As many simple questions, these two also require more than just simple answers. In the following sections we will attempt to construct answers for these questions in small steps. Our starting point will be one of the simplest translations of random walks to the quantum domain --- looking at discrete-time quantum walks taking place both in discrete space and discrete time. Their evolution will be described by an iterative application of a chosen unitary operator, advancing the walk by one step.

The first notions of a discrete-time quantum walk can be traced to Ref.~\cite{AhDaZa93}. The authors considered the spatial evolution of a system controlled by its internal spin-$1/2$ state, defined by the unitary $\op U=\exp(-\ii \op S_z\op P \delta/\hbar)$. The operators $\op P$ and $\op S_z$ correspond to the momentum and the $z$-component of the spin of the particle. The initial state $\ke{\psi(x_0)}(c_+\ke\uparrow+c_-\ke\downarrow)$ under the action of $\op U$ evolves into the state
\begin{equation}
\label{eq:firstQW}
\ke{\Psi}=c_-\ke{\psi(x_0-\delta)}\ke\downarrow+c_+\ke{\psi(x_0+\delta)}\ke\uparrow,
\end{equation}
where $\ke{\psi(x)}$ corresponds to the wave function of the particle centered around position $x$. The evolution described by \eqref{eq:firstQW} will accompany us throughout this whole section as a key ingredient of discrete-time quantum walks --- see e.g. \eqref{eq:translator}. However, we will diverge from \cite{AhDaZa93} on an important point -- the way we measure the system. In the original paper, the authors studied repeated applications of the unitary $\op U$ alongside with the measurement of the spin system and its repeated preparation. As we will see, this course of actions leads to ``classical'' random-walk like evolution (that is why the paper is called \emph{Quantum random walks}). 
The authors have shown that using a well-chosen evolution [similar to \eqref{eq:firstQW} up to the choice of basis] one can steer the system in a desired direction. Together with a choice of a the initial state that has width much larger than $\delta$, one can move this state spatially further than just the distance $\delta$. This is explained as a result of interference and can be used for example to drastically reduce (or amplify) the average number of photons in a cavity, produced by the detection of a single atom after it interacts resonantly with the cavity field.

In quantum walks, the measurement is performed only once, at the end of the evolution. A repetitive measurement process (as the one in \cite{AhDaZa93}) destroys quantum superposition and correlations that emerge during the evolution. Thus, we will use the definitions of quantum walks found in later references \cite{AhAmKeVa01, AmBaNaViWa01}, where we perform the measurement only at the end of the experiment. Similarly, continuous-time quantum walks (see Chapter \ref{sec:ctqw}), introduced by Farhi and Gutmann in \cite{FarhiWalk} involve a unitary (continuous-time) evolution according to the Schr\"{o}dinger equation for some time, followed by a single measurement.

\subsection{Drawing an Analogy from the Classical Case}
\label{sec:DTQWdef}

Let us consider the walk on a linear chain again and attempt to quantize it. We will start with an unsuccessful attempt and then learn from the mistake and find a meaningful way to do it.

In the classical case, a random walk on a linear chain is defined over the set of integers $\mathbb Z$. For a quantum walk, we shall consider a Hilbert space defined over $\mathbb Z$, i.e.~$\mathcal H=\spann(\{\ke x\st x\in\mathbb Z\})$, with the states $\ke{x}$ forming an orthonormal basis. Instead of being using a stochastic matrix, let us now try to describe the evolution by a unitary matrix. This simple ``quantization'' looks as a direct translation of random walks into the quantum world. However, it is easy to see that this model does not work as intended. Just as for the drunkard's walk, we require translational invariance of the unitary evolution. Thus if we start in the state $\ke{x-1}$, in the next step it will turn into $\alpha\ke{x-2}+\beta\ke{x}$ for some (complex) $\alpha$ and $\beta$. Similarly, the state $\ke{x+1}$ will be mapped to the state $\alpha\ke{x}+\beta\ke{x+2}$ with the same $\alpha$ and $\beta$. Note that the two initially orthogonal states $\ke{x-1}$ and $\ke{x+1}$ should remain orthogonal under any unitary evolution. However, this is now  possible only if one of the coefficients $\alpha$ or $\beta$ is zero. We can agree that such an evolution is even simpler than an evolution of the random walker (and quite boring).

The first attempt gave us a hint that using only the position space will not be enough for something interesting. 
Let us then try again and add another degree of freedom -- a \emph{coin space}, describing the direction\footnote{Note that now we are diverging from a classical memory-less random walk, where a walker had ``no idea'' where it came from, or where it would be going in the next step.} of the walker. This additional space will be a set $\Xi=\{\uparrow,\downarrow\}$ so that the state of a particle is described by a tuple $(x,c)\in\mathbb Z\times\Xi$. Because of this addition, we expand the Hilbert space to a tensor product $\mathcal H=\mathcal H_\mathrm{P}\otimes\mathcal H_\mathrm{C}$, where $\mathcal H_\mathrm{P}=\spann (\mathbb Z)$ determines a position of the walker and $\mathcal H_\mathrm{C}=\spann (\Xi)$ is the introduced \emph{coin space}\footnote{Discrete-time quantum walks using coins are also called \emph{coined quantum walks}. Sometimes we will refer to them in this way.} (in this example, the coin could correspond to the spin degree of freedom of a spin-$\frac{1}{2}$ particle). Therefore, $\mathcal H$ is spanned by the orthonormal basis\footnote{When using the states $\ke x\otimes\ke c$ we will often omit the tensor product symbol $\otimes$ to shorten the notation.} $\{\ke x\otimes\ke c\st x\in\mathbb Z,c\in\Xi\}$. How will an initial state $\ke\psinit\in\mathcal H$ evolve? We choose to describe a single evolution step by a unitary evolution operator
\begin{equation}
	\label{eq:coinQW}
	\op U=\op S\,(\iii\otimes\op C).
\end{equation}
It is a composition of $\iii \otimes C$ acting nontrivially only on the coin space, and $S$, which involves the whole Hilbert space $\mathcal H$.
The operator $\op S$ describes spatial translation of the walker, while $\op C$ correspond to coin throwing. Let us look at them in detail.

\paragraph{Translation operator $\op S$.} This operator acts on the Hilbert space $\mathcal H$ as a conditional position shift operator with coin being control qubit. It changes the position of the particle one position up if the coin points up and moves the particle one position down, if the coin points down. It can be written in the form
\begin{equation}
\label{eq:translator}
\op S=\sum_{x\in\mathbb Z}\left( \ke{x+1}\bra{x}\otimes\ke\uparrow\bra\uparrow+\ke{x-1}\bra{x}\otimes\ke\downarrow\bra\downarrow\right).
\end{equation}
We see that the structure of the graph (in this case the line) is reflected in this operator by the allowed transitions in position space only to the nearest neighbours.

\paragraph{Coin operator $\iii \otimes \op C$.} This unitary operator acts nontrivially only on the coin space $\mathcal H_\mathrm{C}$ and corresponds to a ``flip'' of the coin. For the quantum walk on line, $C$ is a $2\times 2$ unitary matrix. A usual choice for $C$ is the \emph{Hadamard coin}\footnote{Quantum walk on line using the Hadamard coin is often called the \emph{Hadamard walk}.}
\begin{equation}
\label{eq:hadamard}
\op C\equiv\op H=\frac{1}{\sqrt{2}}
	\begin{bmatrix}
	 1 & 1\\ 1 & -1\\ 
	\end{bmatrix},
\end{equation}
viewed in a different notation as
\[
\op H: \ke{\uparrow}\mapsto\frac{1}{\sqrt{2}}\left(\ke{\uparrow}+\ke{\downarrow}\right),\qquad
\op H: \ke{\downarrow}\mapsto\frac{1}{\sqrt{2}}\left(\ke{\uparrow}-\ke{\downarrow}\right).
\]
Observe that this coin can be called \emph{unbiased}, since the states $\ke{\uparrow}$ and $\ke{\downarrow}$ of the coin are evenly distributed into their equal superpositions, up to a phase factor. We shall return to the coin later in Sec.~\ref{sec:coins}. In the next Section, we will investigate the evolution of the walker under the influence of the Hadamard coin.

Let us recall that a single step of a coined quantum walk is described by the operator $U$ in \eqref{eq:coinQW}. The state after $n$ steps will thus be described by a vector from $\mathcal H$ as
\begin{equation}
	\label{eq:qevolution}
	\ke{\psi_n}=\op U^n\ke\psinit.
\end{equation}

\subsection{Dispersion of the Hadamard Quantum Walk on Line}
\label{sec:dispersion}

In this section we will investigate the quantum walk in 1D using the Hadamard coin \eqref{eq:hadamard} and compare it to the classical random walk. First, let us argue why we even talk about a connection of the Hadamard walk to the drunkard's walk (for more on the topic see Refs.~\cite{Kendon07,KeTr03,BrCaAm03,KoBuHi06} and Section~\ref{sec:decoherence}). Suppose the state of the system is $\ke{x}\ke{c}$ with $x\in\mathbb Z$ and $c$ being one of the basis states $\ke\uparrow$ and $\ke\downarrow$. Then a single step of the quantum walk gives us
\begin{eqnarray*}
	\op U\ke{x}\ke{c} & = & \op S(\iii\otimes\op H)\ke{x}\ke{c}=\op S(\ke{x}\otimes\op H\ke{c})
	=\frac{1}{\sqrt{2}}\op S\left[\ke{x}\otimes(\ke{\uparrow}\pm\ke{\downarrow})\right]=\\
	 & = &\frac{1}{\sqrt{2}}\left(\op S\ke x\ke\uparrow\pm\op S\ke x\ke\downarrow\right)
	 =\frac{1}{\sqrt{2}}\ke{x+1}\ke\uparrow\pm\frac{1}{\sqrt{2}}\ke{x-1}\ke\downarrow.
\end{eqnarray*}
From this result we may observe that starting in the state $\ke\psinit=\ke x\ke{\uparrow}$ and performing a measurement on the position state space\footnote{Note that we can choose to measure the coin register instead and also end up with $\ket{x+1}$ or $\ket{x-1}$ in the position register with equal probability. Tis is because the two registers become entangled after the application of $U$.} right after an application of $\op U$, the probability of finding $(x+1)$ equals the probability of measuring $(x-1)$. Therefore, if we perform a position measurement after each\footnote{We could imagine performing the measurement only after a certain number of steps or probabilistically, giving us a continuum of behaviors between quantum and classical. We explore this in Section \ref{sec:decoherence}.} application of $U$, we end up with a classical random walk where we shift the position to either side with equal probability $\half$.

\begin{figure}[t!]
\begin{center}
\includegraphics[scale=1]{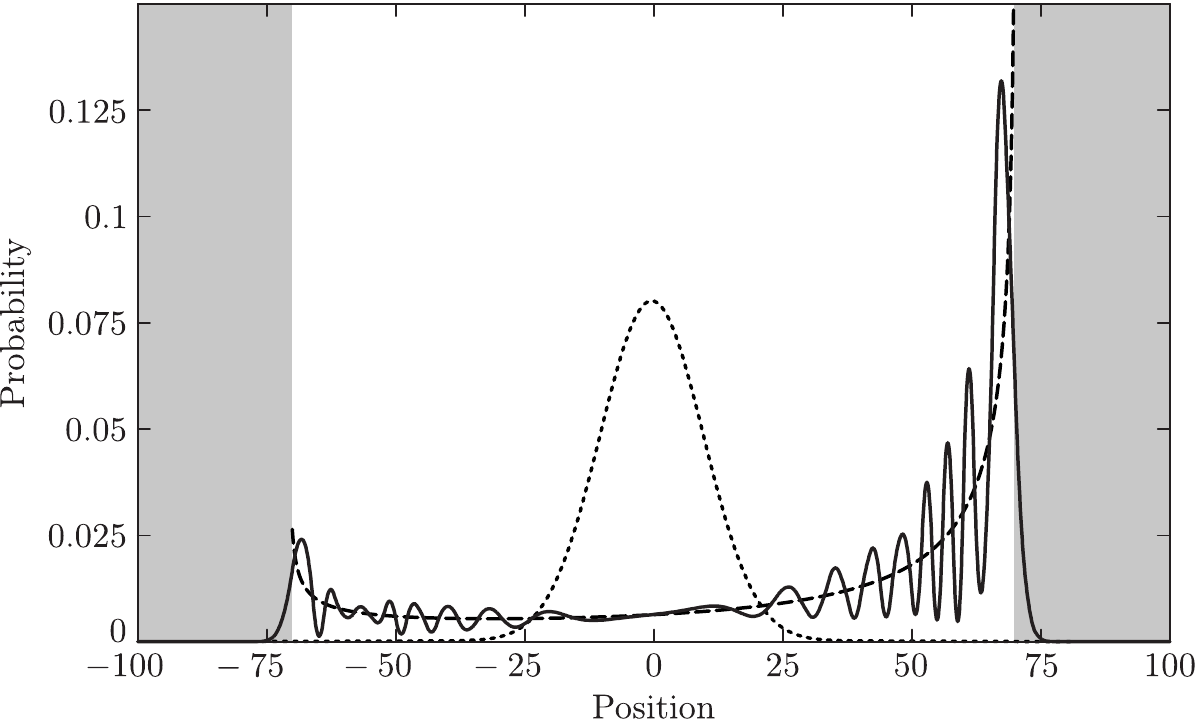}
\end{center}
\caption{\label{fig-3} Comparison of probability distributions after 100 steps for the Hadamard walk on even positions (black line) starting in state $\ke 0\ke\uparrow$ and the drunkard's walk (starting on position 0). Spreading of quantum walk (determined by the variance) is quadratically faster than that of random walk. Hadamard walk can be in region $[-100/\sqrt{2};100/\sqrt{2}]$ approximated by function $P^{100}$ (dashed line) given by Eq.~(\ref{eq:P_slow}) suitable also for evaluation of moments. Outside the region (grey areas), the probability drops exponentially fast. Distribution of the classical walker is depicted by dotted line.}
\end{figure}

However, as we described at the beginning of Chapte \ref{sec:DTQW}, we do not want to measure the system after each step. What will happen if we leave the system evolve for some time and measure the position of the particle then? Let us look at three applications of $U$. After some algebraic manipulation we arrive at
\begin{equation}
	\op U^3\ke 0\ke\uparrow=\frac{1}{\sqrt{8}}\biggl[\ke 3\ke\uparrow+\ke 1\bigl(\ke\downarrow+2\ke\uparrow\bigr)
		-\ke{-1}\ke\uparrow+\ke{-3}\ke\downarrow\biggr].
	\label{U3hadamard}
\end{equation}
This already shows that the amplitudes are not symmetric and start to show interference effects. Let us look at the system after $m$ steps.  
In the final measurement, we do not care about the state of the coin. Thus to get the probability for the particle being on the position $x$ after $m$ steps, we trace the coin register out, and look at
\begin{equation}
	\label{eq:px}	 
	p^m(x)=\no{(\bra x\otimes\bra\uparrow)\op U^m\ke\psinit}^2+\no{(\bra x\otimes\bra\downarrow)\op U^m\ke\psinit}^2.
\end{equation}
Looking at three steps of evolution of the Hadamard walk in \eqref{U3hadamard}, we can read out the values of these probabilities after $n=3$ steps (when starting from $\ke 0\ke\uparrow$):
\[
	p^3(-1)=p^3(3)=p^{3}(-3)=\frac{1}{8},\qquad p^3(1)=\frac{5}{8}
\]
and zero otherwise. Notice the asymmetry between $p^3(1)$ and $p^3(-1)$. We do not observe asymmetrical behavior for classical random walks with an unbiased coin. This is the first difference between quantum and classical random walks that we have seen. The situation is even more interesting for longer evolution times, as depicted in Fig.~\ref{fig-3} where the system is shown after 100 steps. We see that the distribution for the Hadamard quantum walk starting from $\ke 0 \ke \uparrow$ is indeed asymmetrical. What is more interesting, the probability spreads faster than for a classical walk.

We will follow \cite{AmBaNaViWa01,HiBeFe03} to show that the spreading in the case of the quantum walk on a line is linear with time $m$ (the number of steps), meaning that the standard deviation grows as $\sigma\sim m$ as opposed to the classical case where $\sigma\sim\sqrt{m}$ as shown in \eqref{eq:rw_slimak}. Moreover, we can see in Fig.~\ref{fig-3} that the distribution for the quantum walk gets ``close'' to uniform\footnote{When considering decoherence in quantum walks as in \cite{KeTr03}, the distribution of the quantum walker may get even closer to a uniform distribution (see also Section \ref{sec:decoherence}).} in the interval $\left[-\frac{m}{\sqrt{2}},\frac{m}{\sqrt{2}}\right]$.

The method we will use for the analysis is based on switching\footnote{There are also other ways how to deduce the asymptotic behaviour of the quantum walk on line, especially employing the knowledge from the theory of path integrals. Note here, that the Fourier basis used is not normalizable, however, under careful manipulation is very useful.} to the \emph{Fourier basis} in the position space, where
\[
	\ke{\tilde k}=\sum_{x=-\infty}^\infty\e^{\ii kx}\ke x,\qquad k\in[-\pi;\pi].
\]
A converse transformation gives us
\[
	\ke y=\frac{1}{2\pi}\int_{-\pi}^\pi\e^{-\ii ky}\ke{\tilde k}\,\dd k,
\]
with the special case
\begin{equation}
	\label{eq:fourier_zero}
	\ke 0=\frac{1}{2\pi}\int_{-\pi}^\pi\ke{\tilde k}\,\dd k.
\end{equation}
The Fourier basis is useful since the states $\ke{\tilde k}\otimes\ke\uparrow$ and $\ke{\tilde k}\otimes\ke\downarrow$ are eigenvectors of the translation operator $\op S$ corresponding to the eigenvalues $\e^{\mp \ii k}$ respectively.

For the Hadamard walk defined in the previous section, with the Hadamard operator used as the coin operator \eqref{eq:hadamard}, and choosing the initial state $\ke{\psinit^c}=\ke 0\otimes\ke c$ (where $c$ can be either the spin up, or spin down state) and expressing it in the Fourier basis, we find
\begin{equation}
\label{eq:fourier_evolution}
\ke{\psi_m^c}=\op U^m\ke{\psinit^c}=\op U^m(\ke 0\otimes\ke c)=\frac{1}{2\pi}\int_{-\pi}^\pi\ke{\tilde k}\otimes\op M_k^m\ke c\dd k,
\end{equation}
where $M_k$ denotes the $2\times 2$ matrix
\[
	\op M_k=\frac{1}{\sqrt{2}}\begin{bmatrix}
	\e^{-\ii k} & \e^{-\ii k}\\
	\e^{\ii k} & -\e^{\ii k}
	\end{bmatrix}
\]
It turns out that all we really need to know to analyze the evolution is the eigenspectrum of the operator $\op M_k$. Introducing new variables $\omega_k \in[-\pi/2;\pi/2]$ and setting $\sin k=\sqrt{2}\sin\omega_k$, we find that the eigenvalues of $M_k$ are $\lambda_-=\e^{-\ii\omega_k}$ and $\lambda_+=-\e^{\ii\omega_k}$. The corresponding eigenvectors are
\[
	\ke{\phi_-}=\frac{1}{N_-}\begin{bmatrix} 1 \\ \sqrt{2}\e^{\ii(k-\omega_k)}-1 \end{bmatrix} \quad\text{and}\quad
	\ke{\phi_+}=\frac{1}{N_+}\begin{bmatrix} 1 \\-\sqrt{2}\e^{\ii(k+\omega_k)}-1 \end{bmatrix}
\]
with the normalization factors given by
\[
	N_\pm^2=2(1+\cos^2 k\pm\cos k\sqrt{1+\cos^2 k}).
\]
We can expand the expression $\op M_k^m\ke c$ in \eqref{eq:fourier_evolution} in the eigenbasis of the operator $\op M_k$ as
\[
	\op M_k^m\ke c=\bigl(\lambda_-^m\langle\phi_-|c\rangle\bigr)\ke{\phi_-}+\bigl(\lambda_+^m\langle\phi_+|c\rangle\bigr)\ke{\phi_+}.
\]
Inserting it into \eqref{eq:fourier_evolution} we arrive at
\begin{equation}
	\ke{\psi_m^c}=\sum_x\ke x\otimes\left[A^m_c(x)\ke\uparrow+B^m_c(x)\ke\downarrow\right].
	\label{eq:startupprediction}
\end{equation}

\begin{exercise}
Prove the following equalities
\begin{eqnarray*}
A_\uparrow^m(x) &=& \frac{1+(-1)^{m+x}}{2} \left[\alpha^m(x)+\beta^m(x)\right],\\
A_\downarrow^m(x) &=& \frac{1+(-1)^{m+x}}{2} \left[\beta^m(x)-\gamma^m(x)\right],\\
B_\uparrow^m(x) &=& \frac{1+(-1)^{m+x}}{2} \left[\beta^m(x)+\gamma^m(x)\right],\\
B_\downarrow^m(x) &=& \frac{1+(-1)^{m+x}}{2} \left[\alpha^m(x)-\beta^m(x)\right],
\end{eqnarray*}
where
\begin{subequations}
\label{eq:integrale}
\begin{eqnarray}
\alpha^m(x) &=& \int_{-\pi}^\pi\frac{\dd k}{2\pi} \e^{\ii(kx-m\omega_k)},\\
\beta^m(x)  &=& \int_{-\pi}^\pi\frac{\dd k}{2\pi} \frac{\cos k}{\sqrt{1+\cos^2 k}} \e^{\ii(kx-m\omega_k)},\\
\gamma^m(x) &=& \int_{-\pi}^\pi\frac{\dd k}{2\pi} \frac{\sin k}{\sqrt{1+\cos^2 k}} \e^{\ii(kx-m\omega_k)}.
\end{eqnarray}
\end{subequations}
\end{exercise}

\begin{exercise}
Prove that all $a^m(x)$, $b^m(x)$ and $c^m(x)$ are real and that
\[
a^m(x)=(-1)^xa^m(x),\ b^m(x)=-(-1)^xb^m(x),\ c^m(x)=-(-1)^xc^m(x).
\]
\end{exercise}

At this point we may observe the \emph{modularity} property of the walk (previously mentioned in Section \ref{sec:RWproperties}): after an even (odd) number of steps, the walker cannot be on odd (even) positions. 

The results we obtained so far are valid only for states initialized with a computational basis state of the coin -- spin up or down. If we would like the initial state to be completely general, having form
\begin{equation}
	\label{eq:initial_symmetry}
	\ke\psinit = \ke{0}\otimes\left[ \sqrt{q} \ke\uparrow+\sqrt{1-q} \,\e^{\ii\sigma} \ke\downarrow \right]
\end{equation}
parametrized by real parameters $q$ and $\sigma$, the state after $m$ steps will be a superposition of evolved states whose initial coins were set to be either spin-up or spin-down, i.e.
\begin{eqnarray*}
\ke{\psi_m}=\op U^m\ke\psinit &= \sum_x & \ke x\otimes\left\{
\left[ \sqrt{q}A^m_\uparrow(x) +\sqrt{1-q}\e^{\ii\sigma}A^m_\downarrow(x) \right] \ke\uparrow \right.\\
&& \left. + \left[ \sqrt{q}B^m_\uparrow(x) +\sqrt{1-q}\e^{\ii\sigma}B^m_\downarrow(x) \right] \ke\downarrow
\right\}.
\end{eqnarray*}
The probability to find the walker at position $x$ after $m$ steps is then expressed as
\[
p^m(x)=\left| \sqrt{q}A^m_\uparrow(x) +\sqrt{1-q}\e^{\ii\sigma}A^m_\downarrow(x) \right|^2 + \left| \sqrt{q}B^m_\uparrow(x) +\sqrt{1-q}\e^{\ii\sigma}B^m_\downarrow(x) \right|^2.
\]

\begin{exercise}
Show that
\begin{eqnarray}
	p^m(\pm x) &=& \left[\alpha^m(x)\right]^2 + 2\left[\beta^m(x)\right]^2 
			+ \left[\gamma^m(x)\right]^2 \pm (4q-2)\beta^m(x)\left	[\alpha^m(x)\pm\gamma^m(x)\right]\notag\\
	&& \pm 4\sqrt{1-q}\cos\sigma\beta^m(x)\left[\alpha^m(x)\mp\gamma^m(x)\right].
	\label{eq:probabilityonaline}
\end{eqnarray}
\end{exercise}

The integrals in \eqref{eq:integrale} can be approximated by employing the method of stationary phase (see Appendix~\ref{sec:stationary_point}) and we find that the functions $a^m(x)$, $b^m(x)$ and $c^m(x)$ are mostly concentrated within the interval $\left[-\frac{m}{\sqrt{2}};\frac{m}{\sqrt{2}}\right]$ and they quickly decrease beyond the bounding values of this interval. This also holds for the probability $p^m(x)$, which oscillates around the function 
\begin{equation}
\label{eq:P_slow}
P^m(x)=\frac{2m}{\pi(m-x)\sqrt{m^2-2x^2}},
\end{equation}
where we dropped the vanishing part coming from modularity. For details of deriving \eqref{eq:P_slow}, see Appendix~\ref{sec:hadamard_approximation}.

The function $P^m(x)$ allows us to approximately evaluate the averages of position $x$-dependent functions $f^m(x)$ in the $m$-th step of the walk as
\begin{equation}
	\langle f^m(x)\rangle\simeq\frac{1}{2}\int_{-\frac{m}{\sqrt{2}}}^{\frac{m}{\sqrt{2}}}f^m(x)P^m(x)\, \dd x.
	\label{eq:faverage}
\end{equation}
The factor $\half$ comes from modularity (indeed it is easy to check that using \eqref{eq:faverage}, we correctly obtain $\langle 1\rangle=1$. For more interesting $x$-dependent functions we obtain the approximations
\begin{eqnarray*}
\langle x \rangle   &\simeq& m\frac{\sqrt{2}-1}{\sqrt{2}},\\
\langle \no{x} \rangle &\simeq& \frac{m}{2},\\
\langle x^2 \rangle &\simeq& m^2\frac{\sqrt{2}-1}{\sqrt{2}}.
\end{eqnarray*}
The results shows also that the dispersion grows as $\sqrt{\langle x^2 \rangle}\sim m$ for large $m$, increasing linearly with time. This is quadratically faster than the classical random walk \eqref{eq:rw_slimak}, whose dispersion grows with the number of steps as $\sqrt{m}$.

We can use \eqref{eq:probabilityonaline} to find several other properties of the evolution. When tuning the parameter $\sigma$ in the initial coin state \eqref{eq:initial_symmetry}, we can find many possibilities that result in a symmetrical probability distribution $p^m(x)=p^m(-x)$. For example, we can choose $\cos\sigma=0$, which gives a symmetrical final distribution for $q=\half$, i.e.~when the initial state is
\begin{equation}
\label{eq:hadamard_symmetric}
\ke\psinit=\ke 0\otimes\frac{1}{\sqrt{2}}[\ke\uparrow\pm\ii\ke\downarrow].
\end{equation}
Similarly, we can choose $\cos\sigma=\pm 1$, in which case $q=(2\mp\sqrt{2})/4$, corresponding to $\sqrt{q}=\sin\frac{\pi}{8}$ 
or $\sqrt{q}=\cos\frac{\pi}{8}$, again results in a symmetrical distribution.

Instead of looking at symmetrical evolutions, we can just as well try to find the most asymmetrical one. Let us look at the first moment of the probability distribution,
\[
\langle x\rangle=\sum_{x=-\infty}^\infty xp^m(x)\sum_{x=1}^\infty 4x\alpha^m(x)\beta^m(x)\left[(2q-1)+2\sqrt{q(1-q)}\cos\sigma\right].
\]
From Appendix~\ref{sec:hadamard_approximation} we know that $\alpha^m(x)\beta^m(x)$ is positive for our range of positions. This tells us that the maximal right asymmetry can be obtained for $\cos\sigma=1$ and $q=(2+\sqrt{2})/4$, while the maximal left asymmetry can be obtained for $\cos\sigma=-1$ and $q=(2-\sqrt{2})/4$. All these results are in correspondence with the results of \cite{TrFlMaKe03}.

\begin{exercise}
Looking at the mean position is not a good general indicator showing the asymmetry of a distribution. We were able to use it above, as the evolution began in the position zero. In general, the asymmetry of a distribution is given by its \emph{skewness}
\[
\mathrm{Skew}(x)=\left<\left(\frac{x-\langle x\rangle}{\langle x^2\rangle}\right)^3\right>.
\]
Using the approximations \eqref{eq:approximations} given in Appendix \ref{sec:hadamard_approximation}, compute the skewness for the Hadamard walk and find the most asymmetric one [Hint: it should be the initial state choice we already found above].
\end{exercise}

\begin{figure}[t!]
\begin{center}
\includegraphics[scale=0.85]{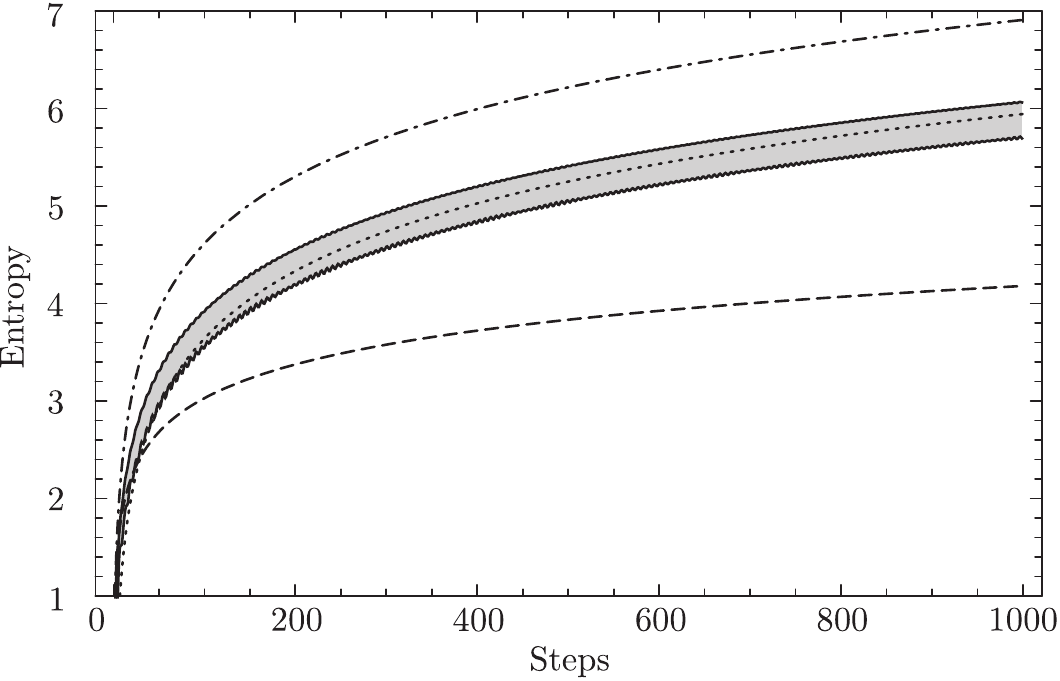}
\end{center}
\vspace{-0.35cm}
\caption{\label{fig-14} Shannon entropy [Eq.~(\ref{eq:entropy})] vs. number of steps taken for probability distributions of different walks, provided the walker starts at position 0 (state $\ke{0}\ke{c}$). The dot-dashed line is an upper boundary for entropy reachable by any walk, the dashed line is the entropy of the drunkard's walk distribution. The shaded area represents the region for entropy values of quantum walks. Its upper boundary is given by symmetric and lower boundary by maximally asymmetric coined quantum walks (see Sec.~\ref{sec:coins} for further information on symmetry of the distribution). The dotted line is the Shannon entropy for the probability distribution coming from the state \eqref{eq:startupprediction}, the Hadamard walk evolution of the initial state �$\ke{0}\ke{\uparrow}$.}
\end{figure}

\begin{example}
We can illustrate the result about fast spreading of the quantum Hadamard walk also by considering the \emph{(Shannon) entropy} defined on distributions as
\begin{equation}
\label{eq:entropy}
S(m)=-\sum_{x=-\infty}^\infty p^m(x)\ln p^m(x),
\end{equation}
on the probabilities $p^m(x)$ for the quantum walk defined by \eqref{eq:px}. This quantity gives us yet another characterization of how the probability of the walker's position is distributed and we shall study its dependence on the number $m$ of steps taken in a given walk. Let us look at what the entropy looks like for certain cases. First, if only one position $x_0$ was possible (and predicted with certainty) for the particle at a particular moment, then $p(x)=\delta_{x,x_0}$ and the sum in \eqref{eq:entropy} would be zero. Second, we can imagine the distribution function evolving so that after $m$ steps it would be uniformly distributed (maximally mixed) between all possible locations in the interval $[-m,m]$, taking modularity into account (excluding unreachable points for both random and quantum walks). For such an evolution we may write
\[
	p^m(x)=\begin{cases}
		\frac{1+(-1)^{m+x}}{m+1} & \text{if $-m\leq x\leq m$,}\\
		0 & \text{otherwise.}
	\end{cases}
\]
The entropy for such a distribution computed from \eqref{eq:entropy} gives us 
\[
	S^\mathrm{max}(m)=\ln(m+1).
\] 
This is, in fact, an upper bound on the entropy that is achievable by any walk on a given graph (see the dot-dashed line in Fig.~\ref{fig-14}).
Third, in the classical drunkard's walk, the probability distribution approaches a normal distribution given by \eqref{eq:pdf_rw}, and its entropy can be easily evaluated as
\[
S^\mathrm{c}(m)=\frac{1}{2}\left(1+\ln\frac{\pi m}{2}\right).
\]
Finally, in the quantum case, we can observe from Fig.~\ref{fig-14} that the entropy is larger than in the classical case given by the dashed line. Even though this entropy depends on the initial coin state, it still tells us that coined quantum walks spread faster than classical random walks also with respect to the Shannon entropy. 
\end{example}

\subsection{Coined Quantum Walks on General Graphs}

Equations \eqref{eq:coinQW}-\eqref{eq:hadamard} describe quantum walks on a line using the unitary update $U$ composed from a coin-diffusing operation and a subsequent translation. The generalization to more general graph structures requires just a slight modification to these equations and only slightly more explanation. We can view the coin degree of freedom $\Xi$ in the 1D walk, given by the two states $\uparrow$ and $\downarrow$, as a coloring of a directed version of the graph. For each $x$, the ``color'' $\downarrow$ was assigned to the directed edges of the graph pointing from $x$ to $x-1$, while the ``color'' $\uparrow$ was assigned to the directed edges of the graph pointing from $x$ to $x+1$. Therefore, even though the graph (line) on which we walked is not directed, each edge can be interpreted as a set of two oppositely directed edges.

Let us present a simple extrapolation of the 1D formalism which allows us to define quantum walks on $d$-regular (undirected) graphs $\mathcal G=(\all,E)$, where $\all$ is the set of vertices and $E\subseteq\all\times\all$ is the set of edges defining the graph. First, we interpret each edge as two oppositely directed edges and then, as described in \cite{AhAmKeVa01}, we assign a ``color'' number from $1$ to $d$ to each of these directed edges in such a way that all the directed edges of one color form a permutation of the vertices. In other words, each vertex has exactly one outgoing and one incident edge with a given color. Such a coloring is always possible using $d$ colors. Thus, in addition to the position space of vertices $\mathcal H_\mathrm{P}$, we expand the Hilbert space with the coin subspace $\mathcal H_\mathrm{C}$ with dimension $d$ as $\mathcal H=\mathcal H_\mathrm{P} \otimes \mathcal H_\mathrm{C}$.

The motion of the walker will be described by the translation operator $\op S$, generalized to
\[
	\op S=\sum_{x\in\all}\sum_{c=1}^d\ke{x\oplus c}\br{x}\otimes\ke{c}\br{c},
\]
where $x\oplus c$ is the vertex accessible from $x$ by the edge with color $c$.

The coin-diffusion $C^{(d)}$ now involves a $d$-dimensional coin space. Furthermore, we can imagine that the coin-flipping operation can be position-dependent 
\begin{equation}
	\label{eq:general_coinQW}
	\op C'=\sum_{x\in\all}\ke x\br x\otimes\op C^{(d)}_x,
\end{equation}
with a different $d\times d$ matrix $C^{(d)}_x$ for different vertices $x$.
The evolution of a general graph quantum walk is then governed by the two-step unitary $\op U=\op S\op C'$. 
Compare this $C'$ to the position-independent (translationally-invariant) coin operator $\iii \otimes C$ from Section \ref{sec:DTQWdef} and observe that we recover it if we choose $\op C_x^{(d)}\equiv\op C$. 

This generalization is often used in the literature, however, there is another (isomorphic) approach called {\em Scattering Quantum Walks}, first introduced by Hillery et al.~\cite{HiBeFe03}, with the proof of the isomorphism provided in Ref.~\cite{AnLu09}. We describe this approach in the following Section. The basic similarities of the two approaches are easier seen after realizing the following points:
\begin{itemize}
\item The condition that all edges with some color $a$ form a permutation of the vertices meansthat for every vertex $x$ and for every color $a$ there is exactly one vertex $x_-$ from which you can get to $x$ by the edge with color $a$ and exactly one vertex\footnote{Vertices $x_+$ and $x_-$ need not be the same; e.g.~in a regular graph with an odd number of vertices and no loops, all (for all vertices and colors) $x_+$ and $x_-$ cannot coincide. If they did, the edges with the same color would connect pairs of vertices and all these pairs would be disjoint, but if the number of vertices is odd, one cannot create this structure.} $x_+$ which is accessible from $x$ by the edge with color $a$. This also means that every state $\ke x\otimes\ke c$ uniquely describes some edge from the directed graph.
\item The coin $\op C_x$ ``determines'' how is the amplitude of particle that came from edge with color $a$ distributed to other edges, while $\op S$ still has the ``trivial'' role of nudging the walker forward.
\end{itemize}
These facts can be translated as essentially describing scattering process, where
\begin{itemize}
\item The incident direction of a particle corresponds to the edge color.
\item A scattering process on a given vertex $x$ corresponds to the action of the coin $\op C_x$. Note that the translation operator $\op S$ is superfluous in this picture.
\end{itemize}
Up to this point, we have described the need for a separate coin in discrete-time quantum walks. Having a coin such as in \eqref{eq:general_coinQW} is viable only for $d$-regular graphs, even though we can generalize it and make it position-dependent. If the underlying graph structure is not regular, a simple description of the coin begins to be difficult and the irregularity of the graph structure becomes problematic as well. There are several approaches to alleviate this problem. One of them is to introduce a position dependent coin with a variable dimension \cite{AhAmKeVa01,ShKeWh03}. However, it means the Hilbert space cannot be factorized into the coin space and the position space anymore. The possible irregularity of the graph structure is in this case embedded in the evolution process effectively acting as an oracle \cite{Kendon06c}. The action of the position-dependent coin, however, has to be in correspondence with the graph structure as well. It is then not such great a leap to start thinking of also embedding the coin operator into the oracle. This second approach is in fact the scattering model of quantum walks we are about to define.

\subsubsection{Scattering Quantum Walks (SQW)}
\label{sec:sqws}


Scattering quantum walks (SQW) describe a particle moving around a graph, scattering off its vertices. The state of the particle lives (is located on) the edges of a graph $\mathcal G=(\all,E)$, defined by a set $\all=\{1,2,\ldots,N\}$ of vertices and a set of edges $E\subseteq \all\times \all$. The Hilbert space $\mathcal H$ for such a walk on $\mathcal G$ is then defined as
\begin{equation}
\label{eq:hilbik}
\mathcal H=\spann(\{\ke{m,l}\st m,l\in \all,ml\in E\}),
\end{equation}
where $ml$ is a short-hand notation for the edge connecting vertices $m$ and $l$. This definition gives us a Hilbert space which is a span of all the \emph{edge states}, which form its orthonormal basis. The state $\ke{m,l}$ can then be interpreted as a particle going from vertex $m$ to vertex $l$. 

\begin{exercise}
Show that the dimension of the Hilbert space for a scattering quantum walk is the same as for a corresponding coined quantum walk.
\end{exercise}

\begin{figure}
\hfill a)
\includegraphics[width=2.5cm]{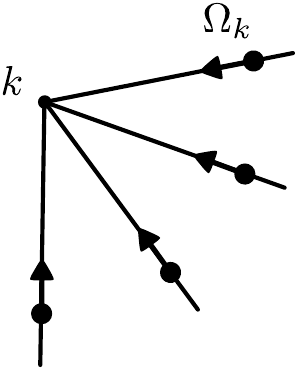}
\hfill b)
\includegraphics[width=2.5cm]{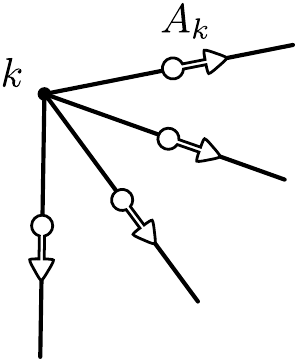}
\hfill c)
\includegraphics[width=2.5cm]{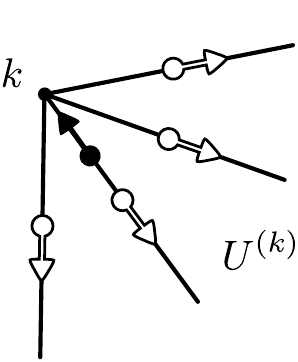}
\hfill\ 
\caption{\label{SQWfigure} 
a) The subspace $\Omega_k$ of walker states entering vertex $k$. b) The subspace $A_k$ of walker states exiting vertex $k$. 
c) The action of the local unitary $U^{(k)} : \Omega_k \rightarrow A_k$ can be viewed as scattering on vertex $k$.}
\end{figure}

Let us look at the structure of the Hilbert space for a SQW. First, we have the subspaces $A_k$ spanned by all the edge-states originating in the vertex $k$,
\[
A_k=\spann(\{\ke{k,m}\st m\in \all,km\in E\}).
\]
Second, we have $\Omega_k$, the subspaces spanned by all the edge-states that end in the vertex $k$,
\begin{equation}
\label{eq:omega}
\Omega_k=\spann(\{\ke{m,k}\st m\in \all,mk\in E\}).
\end{equation}
These subspaces don't overlap, as $A_k\cap A_l=\emptyset$ and $\Omega_k\cap\Omega_l=\emptyset$ for $k\neq l$. Moreover, $\no{\Omega_k}=\no{A_k}$ for all $k$, as the graph $\mathcal G$ is not oriented\footnote{Note that we could describe a SQW coming from an oriented graph as in e.g. \cite{Severini03,Montanaro07}, but not without complications. We thus decide to talk only about QW coming from undirected graphs here.}.
The dynamics of the quantum walk are described by \emph{local unitary evolutions} scattering the walker ``on vertex'' $k$ --- describing the transition from the walker entering vertex $k$ to the walker leaving it. Using our notation for the incoming and outgoing subspaces, the local unitary evolutions act as $\op U^{(k)}:\Omega_k\rightarrow A_k$, as depicted in Figure~\ref{SQWfigure}. 
\begin{example}
The simplest example of a local unitary evolution $U^{(k)}$ for a 1D graph transforms a ``right-moving'' state $\ket{k-1,k}$ (moving from $k-1$ to $k$) into the uniform superposition
$\frac{1}{\sqrt{2}} \ket{k,k-1} + \frac{1}{\sqrt{2}} \ket{k,k+1}$, while a ``left-moving'' state $\ket{k+1,k}$ similarly changes into 
$\frac{1}{\sqrt{2}} \ket{k,k-1} - \frac{1}{\sqrt{2}} \ket{k,k+1}$,
with the minus sign required for unitarity. This transformation corresponds to the Hadamard coin \eqref{eq:hadamard} in DTQW.
\end{example}
We will see more local ``coins'' in the following Section~\ref{sec:coins}, and then a general approach for finding such transformations coming from classical Markov Chains in Section~\ref{sec:qmc}.

The overall unitary $\op U$ describing one step of the SQW acting on the system is then the combined action of the local unitary evolutions:
\[
\op U=\Gamma\bigoplus_{k\in\all}\op U^{(k)},
\]
where $\Gamma$ is just a permutation on the basis elements so that their order would be restored. This can be done as 
\[
\bigcap_{k\in\all}\Omega_k=\bigcap_{k\in\all}A_k
\]
gives the whole computational basis set. In other words, as all $\op U^{(k)}$ act unitarily on disjoint subsets of the Hilbert space, the overall operation $\op U$ they define is also unitary and the restriction of $\op U$ to $\Omega_k$ is just $\op U^{(k)}$. 
When the initial state of the system is $\ke{\psinit}$, the state after $m$ steps is given by $\ke{\psi_m}=\op U^m\ke\psinit$ and the probability of finding the particle (walker) in the state $\ke{k,l}$ is then $\no{\langle k,l|\psi_m\rangle}^2$.

\begin{figure}
\begin{center}
\includegraphics[scale=0.83]{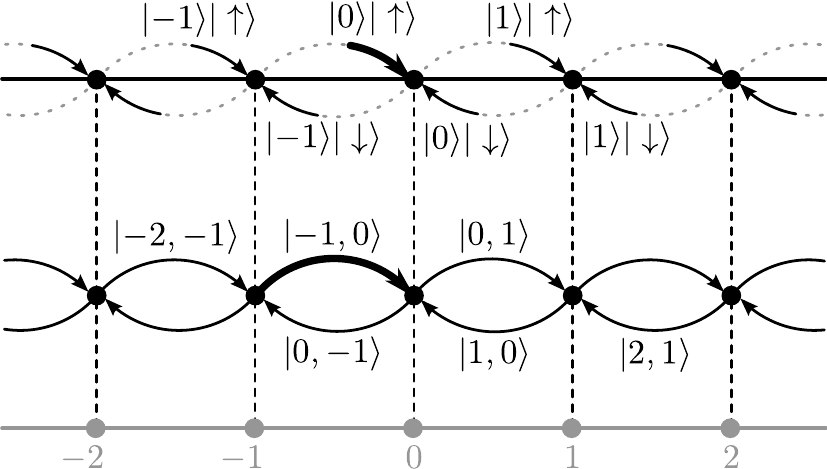}
\end{center}
\vspace{-0.4cm}
\caption{\label{fig:sqw}Correspondence between coined quantum walk (top, for better readability shown without direct product symbol) and scattering quantum walk on a line. For example (thick arrows), state $\ke 0\otimes\ke\uparrow$ represents the walker in coined quantum walk positioned on vertex $0$ with coin pointing ``up'' stating that this was the direction the walker used to get to the vertex. This state corresponds in the scattering quantum walk formalism to the walker going from vertex $-1$ to vertex $0$.}
\end{figure}

Let us compare discrete-time quantum walks to SQW. The action of the coin in discrete-time quantum walks is an analogue of the local unitary evolutions in SQW, transforming a single ``incoming'' state into several ``outgoing'' states from a particular vertex. In addition, the discrete-time quantum walk then requires the action of the translation operator, while in the SQW formalism this is already taken care of by switching the description of the vertex $k$ into the second register as
\begin{equation}
	U^{(k)} \ket{j,k} = \sum_{l=1}^{d} U^{(k)}_{j,k} \ket{k,l}.
\end{equation}
We conclude with a correspondence between the states (see Fig.~\ref{fig:sqw}) of a coined discrete-time QW and a scattering QW, given by 
\[
	\ke x\otimes\ke c \quad \leftrightarrow \quad \ke{x\ominus c,x}.
\]
It means the state ``at'' vertex $x$ with a coin in the state $\ket{c}$ is nothing but a SQW state going ``from'' the vertex $x\ominus c$ ``into'' the vertex $x$.
Moreover, when we talk only about the position of the walker at vertex $x$ in coined quantum walks, it means that we are not interested in the coin state (where is the walker entering $x$ from). In the language of scattering quantum walks this translates to talking about the particle entering vertex $x$, i.e.~when the state of the walker is from subspace $\Omega_x$.

\subsection{More on Coins}
\label{sec:coins}

We have seen that for both coined quantum walks and scattering quantum walks, the dynamics of evolution depends on a set of local unitary operators --- position-dependent coins. There are many choices for them, but some turn out to be much more convenient than others for various uses. First, we will discuss the quantum walk properties arising from using different coins (or initial coin states). Second, we will investigate the usual choices of coins utilized in algorithmic applications.

\subsubsection{Two-dimensional Coins}

\begin{figure}[tb]
\begin{center}
\includegraphics[scale=0.93]{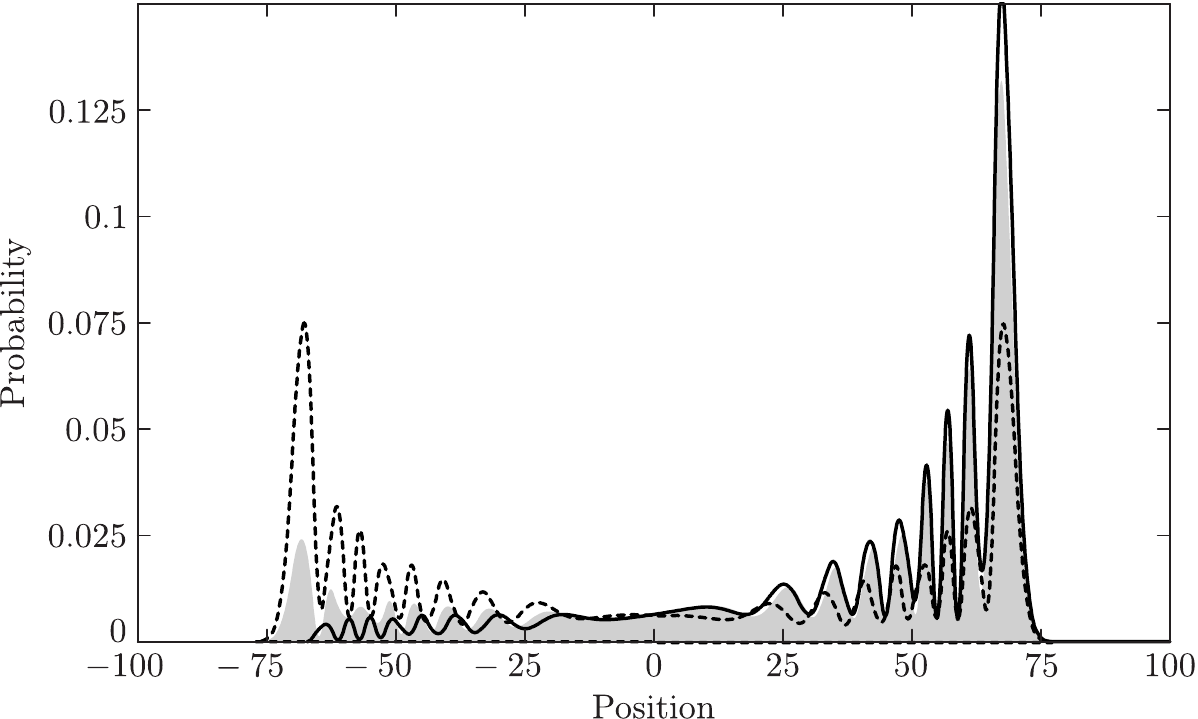}
\end{center}
\vspace{-0.3cm}
\caption{\label{fig:symmetry} Comparison of Hadamard walks after 100 steps, each starting at position 0 and differing only in the initial state of the coin. If the initial coin state is $\ke\uparrow$ (gray shading), the probability distribution is asymmetric. More asymmetry (solid line) is obtained by setting the initial coin state in \eqref{eq:initial_symmetry} as a particular superposition with $\cos\sigma=1$ and $q=\cos^2\pi/8$. On the other hand, symmetrical distributions can be obtained by choosing $\cos\sigma=0$ and $q=1/2$ (dashed line). These are not, however, unique choices, and other are possible.}
\end{figure}

In Sec.~\ref{sec:DTQWdef}, we had the oportunity to notice the need for an additional coin degree of freedom, in order to obtain a non-trivial evolution. We have looked at the quantum walk with the Hadamard coin operator \eqref{eq:hadamard}, resulting in an asymmetrical distribution (see Fig.~\ref{fig-3}). If instead of the Hadamard coin we used the {\em balanced} coin operator 
\begin{equation}
	\op C=\frac{1}{\sqrt{2}}\begin{bmatrix} 1 & \ii\\ \ii & 1\\ \end{bmatrix}, \label{eq:balancedcoin}
\end{equation}
with the coin initial state prepared in the superposition $(\ke\uparrow+\ke\downarrow)/\sqrt{2}$, we would obtain a symmetrical quantum walk.
However, in Sec.~\ref{sec:dispersion} we also saw that the choice of the initial coin state in the Hadamard walk on the line 
determines the final distribution of the walker (see Fig.~\ref{fig:symmetry})
 -- from a right-skewed distribution through a symmetrical distribution to a left-skewed distribution, even though the Hadamard coin itself is \emph{unbiased}.
Under the unitary evolution, the information about the choice of the initial state is thus transferred to the final state (before the measurement).

Is the choice of initial coin state general enough, or do different coin operator choices result in qualitatively different behavior? 
The authors of \cite{TrFlMaKe03} considered this question using similar techniques as in Sec.~\ref{sec:dispersion}, where we switched into the Fourier basis. They found out that the full range of quantum walk behavior (possible with an unbiased coin) on the line can be achieved within the Hadamard walk, combined with choosing the initial coin state.

In these cases the shape of distributions is determined by two factors --- by interference, as is the case for the symmetric Hadamard walk where the initial state is real-valued, or by a combination of probabilities from two mirror-image orthogonal components. Two examples of the latter case are the symmetric Hadamard walk with the initial state \eqref{eq:hadamard_symmetric} and the balanced coin walk \eqref{eq:balancedcoin}.

Although the choice of an {\em unbiased}\footnote{All unbiased coins are equivalent, according to \cite{TrFlMaKe03}. For biased coins, we do not expect to be able to reproduce a symmetric evolution.} coin operator was shown to have only little importance on the line, its impact is significant in other cases. When considering the Hadamard walk on a cycle, the (limiting\footnote{Will be defined in Eq.~(\ref{eq:limiting}).}) distribution depends on the parity of the number of nodes. On the other hand, it can be shown \cite{TrFlMaKe03} that the limiting distribution can be also modified by choosing a different coin and leaving the number of nodes constant.


\subsubsection{General Coins}
Moving away from 1D, we now turn our attention to walks on general graphs with $d$-dimensional vertices, where the coin is described by a $d\times d$ unitary matrix. Not only is the coin space larger, the choice of the coin operator starts to make a difference. For a coined quantum walk on a two-dimensional lattice, a variety of interesting coins were discovered numerically in \cite{TrFlMaKe03}. Each of those (combined with the choice of initial state) affects the characteristics of the walk. The difference between them is mainly in the extent to which the coin can affect them. For further details, see also Ref.~\cite{OmPaShBo06}.

We will now investigate several types of coins commonly used for walks on $d$-regular\footnote{Note that for general $d$-regular graphs which are not symmetric, choosing the coin operator involves taking the directional information into account. For graphs that are not regular, finding a proper coin is even more problematic.} graphs. 
First, a generalization of the Hadamard coin, then a family of unbiased coins related to the Fourier transform, and finally some symmetric coins. 

In two-dimensional coin space, we have looked at the Hadamard coin (which is capable of reproducing all possible behaviors coming from unbiased coins). On $l$-dimensional lattices ($d=2^l$), the Hadamard coin can be generalized to $\op C_{W\!H}=\op H\otimes\op H\otimes\ldots\otimes\op H$. Also called \emph{Walsh-Hadamard operator}, it can be rewritten as
\begin{equation}
\label{eq:WH}
\op C_{W\!H}=\frac{1}{\sqrt{2^l}}\sum_{k=0}^{2^l-1}\sum_{m=0}^{2^l-1}(-1)^{\bar k\odot\bar m} \ke{k}\bra{m},
\end{equation}
where $\bar k\odot\bar m$ is the parity of the bitwise dot product of $l$-bit binary strings representing $k$ and $m$,
\[
\bar k\odot\bar m=\biggl(\sum_{j=0}^{l-1}k_jm_j\biggr)\mod 2.
\]

For some applications, we would like the coin operator to be \emph{unbiased}, i.e. producing equal splitting of the probability of the walker into target vertices. One example of such a coin is the discrete-Fourier-transform (DFT) coin
\[
\op C_{DFT}=\frac{1}{\sqrt{d}}\sum_{\mu=0}^{d-1}\sum_{\nu=0}^{d-1}\exp\frac{2\pi\ii\mu\nu}{d}\ke\nu\br\mu,
\]
which as a matrix has the form
\[
\op C_{DFT}=\frac{1}{\sqrt{d}}\begin{bmatrix}
1 & 1 & 1 & 1 & \ldots\\
1 & \e^{\ii\omega} & \e^{2\ii\omega} & \e^{3\ii\omega} & \ldots\\
1 & \e^{2\ii\omega} & \e^{4\ii\omega} & \e^{6\ii\omega} & \ldots\\
1 & \e^{3\ii\omega} & \e^{6\ii\omega} & \e^{9\ii\omega} & \ldots\\
\vdots & \vdots & \vdots & \vdots & \ddots\\
\end{bmatrix}.
\]
Although unbiased (all elements of the matrix have same magnitude), this coin is \emph{asymmetric} --- different directions of the walker are treated differently, acquiring various phases (all elements of the matrix do not have the same amplitude). Note that for $d=2$ (corresponding to each vertex having two neighbors), the Fourier transform {\em is} the Hadamard transform, so we have $\op C_{DFT}|_{d=2}=\op H$. 

{\em Symmetry} of the coin is also often a desirable property. In general, such coins are written as
\[
\op C=\begin{bmatrix}
-r & t & t & \ldots\\
t & -r & t & \ldots\\
t & t & -r & \ldots\\
\vdots & \vdots & \vdots & \ddots\\
\end{bmatrix}.
\]
The coefficients $t$ and $r$ need to be chosen so that $\op C$ is unitary, obeying the conditions
\begin{eqnarray*}
	\no{r}^2+(d-1)\no{t}^2 & = & 1,\\
	-r^\ast t-t^\ast r+(d-2)\no{t}^2 & = & 0.
\end{eqnarray*}
A specific choice of parameters, biased for all $d\neq 4$, treating the ``return'' direction differently from all others, is 
\begin{equation}
	\label{eq:groverparams}
	t=\frac{2}{d},\qquad r=1-t=1-\frac{2}{d}.
\end{equation}
The coin $\op C_G$ with these coefficients is the asymmetrical coin farthest from the identity \cite{MoRu02}. 
It was used by Grover in \cite{Grover97} in his celebrated quantum algorithm for unstructured search\footnote{Grover's search is the optimal quantum algorithm for the unstructured search problem: given a quantum oracle $R_w = \iii - 2\ket{w}\bra{w}$ acting as $-\iii$ on the target state $\ket{w}$ and as an identity on all other states, the goal is to prepare the marked state $\ket{w}$ with the smallest amount of calls to the oracle $R_w$. For details, see Sec. \ref{sec:grover}.} (see Sec.~\ref{sec:grover}). The role of $C_G$ is a \emph{reflection about the average} state $\ket{s}=\frac{1}{\sqrt{d}}\sum_{j=0}^{d-1}\ke j$. We can see it by observing that
\begin{equation}
	\label{eq:GCbraket}
	\op C_G = -\iii + \frac{2}{d}\sum_{m,n=0}^{d-1}\ke{m}\bra{n} = -\iii + 2\ket{s}\bra{s},
\end{equation}
acting on an input state $\ket{\psi}$ as
$
	\op C_G \ket{\psi} =-\ket{\psi}+2 \braket{s}{\psi} \ket{\psi} 
	= \braket{s}{\psi} \ket{\psi} - (\ket{\psi}-\braket{s}{\psi} \ket{\psi})
$, 
which is the aforementioned reflection of $\ket{\psi}$ about the average state $\ket{s}$.
Besides the algorithm for unstructured search, Grover's coin is used in many other search and walk-based algorithms, such as \cite{ShKeWh03}. For further reference, note that when the Hilbert space dimension is a power of 2  (i.e. $d=2^l$ for some $l$), we can express $C_G$ as
\[
	\op C_G=-\op C_{W\!H}\op R_0\op C_{W\!H}^\dagger=-\id+2\ke s\br s
\]
where $\op R_0=\iii-2\ket{0}\bra{0}$ is the reflection about the state $\ket{0}^{\otimes l}$ and $\op C_{W\!H}$ is the Walsh-Hadamard transformation defined by \eqref{eq:WH}. 
The coin we just defined is the same unitary as used in Sec.~\ref{sec:grover} but you can find its use in many other places as it plays a vital role in (quantum-walk) searches of all kinds.

If we are restricted to lattice of dimension $D$ one obtaines a unique feature --- where other graphs have for each coin state $c$ only one direction, lattices have two. So when in Grover coin Eq.~(\ref{eq:GCbraket}) a large part of the walker was returned to the previous position, on lattices we can modify this coin to move the walker further. When we employ the language of coins, this means, that each coin state is given as $\ke{c,\pm}$ where $c$ determines the direction and the sign determines whether we go ``up'' or ``down''. Grover coin as defined previously would for example change the state $\ke{c,+}$ to
\[
C_G\ke{c,+}=-r\ke{c,-}+t\ke{c,+}+t\sum_{e\neq c}(\ke{e,+}+\ke{e,-})
\]
with $t$ and $r$ being given by Eq.~(\ref{eq:groverparams}) with $d=2D$. The \emph{flip-flop} coin $\op C_{ff}$ that repulses the walker from previous state does not change the direction of the walker to head back, but keeps it the same. That means, that on the state  $\ke{c,+}$ it acts as follows:
\begin{equation}
\label{eq:flip-flop}
\op C_{ff}\ke{c,+}=-r\ke{c,+}+t\ke{c,-}+t\sum_{e\neq c}(\ke{e,+}+\ke{e,-})
\end{equation}
Such coin was used in Ref.~\cite{AmKeRi05} to speed-up the search on lattices. See also Sec.~\ref{sec:discrete_spatial}.


\subsection{Characteristics of Quantum Walks}

The properties of quantum walks given in previous sections suggest that quantum walks could be useful in devising efficient algorithms. From the algorithmic point of view, the properties we have seen so far are quite vague and so to get a more direct feeling about the usefulness of quantum walks, we now turn to some more elaborate properties. We will present them here for discrete-time quantum walks and compare them with their classical analogues introduced in Sec~\ref{sec:RWproperties}. Furthermore, based on Sec.~\ref{sec:sqws}, the readers should be able to translate these concepts also to the language of scattering quantum walks.

\subsubsection{Limiting Distribution and Mixing Time}
\label{sec:discretemixing}

The \emph{mixing time} is a quantity that shall give us qualitatively the same information as the aforementioned Shannon entropy given by \eqref{eq:entropy}, yet taken from a different perspective. In the classical case for connected, non-bipartite graphs, the distribution of random walk always converges (see p.~\pageref{page:convergence} in Sec.~\ref{sec:RWproperties}) to the stationary distribution independent of the initial state. Hence, it is possible to define a {\em mixing time} which tells us the minimal time after which the distribution is $\epsilon$-close to the stationary one as in \eqref{eq:classical mixing}. In the quantum case, such a definition is not straightforward, because in general the $m\rightarrow \infty$ limits of $\op U^m\ke\psinit$ and $p^m(x)$ do not exist. Nevertheless, if we average the distribution over time, in the limit of infinite upper bound on time it does converge to a probability distribution which can be evaluated.

Generalising \eqref{eq:px}, the distribution of the walker position $x$ after $m$ steps, given the initial state is $\ke\psinit$, is given by
\[
p^m(x)=\sum_{c\in\Xi}\no{(\bra{x}\otimes\bra{c}){\op U^m\ke\psinit}}^2.
\]
Formally, then, the \emph{time-averaged distribution} $\bar p^T(x)$ for the walker starting in state $\ke\psinit$ is defined as the average over all distributions up to time $T$
\[
\bar p^T(x)=\frac{1}{T}\sum_{t=0}^{T-1}p^m(x)
\]
The interpretation of this quantity is simple. We start the walk in the state $\ke\psinit$ and let it evolve for time $m$ uniformly chosen from the set $\{0,1,\ldots,T-1\}$. Then the probability of finding the quantum walker at position $x$ is given by $\bar p^T(x)$. The \emph{limiting distribution} (for $T\to\infty$) then is
\begin{equation}
\label{eq:limiting}
\pi(x)\equiv\bar p^\infty(x)=\lim_{T\to\infty}\bar p^T(x).
\end{equation}
For finding the limiting distribution we refer the reader to \cite{AhAmKeVa01}. Here we present only a small portion of their results.

\begin{theorem}
For an initial state $\ke\psinit=\sum_j a_j\ke{\phi_j}$, the limiting distribution is
\begin{equation}
\label{eq:quantum_lim_dist}
\pi(x)=\sum_{i,j\st \lambda_i=\lambda_j}\sum_ca_ia_j^*(\br{x}\otimes\br{c})\ke{\phi_i}\br{\phi_j}(\ke x\otimes\ke c),
\end{equation}
where $c$ is the coin state.
\end{theorem}

If all the eigenvalues are distinct, \eqref{eq:quantum_lim_dist} simplifies to
\[
\pi(x)=\sum_{j}|a_j|^2p_j,
\]
where $p_j=\sum_c|(\br{x}\otimes\br{c})\ke{\phi_j}|^2$ is the probability to measure the initial state in the eigenstate $\ke{\phi_j}$. We may notice that unlike in the classical case, where the limiting distribution did not depend on the initial state, this does not hold anymore in the general quantum case. One of the examples where one still gets a limiting distribution independent of the initial state is for a walk on the Cayley graph of an Abelian group such that all of the eigenvalues are distinct.

\begin{example}
For the Hadamard walk on a cycle with $N$ nodes (vertices), the limiting distribution depends on the parity of the number of nodes $N$. The distribution is uniform only if $N$ is odd \cite{AhAmKeVa01}.
\end{example}

We are ready to define the \emph{mixing time} for a discrete-time quantum walk. It is the smallest time after which the time averaged distribution is $\epsilon$-close to the limiting distribution:
\begin{equation}
\label{eq:quantum mixing}
\mathcal M^\mathrm{q}_\epsilon=\min\{T\st \forall m\geq T,\no{\pi-\bar p^m}_{tvd}\leq\epsilon\},
\end{equation}
where $\no{\ \cdot\ }_{tvd}$ is the total variational distance of distributions $\pi$ and $\bar p^m$ defined by \eqref{eq:TVD}. The quantum mixing time, defined in \eqref{eq:quantum mixing}, tells us the same thing as mixing time $\mathcal M^\mathrm{c}_\epsilon$ in the classical case defined in \eqref{eq:classical mixing}. The difference is that while in the classical case we use the actual distribution to determine $\mathcal M^\mathrm{c}_\epsilon$, in the quantum case we use only the time averaged distribution.

We point out yet another difference between classical and quantum walks which concerns eigenvalues. In the classical case, the difference between the two largest eigenvalues governs the mixing time. In the quantum case where the eigenvalues of the quantum walk unitary step all have amplitude one, we find a different relationship between the elements of the spectrum and the mixing time.

\begin{theorem}
For any initial state $\ke\psinit=\sum_j a_j\ke{\phi_j}$ the total variation distance between the average probability distribution and the limiting distribution satisfies
\[
\no{\pi-\bar p^T}_{tvd}\leq 2\sum_{i,j\st \lambda_i\neq\lambda_j}\frac{|a_i|^2}{T|\lambda_i-\lambda_j|}.
\]
\end{theorem}

\subsubsection{Hitting Time}
\label{sec:discretehitting}

Another important quantity is the \emph{hitting time}. In the classical case, it is the time when we can first observe the particle at a given position \eqref{eq:classical_hitting}. For a quantum walk, the measurement is destructive, not allowing us to meaningfully define the hitting time in the same manner. In \cite{Kempe05}, two ways of defining a hitting time were proposed. One of them is called the \emph{one-shot hitting time} connected with some probability $p$ and time $T$. We say, that a quantum walk has a $(T,p)$ one-shot hitting time between positions $\ke k$ and $\ke j$, if $\no{\br j\op U^T\ke k}^2\geq p$ irrespective\footnote{In the end state, usually any coin state is accepted, while in the initial state the choice of the coin usually respects the topology and symmetry of the graph.} of the coin degree. This quantity tells us that when $p$ is a ``reasonable'' number ($0<p\leq 1$), we need only $\mathrm{poly}(T,p)$ number of steps to reach the vertex $j$ from $k$. In fact, the usual definition \eqref{eq:classical_hitting} is not possible, the an additional parameter $T$ needs to be specified. However, one can use these two values to devise a similar quantity given by \eqref{eq:hitting_analogue}, corresponding to the average number of steps needed when repeating the experiment (each time running it for $T$ steps) until the desired position $j$ is hit.

In the second case, we perform a measurement with two projectors $\Pi_0=\Pi(j)=\ke{j}\bra{j}\otimes\id$ and $\Pi_1=\id-\Pi_0$ after each step, determining whether vertex $j$ is ``hit''. If $\Pi_0$ is measured, the process is stopped, otherwise another step is applied. The action of this operation on a general state $\rho$ can be written as $\Phi_q(\rho)=\Pi_q\op U\rho\op U^\dagger\Pi_q$ for $q=0,1$. This definition allows us to express the probability of reaching the vertex $j$ in the $m$-th step and no sooner, provided the initial state is $\rho_0$, as
\begin{equation}
\label{eq:firsthitprobability}
p^m(j)=\tr[(\Phi_0\Phi_1^{m-1})(\rho_0)].
\end{equation}

\begin{exercise}
The expression \eqref{eq:firsthitprobability} might not be intuitive for everybody. The process of the measurement, instead of using trace-decreasing operations $\Phi_q$, can be also defined as follows. In each step, a projective measurement given by operators $\Pi_0$ and $\Pi_1$ is performed. If we find the walker at position $j$, the process is stopped. Otherwise, the state $\rho$ changes to 
\[
	\rho'=\frac{\Pi_1\op U\rho\op U^\dagger\Pi_1}{\tr[\Pi_1\op U\rho\op U^\dagger]}.
\]
The resulting (normalized) state is then again evolved and a measurement is performed. Show, that this process 
hits the vertex $j$ in the $m$-th step and no sooner with probability given by \eqref{eq:firsthitprobability}.
\end{exercise}

We could readily use the expression for the $m$-th step probability of reaching $j$ \eqref{eq:firsthitprobability}) and plug it into the classical definition \eqref{eq:classical_hitting}. However, the usual choice (in order to correspond to the first definition above) is slightly different. The \emph{concurrent hitting time} for a given probability $p$ is the time $T$ for which the process stops with probability higher than $p$ at a time $m\leq T$. This can be written as
\begin{equation}
\label{eq:qhittingtime}
h_c^\mathrm{q}(j)=\min\left\{T\st \sum_{m=1}^Tp^m(j)\geq p\right\}.
\end{equation}

\begin{example}
Using the quantities defined above, it was shown \cite{Kempe05} that a quantum walk on a hypercube has an exponentially faster hitting time (when traversing from one end to another -- see Fig.~\ref{fig:hypercubefigure}) than the corresponding classical random walk, even though the mixing time might be exponentially large in the number of layers. As was pointed out in \cite{ChildsTreesExp}, however, this does not mean that every classical process is exponentially slower -- there is a classical algorithm capable of traversing the $m$-hypercube in time polynomial in $m$, which is of the same efficiency as that for the quantum walk -- see Exercise \ref{ex:traversing_hypercube} in Sec.~\ref{sec:graphsearching}. For a graph-traversing quantum walk algorithm provably exponentially faster than a classical one, see Sec.~\ref{sec:graphsearching} and \ref{sec:gluedtrees}.
\end{example}

\subsubsection{Absorbing Boundary}
\label{sec:discreteboundary}

We now turn our attention to the Hadamard quantum walk on a line with an {\em absorbing boundary} and compare the results with the classical ones from Sec.~\ref{sec:RWproperties}. In the classical case, we looked at whether a walker starting at position 1 eventually reaches position 0, and found it is so. Following \cite{AmBaNaViWa01}, we now want to compute the probability with which the quantum walker starting in the state $\ke\psinit=\ke 1\otimes\ke\uparrow$ hits an absorbing boundary at position $0$ (after an arbitrary number of steps). Comparing with \eqref{eq:qhittingtime} for the concurrent hitting time, we can also say that we are looking for the smallest $p$ (which we denote $p_{10}$) for which the hitting time is infinite\footnote{This means there is no number of steps that would guarantee the walker has reached position 0 with probability greater than $p_{10}$}. 

The absorbing boundary is a repeated projective measurement with $\Pi_0=\ke{0}\bra{0}\otimes\id$ and $\Pi_1=\id-\Pi_0$, telling us whether the walker reached 0 in a given step or not. 
To find the overall probability of this eventually happening, we will not employ the Fourier transform technique we used before, because we now only look at a semi-infinite line comprised of non-negative positions for the walker. Instead, we shall use the technique of counting paths.

\begin{exercise}
Show that this type of measurement process does not allow the walker to tunnel through the position $0$. i.e.~if she starts at positive position that she cannot pass to negative positions.
\end{exercise}

Let us start with a short analysis of the problem. The walker starts at position $1$ and we want to know, what the amplitude is for him to be at position $0$ after $m$ steps -- let us denote this amplitude $a_{10}(m)$. It can be written as the sum of amplitudes corresponding to all possible classical paths (not going through 0). Each classical path is described by an $m$-tuple $(q_1,q_2,\ldots,q_m)$ with $q_j\in\{\uparrow,\downarrow\}$ standing for the direction the walker went in the $j$-th step. The amplitude for each path is $\pm 2^{-m/2}$, as the Hadamard coin \eqref{eq:hadamard} has only the coefficients $\pm1/\sqrt{2}$. We notice that the coefficient can be negative only in the case we go left in two consecutive steps (let us call this event a \emph{doublet}). In each such case, the amplitude acquires a minus sign. If the path has even number of doublets, the final amplitude for the path is positive and when the path has odd number of doublets, the final amplitude for the path is negative.

We denote the set of all paths of length $m$ with even number of doublets as $A_m^+$ and the set of all paths of length $m$ with odd number of doublets as $A_m^-$. The amplitude $a_{10}(m)$ can now be written as
\[
a_{10}(m)=\sum_{p\in A_m^+}\frac{1}{2^{m/2}}-\sum_{p\in A_m^-}\frac{1}{2^{m/2}}=\frac{1}{2^{m/2}}(|A_m^+|-|A_m^-|),
\]
where $|A|$ denotes the cardinality of set $A$. The probability for the walker to hit the boundary at all is now expressed as
\begin{eqnarray}
p_{10}=\sum_{m=1}^\infty|a_{10}(m)|^2=\sum_{m=1}^\infty\frac{1}{2^m}(|A_m^+|-|A_m^-|)^2. \label{boundaryQ}
\end{eqnarray}
In order to find the coefficients $|A_m^+|-|A_m^-|$, we construct the so called generating function\footnote{
	This is a different concept from the generating functions for the moments of distributions. 
	Although this topic is very interesting from the mathematical point of view, for our purposes we only 
	focus on a few points. For any sequence of numbers $\{a_j\}_{j=0}^\infty$ we can construct the power series at $x=0$,
	\[ g(x)=\sum_{j=0}^\infty a_jx^j. \]
	All we will need is that when we have two generating functions $f(x)$ and $g(x)$, and know 
	the coefficients $\{a_j\}_{j=0}^\infty$ of $g(x)$, then if we can find a functional relation between $f(x)$ and $g(x)$, 
	we can also find a relation between their coefficients $\{a_j\}_{j=0}^\infty$ and $\{b_j\}_{j=0}^\infty$.
} 
for these\footnote{
	In this Section we will always consider coefficients of the type $|A_m^+|-|A_m^-|$ for various conditions, 
	and thus we will use the notion of a generating function in a slightly abusive form just by saying 
	it corresponds to some process instead of always explaining that it corresponds to the coefficients 
	of the type $|A_m^+|-|A_m^-|$ for a given process.
} 
coefficients,
\[ 
	f(x)=\sum_{m=1}^\infty(|A_m^+|-|A_m^-|)x^m. 
\]
If the walker were at position 2 and then got back to 1 by some path (not going to 0), the generating function would be the same. Now if the walker is at position $1$, then moves right to position $2$ and then back to $1$ by some path, we find that the generating function for this process is $xf(x)$. By joining $k$ such paths together we can obtain the generating function for all the paths that get to position $1$ after leaving it exactly $k$-times, obtaining $[xf(x)]^k$. 

There are two ways to get from position $1$ to position $0$. First, we could go left one step. Second, we could leave 1, move around and return to 1, all this $k$-times, and then finally take a step left to 0.When we enter $1$ for the $k$-th time and then go left once more, the amplitude acquires an additional minus sign. Overall, we can write this in the following way,
\[
f(x)=x-x\sum_{k=1}^\infty[xf(x)]^k=x-\frac{x^2f(x)}{1-xf(x)}.
\]
Solving this quadratic equation for $f(x)$ yields
\[
f(x)=\frac{1+2x^2-\sqrt{1+4x^4}}{2x}.
\]
By a simple comparison with \eqref{eq:catalan_genfun}, we find that the generating function $f(x)$ is connected with the generating function $c(x)$ for the Catalan numbers (see Appendix \ref{sec:catalan_numbers}) as
\[
f(x)=x-x^3c\left(-x^4\right)=x+\sum_{k=0}^\infty (-1)^{k+1}C_kx^{4k+3},
\]
where $C_k$ are the Catalan numbers,
\[
C_k=\frac{1}{1+k}\binom{2k}{k}.
\]
In other words, we have
\[
|A_m^+|-|A_m^-|=\begin{cases}
1\text{ \quad for $m=1$,}\\
(-1)^{k+1}C_k\text{ \quad for $m=4k+3$,}\\
0\text{ \quad otherwise.}
\end{cases}
\]
This results in the expression for the probability of eventually hitting the boundary \eqref{boundaryQ}:
\[
p_{10}=\frac{1}{2}+\frac{1}{8}\sum_{k=0}^\infty \frac{C_k^2}{2^{4k}}=\frac{1}{2}+\frac{1}{8}\left(\frac{16}{\pi}-4\right)=\frac{2}{\pi},
\]
using \eqref{eq:catalan_equality} and \eqref{eq:catalan_limit}, obtained for example by employing Stirling's formula.

We found that $p_{10}=2/\pi$ which is different from the random walk case for which the probability is 1 (see p.~\pageref{sec:classicalabsorbingboundary}), meaning the classical random walker will eventually come to the position 0. We see again that interference of amplitudes instead of adding up probabilities plays a crucial role in quantum theory, leading to a different behavior of systems when compared to the classical case.

\begin{exercise}
Show that
\begin{equation}
\label{eq:catalan_equality}
\sum_{k=0}^M\frac{C_k^2}{2^{4k}}=(16M^3+36M^2+24M+5)\frac{C_M^2}{2^{4M}}-4,
\end{equation}
using induction and \eqref{eq:catalan_connection} from Appendix \ref{sec:catalan_numbers}.
\end{exercise}


\subsubsection{Quantum-to-classical Transition and Decoherence}
\label{sec:decoherence}

The definition of discrete-time quantum walks is based mostly on an analogy with classical random walks. So fare, we have done no real attempt to make a canonical quantization of random walks -- this will be done later in Chapter~\ref{sec:qmc}. However, we might think about the opposite processes. Consider now a quantum walk whose evolution is not unitary anymore, but some amount of decoherence comes into play.
Imagine the state unitarily evolves for one step, and then with some small probability $(1-p)$ a projective measurement with projectors $\Pi_x$ gets performed, transforming the state $\rho$ into
\begin{equation}
\label{eq:non-unitary}
\mathcal E(\rho)=p\op U\rho\op U^\dagger+(1-p)\sum_{x}\Pi_x\op U\rho\op U^\dagger\Pi_x.
\end{equation}
The projectors $\Pi_x$ can be either tied to different coin states ($\Pi_\uparrow$, $\Pi_\downarrow$) or work in the position basis of the (coined) walker (when $\Pi_x=\ke x\br x$), or even in both of these (for a SQW, $x$ would correspond to different directed edges). 
This prescription thus states on top of performing a quantum walk, occasionally we measure either the state of the coin, the position of the walker, or the edge state for a SQW, depending on the choice of projectors $\Pi_x$.

Let us look on the quantum walk on a line in the SQW model (see Sec.~\ref{sec:sqws}), concentrating on the Shannon entropy of the walks given by \eqref{eq:entropy}. Consider $\all$ to be the set of all edge-states and $(1-p)$ the probability of performing a measurement after a step of the walk. 
It is clear, that $p=1$ gives an unperturbed unitary evolution. On the other hand, for $p=0$ the evolution is devoid of interference, as it results in a diagonal density matrix. In each step, no matter what the state of the coin might be, the walker moves to the right or left with equal probabilities. The measurements performed after every step thus give us a classically describable evolution, as any superposition that occurs in the coin evolution is broken by the measurement. Such a walk is governed by the same rules as the drunkard's walk. Therefore, in the extreme case of maximal decoherence, a quantum walk becomes classical. This is a conclusion that can be simply extrapolated into any type of graph and a quantum walk on it, with the high-decoherence limit resulting in a classical walk with a transition rule derivable from the corresponding coin operator.

\begin{figure}[t!]
\begin{center}
\vspace{-0.2cm}
\includegraphics[scale=0.83]{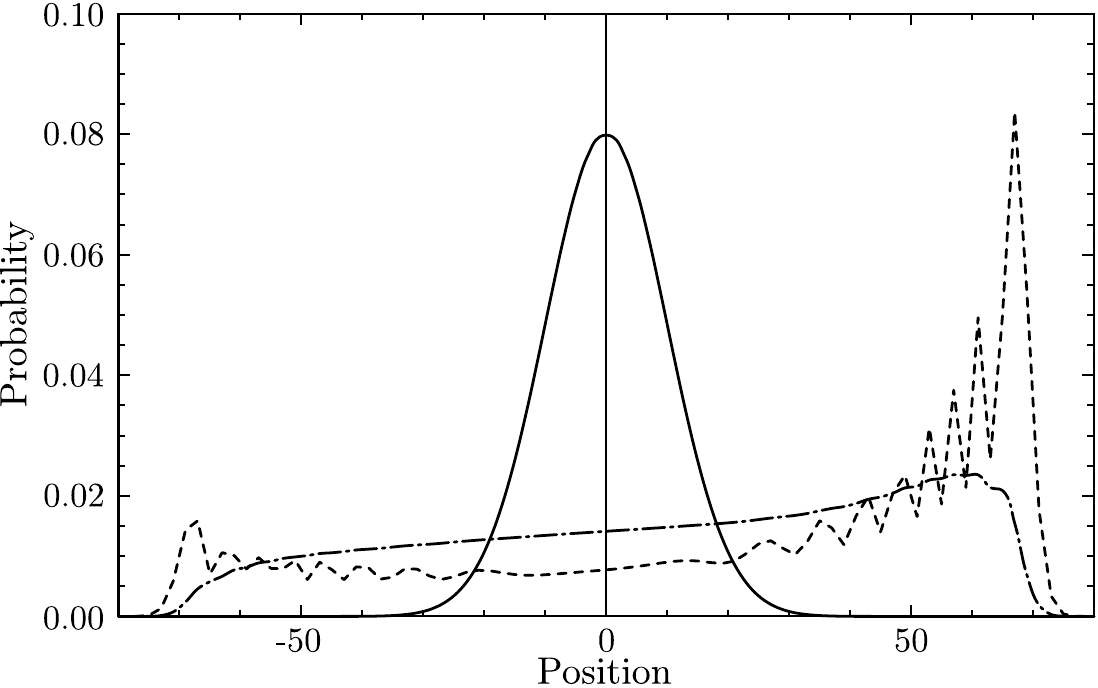}
\end{center}
\vspace{-0.4cm}
\caption{\label{fig:evolutions} Probability distributions for quantum walks on a line with additional decoherence, after 100 steps of evolution. A pure quantum walk with $p=1.0$ (dashed line), a pure classical random walk $p=0.0$ (solid line) and a non-unitary walk with $p=0.96$ leading to a distribution with a maximal Shannon entropy.}
\end{figure}
\begin{figure}[!thp]
\begin{center}
\includegraphics[scale=0.83]{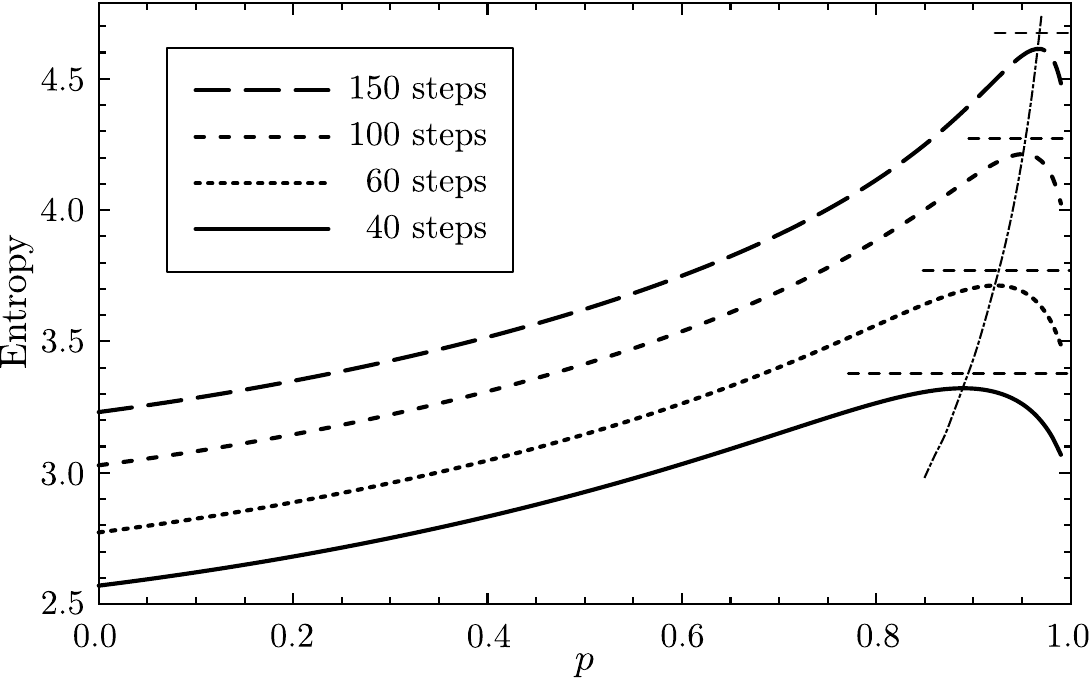}
\end{center}
\vspace{-0.45cm}
\caption{\label{fig:non-unitary_shannon} Shannon entropy of a probability distribution coming from a quantum walk, depending on the unitarity rate $p$, for various numbers of steps taken. Dashed horizontal lines correspond to the maximal achievable entropy of a (uniform) distribution restricted to the position interval $[-x/\sqrt{2};x/\sqrt{2}]$ after taking $x$ steps. The thin dot-dashed line denotes the maximal achievable entropy for a given $p$. Obviously, as a quantum walk without decoherence does not revert to classical side, the maximum of the entropy for $p=1$ goes to infinity.
}
\end{figure}

The situation for general (intermediate) values of $p$ now deserves our attention and is reviewed in \cite{Kendon07}. In short, from Fig.~\ref{fig:evolutions} we see that by varying $p$, we smoothly get from a two-peaked pure quantum-walk distribution to the normal distribution. However, as the computation of Shannon entropy in Fig.~\ref{fig:non-unitary_shannon} suggests, an intermediate point in $p$ gives us an almost uniform distribution, something far from both the purely quantum and purely classical behavior. The point in which Shannon entropy is maximal depends not only on the value of $p$, but also on the number of steps taken. Before reaching this point, the distribution spreads ballistically, while at larger times, the spreading slows down as the distribution approaches a normal distribution and the spreading becomes diffusive. Thus on long timescales, any amount of decoherence leads to the destruction of quantum interference effects. Note that in the limit $p\to 1$, the point of maximal entropy diverges in the number of steps and so the evolution remains (as is expected) ballistic all the time.

We encourage the reader to consult \cite{Kendon07} for a deep review, considering also different types of measurement (showing that it makes a difference whether we measure the coin, position or both), different types of graphs (such as a hypercube or a cycle) and different types of decoherence as well. We conclude that although discrete time quantum walks are not quantized versions of classical random walks, they approach the behavior of classical random walks when undergoing decoherence at an increasing rate. One should keep in mind one additional thing. Even if the decoherence is maximal in both the position and the coin registers ($p=0$ and we measure in every step), the usual unbiased random walk given by \eqref{eq:unbiasedRW} is recovered  only in special cases. The walker always remembers the direction it came from and so it treats all directions equally (as a random walk does) only if the coin is unbiased. Otherwise, the walk obtained by decoherence corresponds to a random walk with memory \cite{KoBuHi06}.

\begin{exercise}
We could also think about what happens to quantum walks under a different type of decoherence. In \cite{KoBuHi06}, every step of a SQW evolution is affected by a random phase-shift on every edge-state. Thus, after evolving the state $\rho$ by the step unitary $\op U$, another unitary $\op\Phi(\hat{\phi})=\mathrm{diag}\,\e^{\ii\hat{\phi}}$ is applied (with a random vector of phase-shifts $\hat{\phi}$). The ``random'' effectively means that such a mapping is described by
\[
\mathcal E(\rho)=\int\pi(\hat{\phi})\op\Phi(\hat{\phi})\op U\rho\op U^\dagger[\op\Phi(\hat{\phi})]^\dagger\, \dd\hat{\phi}\,,
\]
where $\pi$ is the distribution of phases, which we assume to be symmetric and $\hat{\phi}$-components uncorrelated, meaning that the choices of components are independent, i.e. $\pi(\hat{\phi})=\prod_{x\in\all}\pi_0(\phi_x)$. Under this assumption, show that the state after one step of evolution with random phase flips is again described by \eqref{eq:non-unitary} with $p$ dependent on the parameters of noise, still being from interval $[0;1]$, with $p=1$ standing for no noise and $p=0$ for maximal noise described by the choice of phase-shifts sampled uniformly from whole interval $[-\pi;\pi]$.
\end{exercise}

\subsection{Summary}

In this section we have presented the simplest and most na\"\i ve transition from classical random walks to processes we call quantum walks. The definition of these discrete time quantum walks was made on analogy with their classical counterparts rather than by quantization. Nevertheless, there is a closer correspondence between them, as the classical random walk can be obtained from the quantum walk by adding decoherence -- classical random walks are indeed the classical limits of discrete time coined quantum walks (possibly with memory).

On the other hand, quantum walks display a variety of differences from classical random walk, as their spreading is ballistic and not diffusive as is the case for classical random walks. This difference is the result of interference effects and leads to faster mixing times. This in turn can be used to obtain efficient algorithms. The most prominent algorithmic application of discrete-time quantum walks is for searches on graphs, where a quadratic speedup can be obtained if the ``database'' we search is unstructured (Grover's search on a complete graph), but also in many other cases. Although their immediate application doesn't seem apparent, they can in turn be used as subroutines in more complicated tasks such as $k$-subset finding, where they provide a provable speedup over the best possible classical algorithm.

The aim of this section was also to showcase the methods for analysis useful not only in this area -- Fourier transforms (applicable on all Cayley graphs) and counting paths. We looked at the most common graphs: a line, a cycle, a hypercube. They were used to get the reader acquainted with the most usual properties studied in these types of problems -- hitting time, mixing time, reaching an absorbing boundary or finding the (limiting) distribution. We then compared the obtained quantities to their classical counterparts for random walks. The only problem arising when defining these quantities is the different approach to measurement -- whereas in the classical case a measurement does not change the state of the system, in the quantum case measurement destroys the state and leads to the collapse of the wave function. Furthermore, occasional measurement can be used to model decoherence that destroys the superpositions, the key ingredient of quantum walks.
\setcounter{equation}{0} \setcounter{figure}{0} \setcounter{table}{0}\newpage
\section{Quantum Walks and Searches}

In the previous Section we defined discrete-time quantum walks and explored some of their basic properties. Now it is time to show their applications in computational problems. As the title of the section suggests, a large part of the research in this area concerns applying quantum walks for searching. The mixing properties of quantum walks are behind this --- fast mixing leads to a natural suggestion that these walks could perform searches faster than classical algorithms. This approach has led to several important quantum walk search algorithms with provable speedups over their classical counterparts.

The first quantum walk algorithm performing a search for a target vertex was given in Ref.~ \cite{ShKeWh03} for a hypercube. Soon after that, several new algorithms for searching on various graphs emerged, utilizing the symmetric properties of the underlying graphs to analytically describe the unitary evolutions. The common theme resembles the algorithm for unstructured database search by Grover \cite{Grover97}. In fact, the problems of searches on graphs go beyond the framework of amplitude amplification method \cite{BrHoMoTa00} which solves the problem by showing that its evolution is restricted to a two-dimensional subspace of the Hilbert space. On one hand one can use results of Ref.~\cite{AmKeRi05} that approximate the evolution as a restriction to a two-dimensional Hilbert space or one can find higher-dimensional restrictions and describe the evolution within this subspace. This method is described in Sec.~\ref{sec:symmetry} and is applicable for highly symmetrical problems.

Research in the area of quantum walk searches progressed also with Ambainis' algorithm for element distinctness, utilizing quantum walks on a graph with a  more involved structure \cite{Ambainis07} (the vertices of this graph are sets of vertices of the original graph). This specific algorithm was later generalized to the problem of subset (e.g. triangle) finding \cite{ChEi05,MaSaSz05}. Although a general lower bound is still not known in the general case, this approach is the best known up to date. Moreover, it can be modified for the task of solving several other problems such as verification of matrix products \cite{BuSp06} or testing the commutativity of a black-box group \cite{MaNa07}.

\subsection{Grover Search}
\label{sec:grover}

We begin this Section with Grover's search \cite{Grover97}, the algorithm whose power will be used throughout the whole section for comparison and as a point of reference.

In 1997, Lov~K.~Grover introduced \cite{Grover97} a quantum search algorithm for a marked (target) element from an unstructured database. The algorithm is provably quadratically faster than any classical algorithm for the given problem, also in the generalized case soon described in \cite{BoBrHoTa96}. There exists also a fixed point ``version'' of Grover's search --- see Ref.~\cite{GroverFixPoint,GroverFixPoint2} and Appendix~\ref{sec:fixedpoint}. This quadratic speedup comes in the number of oracle calls, where one oracle call gives us only the information whether the questioned-about element is the target one or not. The oracle model is key for Grover's search which serves as a base for the comparison of different algorithms. Therefore, we will now provide a short introduction to oracles and after that we shall analyze Grover's search.

\subsubsection{Oracles and Searches}
\label{sec:oracles_and_searches}

In the previous Sections, we had a few opportunities to notice oracles but left them unexplained. Here we provide some basic information on this topic that concerns Grover's search and further applications.

Within a classical computation, the use of the so-called {\em oracles} (also called black-boxes), is quite abundant. The oracle is a device that performs a specialized task and the number of its calls is used for comparison of efficiencies of algorithms (their query complexity). The algorithmic efficiency is then expressed in the number of calls of such oracle. In the quantum case we can follow a similar path, where we compare different algorithms with respect to the number of oracle calls, where the oracle can now be quantum (i.e. a unitary operation such as a reflection about a vector characteristic for the black box but unknown to us).

A very common task in both classical and quantum setting is a search for a marked/desired element. Here the oracle serves as a device telling us whether the element we are examining is the one we are searching for (and nothing more). This task may be justified in the following way. If we consider a database (structured or non-structured), there are certain kinds of queries for which the database may be viewed as unstructured. For example a phone book (yellow pages) is a database designed so that if you know a name, it is easy to find the corresponding phone number. However, having only the a phone number, it is a tedious task to find the corresponding name, since the phone numbers in the phone book are positioned practically in a random fashion.

Following the previous discussion, we may also construct an oracle that tells us an answer for a query -- given an index of an element from a database, the oracle tells us whether the element accommodates the given query or not. Sticking to the yellow pages analogy, the oracle tells us whether the $k$-th record (name) corresponds to the number we are looking for. The analogy is suitable also to show that such oracle doesn't have to be something magical -- it is a relatively easy task to find the $k$-th name in a phone book -- a well structured database from this point of view.

Let us now define an oracle in a more abstract fashion, denoting the set of all possible choices of queried elements as $\all$ and the subset of this set that accommodates a given query as $\spec\subset\all$. The oracle\footnote{In general, oracle may be any function. For our purposes it will be enough to restrict ourselves to boolean functions.} then is a function $f_\spec(x)$ defined for all $x\in\all$ such that
\begin{equation}
\label{eq:oracle_function}
f_\spec(x)=\begin{cases}
1\text{, \, if $x\in\spec$,}\\
0\text{, \, if $x\in\all\setminus\spec$.}
\end{cases}
\end{equation}
The oracle function $f_\spec$ is determined by the set $\spec$ (\emph{target} or \emph{special set}) and it is not necessary to state this fact in the function definition, yet it will be helpful on occasions. For practical purposes, if $\spec$ contains only one element $k$, we will write it as $f_\spec(x)\equiv f_k(x)$.

In a quantum world, we can define a unitary quantum oracle based upon a classical one $f_\spec$, following the constructions in \cite{BoBrHoTa96}. Given the function $f_\spec$, we can consider a unitary controlled operator $\Ctrl\op V_\spec$ defined for a bipartite (two-register) system, 
\begin{equation}
\label{eq:quantum_oracle}
\Ctrl\op V_\spec: \ke{x}\otimes\ke{m} \mapsto \ke{x}\otimes\ke{m\oplus f_\spec(x)}.
\end{equation}
We can think of the first register as providing a query (asking about the element $x$), while the second subsystem is a qubit to whose value we add (in binary) the result of the oracle function $f_\spec$ evaluated on a state of the first subsystem. This second register is needed for reversibility, as we would like the quantum oracle to be unitary. Such a quantum oracle, no matter what its actual implementation is, is widely used to demonstrate the difference in query complexity between classical and quantum algorithms. 

Observe that the quantum oracle we just defined is a unitary operation, and thus can also act on superpositions. However, if we query it classically (i.e.~if input states are of the form $\ke x\otimes\ke 0$ with $x\in\all$) and measure the result on the second register afterwards, we always get the result of $f_\spec(x)$. Thus, in this classical-like usage, this quantum oracle does not give us any extra power for finding an element from $\spec$. On the other hand, as we will show soon, quantum properties such as interference help us enhance the efficiency of search algorithms -- the only way to achieve a speedup is to use a quantum oracle with quantum inputs.

Finally, there is one interesting feature of the quantum oracle from \eqref{eq:quantum_oracle}. When we intialize the second register to
\[
\ke m=\frac{1}{\sqrt{2}}(\ke 0-\ke 1)\equiv\ke -,
\]
we find that
\[
\Ctrl\op V_\spec\ke{x}\otimes\ke - = (-1)^{f_\spec(x)}\ke{x}\otimes\ke -.
\]
Surprisingly, when the second register is intialized as $\ket{-}$, after the application of $\Ctrl\op V_\spec$ it does not change and thus is not needed in further mathematical considerations. Therefore, we may just as well use
\[
\op R_\spec: \ke x\mapsto (-1)^{f_\spec(x)}\ke x
\]
as our quantum oracle operation.
The operator $\op R_\spec$ can be further written as
\begin{equation}
\label{eq:grover_oracle}
\op R_\spec=\iii-2\sum_{j\in\spec}\ke j\br j,
\end{equation}
flipping the sign of the basis states from $\spec$ and preserving the rest -- it is a \emph{conditional phase-flip.}

As the searches we will study are restricted to graph structures it is also necessary for similar tasks to consider ``graph'' oracles that will be later used. Besides mentioned oracle that marks some vertices we may consider also oracles that give us information about graph structure. We will find use of such oracles later as well. The walk on graph can be viewd also as a query problem, where walker, being on position $j$ queries the \emph{neighbor} oracle to tell him possible neighbors of $j$. Another type of oracle might be called \emph{edge} oracle as it gives just information whether some edge $jk$ exists --- the information is provided in a similar way as in the case of vertex-marking oracle from Eg.~(\ref{eq:quantum_oracle}) with the difference of taking two vertices on input.

\subsubsection{Grover's Algorithm}
\label{sec:grover_search}

\begin{figure}
\begin{center}
\includegraphics{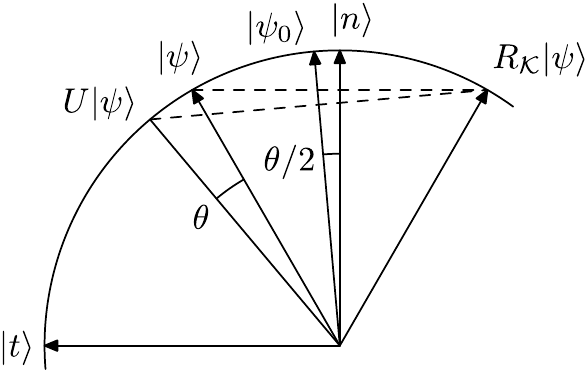}
\end{center}
\caption{\label{fig:grover_evol} The Grover algorithm performs a rotation of the state vector $\ke\psi$ by angle $\theta$ by first reflecting the vector around $\ke n$ and then around $\ke\psinit$.}
\end{figure}

Let us now present the optimal quantum algorithm for unstructured search. Having sets $\all$ and $\spec$ as defined in the previous section, with $N=\no{\all}$ and $k=\no{\spec}$, our goal is to find a state from the marked set $\spec$. We initialize our system in the equal superposition of all states of the computational basis
\begin{equation}
\label{eq:equal_superposition}
\ket{\psi_0}=\ke s=\frac{1}{\sqrt{N}}\sum_{j\in\all}\ke j.
\end{equation}
We define the unitary step operator
\begin{equation}
\label{eq:Grover_unitary}
\op U=\op C_G\op R_\spec,
\end{equation}
where
\begin{equation}
\label{eq:inversion}
\op C_G=-\iii+2\ke s\bra s=-\op R_{s}
\end{equation}
is an \emph{inversion about the average} introduced\footnote{There, we also show how to implement it using the Walsh-Hadamard operation and a reflection about the all-zero state $\ket{0\cdots 0}$.} in Sec.~\ref{sec:coins}, and the operator\footnote{Note, that $\op R_{\{k\}}\equiv \op R_k$, which also justifies Eq.~(\ref{eq:inversion}).} $\op R_\spec$ is defined in \ref{eq:grover_oracle}.
Defining the states
\[
\ke t=\frac{1}{\sqrt{k}}\sum_{j\in\spec}\ke j,\qquad \ke n=\frac{1}{\sqrt{N-k}} \sum_{j\in\all\setminus\spec}\ke j,
\]
simplifies the analysis greatly, as the application of $\op U$ \eqref{eq:Grover_unitary} on a state from the subspace spanned by these two vectors leaves the state in this subspace. In particular, it performs a rotation (see also Fig.~\ref{fig:grover_evol})
\begin{eqnarray*}
\op U\ke n &=& \phantom{-}\cos\theta\ke n+\sin\theta\ke t,\\
\op U\ke t &=& -\sin\theta\ke n+\cos\theta\ke t,
\end{eqnarray*}
where
\begin{equation}
\label{eq:theta_grover}
\cos\theta=\frac{N-2k}{N}.
\end{equation}
Moreover, the uniform superposition initial state $\ket{\psi_0}$ from \eqref{eq:equal_superposition} is also a superposition of the $\ket{t}$ and $\ket{n}$ vectors:
\[
\ket{\psi_0} = \ke s=\sin\frac{\theta}{2}\ke t+\cos\frac{\theta}{2}\ke n.
\]
As a consequence, repeated application of $U$ rotates the initial state towards the state $\ket{t}$ as  
\begin{equation}
\label{eq:grover_result}
\ke{\psi_m}\equiv\op U^m\ke\psinit=\sin(2m+1)\frac{\theta}{2}\ke t+\cos(2m+1)\frac{\theta}{2}\ke n.
\end{equation}
Thus, if we pick the number of applications of $U$ to satisfy $(2\tilde m+1)\theta/2=\pi/2$, we get $\op U^{\tilde m}\ke\psinit=\ke t$ and a subsequent measurement of the system gives us an element from $\spec$. For $k\ll N$, this happens for 
\begin{equation}
\label{eq:grover_steps}
\tilde m\simeq\frac{\pi}{4}\sqrt{\frac{N}{k}}.
\end{equation}
This number of steps is quadratically smaller than the corresponding number of steps we would need when using any  classical algorithm where it is $O(N/k)$.

\begin{figure}
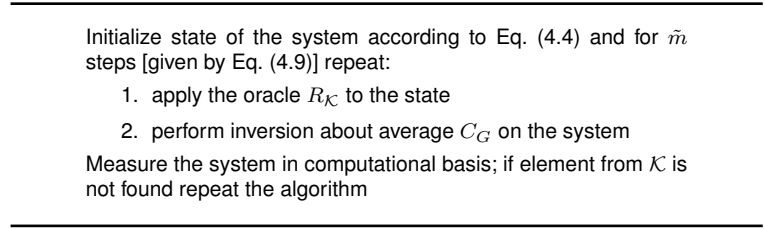

\begin{algorithm}
Initialize state of the system according to Eq.~(\ref{eq:equal_superposition}) and for $\tilde m$ steps [given by Eq.~(\ref{eq:grover_steps})] repeat:
\begin{enumerate}
\item apply the oracle $\op R_\spec$ to the state
\item perform inversion about average $\op C_G$ on the system
\end{enumerate}
Measure the system in computational basis; if element from $\spec$ is not found repeat the algorithm
\end{algorithm}
\caption{\label{fig:grover_alg}Grover algorithm which searches for oracle-selected elements.}
\end{figure}

Note that it is important for this algorithm (see Fig.~\ref{fig:grover_alg}) to start in the equal superposition and to know the number of searched elements $k$ in advance. 
Any deviation from the initial state \eqref{eq:equal_superposition} decreases the probability of successfully identifying the target elements. As a rule of thumb we can say that in order to obtain high enough probabilities, one has to start in a highly superposed initial state (in computational basis). On the other hand, there exist ways to modify the algorithm if the number of searched-for elements is unknown, e.g. by running the algorithm with expecting large $k$ and progressively dividing the expected $k$ after each unsuccessful run.

The Grover algorithm is a special instance of a more general class of amplitude amplification method introduced in Ref.~\cite{BrHoMoTa00}. The algorithm presented there amplifies amplitude $a$ of target state obtainable by quantum algorithm $\mathcal A$. In the classical approach one would measure the system after applying algorithm $\mathcal A$ and obtain target state with probability $|a|^2$. So to find the target state one would have to repeat algorithm $\mathcal A$ for roughly $|a|^{-2}$ times. Approach of the amplitude amplification method repeats the algorithm $\mathcal A$ (adding some intermediate steps) for only $O(|a|^{-1})$ times with measurement process only at the very end to obtain the target with constant probability close to $1$.

\begin{exercise}
A nice conclusion drawn from the previous result, often exploited in experiments verifying the functionality of Grover's search, is obtained by setting $k=\frac{N}{4}$ (meaning that one fourth of the elements from the set $\all$ are target elements). Show that in this case a single application of the unitary $U$ \eqref{eq:Grover_unitary} is needed to reach a target state, as opposed to two oracle queries one needs to make (on average) in the classical case. In particular, this means we can find a single marked element in a list of four by a single application of $U$ (meaning using only a single call to the oracle).
\end{exercise}

\begin{example}
Considering different types of oracles defined at the end of Sec.~\ref{sec:oracles_and_searches} one can notice that the edge oracle can be used together with Grover's search to construct neighbor oracle. If the degree of the graph is much smaller than the number of vertices the time expense of such construction is $O(\sqrt{N})$ as Grover's search has to look through $N$ vertices of the graph to determine which of them are neighbors of given vertex.
\end{example}

\subsection{Searches on Graphs}

One of the first attempts to perform a search with quantum walks used a discrete time quantum walk on a hypercube, with a single type of coin (the Grover coin) for all vertices except the marked one, which had it perturbed to minus identity (effectively acting as minus the Grover coin due to the symmetry of the graph). This work by Shenvi, Kempe and Whaley \cite{ShKeWh03} showed the possibility of quantum speedups in quantum walks on graphs, compared to classical search. The analysis of the hypercube search problem is similar to the one we used for Grover's search \cite{Grover97} in the previous Section, yet the dimension of the subspace into which the evolution is restricted to is larger than the two dimensions of amplitude amplification method of Ref.~\cite{AmKeRi05}. In this way, the approach was unique at the time.

Sarching for marked elements on a graph is the focus of many other works. Ambainis, Kempe and Rivosh \cite{AmKeRi05} , also considered a search on a complete graph with loops, where they found another incarnation of Grover's search. More importantly, they studied searches on regular lattices and found an algorithm with a quadratic speedup over classical search.
However, before that result, in the field of continuous quantum walks (which will be introduced in Sec. 6) Childs and Goldstone in Ref. [44] investigated searching on finite-dimensional lattices, with the special case of a complete graph with loops. The dialogue between continuous and discrete-time quantum walks inspired the quadratically faster (than classical) continuous-time quantum walk search algorithm on regular lattices by Childs and Goldstone \cite{ChildsDirac}.

We will now look at scattering quantum walks, following the definitions in Sec.~\ref{sec:sqws}. We will combine this formalism with searches on graphs, looking at the general algorithm given in Fig.~\ref{fig:graph_search} and compare it to Grover's search (see Fig.~\ref{fig:grover_alg}). In particular we will follow \cite{ReHiFeBu09}, studying scattering quantum walk search on the (symmetric) complete graph. We will call the vertices corresponding to the desired elements from set $\spec$ {\em targets}, while the rest of the (unmarked) vertices will be called {\em normal}. Our goal again is to to find at least one element from set $\spec$ by identifying some target vertex.

\begin{figure}
\begin{algorithm}
\begin{enumerate}
\item Initialize state of the system according to Eq.~(\ref{eq:graph_superposition})
\item for $\tilde m$ steps [given by Eq.~(\ref{eq:graph_optimal})] apply $\op U$
\item measure the position (edge) of the walker; if the resulting edge does not have any target end vertex from $\spec$, restart the algorithm
\end{enumerate}
\end{algorithm}
\caption{\label{fig:graph_search}Searches on graphs usually start in highly superposed initial state in the position basis. After $\tilde m$ iteration the walker is with high probability located at an edge connected to (at least) one target vertex. This algorithm is given for particular case of a search on the complete graph.}
\end{figure}

Considering a vertex $l$, let $\Gamma (l)$ be the set of vertices connected to $l$ by an edge, and if $k\in \Gamma (l)$, let $\Gamma (l;k)$ be the set of vertices connected to $l$ by an edge but excluding $k$. The local unbiased unitary evolutions (coins) corresponding to both the normal and target vertices act as follows
\begin{subequations}
\begin{equation}
\op U^{(l)}\ke{k,l}=-r^{(l)}\ke{l,k}+t^{(l)}\sum_{m\in \Gamma (l;k)}\ke{l,m},
\label{eq:unitary}
\end{equation}
where $r^{(l)}$ and $t^{(l)}$ are reflection and transmission coefficients to be chosen in such way that $\op U^{(l)}$ is unitary. For normal (unmarked) vertices, our choice is the Grover coin \eqref{eq:groverparams} with
\begin{equation}
t^{(l)}=\frac{2}{\no{\Gamma(l)}},\qquad r^{(l)}=1-t^{(l)}.
\label{eq:GC}
\end{equation}
The targed vertices are marked by a special coin, where we choose
\begin{equation}
\label{eq:RC}
r^{(l)}=-\e^{\ii \phi},\qquad t^{(l)}=0,
\end{equation}
\end{subequations}
with a general phase-shift $\e^{\ii \phi}$ (meaning a walker going into a target vertex is reflected completely, gaining some phase shift in the process). Here $\no{\Gamma(l)}=d(l)$ is the degree of vertex $l$, i.e. the number of vertices in the set $\Gamma (l)$.  For both of these choices, the operator $\op U^{(l)}$ is unitary, and, as we shall see in next section, these choices also guarantee that the quantum walk has the same symmetry group as the graph.

\subsection{Symmetry Considerations}
\label{sec:symmetry}

Symmetry plays quite an important role in being able to determine the basic evolution properties of quantum walks. In this Section, we present a general framework of utilizing symmetries, following \cite{ReHiFeBu09}. Suppose we have a graph $\mathcal G=(V,E)$ with vertices of two types\footnote{The number of types of vertices can be arbitrary, but for our purposes two suffices. If the number would be higher, the automorphisms that we define would map vertices of each type onto vertices of the same type.} --- previously called targets and normal vertices. Let $\mathcal{A}$ be the group of automorphisms of the graph that also preserve vertex types.  An automorphism $a$ of $\mathcal G$ is a mapping $a: V\rightarrow V$ such that for any two vertices $v_{1},v_{2} \in V$, there is an edge connecting $a(v_{1})$ and $a(v_{2})$ if and only if there is an edge connecting $v_{1}$ and $v_{2}$.  Each automorphism $a$ induces a unitary mapping $\op U_a$ on the Hilbert space of the graph $\mathcal G$, such that $\op U_{a}\ke{v_{1},v_{2}} =\ke{a(v_{1}),a(v_{2})}$.  Suppose now that $\mathcal{H}$ can be decomposed into $m$ subspaces,
\[
\mathcal{H} = \bigoplus_{j=1}^{m} \mathcal H_{j} ,
\]
where each $\mathcal H_{j}$ is the span of some subset $B_{j}$ of the canonical basis elements and is invariant under $\op U_{a}$ for all $a \in \mathcal{A}$.  We shall also assume that each $\mathcal H_{j}$ does not contain any smaller invariant subspaces. This can always be done --- in the worst case, the subsets $B_j$ are exactly single edge states, while in other cases the subspaces are larger and help us to obtain a considerable reduction of the dimensionality of the problem.

Next, in each invariant subspace we form a vector 
\begin{equation}
\label{eq:eigenone}
\ke{w_{j}} = \frac{1}{\sqrt{d_{j}}} \sum_{\ke{v_{1},v_{2}} \in B_{j}}
 \ke{v_{1},v_{2}}  
\end{equation}
that is the sum of all of the canonical basis elements in the subspace,
with $d_{j}$ the dimension of $\mathcal H_{j}$.  This vector satisfies $\op U_{a}\ke{w_{j}} = \ke{w_{j}}$ for all $a \in \mathcal{A}$. Moreover, it is the only vector in $\mathcal H_{j}$ that satisfies this condition.  With the help of the vectors $\ket{\psi_j}$, we define the space $\subspace= \spann( \{ \ke{w_{j}}\st j=1,2, \ldots, m \})$, and note that $\subspace = \{ \ke\psi \in \mathcal{H}\st \op U_{a}\ke\psi = \ke\psi, \forall a\in \mathcal{A} \}$. The space $\subspace$ has dimension equal to the number of invariant subspaces $B_j$. 

Now suppose that the quantum walk operator $U$ commutes with the automorphisms, i.e. $[\op U,\op U_{a}]=0$ for all $a \in \mathcal{A}$.  This implies that if $\op U_{a}\ke\psi = \ke\psi$, then $\op U_{a}\op U\ke\psi = \op U\ke\psi$. Thus if $\ke\psi \in \subspace$, then $\op U\ke\psi \in \subspace$, meaning the subspace $\subspace$ is closed under the action of the step operator $\op U$. Correspondingly, if the initial state of the walk is in $\subspace$, then we only need to consider states in $\subspace$ to describe the state of the walk at any time. It is useful when the automorphism group is large because $\subspace$ then can have a much smaller dimension than $\mathcal{H}$, simplifying the analysis greatly.

Now let us demonstrate that the unitary operator $\op U$ defined by the local unitary operators in \eqref{eq:unitary} does, in fact, commute with all of the automorphisms of a graph that leave the target vertices fixed. It should be clear that this holds, as the construction of the evolution unitary is based on the structure of the graph. Nevertheless, let us see it directly. If these operators commute when applied to all of the elements of the canonical basis, then they commute. As before, let $\Gamma (v)$ be the set of vertices in $V$ that are connected to the vertex $v$, and if $v^{\prime} \in \Gamma (v)$ then $\Gamma (v; v^{\prime}) = \Gamma (v) \setminus \{v^{\prime}\}$. Finally, let $\no{\Gamma(v)}$ be the number of elements in $\Gamma(v)$. Then we have
\[
\op U_{a} \op U \ke{v_{1},v_{2}} = -r^{(v_2)} \ke{a(v_{2}),a(v_{1})} + t^{(v_2)}\!\!\! \sum\limits_{ v\in \Gamma (v_{2};v_{1})}\!\!\! \ke{a(v_{2}),a(v)} ,
\]
while also knowing that
\[
\op U \op U_{a} \ke{v_{1},v_{2}} = -r^{[a(v_2)]} \ke{a(v_{2}),a(v_{1})} + t^{[a(v_2)]} \!\!\!\!\!\!\!\! \sum\limits_{ v\in \Gamma (a(v_{2}); a(v_{1}))} \!\!\!\!\!\!\!\! \ke{a(v_{2}),v } .
\]
First, note that the reflection and transmission amplitudes in this equation are the same as those in the previous equation, i.e.~$r^{(v_2)}=r^{[a(v_2)]}$ and $t^{(v_2)}=t^{[a(v_2)]}$. This is a consequence of $\no{ \Gamma (v_{2};v_{1})} = \no{  \Gamma (a(v_{2});a(v_{1})) }$ -- the key properties of the local unitaries are conserved as the vertices of some type are mapped to the vertices of the same type, thus also the degree of the vertices is conserved.  Second, we also know that $ \Gamma (a(v_{2});a(v_{1})) = \{ a(v)\st v\in  \Gamma (v_{2};v_{1}) \}$, so that the sums in the two equations are identical.  Therefore, $U_{a} \op U \ke{v_{1},v_{2}}= \op U \op U_{a} \ke{v_{1},v_{2}} $, implying $[\op U, \op U_{a}]=0$.

Symmetries of a graph thus allow us to analyze the evolution under a quantum walk in a reduced subspace, if the initial state belongs to $\subspace$. This is the case for the uniform superposition, which is the reason we use it as our starting point in quantum walk searches. In the next Section, we showcase how this works for the complete graph, where the symmetries dictate that the dimensionality of $\subspace$ is only 4.


\subsection{Search on a Complete Graph}
\label{sec:cg}

\begin{figure}
\begin{center}
\includegraphics{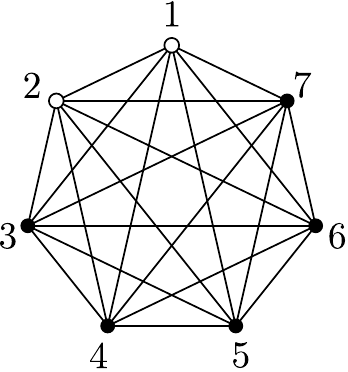}
\end{center}
\caption{\label{fig:CG}An example of a complete graph with $N=7$ vertices out of which $k=2$ are targets (white ones). A solution for the scattering-quantum-walk search on such graph leads to a reduction in dimensionality of the problem to only four dimensions.}
\end{figure}

Let us now consider a specific search problem where we will show basic manipulations and methods used to solve these problems. We consider a complete graph with $N$ vertices comprising the set $\all$ (see Fig.~\ref{fig:CG}). This is a specific graph with each vertex connected to all of the other vertices by an edge. The graph has thus $N(N-1)/2$ edges which define the Hilbert space of dimension $\dim\mathcal H =N(N-1)$ for a scattering quantum walk, according to \eqref{eq:hilbik}. Let $k$ be the number of special vertices. Recalling what was said in Sec.~\ref{sec:symmetry}, the elements of the set $\spec$ of these target vertices can be labeled as $j=1,2,\ldots, k$ and corresponding local unitary evolutions will be defined by \eqref{eq:unitary} and \eqref{eq:RC}. We label the normal vertices $j=k+1,k+2,\ldots, N$, and define their local unitary evolutions  by \eqref{eq:unitary} and \eqref{eq:GC}. The transmission and reflection coefficients for all normal vertices $j$ are the same, with $\no{\Gamma(j)}=N-1$.

There are four types of edge states here: those directed from normal vertices to special vertices, from special vertices to normal vertices, connecting normal vertices and connecting special vertices. Because of the symmetry of the complete graph, these four types also define the basis vectors of the subspace $\subspace$ according to \eqref{eq:eigenone}:
\begin{eqnarray*}
\ke{w_{1}} & = & \frac{1}{\sqrt{k(N-k)}}\sum_{a=k+1}^{N}\sum_{b=1}^k \ke{a,b},  \nonumber \\
\ke{w_{2}} & = & \frac{1}{\sqrt{k(N-k)}}\sum_{a=1}^{k}\sum_{b=k+1}^N \ke{a,b},
\end{eqnarray*}
\begin{eqnarray*}
\ke{w_{3}} & = & \frac{1}{\sqrt{(N-k)(N-k-1)}}\sum_{a=k+1}^{N}\sum_{\substack{b=k+1\\b\neq a}}^{N} \ke{a,b}, \nonumber \\
\ke{w_{4}} & = & \frac{1}{\sqrt{k(k-1)}}\sum_{a=1}^{k}\sum_{\substack{b=1\\b\neq a}}^{k} \ke{a,b}.
\end{eqnarray*}
These are equal superpositions of all edge states directed from normal to special, from special to normal, connecting only normal and connecting only special vertices. The unitary evolution of a uniform superposition of edge states, as shown in Sec.~\ref{sec:symmetry}, happens within the subspace $\subspace$ spanned by the four vectors $\ke{w_k}$, $k=1,2,3,4$, and is given by a $4\times 4$ matrix
\begin{equation}
\label{eq:US}
\op U \big|_{\subspace} = \begin{bmatrix} 0 & q & s & 0 \\ \e^{\ii\phi} & 0 & 0 & 0 \\ 0 & s & -q & 0 \\ 0 & 0 & 0 & \e^{\ii\phi}
\end{bmatrix},
\end{equation}
where
\begin{eqnarray*}
q & = & -r+t(k-1)=-1+\frac{2k}{N-1},\nonumber\\
s & = & \sqrt{1-q^2}=t\sqrt{k(N-k-1)}.\nonumber
\end{eqnarray*}
Note that the subspace spanned by the vectors $\ke{w_k}$, $k=1,2,3$ is decoupled from the subspace spanned by the vector $\ke{w_4}$. Indeed, the vector $\ke{w_4}$ (a superposition of edge states from target to target vertices) is invariant under the step unitary, up to a phase factor.

In analogy with Grover's search [see \eqref{eq:equal_superposition}], we now take the equal superposition of all edge states
for our initial state:
\begin{equation}
\label{eq:graph_superposition}
\ke\psinit=\frac{1}{\sqrt{N(N-1)}}\sum_{\ke{j,l}\in\mathcal H}\ke{j,l}.
\end{equation}
This is on one hand a necessity in order for the search to be efficient, but on the other hand it is a state where no prior information about any target vertex is used, as we do not expect to know it beforehand. We can write $\ket{\psi_0}$ as a superposition of the states $\ke{w_k}$ (thus it belongs to $\subspace$) as
\begin{eqnarray*}
\ke\psinit & = & \sqrt{\frac{k(N-k)}{N(N-1)}}\left(\ke{w_1}+\ke{w_2}\right)+\\
& & + \sqrt{\frac{(N-k)(N-k-1)}{N(N-1)}}\ke{w_3}+\sqrt{\frac{k(k-1)}{N(N-1)}}\ke{w_4}.
\end{eqnarray*}

We shall analyze what happens for two choices of the phase-shift on the target vertices: 
$\phi = 0$ and $\phi = \pi$.  First, we start with a bad choice. For $\phi=0$ (target vertices do not scatter, just reflect without a phase shift), we find that the initial state can be written as a superposition of eigenstates of $\op U$ with an eigenvalue equal to unity, in particular,
\begin{eqnarray*}
\ke{\tilde u_0} & = & {\frac{1}{\sqrt{3+q}}}\begin{bmatrix}
\sqrt{1+q} \\ \sqrt{1+q} \\ \sqrt{1-q} \\ 0\end{bmatrix}, \nonumber \\
\ke{\tilde u_0'} & = & (0,0,0,1)^{T} ,
\end{eqnarray*}
and the initial state can be expressed in terms of them as
\[
\ke\psinit = \sqrt{\frac{(N-v)(N+v-1)}{N(N-1)}}\ke{\tilde u_0} + \sqrt{\frac{v(v-1)}{N(N-1)}}\ke{\tilde u'_0}.
\]
This, however, means that the initial state is an eigenstate of the step unitary $\op U$ as well. In this case, the quantum walk gives us no advantage over a classical search, as the measurement in any time gives us only a random edge state.

Second, we pick the value of $\phi=\pi$, and see that the behavior of the quantum walk is quite different. The state of the walk after $m$ steps is derived by employing the decomposition formula
\begin{equation}
\label{eq:evolution}
\ke{\psi_m}=\sum_{\lambda}\left(\lambda^m\langle \mu_\lambda\ke\psinit\right)\ke{\mu_\lambda},
\end{equation}
where $\ke{\mu_\lambda}$ are the orthogonal eigenvectors corresponding to the eigenvalue $\lambda$ of the unitary operator $\op U$. In the limit $N\gg k \geq 1$ we find that
\begin{equation}
\label{eq:result}
\op U^m\ke\psinit\simeq\frac{1}{\sqrt{2}}\begin{bmatrix}
\phantom{-} \sin (2m+1)\frac{\theta}{2}\\
-\sin (2m-1)\frac{\theta}{2}\\
\sqrt{2}\cos m\theta\\
0
\end{bmatrix}, 
\end{equation}
where
\begin{equation}
\tan\theta = \frac{\sqrt{k(2N-k-2)}}{N-k-1}.
\end{equation}
We find that the probability amplitudes for edge states not connected to special vertices (the third component in the vector) are approximately equal to zero when $\theta m=\pi/2$. If we measure the walker after such a choice of steps, with probability close to unity we will find an edge connected to one of the target vertices.  Therefore, the number of steps needed to find one of the target vertices with reasonable probability is of the order $\order (\sqrt{N/k})$ for large $N$, in particular
\begin{equation}
\label{eq:graph_optimal}
\tilde m\simeq\frac{\pi}{2\sqrt{2}}\sqrt{\frac{N}{k}}.
\end{equation}
This is a quadratic speedup over any classical algorithm that needs at least $\order (N/k)$ steps to do the task when searching an unstructured database.

It should be noted that for the case $k=1$, the vector $\ke{w_4}$ is not defined and the dimension of the problem is reduced to three. In this special case, the analysis becomes even simpler and the quadratic speedup result remains valid.

To put the problem into perspective, let us compare it with the results of \cite{ShKeWh03}, studying search on a hypercube. They present an algorithm for finding a single target vertex with success probability of approximately $1/2$, which is in contrast with our finding an edge with a special state with probability close to unity. In our case, the probability is (almost) equally split between two possible sets of edge states, those \emph{leaving} the special vertex and those \emph{entering} the special vertex (states $\ke{w_1}$ and $\ke{w_2}$). If we would reformulate our search in the coined quantum walk language, we would find that our results correspond to either finding the particle on the special vertex (with an arbitrary coin state, i.e.~our state $\ke{w_1}$) or finding it on one of the neighboring vertices with the coin pointing to the special one (our state $\ke{w_2}$).

In \cite{AmKeRi05}, a quantum walk that performs an exact Grover search is discussed. The graph the quantum walk is performed on is a complete graph that has loops added to each vertex. The coins for normal vertices are chosen to be of the Grover type. However, the coins for the special vertices are ``minus'' the Grover coins (adding a phase shift of $\pi$). This leads to almost the same evolution as in the case of Grover's search; the only difference is that one step of Grover's algorithm corresponds to two steps of the quantum walk.

Finally, let us summarize the situation as it presently stands. Quantum-walk searches have two figures of merit, the number of steps necessary to find a special vertex and the probability of finding it after a specific number of steps.  On a complete graph without loops [the result we obtained in \eqref{eq:result}) one needs $\sqrt{2}$-times as many steps as in the Grover search for the corresponding problem (a search within $N$ elements), and after this many number of steps the probability of finding the special vertex is equal to unity. On a complete graph with loops \cite{AmKeRi05}, twice as many steps as in the corresponding Grover search are required, and the probability of finding the special vertex is again equal to unity. A rigorous comparison of these properties on a hypercube can be found in Ref.~\cite{PoGaKiJe08}, where adding loops to the graph again results in the increase if the necessary number of walk steps by a factor of $\sqrt{2}$. The result holds also for the complete graph and explains the differences stated.

\begin{exercise}
Find the evolution of an equal superposition initial state for a quantum scattering walk on a complete graph with loops using the Grover coin on normal vertices and minus the Grover coin on target vertices.
\end{exercise}

\subsubsection{Oracle Controlled Evolution}
\label{sec:oracles}

In this Section we will make a small detour from the particular results obtained in previous pages. We stated that quantum walk search is faster than the classical one, yet we have not really supported the statement. In order to be able to make such a comparison more rigorously we will use oracles defined in Sec.~\ref{sec:oracles_and_searches} and the number of their calls as a mean of comparison. Oracles are the bearers of information about searched-for targets and they provide a way to quantify the resources needed to perform a search. It is the number of times we call an oracle that tells us how efficient the search is. Let us have a closer look on how oracles fit into the problems of quantum-walk searches.

In particular, the oracle $\Ctrl\op V_f$ we will use is given by \eqref{eq:quantum_oracle} from Sec.~\ref{sec:oracles_and_searches}. On one hand, when marking target elements, it can be used to count the resources used in a search algorithm such as Grover's search. On the other hand, we have devised and discussed quantum walks with position-dependent coins (scattering quantum walks) in the previous Sections. If we want to compare the necessary resources, we need to connect the quantum oracle concept with the notion of a coin for a quantum walk. Just as in scattering quantum walks the state of the walker is given by two vertices (an edge connecting them), we will use the oracle in two steps. In the first step, the information about one of the vertices contained in the state of the walker will be extracted to one ancillary system. This information will be then used as the input to the oracle. The oracle will tell us whether the vertex is a target or not, adding this information to the second ancillary system. Finally, this information will be used to determine what type of coin will be used on a given vertex.

In particular, if $l$ is a vertex of the graph, $f(l)=0$ corresponds to a normal vertex and $f(l)=1$ corresponds to a target vertex, i.e. the type of vertex we are trying to find.  If $l$ is a normal vertex, the local unitary operator corresponding to it will be denoted by $\op U_{0}^{(l)}$, and if it is a target vertex the local unitary will be denoted by $\op U_{1}^{(l)}$.

Our quantum circuit will act on a tensor product of the Hilbert space $\mathcal H$ for the quantum walk \eqref{eq:hilbik}, the Hilbert space for vertices, $\mathcal{H}_{v}$, and a qubit Hilbert space, $\mathcal{H}_{2}$. The vertex space is
\begin{equation}
\mathcal{H}_{v} = \ell^2(\{\ke{l}\st l\in V\}).
\end{equation}
We now define an operator $\CtrlU$ acting on in the following way on the edge register and the two ancillary ones as 
\begin{equation}
\label{eq:walk}
\CtrlU(\ke{k,l}\otimes\ke l\otimes\ke c)=\left(\op U^{(l)}_{c}\ke{k,l}\right)\otimes\ke l\otimes\ke c.
\end{equation}
This equation does not completely specify the actions of $\CtrlU$. In particular, it does not specify its action on states of the form $|k,l\rangle \otimes |l^{\prime}\rangle \otimes |c\rangle$, where $l\neq l^{\prime}$, but we will only need to consider its action on states of the form given in the previous equation as it can always be expanded to a unitary operator on the whole Hilbert space.

\begin{figure}
\begin{center}
\includegraphics{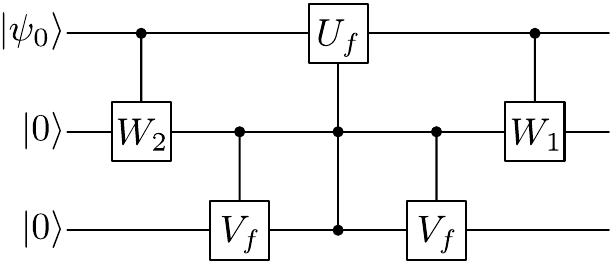}
\end{center}
\caption{\label{fig:circuit}
A logical circuit (network) that implements a single step of a scattering quantum walk search, making use of the quantum oracle $\Ctrl$. The first input corresponds to a quantum walker originally prepared in the state $\ke\psinit$. The second input represents a vertex
state, while the third input represents an ancillary qubit.}
\end{figure}

The quantum circuit that implements one step of our quantum walk search is given in Fig.~\ref{fig:circuit}. The first input stands for a state of the quantum walker (i.e.~any superposition of edge states), the second for a vertex state, and the third for an ancillary qubit.  The input state is $\ke{\psi_\mathrm{init}}=\ke\psinit\otimes\ke 0\otimes\ke 0$ where $\ke\psinit$ is a general state in the Hilbert space $\mathcal{H}$ of edge states,
\[
\ke\psinit=\sum_{ml\in E}a_{ml}\ke{m,l},\qquad\sum_{ml\in E}|a_{ml}|^2=1.
\]
The state $|0\rangle$ in the second slot (input) in Fig.~\ref{fig:circuit} is one of the vertex states, which, besides labeling a particular vertex, will also serve as a reference ``zero'' state.  First, we apply the operator $\CtrlWtwo$ which maps the state $|m,l\rangle \otimes |0\rangle$ in $\mathcal{H} \otimes \mathcal{H}_{v}$ to $|m,l\rangle \otimes |l\rangle$.  Such a unitary can be implemented, e.g.~as presented in \cite{BoBrHoTa96} or \cite{BrBuHi01}.  After this  operator is applied, we get
\[
\ke{\psi_0}\mapsto\sum_{ml\in E}a_{ml}\ke{m,l}\otimes\ke{l}\otimes\ke 0\equiv\ke{\psi_1}.
\]
Next, we apply the quantum oracle \eqref{eq:quantum_oracle} to the vertex state and the qubit, yielding
\[
\ke{\psi_1}\mapsto\sum_{ml\in E}a_{ml}\ke{m,l}\otimes\ke{l}\otimes\ke{f(l)}\equiv\ke{\psi_2}.
\]
Now we can apply the $\CtrlU$ operator (which coin gets applied is controlled by the last two registers) from \eqref{eq:walk}to the state, producing
\[
\ke{\psi_2}\mapsto\sum_{ml\in E}a_{ml}\left(\op U_{f(l)}^{(l)}\ke{m,l}\right)\otimes\ke{l}\otimes\ke{f(l)}\equiv\ke{\psi_3}.
\]
Both of the ancillary systems acted as controls, and they need to be reset before we can make another quantum walk step.  This erasure is the task of the remaining two gates in the circuit.  Since the local unitary operator $\op U_{f(l)}^{(l)}$ acts only on the edge space and maps $\Omega_l$ to $A_l$ as given in Sec.~\ref{sec:sqws}, the state $\ket{\psi_3}$ can be rewritten as
\[
\ke{\psi_3}=\sum_{lm\in E}b_{lm}\ke{l,m}\otimes\ke{l}\otimes\ke{f(l)}.
\]
A second application of the quantum oracle on the last two register resets the last qubit state to $|0\rangle$. However, to reset the vertex state in the second register, we cannot use the $\CtrlWtwo$ operation as before, since the information contained in the edge state about the vertex state has moved from the second to the first position. Therefore, we need the operator $\CtrlWone$, which would map $|l,m\rangle \otimes |l\rangle$ in $\mathcal{H} \otimes \mathcal{H}_{v}$ to $|l,m\rangle \otimes |0\rangle$, giving
\[
\ke{\psi_3}\mapsto\sum_{lm\in E}b_{lm}\ke{l,m}\otimes\ke{0}\otimes\ke{0}.
\]
This $\CtrlWone$ operation can be constructed in the same manner as operation $\CtrlWtwo$ in \cite{BoBrHoTa96}, using now the first of the edge states as a control.

We have thus performed one step of the walk and reseted the ancillas, so that the circuit can be applied again to perform additional steps of the walk.
The oracle in these problems is a resource -- giving us additional information every time we call it. The number of necessary oracle calls then tells us how efficient our algorithm is. Here we see that at most $2m$ oracle calls are needed in a quantum walk search algorithm that requires $m$ walk steps.

\subsection{Other Examples of Searches on Graphs}
\label{sec:othersearches}

The complete graph search example from Sec.~\ref{sec:cg} is just one of many possibilities. In \cite{ReHiFeBu09}, the reader can find a number of possibilities for the choice of graph suitable for quantum walk search (with at most quadratic speedups). On one hand these examples are artificial, however one should bear in mind the results of Sec.~\ref{sec:oracles}, that the oracle determining the selected elements is independent of the choice of the graph we make. Hence, whatever choice of graph with a proper number of vertices we make, we can always use the same oracle to perform the search. It is however crucial to make a good graph choice, as the efficiency of the walk algorithm is dependent on this choice. Indeed, taking a circular graph clearly cannot give us any means to speed up classical search, as the ``information'' about the target has to ``travel'' at least along half of the vertices (the amplitude needs to rise significantly at the target vertex). On the other side of the spectrum of graph choices is the complete graph which makes the search quadratically faster than in the classical case.

\begin{figure}
\begin{center}
a) \includegraphics[height=5cm]{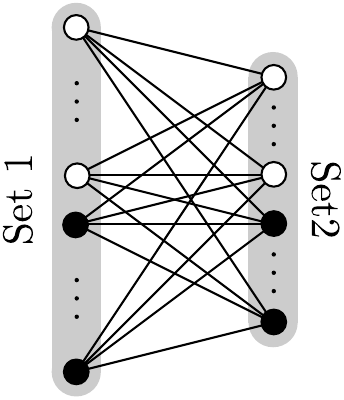}\hskip1cm
b) \includegraphics[height=6cm]{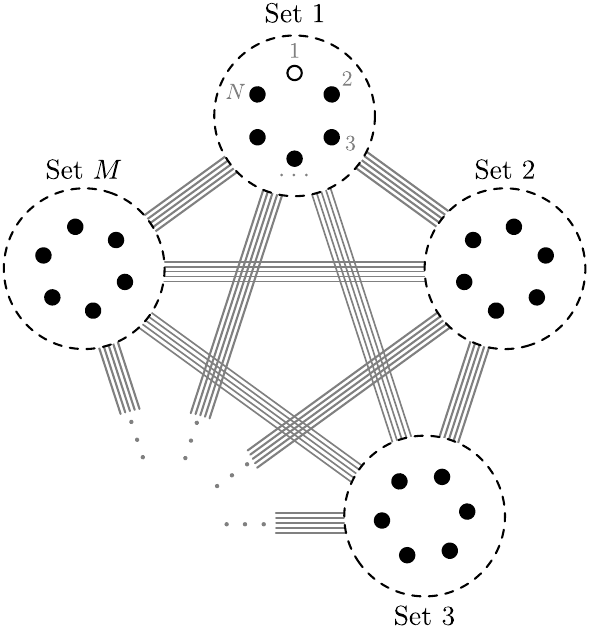}
\end{center}
\caption{\label{fig:othergraphs} Two types of graphs suitable for quantum-walk searches: a) A bipartite graph consisting of two sets of vertices where each vertex from one set is connected to all of the vertices from the other set, while there are no connections within a set. There may also be a different number of target vertices in each set. b) An $M$-partite complete graph consisting of $M$ sets of vertices, where each set contains $N$ vertices. There exists an edge for every pair of vertices not belonging to the same set, while there is no edge connecting any two vertices within the same set. Here we look for one special vertex in one of the sets.}
\end{figure}

One interesting example of a graph to search on is the complete bipartite graph depicted in Fig.~\ref{fig:othergraphs}a. It is a graph with vertices belonging to two sets. Each vertex belongs to a particular set, is connected to all the vertices belonging to the other set, but not with any vertices within the same set. Another graph of interest is the complete $M$-partite graph (Fig.~\ref{fig:othergraphs}b). This graph is composed of $M$ sets, each containing $N$ vertices. Here the vertices from one set are connected to all the vertices from any other set, but again, vertices within a set are not connected. Clearly, taking $M=2$ we recover the complete bipartite graph and taking $N=1$ we recover the complete graph. In all of these cases, a quadratic speedup (in the number of oracle calls) over the best classical search algorithm is possible.

Until now, we had the opportunity to study quantum-walk searches where target vertices were marked by an additional phase-shift, utilizing the oracle from \eqref{eq:quantum_oracle} -- the oracle tells us which vertices are special, while the choice of the graph is ours. There are, however, also examples where the goal of the search is to find a distinctive topological feature of the graph. In those cases, we may also employ an oracle, albeit in a different form -- giving us the information about the neighbors that the walker can visit. This also means, that the structure of the graph is now a part of the oracle and not our choice anymore.

\begin{figure}
\begin{center}
\includegraphics[scale=0.8]{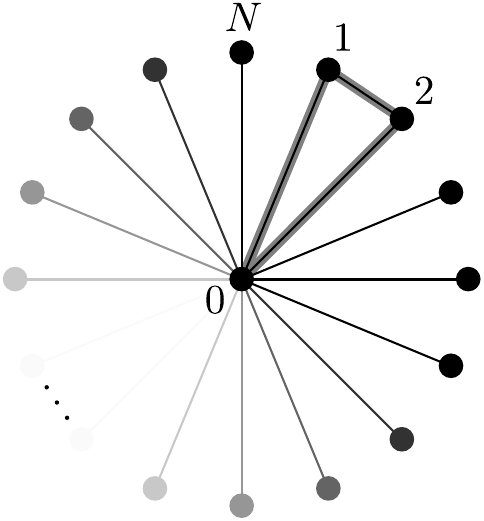}
\end{center}
\caption{\label{fig:defect} A star graph having two arm vertices connected; these vertices are purely transmissive. Other arm vertices are purely reflective and the central vertex is Grover-like. The evolution, when starting from equal superposition, gives us again a speedup in locating the extra edge with high probability and the state can be localised almost entirely on the grey triangle.}
\end{figure}

Let us now take a look at the graph depicted in Fig.~\ref{fig:defect} (see also Ref.~\cite{FeHi+10}) -- a star graph with $N$ spikes, where two of the spike vertices (say $1$ and $2$) are connected. The Hilbert space is again defined as a span of all edge states of the graph. We choose the central node labeled 0 to have a Grover-like coin, obeying \eqref{eq:GCbraket} with $t=2/(N+1)$. 
The local unitary evolution for the outside vertices is the same as before (without a phase-shift), except for the two connected vertices $1$ and $2$, which obviously have two outgoing edges. There we set the local unitary evolution $\op U_0$ to 
\begin{eqnarray*}
\op U_0\ke{0,1}=r_0\ke{1,0}+t_0\ke{1,2}, &\qquad& \op U_0\ke{0,2}=r_0\ke{2,0}+t_0\ke{2,1},\\
\op U_0\ke{2,1}=t_0\ke{1,0}-r_0\ke{1,2}, &\qquad& \op U_0\ke{1,2}=t_0\ke{2,0}-r_0\ke{2,1},
\end{eqnarray*}
with $t_0=\sqrt{1-r_0^2}$ and $r_0$ chosen from the interval $[-1;1]$, with $r_0=-1$ corresponding to Grover's search (``marked'' edges get a $-1$ phase), $r_0=1$ giving a repetitive application of Grover's coin without any phase-flips and $r_0=0$ for purely transmitting vertices. The quantum walk evolution is based on an oracle that determines the neighbors of the vertex we enter as an input. We further suppose that the (name of the) central vertex is known to us beforehand.

To classically find the extra edge, we would have to sift through the outlying vertices to find one that has two neighbors.
Note that looking at a typical vertex with only 1-neighbor, we only get the knowledge that we missed. 
Thus, this is unstructured search, which classically requires $O(N)$ calls to the oracle giving out the neighbors of a given vertex. Again, as we will see, we can obtain a quadratic improvement but now without using any phase-flips. The results of Sec.~\ref{sec:symmetry} once again give us a way to find an invariant subspace, which is now five-dimensional, spanned by the vectors
\begin{eqnarray*}
\ke{w_1} &=& \frac{1}{\sqrt{2}}(\ke{0,1}+\ke{0,2}),\\
\ke{w_2} &=& \frac{1}{\sqrt{2}}(\ke{1,0}+\ke{2,0}),\\
\ke{w_3} &=& \frac{1}{\sqrt{N-2}}\sum_{j=3}^N\ke{0,j},\\
\ke{w_4} &=& \frac{1}{\sqrt{N-2}}\sum_{j=3}^N\ke{j,0},\\
\ke{w_5} &=& \frac{1}{\sqrt{2}}(\ke{1,2}+\ke{2,1}).
\end{eqnarray*}
Note, that it is desirable to end in $\ke{w_5}$ or at least in the states $\ke{w_1}$ or $\ke{w_2}$. Meanwhile, the probability of ending in the states $\ke{w_3}$ and $\ke{w_4}$ should be as small as possible, as these states correspond to edges not connected to either of the vertices $1$ or $2$. 

The choice of initial state,
\begin{equation}
\label{eq:init_extra_edge}
\ke\psinit=\frac{1}{\sqrt{2N}}\sum_{j=1}^N(\ke{0,j}-\ke{j,0}),
\end{equation}
will lead us to the evolution
described by the state after $m$ steps,
\[
\ke{\psi_m}=\op U^m\ke\psinit\simeq\frac{(-1)^m\Delta}{\sqrt{2}}
\begin{bmatrix}
\sin m\Delta\sqrt{t}\\
\sin m\Delta\sqrt{t}\\
\phantom{-}\Delta^{-1}\cos m\Delta\sqrt{t}\\
-\Delta^{-1}\cos m\Delta\sqrt{t}\\
-\frac{t_0}{1-r_0}\sin m\Delta\sqrt{t}\\
\end{bmatrix},
\]
where
\[
\Delta=\sqrt{\frac{2(1-r_0)}{3-r_0}}.
\]
We chose the initial state \eqref{eq:init_extra_edge} in such a way that the contributions from different eigenvectors do not cancel in the components for states $\ke{w_3}$ and $\ke{w_4}$ (which would happen if the intitial state was the equal superposition), but rather add up.

The different choices of $r_0$ result in various behavior of the walk. For $r_0=0$ (when the vertices $1$ and $2$ are purely transmitting), we get a quantum walk algorithm taking $\opt=\pi\Delta^{-1}\sqrt{N/8}$ steps, 
ending up with the walker equally distributed on the triangle $012$, with equal probability of $1/3$ on every edge of this triangle.

When increasing $r_0$, the probability to find the walker in the state $\ke{w_5}$, i.e.~on the edge $1$--$2$, rises. It might seem that taking $r_0\to 1$ will give us the walker positioned entirely on the extra edge. However, the parameter $\Delta$ depends on $r_0$ as well. Taking $r_0\to 1$ gives $\Delta\to 0$, increasing the required number of steps $\opt$ necessary to assure us the maximum probability of success. Evidently, $r_0=1$ causes the extra edge to become ``invisible'', making the vertices $1$ and $2$ no different from others -- in this case, the evolution is trivial (up to a global phase).

On the other hand, taking $r_0=-1$ is a good choice, as it results in $\Delta=1$. In this case, the extra edge is again not ``visible'', but the vertices $1$ and $2$ get a phase-shift $\pi$ leading to Grover's search on a star graph with $k=2$ target vertices. Note that for every step of Grover's algorithm we need two steps of the walk on the star graph, so the number of steps needed $\opt=\pi\sqrt{N/8}$ corresponds to Grover's result in \cite{Grover97}.

Let us go back to the most natural $r_0=0$ case. There is no active phase-shift occurring anywhere in the graph, yet for the Grover search a phase-flip is a very important element. Where is then the place where something like that occurs in this case? The graph in Fig.~\ref{fig:defect} is not bipartite. The part of the walker leaving the central vertex (outgoing edges) and the part entering it (incoming edges) have a way to interfere via the extra edge. Notice that in the initial state \eqref{eq:init_extra_edge} has the incoming and outcoming parts initialized with opposite signs. Therefore, the directly reflected amplitude and the amplitude transmitted ``around'' the triangle can have different signs, combining as if a part that was reflected from one of the special vertices gained a $\pi$ phase-shift. Thus, in this case a phase-shift action of a unitary was replaced by an interference effect of phase-shifted (otherwise) non-interfering parts of the walker.

\begin{example}
\label{ex:star_bound}
Search on the star graph with an extra edge can be also used as a lower bound for a triangle-finding algorithm --- see Sec.~\ref{sec:cliques}. In the triangle problem for a graph $G$ with $N$ vertices, we are supposed to find out whether it contains a triangle. The graph is given to us via an oracle $O_G \ket{x,y}\ket{z} = \ket{x,y}\ket{z\oplus E(x,y)}$ that holds the information whether vertices $x$ and $y$ have an edge between them. A clever quantum algorithm for this was given by Magniez et al.~\cite{MaSaSz05}, taking $O(N^{\frac{13}{10}})$ queries of the edge oracle. \dots However, we do not know whether this is an optimal algorithm. The best lower bound for this problem comes from the star graph. Finding the extra edge between a pair of spike vertices certifies the existence of the triangle. Any quantum algorithm for this unstructured problem has to take $O(\sqrt{N(N-1)}$ queries of oracle $O_G$, as there are $N(N-1)$ possible edges between the $N$ spike vertices in the star graph. Thus, the best algorithm for the triangle problem can not take fewer than $O(N)$ edge oracle $O_G$ queries. Note that if we have an oracle giving us a list of neighbors instead, we can find the special vertex in $\sqrt{N}$ queries using Grover's search with unit cost of asking for its neighbors, as was done above in this Section.
\end{example}

\subsection{Abstract Search Algorithm and Spatial Search}
\label{sec:discrete_spatial}

All the kinds of search algorithms introduced previously were analyzed in the same way. First the unitary $\op U$ is spectrally decomposed and then Eq.~(\ref{eq:evolution}) is used to find the evolution of the initial state. This approach is quite universal, however sometimes such analysis is overcomplicated and some other approaches can be used. One such approach was presented in Ref.~\cite{AmKeRi05} --- abstract search algorithm. It is not as universal, yet it is still quite broad to help solve many search problems easier than by determining the whole evolution. This approach to searches view them from a different perspective than the generalized approach of amplitude amplification \cite{BrHoMoTa00}.

Suppose you have evolution driven by unitary $\op U=\op V\op R$, where $\op R$ is controlled phase-flip on a single target element $\ke t$ given by Eq.~(\ref{eq:grover_oracle}) and $\op V$ is real unitary operation with a unique (real) eigenvector $\ke\psinit$ with eigenvalue $1$.

As $\op V$ is real unitary matrix, its non-$\pm 1$ eigenvalues $\lambda_j^\pm$ come in pairs of complex conjugate numbers $\e^{\pm\ii\theta_j}$. The eigenvector with eigenvalue $1$ is $\ke\psinit$; let the complex eigenvalues $\lambda_j^\pm$ have eigenvectors $\ke{\phi_j^\pm}$. One can show that $\ke{\phi_j^+}^*=\ke{\phi_j^-}$. Finally, let the eigenstates of eigenvalue $-1$ be $\ke{\rho_k}$. Setting $a_j^\pm=\br{\phi_j^\pm} t\rangle$ one can also see that the global phase of states $\ke{\phi_j^\pm}$ can be set so that $(a_j^+)^*=a_j^-\equiv a_j$. From this it follows that the expansion of state $\ke t$ in the basis of $\op V$, which we will need, is
\begin{equation}
\label{eq:spatial_expansion}
\ke t=a\ke\psinit+\sum_j a_j(\ke{\phi_j^+}+\ke{\phi_j^-})+\sum_k a_k \ke{\rho_k}.
\end{equation}
This expansion is used to analyze the operator $\op U$. Taking $\theta_{\min}$, the smallest phase of the eigenvalues of $\op V$ one can find \cite{AmKeRi05} that the most important eigenvalues of $\op U$ are $\e^{\pm\ii\alpha}$, where\footnote{Symbol $\Theta(g)$ means that the function $f$ is bounded both above and below by $g$ asymptotically, i.e.~there exist positive numbers $a$ and $b$ such that $a g(x)\leq f(x) \leq b g(x)$.}
\[
\alpha=\Theta\left(a \left( \sum_j \frac{a_j^2}{1-\cos\theta_j}+ \frac{A^2}{4} \right)^{-1/2}\right),
\]
where the sum goes only over the complex eigenvalues indices and
\[
A=\sqrt{\sum_k a_k^2}
\]
is the contribution from $-1$ eigenvalue expansion coefficients. When $\theta_{\min}$ is small, which is ususal, then also $\alpha$ is small. Let us set also states
\[
\ke{\alpha^\pm}=\frac{1}{\sqrt{2}}(\ke{\alpha}\pm\ke{-\alpha}),
\]
where $\ke{\pm\alpha}$ are eigenvectors of $\op U$ corresponding to eigenvalues $\e^{\pm\ii\alpha}$. If $\alpha<\theta_{\min}/2$, then the initial state $\ke\psinit$ is close to the state $\ke{\alpha^-}$, in particular
\begin{subequations}
\label{eq:spatial_approx}
\begin{equation}
|\br\psinit \alpha^-\rangle|\geq 1-\Theta\left(\alpha^4\sum_j\frac{a_j^2}{a^2}\frac{1}{(1-\cos\theta_j)^2}\right)-\Theta\left(\frac{A^2\alpha^4}{a^2}\right),
\end{equation}
while the target state $\ke t$ falls out to be close to the state $\ke{\alpha^+}$,
\begin{equation}
\label{eq:spatial_approx_final}
|\br t \alpha^+\rangle|^2=\Theta\left(\min\left\{\left(\sum_j a_j^2\cot^2\frac{\theta_j}{4}\right)^{-1/2},1\right\}\right).
\end{equation}
\end{subequations}

This means, that one can use this general procedure to find the approximate evolution of the system: if we apply $m$-times operator $\op U$ on the initial state, keeping in mind approximations of Eqs.~(\ref{eq:spatial_approx}), we find
\[
\op U^m\ke\psinit\simeq\op U^m\ke{\alpha^-}=\op U^m\frac{1}{\sqrt{2}}(\ke\alpha-\ke{-\alpha})=\frac{1}{\sqrt{2}}\left(\e^{\ii m\alpha}\ke\alpha-\e^{-\ii m\alpha}\ke{-\alpha}\right).
\]
When we take $m\alpha=\pi/2$ we find that
\[
\op U^m\ke\psinit\simeq\frac{\ii}{\sqrt{2}}(\ke\alpha+\ke{-\alpha})=\ii\ke{\alpha^+}.
\]
So the method tells us, that if we start in state $\ke\psinit$ and apply operation $\op U$ for $\lfloor\pi/2\alpha\rfloor$-times, Eq.~(\ref{eq:spatial_approx_final}) will tell us, how close we are to the state $\ke t$.

\begin{example}
Let us consider a lattice with dimension $d\geq 3$ with $N$ vertices arranged as $\sqrt[d]{N}\times\ldots\times\sqrt[d]{N}$ with periodic boundary conditions. Then we can use flip-flop grover coin from Eq.~(\ref{eq:flip-flop}) in quantum walk search given by algorithm in Fig.~\ref{fig:graph_search} to find one marked vertex in (optimal) time $O(\sqrt{N})$ starting from initial state of the equal superposition on all sites --- such initial state can be constructed from localized state with time expense of $O(\sqrt[d]{N})$. Eqs.~(\ref{eq:spatial_approx}) tell us that initial state is almost $\ke{\alpha^-}$ and the probability of success is constant, i.e.~few repetitions of the algorithm suffice to find targeted vertex.

Interestingly continuous-time quantum walks for a long time did not succeed to find an efficient continuous-time alternative to this discrete-time algorithm. Only after introducing spin degree of freedom into the walk researches suceeded to get the same limits.
\end{example}

\begin{example}
Spatial search on two-dimensional lattice of $\sqrt{N}\times\sqrt{N}$ vertices with periodic boundary can be analyzed with the abstract search algorithm as well, yet the results are not as optimistic as in higher dimensions. For such quantum walk there is a $T=O(\sqrt{N\log N})$ such that after $T$ steps the probability to determine the target vertex is $p=O(1/\log N)$ \cite{AmKeRi05}. By using the method of amplitude amplification (see Sec.~\ref{sec:grover_search}) one can obtain constant probability of success with the running time of the algorithm of $O(\sqrt{N}\log N)$. Deeper analysis of the abstract search algorithm and introduction of properly chosen ancillary system and allowed Tulsi \cite{Tulsi08} to boost the probability of succes to constant thus reducing the running time of the algorithm even more to $O(\sqrt{N\log N})$ steps.
\end{example}

\subsection{Subset Finding and Related Problems}
\label{subsetfinding}

Probably the most ingenious algorithm showing the usefulness of discrete-time quantum walks is their application to various subset finding problems. This range of algorithms is based on Grover-like evolution on specially constructed graphs allowing for better efficiency of these algorithms. The history begins in 2003 when Ambainis \cite{Ambainis07} gave an algorithm for element $k$-distinctness. It determines whether a given set contains $k$ elements with the same assigned value provided by an oracle, and finds such a set if there is one. Building on this work, Magniez, Santha and Szegedy \cite{MaSaSz05} then provided a triangle-finding algorithm, deciding whether a given graph contains a triangle. This approach was generalized in \cite{ChEi05}, where an algorithm for subset-finding was provided. A better efficiency was then provided in the updated version of \cite{MaSaSz05}, where the authors performed a deeper algebraic analysis of the algorithm. Here we present the algorithm in its most up-to-date and efficient form.

\subsubsection{Algorithm for $k$-subset Finding}

Let us discuss the algorithm for the following problem, introduced in \cite{Ambainis07}. Consider a set $\all$ with $N$ elements combined with values from a finite set $\mathcal R$ assigned by a function $f:\all\to\mathcal R$. 
The problem is to determine, whether a $k$-subset with a given \emph{property} $\mathcal P\subset (\all\times\mathcal R)^k$ exists in the set $\all$. For example, in the {\em collision problem}, we are given a list of vertices and their assigned colors, and we're asked to find two vertices ($k=2$) of the same color (if they exist). Another example is the {\em triangle problem}: given a list of edges in a graph, determine whether we can find a set of 3 edges (here $k=3$) that form a triangle (if such a set exists). 

The information hidden in the function $f$ is given to us in the form of a classical or quantum oracle. Evaluating whether a $k$-subset has a given property $\mathcal P$ (such as: do these three edges form a triangle) should be simple once we know the values of $f$ on the vertices of the $k$-subset. This allows us to determine the efficiency of a given algorithm using query complexity -- counting the number of oracle calls required to find the desired $k$-subset with a given property.

The aim of the algorithm is to output a subset $\spec=\{x_1,x_2,\ldots,x_k\}\subset\all$ such that the $l$-tuple $((x_1,f(x_1)),(x_2,f(x_2)),\ldots,(x_l,f(x_l)))\in\mathcal P$ if it exists,
when we are given the set $\all$ of elements, a description of an easily computable property $\mathcal P$ and the oracle computing the function $f$ on vertices. If there is none such $l$-tuple, the algorithm should say so. In the classical case, it we have to query the oracle $O(N)$ times to determine the solution to the problem, as the very last element we query could be the one that completes a $k$-tuple with the property $\mathcal P$.

\begin{figure}
\begin{center}
\includegraphics{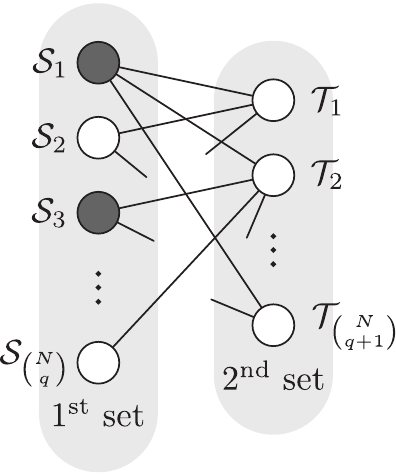}
\end{center}
\caption{\label{fig:distinctness}The algorithm for subset finding performs a walk on a bipartite graph whose vertices are identified with subsets of $\all$ having either $q$ elements (on the left), or $q+1$ elements (on the right). The vertices are connected only if the sets they correspond to differ in exactly one element. The evolution on the graph is given by the Grover coin on most vertices, combined with phase-flips on vertices ($q$-subsets) containing $k$-subsets with elements having with property $\mathcal P$.}
\end{figure}

The $k$-subset finding quantum algorithm is a quantum walk on a specially constructed bipartite graph, depicted in Fig.~\ref{fig:distinctness}. The vertices on the left of this graph correspond to all the possible $q$-element subsets $\mathcal S_j$ of $\all$, while the vertices on the right label all the possible $(q+1)$-element subsets $\mathcal T_j$ of $\all$. Two vertices are connected only if the corresponding sets differ in exactly one element. In the following, we will not make any further distiction between the vertex and its corresponding set, i.e.~$\mathcal S$ will be both the set and the vertex it determines.

The size of the parameter $q$ shall be determined later to provide the best efficiency. The simplest choice would be $q=k$, but that does not give much benefit over straightfoward Grover's search over $k$-tuples, looking whether they have the property $\mathcal P$. It will be much better to think of rather large subsets $q=N^{\mu}$ with $\mu<1$. 

The Hilbert space the algorithm runs in will be spanned by the orthonormal basis
\begin{eqnarray*}
\ke{\mathcal S}\otimes\ke r\otimes\ke j &\equiv& \ke{\mathcal S,r,j}\text{, where } |\mathcal S|=q, \textrm{ and }j\not\in\mathcal S,\\
\ke{\mathcal T}\otimes\ke r\otimes\ke j &\equiv& \ke{\mathcal T,r,j}\text{, where } |\mathcal T|=q+1, \textrm{ and } j\in\mathcal T,
\end{eqnarray*}
whose second ({\em data}) register $r$ stores information about $\mathcal S$ or $\mathcal T$ obtained from the oracle. This part of the system is just an ancillary subspace of suitable dimension. We will also use the notation $f(\mathcal S)=\{f(j):j\in\mathcal S\}$ for simplicity. The third register holding $\ke j$ functions somewhat like a coin -- the state $\ket{j}$ unambiguously points to some neighbor of a subset, so we can use it for moving between sets of the type $\mathcal S$ and $\mathcal T$. For the vertex $\mathcal S$ from the first set it is the element that shall be added to get to vertex $\mathcal T$ from the second set (that is why $j \notin \mathcal S$). On the other hand, for vertex $\mathcal T$ from the second set, the element $j\in \mathcal T$ can be removed to get to $\mathcal S$.

The graph structure and the Hilbert space we introduced are very useful, as they require only a single query of the oracle when using the translation operator $\op S$ between the vertices:
\begin{eqnarray*}
\op S\ke{\mathcal S,f(\mathcal S),j} &=& 
	\ket{\mathcal S\cup\{j\}} \ket{f(x_1),\dots,f(x_q),f(j)}\ket{j}=
	\ke{\mathcal S\cup\{j\},f(\mathcal S\cup\{j\}), j},\notag\\
\op S\ke{\mathcal T,f(\mathcal T),j} &=& 
	\ke{\mathcal T\setminus\{j\},f(\mathcal T\setminus\{j\}), j}.
\label{eq:collisionS}
\end{eqnarray*}
This operation changes only the element $j$, so its addition to the set $\mathcal S$ requires a single call of the oracle $f$ to determine $f(j)$. On the deletion of the element $j$, a single call of the oracle is also needed to clean up the register previously holding $f(j)$.

The translation operation handles states that have the oracle information stored in the data register. Before starting the quantum walk, this register needs to be initialized, which takes $q$ calls to the oracle. 

The overall evolution is specified as
\[
\left[(\op S\op C_G)^{2\tau_1}\op P\right]^{\tau_2},
\]
where $\op C_G$ is the standard Grover coin from \eqref{eq:GCbraket}, with $d=N-q$ for the first set (the number of elements we can possibly add to the chosen set $\mathcal S$) and $d=q+1$ for the second set (the number of elements we can remove from the chosen set $\mathcal T$). In particular,
\begin{eqnarray*}
\op C_G\ke{\mathcal S,f(\mathcal S),j} &=& -\ke{\mathcal S,f(\mathcal S),j} + \frac{2}{N-q}\sum_{l\in\all\setminus\mathcal S}\ke{\mathcal S,f(\mathcal S),l},\notag\\
\op C_G\ke{\mathcal T,f(\mathcal T),j} &=& -\ke{\mathcal T,f(\mathcal T),j} + \frac{2}{q+1}\sum_{l\in\mathcal T}\ke{\mathcal T,f(\mathcal T),l}.\\
\label{eq:collisionCG}
\end{eqnarray*}
This operation only affects the coin state (last register), thus it does not require a query to the oracle. After flipping the coin (chosing which element to add/remove), we apply the translation operator $S$. When this happens $2\tau_1$ times, we use the conditional phase-flip operator $P$:
\begin{equation*}
\label{eq:collisionP}
\op P\ke{\mathcal S,f(\mathcal S),j}=\begin{cases}
-\ke{\mathcal S,f(\mathcal S),j}\text{ if some subset of $\mathcal S$ has the property $\mathcal P$,}\\
\phantom{-}\ke{\mathcal S,f(\mathcal S),j}\text{ otherwise.}
\end{cases}
\end{equation*}
The construction that will be presented in a moment is such that $\op P$ doesn't need to be specified for the states from the second set, we may still assume it acts as identity there. Finally, the algorithm consists of repeating the ``walk $2\tau_1$ times, do a conditional phase flip'' combination $\tau_2$ times (see also Fig.~\ref{alg:collision}).

\begin{figure}
\begin{algorithm}
\begin{enumerate}
\item prepare initial state $\ke\psi=\ke\psinit$ according to Eq.~(\ref{eq:collision_init})
\item for $\tau_2$ steps repeat:
\begin{enumerate}
\item for $2\tau_1$ times apply $\op S\op C_G$ on the state $\ke\psi$
\item aplly operation $\op P$ on state $\ke\psi$
\end{enumerate}
\item measurement should with high probability return $\mathcal S$ such that colliding elements are in the set
\end{enumerate}
\end{algorithm}
\caption{\label{alg:collision}Algorithm for distinctness problem repeats Grover-like evolution for time $\tau_1$ without applying phase-flip $\op P$. The phase-flip is applied only afterwards and then the process is repeated for $\tau_2$ times.}
\end{figure}

The initial state is chosen as an equal superposition of all subsets of size $q$, with their data registers initialized, and the third register (the walker) uniformly spread through the available positions not the corresponding set $\mathcal S$ from the first register,
\begin{equation}
\label{eq:collision_init}
\ke\psinit=\frac{1}{\sqrt{c}}\sum_{\substack{\mathcal S\in\all\\ |\mathcal S|=q}}\sum_{j\in\all\setminus\mathcal S}\ke{\mathcal S,f(\mathcal S),j},
\end{equation}
with the normalization constant $c={N\choose q}(N-q)$. The choice of this state on one hand makes no assumptions about the target vertex we shall end in. To assure we end up in a state of the type $\ket{\mathcal S,f(\mathcal S),j}$, the operation $\op S\op C_G$ is repeated an even number of times.

Up to now, we decided what the initial state will be and how the evolution proceeds. It is time to specify $\tau_1$ and $\tau_2$. A quite lengthy derivation in \cite{ChEi05} shows a choice of these parameters leading to an evolution that transforms the initial state $\ke\psinit$ to a state with $\mathcal S$ containing a set with property $\mathcal P$ with high probability, if such a set it exists. It is
\begin{equation}
\label{eq:collision_tau}
\tau_1=\left\lfloor\frac{\pi}{2}\sqrt{\frac{q}{k}}\right\rceil,\qquad\tau_2=\left\lfloor\frac{\pi}{4}\left(\frac{N}{q}\right)^{k/2}\right\rceil,
\end{equation}
where $\lfloor x\rceil$ means the closest integer to $x$. These required repetition numbers are approximations under the assumptions $N,q\gg k\geq1$. Putting the things together and setting $q=N^\mu$, one finds (see also Tab.~\ref{tab:efficiencies}) that the efficiency of the algorithm is
\begin{equation}
\label{eq:subset_efficiency}
O(q+2\tau_1\tau_2)=O\left(N^\mu+N^{\frac{1}{2}[(1-k)\mu+k]}\right).
\end{equation}
This efficiency is smallest (with respect to the choice of $\mu$) when both terms have the same exponents, i.e. for $\mu=k/(k+1)$. The efficiency of this quantum algorithm then is $O(N^{\frac{k}{k+1}})$ of oracle calls. Again, we stress that here the process of determining whether the set $\mathcal S$ contains a subset with the property $\mathcal P$ is considered to be ``fast'', i.e.~it requires no resources\footnote{It doesn't require calling the oracle, only some extra computational resources, such as for the collision problem we need to sort the list $f(x_1),\dots,f(x_q)$ and see if any two of the values match.}. 

Finally, we can compare the query complexity $O\left(N^\frac{k}{k+1}\right)$ of the quantum subset-finding algorithm to the best classical one, where $O(N)$ queries are required both on average and in the worst case if there exists a collision of two elements. If there is no collision, one has to query the oracle $N$ times in order to find, that there is no collision.

\begin{example}
When we set $k=1$ in \eqref{eq:subset_efficiency}, we get an algorithm with query complexity $O\big(N^\mu+$ $N^\frac{1}{2}\big)$. We are free to choose $\mu$ from the interval $[0;1/2]$ where the efficiency remains the same as in Grover's search. In fact, by setting $\mu=0$ we make the vertices singleton sets and the algorithm becomes Grover's search by a quantum walk on a complete graph as in Sec.~\ref{sec:cg}.
\end{example}

\subsubsection{Algorithm for Finding $k$-cliques in Graphs}
\label{sec:cliques}

The following is an extension of the above results, also presented in \cite{Ambainis07,ChEi05}, dealing with a problem of finding complete subgraphs (cliques) of size $k$ in a graph\footnote{Not to be confused with the bipartite graph from Fig.~\ref{fig:distinctness} which just shows a process of construction of the walk performing the search for the clique.} $\mathcal G$ with $N$ vertices. The oracle in this problem is a device that takes two vertices $j$, $m$ of the graph $\mathcal G$ as input and answers whether the pair $jm$ is an edge of $\mathcal G$.

Again, the quantum walk algorithm will work with subsets of $\mathcal G$'s vertices containing $q$, resp.~$q+1$ vertices. 
There are, however, a few necessary changes to what we saw in the previous Section. First, the costs of initialization and application of $\op S$ increase. When initializing the system, we need to know the information (stored in the ancillary data register) about the edges between the vertices contained in the subset, which now requires $O(q^2)$ queries to the oracle. Also, when performing a translation with the operator $\op S$, the addition (or removal) of a vertex $j$ requires $O(q)$ queries to the oracle for the potential edges connected to the vertex $j$. This does not change the exponent $\mu$, but raises the query complexity to $O\big(N^\frac{2k}{k+1}\big)$ oracle calls (see Tab.~\ref{tab:efficiencies}).

This is not the best that one can do, and the efficiency can be improved further by an ingenious approach from \cite{Ambainis07,ChEi05}. Instead of searching for $k$-cliques, the crucial idea is to look for only $(k-1)$-cliques. In particular, we search for subsets with $(k-1)$ elements that fulfill the property that these elements form a $(k-1)$-clique (this changes $\tau_2$, see Tab.~\ref{tab:efficiencies}). At the same time, we also want all these vertices to be connected to one other vertex. This is a ``redefinition'' of the property $\mathcal P$ which now has an impact on the complexity of performing the operation $\op P$. Effectively, $P$ works as an oracle that takes some $\mathcal S$ on input (and either lets it be or gives it a $-1$ phase) and we will thus call it oracle$_2$ as opposed to the oracle giving us information about edges of graph $\mathcal G$ which, for the time being, we shall call oracle$_1$.

\begin{table}
\begin{center}
\begin{tabular}{c||c|c|c}
& $k$-subset & $k$-clique & recursive $k$-clique\\
\hline\hline
initialization & $q$ & $q^2$ & $q^2$\\
$\op S$ & 1 & $q$ & $q$\\
$\op C_G$ & 0 & 0 & 0\\
$\op P$ & 0 & 0 & $\sqrt{N}\times N^{\frac{k-1}{k}}$\\
\hline
$\tau_1$ & $\sim\sqrt{q}$ & $\sim\sqrt{q}$ & $\sim\sqrt{q}$\\
$\tau_2$ & $\sim\left(\frac{N}{q}\right)^{\frac{k}{2}}$ & $\sim\left(\frac{N}{q}\right)^{\frac{k}{2}}$ & $\sim\left(\frac{N}{q}\right)^{\frac{k-1}{2}}$\\
\hline\hline
query complexity & $q+2\tau_1\tau_2$ & $q^2+2q\tau_1\tau_2$  & $q+\tau_2\left(2q\tau_1+\sqrt{N}\times q^\frac{k-1}{k}\right)$\\
\hline
optimal $\mu$ & $\frac{k}{k+1}$ & $\frac{2k}{k+1}$ & $\frac{5k-2}{2k+4}={1.3}$ for $k=3$\\
&&& \qquad\,$\frac{2(k-1)}{k}$ for $k>3$\\
\end{tabular}
\end{center}
\caption{\label{tab:efficiencies} Summary of the query complexity for quantum walk algorithms for $k$-subset finding, $k$-clique finding and recursive $k$-clique finding. The algorithms involve a walk on subsets of size $q=N^\mu$. The bottom part of the table shows the summary complexity and the best choice for the exponent $\mu$, resulting in query complexity $N^\mu$.}
\end{table}

The oracle$_2$ can be implemented by an algorithm that performs a search for a vertex that is fully connected to some $(k-1)$-clique within $\mathcal S$. Whether some vertex $j$ has this property shall be provided to us by another oracle$_3$. If we had it at hand, then searching for the special vertex could be done by a Grover search on $O(N)$ vertices (elements), having to query oracle$_3$ $O(\sqrt{N})$ times.

We are able to construct (at least in principle) this oracle$_3$, having some vertex $j$ and a subset $\mathcal S$ as input. We know how to search for a $(k-1)$-clique within the set $\mathcal S$, having $q$ elements. The search for this $(k-1)$ clique connected to $j$ is application of a standard $(k-1)$-subset finding procedure within $\mathcal S$, where the oracle$_1$ sets the property $\mathcal P$ for this sub-search. Checking whether some $(k-1)$-element subset of $\mathcal S$ connected to $j$ is a clique requires no additional queries to oracle$_1$, as this information is already stored in the ancillary data register state corresponding to $\mathcal S$. The oracle$_3$ thus needs to call oracle$_1$ $O\big(q^\frac{k-1}{k}\big)$ times. Knowing the complexity of oracle$_3$, we can determine 
the complexity of calling oracle$_2$, i.e.~of the operation $\op P$ -- it is $O\big(\sqrt{N}\times q^\frac{k-1}{k}\big)$.
Putting it all together, the combined algorithm for $k$-subset finding requires 
\[
O\left(q+\tau_2\left(2q\tau_1+\sqrt{N}\times q^\frac{k-1}{k}\right)\right)
\]
calls to the\footnote{From this point on oracle$_1$ is again called only ``the oracle''.} oracle$_1$. 

Let us look at the result for various $k$. For $k=2$ (we look for an edge) the optimal efficiency is that of Grover's search on the $N(N-1)$ possible edges, i.e.~$O(N)$ calls to the oracle. However, for $k>2$, the presented algorithm is more efficient than direct Grover's search for $k$-cliques which requires $O(N^{k/2})$ calls to the oracle. First, a brute force querying of
all the edges and then performing a search on the received information has query complexity $\Omega(N^2)$, besting direct Grover's search for $k>4$. However, the presented algorithm does even better. For $k=3$ (triangle finding) the recursive algorithm given above can find a triangle with $O(N^{1.3})$ calls to the oracle (although the lower bound for this problem is so far only $O(N)$ - see Example \ref{ex:star_bound}). For $k>3$, its complexity goes as $O\big(N^\frac{2(k-1)}{k}\big)$, but again, it is not known whether this is the optimum.

Note that all these algorithms were presented under the assumption that there is exactly one solutions or none at all. If there would be more solutions, we could use the approach given in \cite{Ambainis07}, which preserves the efficiencies and speedups over the classical case. There is a number of discrete quantum walk algorithms that are more efficient than the best possible classical ones and are based on the algorithm of $k$-subset finding. Two of the examples are the verification of matrix products \cite{BuSp06} and testing the commutativity of a black-box group \cite{MaNa07}.

\subsection{Summary}

In this Chapter, we explored the potential of discrete time quantum walks, focusing on search algorithms. We have seen they offer enough potential for devising new and more efficient algorithms for several problems --- searches on hypercube \cite{ShKeWh03}, complete graph \cite{ReHiFeBu09}, lattices \cite{AmKeRi05} or for anomalies in symmetry \cite{FeHi+10}, collision problem \cite{Ambainis07}, finding triangles in graphs \cite{MaSaSz05}, verifying matrix products \cite{BuSp06} or testing the commutativity of a black-box group \cite{MaNa07}. 

Viewed in a slightly abstract manner, quantum walks can be used in many oracle problems spanning from unstructured search to searching for graph substructures. Separating the oracle (which holds information about a set of elements) and the graph underlying the quantum walk allows one to make good choices resulting in clever algorithms as the one for $k$-subset finding. The simpler oracle algorithms can be subsequently employed as subroutines in more elaborate algorithms. We have looked at $k$-subset finding, where the search is performed on subsets of the set of elements rather then on elements themselves.
This algorithm was used as a subroutine in the algorithm that finds $k$-cliques in a graph. 
\setcounter{equation}{0} \setcounter{figure}{0} \setcounter{table}{0}\newpage
\section{Quantizing Markov Chains}
\label{sec:qmc}

In this Chapter, we will look at how to obtain discrete quantum walks
from {\em any Markov Chain}, which will result in a quantum speedup
for many classical algorithms. It is hard to quantize Markov chains that are not regular. In particular, we would have to define a different ``coin'' at each vertex, which presents encoding difficulties. That's why Szegedy \cite{Szegedy} took a different approach, using the state of the quantum walk itself as the basis for diffusion instead of an external coin register. Taking this route again results in a quantum walk that has some kind of ``memory'', as the unitarity of each transformation implies dependence on where we came from, in contrast to classical random walks.

We start with the definition of quantum walks on systems with two registers, analyze their spectra and prove some speedup results for hitting times. We then turn our attention to sampling (Monte-Carlo Markov Chain) algorithms. Finally, we take a look at quantum Metropolis sampling \cite{qmetropolis,qmetropolis2}. 


\subsection{Walks on Two Registers}
Let us recall a discrete-time quantum walk
on a regular degree-$d$ graph which uses the Grover coin $C_G$.
The state of the system is contained in two registers (vertex, coin), with Hilbert space of dimension $Nd$. When we start walking at vertex $\ket{x}$ with the coin in the state $\ket{c}$, a step of the walk results in
\be
	\ket{x}\ket{c}_{coin} 
	\stackrel{\textrm{diffuse}}{\longrightarrow} 
	\ket{x}\sum_{\tilde{c}=1}^{d}(C_G)_{c\tilde{c}}\ket{\tilde{c}}_{coin}
	\stackrel{\textrm{shift}}{\longrightarrow} 
	\sum_{\tilde{c}=1}^{d} (C_G)_{c\tilde{c}} \ket{x\oplus \tilde{c}}\ket{\tilde{c}}_{coin},
	\label{TRcoin}
\ee
a superposition over the neighbors of $\ket{x}$ in the vertex register, with corresponding states of the coin in the coin register.
In Section \ref{sec:sqws}, we viewed it as a scattering quantum walk in a system with two registers containing a target vertex and a source vertex. The Hilbert space for such a walk has dimension $\C^N\otimes\C^N$.  
When a scattering walk starts in the state that ``moves'' from vertex $\ket{x\oplus c}$ towards the vertex $\ket{x}$, one step brings it to
\be
	\ket{x}\ket{x\oplus c} 
	\stackrel{\textrm{diffuse}}{\longrightarrow} 
	\ket{x} \sum_{\tilde{c}=1}^{d}(C_G)_{c\tilde{c}}\ket{x \oplus \tilde{c}}
	\stackrel{\textrm{swap registers}}{\longrightarrow} 
	\sum_{\tilde{c}=1}^{d} (C_G)_{c\tilde{c}} \ket{x\oplus \tilde{c}} \ket{x},
	\label{TRscatter}
\ee
a superposition of states originating in $\ket{x}$ and ``going'' towards the neighbors of $\ket{x}$. Note that the diffusion in \eqref{TRcoin} and \eqref{TRscatter} is governed by the same $d\times d$ diffusion matrix $C_G$.
It turns out we can express the unitary operator for the Grover diffusion step as a reflection operator on the whole two-register Hilbert space. Let us define the state
\be
	\ket{\alpha_x} &=& \ket{x} \sum_{\tilde{c}=1}^{d} 
		\frac{1}{\sqrt{d}} \ket{x\oplus \tilde{c}}.
\ee 
The reflection of the state $\ket{x}\ket{x\oplus c}$ about $\ket{\alpha_x}$ is
\be
	\left(2\ket{\alpha_x}\bra{\alpha_x}-\iii\right) \ket{x}\ket{x\oplus c}
		&=& \frac{2}{\sqrt{d}} \ket{\alpha_x} - \ket{x}\ket{x\oplus c} \\
		&=& \ket{x}\left[ \left(\frac{2}{d}-1\right)\ket{x\oplus c}
		+		
		\frac{2}{d}\sum_{\tilde{c}\neq c} \ket{x\oplus \tilde{c}}\right]\\
		&=& \ket{x}
		\sum_{\tilde{c}=1}^{d}(C_G)_{c\tilde{c}}\ket{x \oplus \tilde{c}},
\ee
recalling the definition of the Grover coin $C_G$ from \eqref{eq:GCbraket}, where we analyzed coins for discrete-time walks.
Not restricting ourselves to a specific vertex $x$, 
we now define the projector
\be
	\Pi_{G} = \sum_{x} \ket{\alpha_x}\bra{\alpha_x}.
\ee
A generalization of the above computation shows that the reflection
$2\Pi_G -\iii$
is exactly the diffusion step on the whole two-register space of our scattering quantum walk.
The scattering quantum walk with the Grover coin on a $d$-regular graph can thus conveniently be written as
\be
	W_{G} = S (2\Pi_{G} - \iii),
\ee
where $S = \sum_{x,y} \ket{x,y}\bra{y,x}$ is an operator swapping the two registers.
Such two-register scattering quantum walks (SQW) on regular graphs have a connection to classical walks whose next step chooses uniformly at random among the neighbors of each vertex. It remains a problem to concisely describe quantum walks that correspond to asymmetric classical walks, or to walks with a general stochastic\footnote{each row sums to 1} transition matrix $P_{x,y}$.

In the case the diffusion is different at different vertices, we will again utilize a system with two-registers (vertex,vertex), related to scattering quantum walks. 
Using Szegedy's generalization \cite{Szegedy}, instead of reflecting about the uniform states $\ket{\alpha_x}$, we now define a state
\be
	\ket{\phi_x} &=& \ket{x} \otimes \left(\sum_{y\in X} \sqrt{P_{x,y}} \ket{y}\right) \label{bipartphi}
\ee
for every vertex $x$. We then choose the unitary for the diffusion to be a reflection about the subspace spanned by all the states $\ket{\phi_x}$, i.e.
\be
	R_1 = 2 \left(\sum_{x\in X} \ket{\phi_x}\bra{\phi_x}\right) - \iii. \label{tworegisterreflection}
\ee
As for the SQW, the diffusion step is followed by $S$, a swap of the registers.
Applying this twice, we get Szegedy's quantization of a Markov chain with a transition matrix $P$. The (composed) step of the walk is then  
\be
	W = S R_1 S R_1 = R_2 R_1, \label{MCqwalk}
\ee
a product of two reflections, with $R_2$ given by 
\be
	\ket{\psi_y} &=& \left(\sum_{x\in X} 
			\sqrt{P_{x,y}}\ket{x}\right) \otimes \ket{y}, \label{bipartpsi}\\
	R_2 &=& 2 \left(\sum_{y\in Y} \ket{\psi_y}\bra{\psi_y}\right) - \iii.
\ee

We measure the position of the walker in a two-register state $\ket{\psi}$ by measuring only the first register. The probability of finding the walker at vertex $x$ is thus 
\be
	p(x) = \bra{\psi}\left( \ket{x}\bra{x}\otimes \iii \right)\ket{\psi}.
\ee

Note that for a symmetric Markov chain (with $P_{x,y} = P_{y,x}$), the only
state invariant under both of these reflections (and thus under $W=R_2 R_1$) is
\be
	\ket{\psi} = \frac{1}{\sqrt{N}}\sum_{x\in X}\ket{\phi_x} 
	= \frac{1}{\sqrt{N}} \sum_{y\in Y}\ket{\psi_y}
	= \frac{1}{\sqrt{N}} \sum_{x\in X}\sum_{y \in Y} \sqrt{P_{x,y}} \ket{x}\ket{y},
\ee
a state with probability $p(x)=\frac{1}{\sqrt{N}}$ for each $x$ (because $\sum_{y\in Y} P_{x,y} = 1$). 
This quantum state corresponds to the uniform-superposition stationary state 
of the classical Markov chain, and obeys the detailed balance equations
\be
	P_{x,y} \, p(x) = P_{y,x} \, p(y).
\ee


\subsection{The Spectrum of the Walk}
\label{sec:walkspectrum}
We have written the quantum walk \eqref{MCqwalk} as a product of two reflections.
Let us investigate the ``mixing'' properties of this type of walk and how
we can connect them to the mixing properties of the original Markov Chain. 

In many classical algorithms, Markov Chains are used for their fast mixing towards their stationary states.
Do quantum walks bring anything new to the table? In what sense do unitary walks mix? We have discussed
mixing previously in Section \ref{sec:discretemixing}, where we defined mixing towards
a limiting distribution in a time-averaged sense. We now want to look at how fast (and how close) 
we are getting towards the state that encodes the stationary distribution. 
For this, we will need to understand the spectrum of the quantum walk unitary operator.
Let us start by investigating the action of two reflections. It will be helpful to recall Jordan's lemma.
\begin{lemma}[Jordan '75]
For any two Hermitian projectors $\Pi_1$ and $\Pi_2$, there exists an
orthogonal decomposition of the Hilbert space into one dimensional
and two dimensional subspaces that are invariant under both $\Pi_1$ and $\Pi_2$.
Moreover, inside each two-dimensional subspace, $\Pi_1$ and $\Pi_2$ are
rank-one projectors.
\end{lemma}
The two reflections $R_1$ and $R_2$ reflect around the subspaces defined by the projectors $\Pi_1$ and $\Pi_2$. Jordan's lemma implies that we can rewrite the Hilbert space as an orthogonal sum of 1D and 2D subspaces invariant under $\Pi_1$ and $\Pi_2$. Consequently, these subspaces are also invariant under the reflections $R_1$ and $R_2$. Moreover, within each 2D subspace, two reflections compose into a rotation.

Szegedy proved a spectral theorem for the quantum walk $W=R_2 R_1$. We now present a slightly different version, including the 1D and 2D invariant space decomposition intuition from Jordan's lemma. 

\begin{theorem}
Consider two Hermitian projectors $\Pi_1$, $\Pi_2$ and the identity operator
$\iii$. 
The unitary operator $(2 \Pi_2-\iii)(2 \Pi_1-\iii)$ has eigenvalues $e^{\pm
i 2 \theta_j }$,  $0<\theta_j <\frac \pi 2$  in the two-dimensional
subspaces $S_i$ invariant under $\Pi_1$ and $\Pi_2$, and it has eigenvalues
$\pm 1$ in the one-dimensional subspaces invariant under $\Pi_1$ and $\Pi_2$.
\end{theorem}

The following proof comes from \cite{ChildsNotes}. The 1D invariant subspaces are spanned by the common eigenvectors of $\Pi_1$ and $\Pi_2$,
making the product of two reflections either $\iii$ or $-\iii$.  
What is more interesting, in each 2D subspace invariant under the two projectors, a product of two reflections is a rotation by $2\theta_j$, the angle between the axes of the two bases of the 2D subspace related to $\Pi_1$ and $\Pi_2$. 
We can find the Hilbert space decomposition 
and the angles $\theta_j$ with the help of the Hermitian {\em discriminant matrix} with entries
\be
	D_{x,y} = \sqrt{P_{x,y} P_{y,x}}
\ee
of the Markov chain. In the case of a symmetric chain, it is equal to the matrix $P$ itself.
Let the eigenvectors of $D$ be $\ket{\lambda_j}$ with eigenvalues $\lambda_j$. 
Define an isometry from $\C^n$ to $\C^n \otimes \C^n$ by
\be
	T = \sum_{x} \ket{\psi_x}\bra{x} 
	= \sum_{x,y} \sqrt{P_{x,y}}\ket{x}\ket{y}\bra{x}.
\ee
We claim that each two-dimensional invariant subspaces can be constructed as  
$\textrm{span}\{\kets{\tilde{\lambda}_j},S\kets{\tilde{\lambda}_j}\}$,
where the vectors $\kets{\tilde{\lambda}_j}=T\ket{\lambda_j}$ come from the eigenvectors of $D$
with the help of the isometry $T$. To see that this is indeed so, we first observe that
\be
	T T^\dagger &=& \sum_{x,y} \ket{\psi_x}\braket{x}{y}\bra{\psi_y} 
			= \sum_{x} \ket{\psi_x}\bra{\psi_x} = \Pi_1, \\
	T^\dagger T &=& \sum_{x,y} \ket{x}\braket{\psi_x}{\psi_y}\bra{y} 
			= \sum_{x,y,w,z} \sqrt{P_{x,w} P_{y,z}} \ket{x}\braket{x}{y}\braket{w}{z}\bra{y}
			  \nonumber\\
			&=& \sum_{x,w} P_{x,w} \ket{x}\bra{x} = \sum_{x}\ket{x}\bra{x} = \iii, \\
	T^\dagger S T &=& \sum_{x,y} \ket{x}\bra{\psi_x}S\ket{\psi_y}\bra{y}  
					= \sum_{x,y,z,w} \sqrt{P_{x,z}} \sqrt{P_{y,w}}
							\bra{x}\bra{x}\bra{z} 
							S  \ket{y}\ket{w}\bra{y}  \nonumber\\
					&=& \sum_{x,y} \sqrt{P_{x,y} P_{y,x}}
								\ket{x}\bra{y} = D,
\ee
and then check that each subspace $\textrm{span}\{\kets{\tilde{\lambda}_j},S\kets{\tilde{\lambda}_j}\}$ is indeed invariant:
\be
	R_1 \kets{\tilde{\lambda}_j} &=& (2\Pi_1 - \iii) T \ket{\lambda_j} 
			= T\ket{\lambda_j} = \kets{\tilde{\lambda}_j}, \\
	R_1 S \kets{\tilde{\lambda}_j} &=& (2\Pi_1 - \iii) ST \ket{\lambda_j}
		= (2TT^\dagger - \iii) ST \ket{\lambda_j} \nonumber\\
		&=& (2 T D - ST) \ket{\lambda_j} = 2\lambda_j \kets{\tilde{\lambda}_j} 
					- S\kets{\tilde{\lambda}_j}, \\
	R_2 \kets{\tilde{\lambda}_j} &=& (SR_1 S) \kets{\tilde{\lambda}_j} 
			= 2\lambda_j S\kets{\tilde{\lambda}_j} 
					- \kets{\tilde{\lambda}_j}, \\
	R_2 S \kets{\tilde{\lambda}_j} &=& S R_1 S S \kets{\tilde{\lambda}_j}
			= S R_1 \kets{\tilde{\lambda}_j} = S \kets{\tilde{\lambda}_j}.
\ee
This in turn helps us relate the eigenvalues of $W=R_2 R_1$ to the
eigenvalues of $D$. Jordan's lemma implies the action of $W$ is a rotation
in each 2D subspace, so its eigenvalues are $e^{\pm i \theta_j}$,
where $\theta_j$ is the angle between the $+1$ eigenvectors of $\Pi_1$ and $\Pi_2$
in this subpspace. Let us calculate these $\theta_j$. First, recall that $\Pi_1 \kets{\tilde{\lambda}_j}=\kets{\tilde{\lambda}_j}$ and $\Pi_2 = S\Pi_1 S$, 
which means that $T^\dagger \Pi_2 T =T^\dagger S T T^\dagger S T =D^2$. Thus
\be
	\cos \theta_j
	= \frac{
	\bras{\tilde{\lambda}_j}\Pi_2 \Pi_1 \kets{\tilde{\lambda}_j}
	}{
	\sqrt{
		\bras{\tilde{\lambda}_j}\Pi_2\kets{\tilde{\lambda}_j}
		\bras{\tilde{\lambda}_j}\Pi_1\kets{\tilde{\lambda}_j}
		}
	}
	= \sqrt{
	\bras{\lambda_j} T^\dagger \Pi_2  T \kets{\lambda_j}
	}
	=
	\sqrt{
	\bras{\lambda_j} D^2 \kets{\lambda_j}
	}
	= \lambda_j.
\ee
The eigenvalues of $W$ are thus $\pm 1$ in the 1D invariant subspaces and 
\be
	e^{\pm i2\theta_j} 
	= e^{\pm i2\arccos \lambda_j}
	= (2\lambda_j^2 - 1) \pm i2\lambda_j\sqrt{1-\lambda_j^2},
\ee
in the 2D invariant subspaces (this is a form seen in Szegedy's papers).

\subsection{Speeding up Searching for Marked Vertices}

With an understanding of the spectrum of the quantum walk unitary, we now turn our attention
to possible square root speedups over its classical counterpart. 
While the number of necessary classical Markov Chain steps depends on the gap $\delta$ of the classical 
chain as $f(\epsilon,\delta)$. Szegedy has shown that the quantum walk will need to only take $f(\sqrt{\epsilon},\sqrt{\delta})$ steps.
The speedups thus appear in shorter hitting times when searching, and in the later sections we will see them 
in shorter required times to get within distance $\epsilon$ of the stationary distribution. 

Let us look at a rather generic search algorithm based on a Markov chain $P$ (with spectral gap $\delta$) on set $X$, with a marked subset $M$. The goal is to find one of the vertices of the subset $M$. We will see that a quantum version of this algorithm has a ``square-root speedup'' over its classical counterpart. This section also follows \cite{ChildsNotes}.

First, we start with a classical algorithm, and modify the Markov chain so that it stays in a marked vertex $x\in M$ once we hit it. For this, we choose the transition matrix to be 
\be
	P'_{x,y} = \begin{cases}
		0 & x\in M, x\neq y, \\
		1 & x\in M, x=y, \\
		P_{x,y} & x\notin M.
	\end{cases}
\ee
It has block form
\be
	P' = \left[
		\begin{array}{cc}
		P_M & 0 \\
		B & \iii
		\end{array}
	\right],
\ee
where $P_M$ corresponds to the rows/columns of the original $P$ that are not marked (in $M$).
We will now show that when the fraction of marked vertices is $\frac{|M|}{N} \leq \ep$ and the second largest eigenvalue of $P$ is lower than $(1-\delta)$, the classical hitting time is lower bounded by $O\left(\frac{1}{\delta\ep}\right)$. The decision problem (is there a marked vertex at all) complexity will also have the same lower bound.

First, let us analyze the classical chain. When we take $t$ steps of the walk, we get
\be
	(P')^t = \left[
		\begin{array}{cc}
		P_M^t & 0 \\
		B+BP_M+BP_M^2 + \cdots + BP_M^t  & \iii
		\end{array}
	\right]
	= \left[
		\begin{array}{cc}
		P_M^t & 0 \\
		B\frac{P_M^t-\iii}{P_M-\iii} & \iii
		\end{array}
	\right].
\ee
We start in the uniform distribution over unmarked vertices (by choosing a random unmarked vertex as the starting point). Denoting 
$
     \ket{o} = \frac{1}{\sqrt{N-|M|}}\sum_{x\notin M} \ket{x},
$
we can express the probability of {\em not} reaching a marked vertex
after $t$ steps as
\be
	p^-_t = \bra{o}P_M^t \ket{o} \leq \norm{P_M^t} = \norm{P_M}^t,
	\label{uniformunmarked}
\ee
using the operator norm (largest eigenvalue) of the matrix. When
$\norm{P_M} \leq 1-\Delta$, 
\be
	p^+_t = 1-p^-_t \geq 1-\norm{P_M}^t \geq 1-(1-\Delta)^t.
		\label{MChitprob}
\ee
It is then enough to take $t = O\left(\frac{1}{\Delta}\right)$ to ensure an $\Omega(1)$ success probability.

Finally, we can relate $\norm{P_M}$ to the second largest eigenvalue $(1-\delta)$ of the original transition matrix $P$, and the fraction $\frac{|M|}{N} \leq \ep$ of marked vertices.
The original matrix $P$ is symmetric, so its principal eigenvector with eigenvalue $1$ is $\ket{s} = \frac{1}{\sqrt{N}} \sum_{x} \ket{x}$. 
Let $\ket{w} \in \C^N$ be the principal eigenvector of the matrix $P_M$ (which has size $N-M$), padded with zeros on the marked vertices.

Using the Cauchy-Schwartz inequality, we can upper bound the overlap
\be
	|\braket{s}{w}|^2 = |\bra{s}\Pi_{x\notin M} \ket{w}|^2
	\leq \norm{\Pi_{x\notin M}\ket{s}}\cdot \norm{\ket{w}}^2
	= \frac{N-|M|}{N} = 1-\ep.
\ee
Next, using the eigenvectors $\ket{\lambda}$ of $P$, we can express
\be
	\ket{w} &=& \braket{s}{w}\ket{s} 
				+ \sum_{\lambda\neq 1}\braket{\lambda}{w}\ket{\lambda}, \\
	\norm{P_M}^2 &=& \norm{P\ket{w}}^2 = 
		|\braket{s}{w}|^2 + \sum_{\lambda\neq 1}|\braket{\lambda}{w}|^2 \lambda^2 \\
		&\leq&
		|\braket{s}{w}|^2 + (1-\delta)^2 \sum_{\lambda\neq 1}|\braket{\lambda}{w}|^2  \\
		&=&
		|\braket{s}{w}|^2 [1- (1-\delta)^2] + (1-\delta)^2 \sum_{\mathrm{all\,}\lambda}|\braket{\lambda}{w}|^2  \\
		&\leq&
		(1-\ep) (2\delta - \delta^2)+ (1-\delta)^2 \\
		&\leq&
		1-2 \ep \delta,
\ee
because the first eigenvalue of $P$ is 1 and all the other eigenvalues are upper bounded by $(1-\delta)$.
Therefore, $\norm{P_M} \leq \sqrt{1-2\delta\ep} \leq 1- \delta \ep$.
Together with \eqref{MChitprob}, this means 
we need to take $O\left(\frac{1}{\delta\ep}\right)$ steps of the walk to get a constant success probability for this classical random walk search algorithm.

Let us now look at a quantum walk algorithm for the same task. If we start in a uniform superposition and measure whether we get a marked vertex, success is unlikely and we end up in a uniform superposition over unmarked vertices. However, we can now perform phase estimation of the unitary quantum walk $W$ instead. We have seen how we can relate its eigenvalues to the discriminant matrix, which in this case is 
\be
	D = \left[
	\begin{array}{cc}
		P_M & 0\\
		0 & \iii
	\end{array}
	\right].
\ee
If there are no marked vertices, $\ket{o}$ \eqref{uniformunmarked} is the principal eigenvector of the quantum walk $W$ with eigenvalue one, and phase estimation \cite{NielsenChuang} of $W$ returns $0$. On the other hand, if marked vertices exist, $\ket{o}$ lives in the ``busy'' subspace of the quantum walk (corresponding to eigenvalues $e^{\pm i2\theta_j}$,
where $\lambda^M_j=\cos \theta_j$ are the eigenvalues of $P_M$).
Phase estimation of $W$ would thus return a phase greater than
\be
	\phi_0 = 2\, \textrm{min}_j |\theta_j| \geq 2\, \textrm{min}_j \arccos \lambda^M_j 
	\geq 2\, \textrm{min}_j \sqrt{1-\lambda^M_{j}}
	=
	2 \sqrt{1-\norm{A}} \geq 2 \sqrt{\delta\ep}.
\ee
Phase estimation of a unitary $W$ with precision $\phi_0$ takes $O\left(\phi_0^{-1}\right)$ evaluations of $W$, so a quantum walk algorighm deciding between no marked vertices and a fraction of $\ep$ marked vertices will take $O\left(\frac{1}{\sqrt{\delta\ep}}\right)$ steps, a quadratic speedup over its classical counterpart described above.
This is the result of Szegedy's $\delta\ep$-paper \cite{Szegedy}. 

Note that we did not claim the classical algorithm was optimal for a particular search problem -- we only compared classical and quantum versions of these generic approaches. Furthermore, we have analyzed a quantum walk algorithm that only {\em decides} whether marked vertices exist, and does not output one. However, by adding an additional layer of ``marking'' vertices, we could obtain a marked-vertex-identifying algorithm with at most logarithmic overhead. On the other hand, sampling from the set of marked vertices (or according to some probability distribution) is a different and interesting problem, and we will look at it in the following Sections.

\subsection{Walks and Sampling}

We have seen that quantization of classical Markov chains has been crucial in the design of efficient quantum algorithms for a wide range of search problems \cite{Santha} that outperform their classical counterparts. 
We now extend the scope of use of quantum walks (quantized Markov chains) beyond search problems. They can be employed to speed up sampling from probability distributions. This results in a variety of quantum algorithms, including quantum simulated annealing, fully polynomial-time quantum approximation schemes for partition functions, and the quantum Metropolis algorithm.

Sampling from the stationary distributions of Markov Chains is a strong classical algorithmic tool, useful for counting (\#P) problems, as described in Section \ref{sec:MCMC}. Classically, Aldous has shown \cite{Aldous} that the mixing time (the number of steps guaranteeing closeness to the stationary distribution) for a Markov Chain is related to its spectral gap $\delta$ and the minimum of the distribution $\pi_{*}$ as
$O(\delta^{-1} \log 1/\pi_{*})$. The question is whether we can do better using quantum walks. Richter \cite{Richter1}, introducing decoherence into quantum walks generated a classically converging Markov Chain, and proved that on a periodic lattice $\mathbb{Z}_n^d$, the mixing time gets a square root speedup (with respect to the spectral gap) to $O(\sqrt{\delta^{-1}} \log 1/\pi_{*})$. 
Another interesting application of quantization of classical MC's is simulated annealing which involves
Markov Chains whose stationary distributions correspond to Gibbs distributions at particular temperatures.
Bringing quantum mechanics to the picture, Somma et al. \cite{Somma2} proposed a quantum simulation of classical annealing, using quantized Markov chains, and relying on the quantum Zeno effect and phase estimation. They sequentially prepare coherent state encodings of the stationary distribution of the MC's, using a simulation of projective measurements. This results in a quantum algorithm with an $O(\sqrt{\delta^{-1}})$ dependence on the spectral gap (again a square root speedup), with some additional factors. 
We discuss these two approaches to sampling in Section \ref{sec:MixDecohere}, describe the Monte-Carlo Markov Chain (MCMC) method in detail in \ref{sec:MCMC}, look at a general approach to its quantization \cite{NW08} in \ref{sec:quantizeMCMC} and showcase some of its applications and speedups. We then turn our attention to the quantum Metropolis algorithm \cite{qmetropolis} in Section \ref{sec:Metropolis}.



\subsubsection{Speeding up Mixing Using Quantum Walks}
\label{sec:MixDecohere}

We have seen that sampling from stationary distributions of Markov Chains has very useful applications. 
How could one speed up the preparation of these states using quantum walks? Unitarity prohibits us from
talking about mixing in the classical sense. However, according to Richter \cite{Richter1} we can use a quantum walk to get a random process by combining \begin{enumerate}
\item time evolution according to the transition rule $U$ (discrete with $U^t$ or continuous with $U(t)$), and 
\item a measurement in the computational basis, evaluated at a time chosen randomly according to a measurement rule, given by a probability distribution $\omega_T$ (e.g. the uniform distribution $\bar{\mu}_T=\frac{1}{T}\chi_{[0,T]}$, the delta-function distribution $\delta_T(t)=\delta(t-T)$, or other). 
\end{enumerate}
Thus, we can view quantum walks as the pair $\langle U,\omega_T\rangle$. Repeating the unitary evolution and random-time measurement many times gives us an algorithm which probabilistically outputs\footnote{For each sample, we pick some simple initial state (or the previously measured vertex), let the quantum walk run for some time, and measure. With this measurement, the superposition randomly collapses to a vertex, producing a sample.} a vertex -- the probability distribution for this output vertex converges to a stationary distribution just as its classical Markov Chain counterpart did. Let us ask how fast this happens.

The Aldous theorem for a classical reversible, ergodic Markov Chains with stationary distribution $\pi$ and spectral gap $\delta$ states that the mixing time obeys
\be
	\frac{1}{\delta} \leq \tau_{mix}\leq \frac{1}{\delta}\log \frac{1}{\pi_*},
\ee
where $\pi_*=\min_x \pi(x)$.
Could we speed up this dependence on the spectral gap in a square root fashion by using quantum walks?
Richter \cite{Richter1} showed that on a periodic lattice $\mathbb{Z}_n^d$, the mixing time 
for the quantum decohering walk gains a square root speedup (with respect to the spectral gap), 
as it converges with mixing time $O(\sqrt{\delta^{-1}} \log 1/\pi_{*})$. It remains open whether this speedup can be achieved in general, for (m)any quantized Markov Chains.

Classical simulated annealing \cite{KiGeVe83, Cerny85} imitates the process where a metal is heated to a high temperature and then slowly cooled down. This is supposed to allow thermal excitations to jump out of local minima, letting the system end up in a low-energy state at the end of the process. The thermalization is modeled by a Markov Chain, whose stationary distribution corresponds to the Gibbs distribution at a given temperature. Again, we recall the Aldous theorem which says the mixing time for a MC scales with the spectral gap $\delta$ as $O(\delta^{-1})$. Could we speed this process up using quantized Markov Chains? In \cite{Somma2}, Somma et al. use a sequence of quantized (two-register) Markov chains, but instead of preparing probability distributions $\pi_i$, they look at coherent states $\ket{\pi_i} = \sum_{x} \sqrt{\pi_i(x)} \ket{x}$ encoding them. When we have access to the state $\ket{\pi_i}$, could we use it to prepare the next state $\ket{\pi_{i+1}}$? As we lower the temperature slowly, the states are close. Thus, it is likely that a projection onto $\ket{\pi_{i+1}}$ would succeed. This projective measurement is simulated by phase estimating the walk operator $W_{i+1}$ for the next quantum MC, as $\ket{\pi_{i+1}}$ is its eigenvector with eigenvalue 1. The quantum Zeno effect \cite{NielsenChuang} and large overlaps $\braket{\pi_i}{\pi_{i+1}}$ are responsible for the high success probability of each step, and consequently, the overall procedure. The resulting algorithm \cite{Somma2} requires $O(\sqrt{\delta^{-1}})$  implementations of a step of a quantum MC (with some additional factors).

\subsubsection{Markov Chain Monte Carlo (MCMC) Methods}
\label{sec:MCMC}

%
%

Sampling from stationary distributions of a sequence of Markov chains, combined with simulated
annealing (progressive lowering of a temperature parameter) lies at the heart of 
many important classical approximation algorithms. These methods are in general called {\em Markov Chain Monte Carlo} (MCMC).
Some out of the many examples include the approximation algorithms for the volume of convex
bodies \cite{Vempala}, the permanent of a non-negative matrix \cite{Vigoda},
and the partition function of statistical physics models such as the Ising
model \cite{Jerrum2} and the Potts model \cite{Vazirani}. Each of these algorithms is a {\em fully
polynomial randomized approximation scheme} (FPRAS), outputting a random number
$\hat{Z}$ within a factor of $(1\pm \ep)$ of the real value $Z$, with probability 
greater than $\frac{3}{4}$, i.e.
\begin{equation}
	\Pr\big[ (1-\epsilon)Z \le \hat{Z} \le (1+\epsilon)Z \big] \ge \frac{3}{4},
	\label{fpras}
\end{equation}
in a number of steps\footnote{Note that these FPRAS can not be used to {\em solve} counting problems when $Z$ is exponentially large (in the problem size). The precision $\ep$ would then have to be exponentially small, and the number of required algorithmic steps would grow exponentially.} polynomial in $1/\ep$ and the problem size. 
				
The classical algorithms can be outlined as follows.
Consider a physical system with state space $\Omega$ and an energy function
$E: \Omega\rightarrow \R$, assigning each state $\sigma\in\Omega$
an energy $E(\sigma)$.	
Our task is to estimate the Gibbs partition function
\begin{eqnarray}
		         Z(T) = \sum_{\sigma \in \Omega} e^{-\frac{E(\sigma)}{kT}}
		         \label{partition}
\end{eqnarray}
at a low final temperature $T_F$. The value of $Z$ at zero temperature is interesting, as it is equal to the number of the system configurations with zero energy\footnote{This relationship is used e.g. in the algoritm \cite{Vigoda} for approximating the permanent of a non-negative matrix -- one can find the value of the permanent by counting the number of perfect matchings of a particular bipartite graph, which in turn is equal to the zero-temperature partition function of a certain spin system.}. This could be a solution to a counting problem, such as providing us the number of graphs with a particular property. 

The partition function $Z(T)$ encodes the 
		thermodynamical properties of the system in equilibrium at temperature $T$,
		where the probability of finding the system in state $\sigma$ is given by the Boltzmann distribution
		\begin{eqnarray}
		         \pi(\sigma) = \frac{1}{Z(T)}\, e^{-\frac{E(\sigma)}{kT}}.
		         \label{boltzmann}
		\end{eqnarray}
It is hard to estimate $Z(T)$ directly -- using a single Markov Chain (whose stationary distribution is a low temperature Gibbs state) and letting it thermalize could take a very long time. The schemes we want to speed up thus attempt to reach the low-temperature thermal states in several stages. Consider a sequence of decreasing temperatures 
$T_0 \geq T_1 \geq \dots \geq T_{\ell},$
where $T_0$ is a very high starting temperature and $T_{\ell}=T_F$ is the desired
final temperature. The final partition function $Z(T_F)$ can then be expressed as a telescoping product
\begin{eqnarray}
			Z(T_F) &=& Z_0 \, \frac{Z_1}{Z_0} \cdots \frac{Z_{\ell-1}}{Z_{\ell-2}} \frac{Z_\ell}{Z_{\ell-1}} 
			= 
			Z_0 \underbrace{ \left(\alpha_0 \alpha_1 \cdots \alpha_{\ell-2} \alpha_{\ell-1} \right)}_{\alpha}\,,
			\label{telescope}
\end{eqnarray}
where $Z_i = Z(T_i)$ stands for the Gibbs partition function at temperature
$T_i$ and $\alpha_i = Z_{i+1}/Z_{i}$ is the ratio of successive $Z$'s. 
		To start, it is easy to calculate the partition function $Z_0 = Z(T_0)$ at high
		temperature (its value is the volume of the state space). 
		Next, for each $i$, we can estimate the ratio $\alpha_i$ by sampling from a distribution that is sufficiently close to the
		Boltzmann distribution $\pi_i$ \eqref{boltzmann} at temperature $T_i$. 
		This is possible by using a rapidly-mixing Markov chain $P_i$ whose stationary distribution is equal to the 
		Boltzmann distribution $\pi_i$.
		
		To be efficient, these classical schemes require that
		\begin{enumerate}
		\item we use a cooling schedule such that the resulting ratios $\alpha_i = Z(T_{i+1})/Z(T_i)$ are lower bounded by a constant $c^{-1}$ (to simplify the presentation, we use $c=2$ from now on), 
		\item the spectral gaps of the Markov chains $P_i$ are bounded from below by $\delta$.
		\end{enumerate}
		The time complexity of such FPRAS, i.e., the number of times we have to invoke an update step
		for a Markov chain from $\{P_1,\ldots,P_{\ell-1}\}$, is
		\begin{equation}
		         \tilde{O}\left(\frac{\ell^2}{\delta \cdot \epsilon^2}\right)\,,
		         \label{classicalfpras}
		\end{equation}
		where $\tilde{O}$ means up to logarithmic factors.
		
We will now follow the presentation in \cite[Section 2.1]{Vazirani}
and describe the classical approximation schemes in more detail,
starting with the partition function as a telescoping product \eqref{telescope}.
At $T_0=\infty$, the partition function $Z_0$ is equal to
\begin{eqnarray}
   Z_0 = |\Omega|,
\end{eqnarray}
the size of the state space. On the other hand, for each $i=0,\dots,\ell-1$, we
can estimate the ratio
\begin{eqnarray}
	\alpha_i = \frac{Z_{i+1}}{Z_i}
	\label{alpharatio}
\end{eqnarray}
in \eqref{telescope} by sampling from the Boltzmann distribution $\pi_i$ as follows.  Let $X_i \sim \pi_i$ denote a random state chosen according to the Boltzmann distribution $\pi_i$, i.e., 
\begin{equation}
	\Pr(X_i=\sigma)=\pi_i(\sigma)\,.
\end{equation}
Define a new random variable $Y_i$ by 
\begin{equation}
	Y_i = e^{-(\beta_{i+1}-\beta{i})\, E(X_i)},
\label{yvariable}
\end{equation}
where $\beta_i = (kT_i)^{-1}$ is the inverse temperature ($k$ is the Boltzmann constant). 
This $Y_i$ is an unbiased estimator for $\alpha_i$ since 
\begin{eqnarray}
	\EE{Y_i} 
	& = & \sum_{\sigma \in \Omega} \pi_i(\sigma) \, e^{-(\beta_{i+1}-\beta{i})\, E(\sigma)} 
	 =   
	\sum_{\sigma \in \Omega} \frac{e^{-\beta_i E(\sigma)}}{Z_i} \, e^{-(\beta_{i+1}-\beta{i})\, E(\sigma)} \\
	& = & 
	\sum_{\sigma \in \Omega} \frac{e^{-\beta_{i+1}\, E(\sigma)}}{Z_i}
	= 
	\frac{Z_{i+1}}{Z_i} = \alpha_i.
\label{expecty}
\end{eqnarray}
Assume now that we have an algorithm for generating states $X_i$ according to $\pi_i$. We draw 
$
	m:=64\ell/\ep^2
$
samples of $X_i$ and take the mean $\overline{Y}_i$ of their corresponding estimators $Y_i$.
Then, assuming $\frac{1}{2}\le\alpha_i\le 1$, the mean $\overline{Y}_i$ satisfies
\begin{eqnarray}
	\frac{\VV{\overline{Y}_i}}{\left(\EE{\overline{Y}_i}\right)^2} 
	=
	\frac{\ep^2}{64\ell}\, \frac{\VV{Y_i}}{\left(\EE{Y_i}\right)^2}
	\leq
	\frac{\ep^2}{16 \ell}\,.
\end{eqnarray}
We can now compose such estimates of $\alpha_i$. 
Define a new random variable $\overline{Y}$ by 
\begin{equation}
	\overline{Y} = \overline{Y}_{\ell-1}\overline{Y}_{\ell-2}\cdots\overline{Y}_0
\end{equation}
Since all $\overline{Y}_i$ are independent, we have
\begin{eqnarray}
	\EE{\overline{Y}}
	&=&
	\EE{Y_{\ell-1}} \EE{Y_{\ell-2}} \cdots \EE{Y_{0}} 
	= 
	\alpha_{\ell-1} \alpha_{\ell-2} \cdots \alpha_{0} = \alpha,
\end{eqnarray}
Moreover, $\overline{Y}$ has the property
\begin{eqnarray}
	\frac{\VV{\overline{Y}}}{\left(\EE{\overline{Y}}\right)^2} 
	 &=& 
		\frac{\EE{\overline{Y}^2_{\ell-1}}\cdots\EE{\overline{Y}^2_{0}} -
				\EE{\overline{Y}_{\ell-1}}^2\cdots\EE{\overline{Y}_{0}}^2}		         
		{\EE{\overline{Y}^2_{\ell-1}}^2\cdots\EE{\overline{Y}_{0}}^2}
		\nonumber \\
	 &=& 
		         \left(
		                 1 +
		                 \frac{\VV{\overline{Y}_{\ell-1}}}{\left(\EE{\overline{Y}_{\ell-1}}\right)^2}
		         \right)
		         \cdots
		         \left(
		                 1 +
		                 \frac{\VV{\overline{Y}_{0}}}{\left(\EE{\overline{Y}_{0}}\right)^2}
		         \right)- 1 \nonumber\\
	&\le&
		                 \left(e^{\ep^2/16\ell}\right)^\ell - 1 
	\le
		                 \epsilon^2/8\,, \label{use2} 
\end{eqnarray}
where we used $1+x\le e^x$ (true for all $x$) and $e^x-1\le 2x$ (true for all $x\in[0,1])$ in the last two steps, respectively.
Chebyshev's inequality now implies that the value of
$\overline{Y}$ is in the interval $[(1-\epsilon)\alpha, (1+\epsilon) \alpha]$
with probability at least $\frac{7}{8}$.
		
Of course, we are not able to obtain perfect samples $X_i$ from $\pi_i$.  Assume now that we have $X_i'$ that are from a distribution with a variation distance from $\pi_i$ smaller than 
$
	d:= \epsilon^2/(512\ell^2)$.
Let $\overline{Y}'$ be defined as $\overline{Y}$ as above, but instead of $X_i$ we use $X_i'$.  Then, with probability at least $\frac{7}{8}$, we have $\overline{Y}=\overline{Y}'$.  To derive this, observe that the algorithm can be thought to first take a sample from a product probability distribution $\pi$ on the $(m\ell)$-fold direct product of $\Omega$.  We denote the probability distribution in the case of imperfect samples by $\pi'$.  The total variation distance between $\pi$ and $\pi'$ is then bounded from above by 
\begin{equation}
d \cdot m \cdot \ell = \frac{\epsilon^2}{512\ell^2} \cdot \frac{64\ell}{\epsilon^2} \cdot \ell = \frac{1}{8}\,.
\end{equation}
Therefore, $\overline{Y}'$ is in the interval $[(1-\epsilon)\EE{Y},(1+\epsilon)\EE{Y}]$ with probability at least $\frac{3}{4}$.

We obtain the samples $X'_i$ by applying rapidly mixing Markov chains $P_i$ whose limiting distributions are equal to $\pi_i$.  
Constructing such Markov chains is a hard task, but it has been done for 
the Ising model \cite{Jerrum2} and the Potts model \cite{Vazirani}, resulting in FPRAS for the partition functions for these models.

Thus, using a sequence of rapidly-mixing Markov Chains, it is possible to prepare low-temperature Gibbs states. Not only that, sampling from these states and estimating the variable \eqref{yvariable} allows us to estimate partition functions by a telescoping product \eqref{telescope}, because of \eqref{expecty}.
Note that for each sample from a state close to the thermal state for a given MC, we need to prepare it using a sequence of MC's. The run is then discarded. Could we somehow ``reuse'' the mixing we have done so far using quantum mechanics? It turns out it is possible, and that estimating the ratios $\alpha_i$ using \eqref{yvariable} is possible by phase estimation of a certain quantum walk operator. The method described in the next section, developed by Wocjan et al. in \cite{WA08, NW08} is related to the one used by Somma \cite{Somma2}, and brings an additional square root speedup in the approximation precision parameter.

\subsubsection{Quantizing MCMC Methods for Approximating Partition Functions}
\label{sec:quantizeMCMC}

In Section \ref{sec:MCMC}, we have seen how classical MCMC methods work and that they require $\tilde{O}\left(\frac{\ell^2}{\delta \cdot \epsilon^2}\right)$ steps \eqref{classicalfpras}, where $\delta$ is the lower bound on the spectral gap of the MC's involved and $\ep$ is the desired approximation precision for the FPRAS.
Let us now turn to the quantum world and put quantum walks to use, following \cite{NW08}. We will see that the resulting quantized algorithm
gets a square root speedup in both the parameter $\delta$ and $\epsilon$. Let us summarize the result:
\begin{theorem}
		\label{mainresult}
Consider a classical FPRAS for approximating the Gibbs partition function of a physical system
		at temperature $T_F$, satisfying the above conditions. Then, there exists a fully polynomial quantum approximation scheme that uses
		\begin{equation}
		   \tilde{O}\left(\frac{\ell^2}{\sqrt{\delta} \cdot \epsilon}\right)
		\end{equation}
		applications of a controlled version of a quantum walk operator from
		$\{W(P_1),\ldots,W(P_{\ell-1})\}$.
		\end{theorem}
		
\begin{figure}
	\begin{center}
		\includegraphics[width=3in]{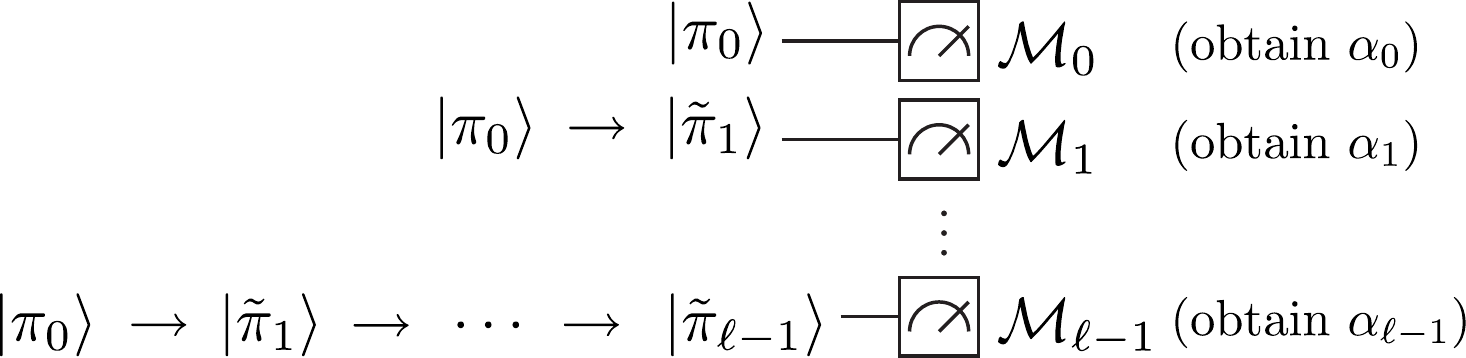}
	\end{center}
\caption{Structure of the quantum algorithm.}
\label{figstructure}
\end{figure}

The reduction in complexity (in comparison to the
classical FPRAS \eqref{classicalfpras}) for the quantum algorithm we will now look at is twofold. First, the factor $1/\delta$ is reduced to $1/\sqrt{\delta}$ by using quantum
walks instead of classical Markov chains (for preparing the distributions to sample from), and utilizing the quadratic relation between spectral and phase gaps.
This relation is the basis of success for many quantum search algorithms based on quantum walks \cite{Santha}.
Thus, instead of letting a Markov chain $P_i$ mix towards its stationary distribution $\pi_i$,
we choose to approximately prepare the state 
\begin{eqnarray}
	|\pi_i\> = \sum_{\sigma \in \Omega} \sqrt{\pi_i(\sigma)} \ket{\sigma}
\end{eqnarray}
a {\em coherent encoding} of the Boltzmann distribution. 
The preparation method \cite{WA08} is based on Grover's $\frac{\pi}{3}$-fixed-point search \cite{GroverFixPoint,GroverFixPoint2}, described in Appendix~\ref{sec:fixedpoint}, efficiently driving the state $\ket{\pi_0}$ towards the desired state $|\pi_i\>$ 
through a sequence of intermediate states. 

Second, we speed up the way to determine the ratios $\alpha_i$.
Instead of using classical samples from the stationary distribution $\pi_i$ of a Markov chain $P_i$, 
we approximate $\alpha_i$ by phase-estimating a certain unitary (related to quantum walks) on the state $\ket{\pi_i}$. This results in the reduction of the factor
$1/\epsilon^2$ to $1/\epsilon$. 

The structure of the algorithm is depicted in Fig. \ref{figstructure}.
It consists of successive approximate preparations of $\ket{\pi_i}$ followed by 
a quantum circuit outputting a good approximation to $\alpha_i$ (with high probability).
We will now show how to quantize the the classical algorithm, assuming that we can take perfect samples $X_i$ from $\pi_i$. The interested reader is invited to look into \cite{NW08} for the full quantum algorithm which deals with the fact that samples from $\pi_i$ can be taken only approximately, as the states we can prepare are only approximations (but good ones) to $\ket{\pi_i}$. However, the errors can be handled and collected in such a way, that the result of Theorem \ref{mainresult} holds, giving a fully quantum FPRAS with a double (in $\delta$ and $\ep$) square root speedup over the classical method.

To estimate the ratios $\alpha_i$ in \eqref{telescope}, the classical algorithm generates 
random states $X_i$ from $\pi_i$ and computes the mean $\overline{Y}_i$ of the random variables $Y_i$.
The process of generating a random state $X_i$ from $\pi_i$ is equivalent to preparing the mixed state
$\rho_i = \sum_{\sigma\in\Omega} \pi_i(\sigma) |\sigma\>\<\sigma|$.
Instead of this, we choose to prepare the pure states
\begin{equation}
	|\pi_i\> = \sum_{\sigma\in\Omega} \sqrt{\pi_i(\sigma)} |\sigma\>\,.
\end{equation}
We call these states {\em quantum samples} since they coherently encode the probability 
distributions $\pi_i$. Let us for now assume that we can prepare these exactly and efficiently.

The random variable $Y_i$ can be viewed as the outcome of the measurement of the observable
\begin{equation}
	A_i = \sum_{\sigma\in\Omega} y_i(\sigma) |\sigma\>\<\sigma|
\end{equation}
where $y_i(\sigma) = e^{-(\beta_{i+1}-\beta_i) E(\sigma)}$, for the state $\rho_i$.
This interpretation implies that to estimate $\alpha_i$ 
classically, we need to estimate the expected value ${\rm Tr}(A_i \rho_i)$ by repeating the above measurement 
many times and outputting the mean of the outcomes.
However, we can quantize this process.  We add an ancilla qubit to our quantum system in which the quantum samples $|\pi_i\>$ live.  For each $i=0,\ldots,\ell-1$, we define the unitary
\begin{equation}
V_i = \sum_{\sigma\in\Omega} \ket{\sigma}\bra{\sigma} \otimes
\left(
\begin{array}{cc}
  \sqrt{y_i(\sigma)}   & \sqrt{1-y_i(\sigma)} \\
- \sqrt{1-y_i(\sigma)} & \sqrt{y_i(\sigma)}
\end{array}
\right)\,.
\end{equation}
This $V_i$ can be efficiently implemented, as it is a rotation on the extra qubit controlled
by the state of the first tensor component.
Let us label
\begin{equation}
	\ket{\psi_i} = V_i \big( |\pi_i\> \otimes |0\> \big).
\label{psistate}
\end{equation}
Consider now the expected value of the projector
\begin{equation}
	P = \iii \otimes \ket{0}\bra{0}
\label{projector}
\end{equation}
in the state $\ket{\psi_i}$. We find
\begin{equation}
	\<\psi_i|P|\psi_i\> = \<\pi_i|A_i|\pi_i\> = \alpha_i\,.
\end{equation}
We now show how to speed up the process of estimating $\alpha_i$ 
with a method that generalizes quantum counting \cite{BrassardHoyerTapp}. 
Assuming efficient preparation of $\ket{\pi_i}$ implies that we can also efficiently implement the reflections
\begin{equation}\label{eq:reflectPii}
	R_i = 2|\pi_i\>\<\pi_i| - \iii\,.
\end{equation}
Thus, we arrive at the existence of a quantum FPRAS for estimating the partition function, 
assuming efficient and perfect preparation of $\ket{\pi_i}$, summed in Theorem \ref{th:perfectZ}:
\begin{theorem}\label{th:perfectZ}
	There is a fully polynomial quantum approximation scheme $\cA$ for the partition function $Z$.  
	Its output $Q$ 	satisfies
	\begin{equation}
		\Pr\big[(1-\epsilon) Z \leq Q \leq (1+\epsilon) Z \big] \geq \frac{3}{4}\,.  
	\end{equation}
	For each $i=0,\ldots,\ell-1$, the scheme $\cA$ uses $
		O\left(\log \ell \right)$
	perfectly prepared quantum samples $|\pi_i\>$, and applies the controlled-$R_i$ operator $
		O\left( \frac{\ell}{\ep} \log \ell \right)$
		times, where $R_i$ is as in (\ref{eq:reflectPii}).
\end{theorem}
	   
The proof of Theorem~\ref{th:perfectZ} builds on the following three technical results.

\begin{lemma}[Quantum ratio estimation]\label{lem:generalPE}
Let $\ep_{pe}\in (0,1)$. For each $i=0,\ldots,\ell-1$ there exists a quantum approximation scheme $\cA'_i$ for $\alpha_i$. Its output $Q'_i$ satisfies
$
\Pr\big[(1-\ep_{pe}) \alpha_i \leq Q'_i \leq (1+\ep_{pe}) \alpha_i \big] \geq \frac{7}{8}.
$
The scheme $\cA'_i$ requires one copy of the quantum sample $|\pi_i\>$ and invokes the controlled-$R_i$ operator
$O\left( \ep^{-1}_{pe}\right)$ times, where $R_i$ is as in (\ref{eq:reflectPii}).
\end{lemma}
		
We can boost the success probability of the above quantum approximation scheme for the ratio $\alpha_i$ 
by applying the \emph{powering lemma} from
\cite{Valiant}, which we state here for completeness:
		
\begin{lemma}[Powering lemma for approximation schemes]\label{lem:powering}
Let $\cB'$ be a (classical or quantum) approximation scheme whose estimate $W'$ is within $\pm\ep_{pe} q$ to some value $q$ with probability $\frac{1}{2}+\Omega(1)$.  Then, there is an approximation scheme $\cB$ whose estimate $W$ satisfies
$
\Pr\big[(1-\ep_{pe}) q \leq W \leq (1+\ep_{pe}) q \big] \geq 1- \delta_{boost}\,.
$
It invokes the scheme $\cB'$ as a subroutine $O\left(\log \delta_{boost}^{-1} \right)$ times.
\end{lemma}

Lemma \ref{lem:powering} ensures we can get precise estimates of $\alpha_i$, which we can than compose 
into an approximation for the partition function \eqref{telescope}.
		
\begin{lemma}[Composing ratio estimates]\label{lem:boundProdEstimates}
Let $\epsilon>0$. Assume we have approximation schemes $\cA_0,\cA_1,\ldots,\cA_{\ell-1}$ such that their estimates $Q_0,Q_1,\ldots,Q_{\ell-1}$ satisfy
$\Pr\left[ \left|Q_i - \alpha_i\right|\leq \frac{\ep\alpha_i}{2\ell} \right] \ge 1 - \frac{1}{4\ell}$.
Then, there is a simple approximation scheme $\cA$ for the product $\alpha=\alpha_0 \alpha_1 \cdots \alpha_{\ell-1}$. 
The result $Q=Q_0 Q_1 \cdots Q_{\ell-1}$ satisfies
\begin{equation}\label{eq:multApprox}
\Pr\big[ (1-\ep) \alpha \le Q \le (1+\ep) \alpha \big] \ge \frac{3}{4}\,.
\end{equation}
\end{lemma}
		
We are now finally ready to prove Theorem \ref{th:perfectZ}. For each $i=0,\ldots,\ell-1$, we can apply Lemma~\ref{lem:generalPE} with the state $|\psi_i\>$ (\ref{psistate}) 
and the projector $P$ (\ref{projector}). This gives us a quantum approximation scheme for $\alpha_i$.
Note that to prepare $|\psi_i\>$, it suffices to prepare $|\pi_i\>$ once. Also, to realize a controlled reflection 
around $|\psi_i\>$, it suffices to invoke the controlled reflection around $|\pi_i\>$ once. 

We now use the reflection $2|\psi_i\>\<\psi_i|-\iii$ and set $\epsilon_{pe}=\epsilon/(2\ell)$ in Lemma~\ref{lem:generalPE}.
With these settings, we can apply Lemma \ref{lem:powering} to the resulting approximation scheme 
for $\alpha_i$ with $\delta_{boost}=1/(4\ell)$. This gives us approximation schemes $\cA_i$ outputting $Q_i$ with 
high precision and probability of success that can be used in Lemma~\ref{lem:boundProdEstimates}. 
The composite result $Q=Q_0 \cdots Q_{\ell-1}$ is thus an approximation 
for $\alpha=\alpha_0 \cdots \alpha_{\ell-1}$ 
with the property
$
	\Pr\big[ (1-\ep) \alpha \le Q \le (1+\ep) \alpha \big] \ge \frac{3}{4}
$.

Finally, we obtain the estimate for $Z$ by multiplying $Q$ with $Z_0$.
Let us summarize the costs from Lemmas \ref{lem:generalPE}-\ref{lem:boundProdEstimates}.
For each $i=0,\dots,\ell-1$, this scheme uses $\log \delta_{boost}^{-1} = O(\log \ell)$ copies of the state $\ket{\pi_i}$, and invokes $\left(\log \delta_{boost}^{-1}\right) \ep_{pe}^{-1} = O\left(\frac{\ell}{\ep} \log \ell\right)$
reflections around $\ket{\pi_i}$.

Finally, this is where quantum walks come into play.
So far, we have assumed that we can prepare the quantum samples $|\pi_i\>$ 
and implement the controlled reflections $R_i=2|\pi_i\>\<\pi_i|-\id$ about these states 
perfectly and efficiently. These assumptions can be released and accomplished approximately 
with the help of quantum walk operators.  The errors arising from these approximate procedures 
do not significantly decrease the success probability of the algorithm.

Using the fact that the consecutive states $|\pi_i\>$ and $|\pi_{i+1}\>$ are close, we can utilize Grover's $\frac{\pi}{3}$ fixed-point search \cite{GroverFixPoint} to drive the starting state $\ket{\pi_0}$ towards the desired state $|\pi_i\>$ through multiple intermediate steps\footnote{Compare this to the projections of $\ket{\pi_i}$ onto $\ket{\pi_{i+1}}$ and the quantum Zeno effect in \cite{Somma} discussed in Section \ref{sec:MixDecohere}.}. Moreover, to be able to perform this kind of Grover search, we have to be able to apply selective phase shifts of the form $S_i = \omega |\pi_i\>\<\pi_i| + (\iii - |\pi_i\>\<\pi_i|)$ for $\omega=e^{i\pi/3}$ and $\omega=e^{-i\pi/3}$. There is an efficient way to apply these phase shifts  approximately, based on quantum walks and phase estimation \cite{WA08}. 

The important condition for this method to work, the overlap of two consecutive quantum samples $|\pi_i\>$ and $|\pi_{i+1}\>$ has to be large. This is satisfied when $\alpha_i=Z_{i+1}/Z_i$ is bounded from below by $\frac{1}{2}$, since
\begin{eqnarray}
	|\<\pi_i|\pi_{i+1}\>|^2
	 = 
	\left|
	\sum_{\sigma\in\Omega} \frac{\sqrt{e^{-\beta_i E(\sigma)} \, 
	e^{-\beta_{i+1} E(\sigma)}}}{\sqrt{Z_i \, Z_{i+1}}} \right|^2 
	 \ge 
	\left|
	\frac{\sum_{\sigma\in\Omega} \, e^{-\beta_{i+1}E(\sigma)}}{\sqrt{2 Z_{i+1}} \, \sqrt{Z_{i+1}}} \right|^2 
	= 
	\frac{1}{2}\,. \label{condition2ZZ}
\end{eqnarray}
We can then use additional ancilla qubits and the phase estimation of a quantum walk operator to implement the selective phases required for Grover's fixed-point search, with
$O\left( \frac{\ell}{\sqrt{\delta}} \log^2 \ell \right)$ applications of a quantum walk operator.
Second, the reflections $R_i=2|\pi_i\>\<\pi_i|-\id$ can to done approximately and effectively, again using phase estimation of a controlled-quantum walk operator. 
Altogether, with proper choices for the precision for the phase estimations required in the state preparation and reflections, 
the resulting cost of this scheme (the number of times we have to invoke the controlled 
quantum walk operators) is $
	\tilde{O}
		\left(
			\frac{\ell^2}{\ep \sqrt{\delta}}
		\right)$.
It remains open how to quantize FPRAS which are based on sequences of MC's which do not obey the condition \eqref{condition2ZZ}, or use adaptive steps (not knowing the range into which $\alpha_i$ will fall in advance).

\subsubsection{Quantum Metropolis Sampling}
\label{sec:Metropolis}

The preparation of ground states and (sampling from) Gibbs states is generally a hard task, as finding them is related to optimization problems. However, for classical systems (Hamiltonians), the Metropolis algorithm, described in Section \ref{sec:metropolisandsuch}, is widely used and often efficient. When we want to use it to prepare Gibbs (thermal) states of a system, the strategy is to perform a random walk on the states of a system, changing a few local parameters in each step. There is a simple rule for accepting or rejecting the change, depending on the difference in energies of the two states and on the system temperature, which we slowly cool down.

The big question is whether we could we use a quantum computer and quantum walks to prepare Gibbs states of {\em quantum} systems, especially those plagued by the sign problem (and thus unfit for Quantum Monte Carlo methods). We would thus like to prepare a sample from the eigenstates of a quantum system (according to their energies). However, the energy eigenbasis is now not equal to the computational basis -- and making a Markov Chain jumping between the (unknown) energy eigenstates, according to a Metropolis rule, is a hard task. The most serious obstacle to quantizing the Metropolis algorithm is the necessity of rejecting a state change, when the energy of the proposed new state is much larger than the energy of the original state. Is a return from the undesired quantum state to the original (but uknown) one possible and could we do it coherently? Recently, Temme et al. \cite{qmetropolis} have discovered a way of doing exactly that, using a property of two reflections that is essential to quantum walks, but also found uses in quantum complexity theory \cite{MWqma, FastQMA}. Later, Yung et al. \cite{qmetropolis2} made the algorithm even more quantum, providing a square root speedup in the dependence of the convergence time on $\delta$, the gap of the associated classical Markov Chain.

Let us now briefly sketch the principle of the Quantum Metropolis algorithm due to Temme et al. \cite{qmetropolis}. For simplicity, let us assume we would like to prepare the Gibbs state of a system of $n$ 2-level particles, i.e. Ising spins. We would like to set up a rapidly mixing Markov chain, which samples from the configurations $x$ with the corresponding probabilities
\be
 \pi(x) = \frac{e^{-\beta E_x}}{Z}. 
\ee
We start from a random state and run the Markov chain for many steps. In each step, we start with some state $x$ with energy $E_{x}$, flip a few (a local change) randomly selected spins, and obtain a state $y$. We accept this move if the energy decreases. On the other hand, if the energy $E_{y}$ of the new state is higher, we accept the move only with probability $e^{-\beta (E_{y}-E_{x})}$.
This Markov chain obeys detailed balance, because the following is true when $E_y > E_x$:
\be
	\pi(x) = \pi(y) e^{-\beta(E_x-E_y)}.
\ee
Each step of the quantum version of the Metropolis algorithm now takes as input an energy eigenstate $\ket{\psi_i}$ and applies a random local unitary $C$ to it, producing $\sum_{k} x_{k}^{i} \ket{\psi_k}$, some superposition of energy eigenstates. To be able to decide whether we want to take a step or not, let us also add two extra energy-labeling registers in the state $\ket{E_i}\ket{0}$. Without touching the register with $\ke{E_i}$, we can use phase estimation on the first and third register to produce the label $E_k$, giving
\be
   \sum_{k} x_{k}^{i} \ket{\psi_k}\ket{E_i}\ket{E_k}. \label{QMwithElabel}
\ee
We would now like to accept the states with energies $E_k > E_i$ with probability $e^{-\beta(E_k-E_i)}$. However, a direct measurement of the energy register would collapse the superposition, and this would disallow us to return back to the state $\ket{\psi_i}$. We can work around this obstacle by adding another ancilla register denoting the acceptance of the move, using a unitary $F$ on the energy registers and the acceptance ancilla to locally transform \eqref{QMwithElabel} into
\be
   \underbrace{\sum_{k} x_{k}^{i} \sqrt{f_k^i} \ket{\psi_k}\ket{E_i}\ket{E_k}}_{\kets{\psi_i^+}}\ket{1}
   + \underbrace{\sum_{k} x_{k}^{i} \sqrt{1-f_k^i} \ket{\psi_k}\ket{E_i}\ket{E_k}}_{\kets{\psi_i^-}}\ket{0}, \label{QMwithAccept}
\ee
where $f_{k}^{i} = \mathrm{min}\left(1, e^{-\beta(E_k-E_i)}\right)$ corresponds to the Metropolis rule \eqref{eq:transition_rate} for the Gibbs distribution\footnote{Note that the Metropolis algorithm is general, not restricted to use the Gibbs distribution.}. If we now perform a projective measurement $Q$ on the  acceptance register and obtain 1, we get exactly what we wanted, as the amplitudes $x_{k}^{i}\sqrt{f_k^i}$ correspond to the classical Metropolis rule transition probabilities $|x_k^i|^2 f_k^i$. However, if we measure 0 in the acceptance register, we project into a state that we did not want. Nevertheless, there is a way of undoing this, similar to the procedure used in QMA amplification \cite{MWqma, FastQMA}. 
The projective measurement of the acceptance register gave us the state $\kets{\psi_{i}^-}\ket{0}$.
The key observation is that this projective measurement left the state  within the 2D subspace spanned by $\kets{\psi_{i}^+}\ket{1}$ and $\kets{\psi_{i}^-}\ket{0}$, which has a different basis as well:
\be
   \ket{\varphi_{+}} &=& \kets{\psi_i^+}\ket{1} + \kets{\psi_i^-}\ket{0}. \\
   \ket{\varphi_{-}} &=& \frac{\kets{\psi_{i}^+}\ket{1}-\ket{\varphi_+}\braket{\varphi_+}{\psi_i^+}\ket{1}}{\sqrt{1-|\braket{\varphi_+}{\psi_i^+}\ket{1}|^2}}   \label{QMotherbase}
\ee
If we could devise a projective measurement according to the projector  $P = \ket{\varphi_{+}}\bra{\varphi_{+}}$, we could get back into business (obtain the state $\kets{\psi_i^+}\ket{1}$) in the following way. Because the 2D subspace we are in is invariant under the projectors $P$ and $Q$, performing alternating projective measurements $P$ and $Q$ (as in the Marriott-Watrous scheme) gives a rotation in this 2D subspace. More than that, it will eventually result in the $+1$ eigenstate  of $Q$, which is our desired state $\kets{\psi_i^{+}}\ket{1}$. 

The last necessary ingredient is the projective measurement according to $P$. It can be performed as follows. We uncompute the unitary $F$, uncompute the energy label $E_k$, undo the local mixing unitary $C$ and uncompute the energy label $E_i$. We can now project onto $\ket{0}\ket{0}\ket{0}$ on the ancilla registers, and again compute $E_i$, apply $C$, compute $E_k$ and apply $F$. Observe that on the state $\ket{\varphi_{+}}$ this projection acts as an identity, while the state $\ket{\varphi_{-}}$ is in its kernel. 
The required number of steps for applying this rejection (and state repair) procedure is worked out thoroughly in \cite{qmetropolis}, resulting in a quantum algorithm for preparing the Gibbs states of quantum systems. The required number of steps of the algorithm again depends on the inverse of the spectral gap $\delta$ of the Markov Chains involved. We believe natural systems thermalize easily, but it is unlikely that MC's whose Hamiltonians correspond to hard computational problems have gaps that scale favorably (as inverse polynomials) with the system size. 

The quantum Metropolis algorithm receives a square root speedup (exchanging $\delta$ for $\sqrt{\delta}$) in Yung et al.'s work \cite{qmetropolis2}, where the authors use quantized (two-register) MC's and quantum simulated annealing instead of the random walk described in this Section which is classical at heart, although running on quantum states and combined with a quantum evaluation and recovery procedure.


\subsection{Summary}
In this Chapter, we have seen how to implement a quantum walk corresponding to a particular (classical) Markov Chain given by its transition matrix. Instead of appending a coin to the position of a walker, we did it with a system with two-registers both having ranges on the vertices (positions) of the underlying graph. A step of the quantum walk is then made by using one of the registers to govern the diffusion (according to \eqref{tworegisterreflection}) in the other one. We then swap the registers and continue. 

The unitary quantum walk we obtain has nice spectral properties, relating the spectra of the Quantum MC to the original spectra of the underlying MC's as seen in Section \ref{sec:walkspectrum}. This underlies the quadratic speedups for many algorithms, including phase estimation. Taking it further, Quantum MC's form the basis of sampling algorithms, culminating with the quantum Metropolis algorithm \cite{qmetropolis}.

\setcounter{equation}{0} \setcounter{figure}{0} \setcounter{table}{0}\newpage
\section{Continuous Time Quantum Walks}
\label{sec:ctqw}

In this Chapter we will look at a different approach to quantum walks, with a system undergoing continuous-time evolution according to the Schr\"{o}dinger equation, governed by a fixed ``hopping'' Hamiltonian related to an underlying graph. This is a more natural description related to the dynamics of a single excitation in condensed-matter systems in physical systems. We start with a review of the development of continuous-time quantum walks, discuss their behavior, and present several algorithms (one with an exponential speedup) utilizing graph symmetries. We then show how continuous-time quantum walks (CTQW) can be used as a universal model for quantum computation, and wrap up with the comparison (and connection) of continuous-time to discrete-time quantum walks.


\subsection{Quantizing Continuous Random Walks}

So far, we have investigated classical and quantum walks on graphs in {\em discrete} time. They are described by a particular update rule -- a transition matrix (classical Markov Chains), a unitary coin-toss and shift matrices (usual discrete-time quantum walks), or a product of two reflections (quantized Markov chains).
Where classical walks involve the evolution of probabilities, quantum discrete-time walks involve the evolution of amplitudes (of quantum states) and require an extra coin register to ensure unitary evolution. They describe discrete quantum diffusion processes (retaining at least some degree of coherence) on graphs. We will now look for a quantum process described by a continuous-time equation similar to the diffusion equation, and call what we find  a {\em continuous-time quantum walk}. In contrast to discrete-time quantum walks, there will be no need for the extra register (``coin'' or spin degree of freedom), but using one can be helpful, as shown by the algorithm for spatial search in Section \ref{ContinuousSearchSection}. 

Let us consider a graph $G=(V,E)$ where $V$ is the set of its vertices and $E$ the set of its edges. 
Our goal si to model a continuous-time diffusion process on this graph. 
We start with a classical process, where in each small time-step, vertex $j$ leaks probability to its neighbors (there are $\textrm{deg}(j)$ of them), while collecting some probability back from them. We can describe this process by the diffusion equation
\be
	\deriv{}{t}p_j(t) = \sum_{k\in V} L_{j,k} p_k(t), \label{diffconttime}
\ee
where $L$ is the Laplacian of $G$, given by
\be
	L_{j,k} = \begin{cases}
			-\textrm{deg}(j) & j=k,\\
			1 & (j,k)\in E,\\
			0 & \textrm{otherwise}.\\
	\end{cases}
\ee
This matrix respects the graph structure, and is a representation of the discrete Laplace operator $\nabla^2$. For example, on a cycle of length 5, the Laplacian is
\be
	L_{\circ_5} = \left[
		\begin{array}{rrrrr}
			-2 &  1 &  0 &  0 &  1\\
			 1 & -2 &  1 &  0 &  0 \\
			 0 &  1 & -2 &  1 &  0 \\
			 0 &  0 &  1 & -2 &  1 \\
			 1 &  0 &  0 &  1 & -2 
		\end{array}
	\right].
\ee

We will now define a quantum process involving the evolution of amplitudes (instead of probabilities), described by an equation similar to \eqref{diffconttime}. The matrix $L$ is Hermitian, so we can also think of it as a Hamiltonian in the Schr\"odinger equation as
\be
	i\deriv{}{t}\ket{\psi(t)} = L \ket{\psi(t)}.
\ee
The essential difference from \eqref{diffconttime} is the appearance of the imaginary $i$ (inducing unitary evolution), and the interpretation of $\ket{\psi(t)}$ as the amplitudes of the system's wavefunction. We will soon see this has interesting consequences, just as using unitary update rules gave discrete quantum walks their rich structure.

This formulation (using the Laplacian matrix) was used by Farhi and Gutmann in the first paper on continuous-time quantum walks  \cite{FarhiWalk}, where they investigated the transition/reflection properties of binary trees.
We could also choose a different matrix for our Hamiltonian. One example also containing the information about the graph is the adjacency matrix\footnote{also with size $|V|\times|V|$} of $G$ 
\be
	A_{j,k} = \begin{cases}
			1 & (j,k)\in E,\\
			0 & (j,k)\notin E.\\
	\end{cases}
\ee
It is often suitable to use as the Hamiltonian, but the analogy with the classical diffusion is now lost. It is not always clear how to use $A$ as a generator of a continuous time {\em classical} random walk, where one may need to use its ``lazy'' version generated by $\iii-A$.


\subsubsection{Walking in 1D and Mixing}
\label{sec:continuouswalk1D}
The discrete quantum walk in 1D behaves quite differently than the classical drunken-sailor walk. The continuous version of the 1D walk resembles the quantum walk we have seen in Section \ref{sec:dispersion}. How fast the probability is spreading is proportional to $t$ instead of the classical $\sqrt{t}$. We will now show this, and then discuss the sense in which we can talk about mixing for continuous-time quantum walks.

The system we will investigate is a cycle of length $N$ (periodic boundary conditions). The behavior of the continuous-time quantum walk on a cycle is similar for both basic choices of the Hamiltonian --- the Laplacian or the adjacency matrix, as they are related by $L=A-2\iii$. For simplicity, let us then choose
\be
 	H_{N}= - A_{N},
\ee
the negative of the adjacency matrix of the cycle, acting on the $N$ basis states $\ket{x}$ as
\be
	H_{N} \ket{x} &=& -\ket{x-1} - \ket{x+1}, \quad \textrm{for } 2\leq x\leq N-1, \\
	H_{N} \ket{1} &=& -\ket{N} - \ket{2}, \nonumber\\
	H_{N} \ket{N} &=& -\ket{N-1} - \ket{1}. \nonumber
\ee
One obvious eigenvector of $H_{N}$ is the uniform superposition over all states $\ket{x}$. 
\begin{exercise}
Show that all of the eigenvectors of $H_{N}$ are plane waves, i.e.  
\be
	\ket{\phi_k} = \frac{1}{\sqrt{N}} \sum_{x=1}^{N} e^{i p_k x} \ket{x}, \label{cyclevecs}
\ee
with the corresponding eigenvalues (energies)
\be
	E_k = - \left( e^{-ip_k} + e^{ip_k} \right) = - 2 \cos p_k. \label{cycleeigs}
\ee
Show also that the periodic boundary conditions constrain the momenta to values
\be
	p_k = \frac{2\pi k}{N},
\ee
for $0 \leq k \leq N-1$, or alternatively, $-\lfloor \frac{N}{2} \rfloor \leq k \leq \lceil \frac{N}{2} \rceil$. 
\end{exercise}

Let us start in a state concentrated at vertex $x$ and let it evolve for time $t$ according to the Schr\"odinger equation with Hamiltonian $H_N$. Using an expansion in terms of the eigenvectors, the amplitude at vertex $y$ at time $t$ is 
\be
	\bra{y} e^{-iH_N t} \ket{x} &=& \bra{y} e^{-iH_N t} \left(\sum_{k=0}^{N-1}\ket{\phi_k} \bra{\phi_k}\right)\ket{x} \\
	&=&
	\sum_{k=0}^{N-1}  e^{-i (-2 \cos p_k) t} \braket{y}{\phi_k} \braket{\phi_k}{x}  \\
	&=&
	\frac{1}{N} \sum_{k=0}^{N-1}  e^{i 2t \cos p_k} e^{i p_k (y-x)}. \label{computecontinuoustransition} 
\ee
For large $N$, we can approximate the sum by an integral, and obtain \cite{ChildsThesis} the Bessel function of the first kind of order $(y-x)$ as
\be
	\bra{y} e^{-iH_Lt} \ket{x} 
	&\approx&
	i^{(y-x)} J_{y-x}(2t), \\
	p_{y-x}(t) &\approx& \left| J_{y-x}(2t) \right|^2 \label{BesselApprox}
\ee

\begin{figure}
\begin{center}
\includegraphics[width=5.3in]{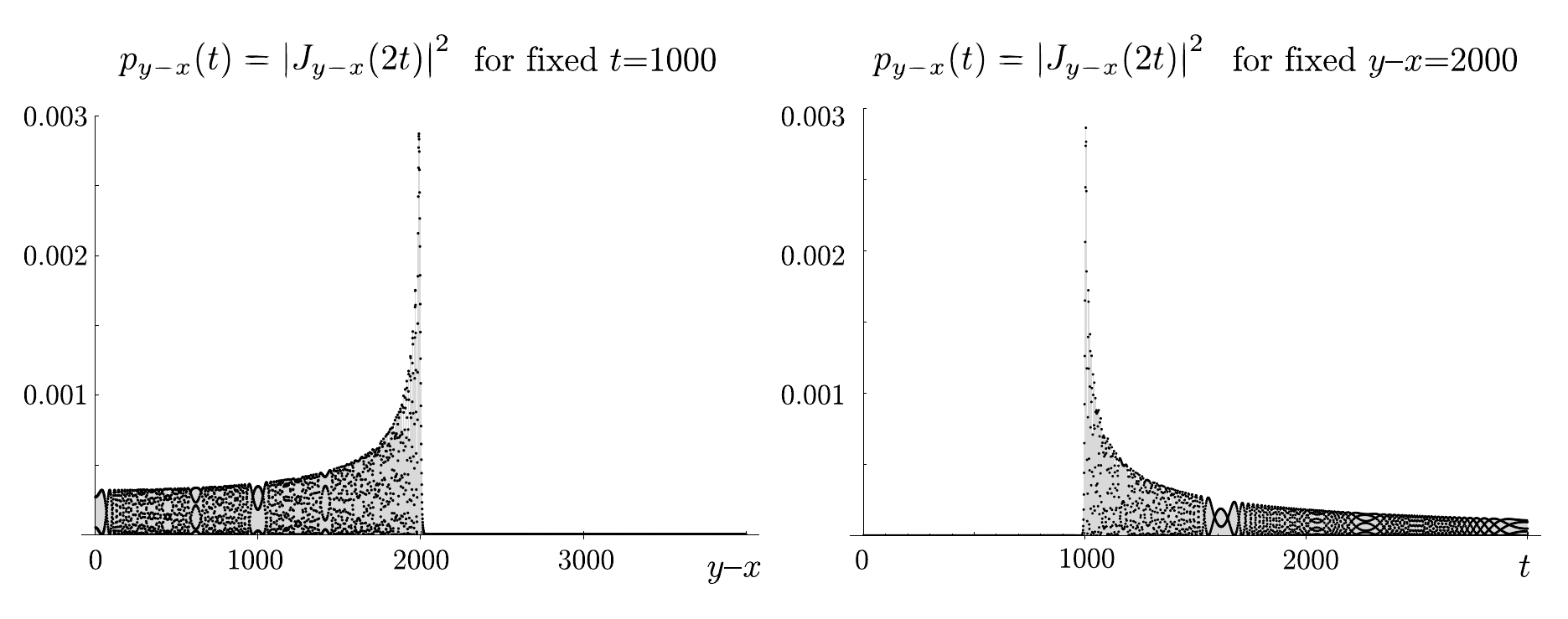}
\end{center}
\caption{Plotting the probabilities of being at vertex $y$ when starting at $x=0$ for a continuous-time quantum walk on an infinite line, approximated with the help of Bessel functions as $p_{y-x}(t)\approx |J_x(2t)|^2$. Note that at time $t=1000$, the wavefront of the walk is at position $y-x=2000=2t$, agreeing with the speed of spreading equal to 2 distance points per unit of time, coming from the largest eigenvalue of $H$, which is 2.
}\label{besselfigure}
\end{figure}

We can now look at the asymptotics of the Bessel function \cite{NumericalRecipes}. For $2t$ large, but still obeying $2t \ll y-x$, we have $J_{y-x}(2t) \approx \frac{t^{y-x}}{(y-x)!}$, which is exponentially small. However, for $2t \approx y-x$, the values of the Bessel function start to rise, and $J_{y-x}(y-x)$ is on the order of $(y-x)^{-\frac{1}{3}}$. Furthermore, for $2t > y-x$ the function $J_{y-x}(2t)$ becomes qualitatively a cosine wave with amplitude decreasing as $\frac{1}{\sqrt{2t}}$. 
For illustration, we plot the probabilities arising from the transition amplitudes for a fixed large time $t$, and for a fixed large distance $y-x$ in Figure \ref{besselfigure}. 

There is a significant probability of transition $x\rightarrow y$ for times of order $2t \approx y-x$, which corresponds to the wavefront moving with a constant speed of the walker\footnote{Note that the speed of spreading for the discrete quantum walk in 1D is proven to be $\frac{1}{\sqrt{2}}$.} equal to 2. Just as for the discrete quantum walk, the average distance of the walker
for the continuous quantum walk on a line rises linearly with time. Contrast this to the classical drunkards' walk, in which the average distance rises as a square root of the number of steps (or equivalently, time). 
The similarities between continuous and discrete quantum walks are discussed in more detail in Section \ref{sec:connectthewalks}. We refer the reader to Figure \ref{walkcomparisonfigure}, where we compare the probabilities of finding the walker at a given vertex for both types of quantum walks in 1D.

The evolution according to the Schr\"odinger equation -- and thus also the continuous quantum walk -- is unitary. There is no mixing towards a stationary distribution -- a wavefunction does not change with time only if we start in an eigenstate. However, for finite graphs we can again talk about its mixing in a time-averaged sense, just as we did for discrete-time quantum walks in Section \ref{sec:discretemixing}, and follow Aharonov et al. \cite{AhAmKeVa01}. The probability of being at vertex $y$ at time $t$ when starting from vertex $x$ is
\be
	p_t (x\rightarrow y) = \left| \bra{y} e^{-iHt} \ket{x} \right|^2,
\ee
and this probability does not converge. However, when we pick a random time $t$ between $0$ and some large $T$, we can talk about a {\em time-averaged distribution}
\be
	\bar{p}_T (x\rightarrow y) = \frac{1}{T} \int_{0}^{T} \left| \bra{y} e^{-iHt} \ket{x} \right|^2\, dt.
\ee
In the $T\rightarrow \infty$ limit, this gives rise (and converges) to the {\em limiting distribution}
\be
	\pi(x\rightarrow y) &=& \lim_{T\rightarrow \infty} \bar{p}_T (x\rightarrow y).
\ee
This limiting distribution can be computed using the expansion to energy eigenstates $\ket{\phi_k}$, similarly to what we did in \eqref{computecontinuoustransition} as
\be
	\pi(x\rightarrow y) &=& \lim_{T\rightarrow \infty} \int_{0}^{T} \left| \bra{y} e^{-iHt} \ket{x} \right|^2\, dt  \\
	&=&  \lim_{T\rightarrow \infty} \int_{0}^{T} \sum_{k,m} 
		\braket{y}{\phi_k} \bra{\phi_k}e^{-iHt}\ket{x}  
		\braket{x}{\phi_m} \bra{\phi_m}e^{-iHt}{y}\ket{y} \,dt \\
	&=& \sum_{k,m}\braket{y}{\phi_k} \braket{\phi_k}{x} \braket{x}{\phi_m} \braket{\phi_m}{y}
	   \underbrace{ \lim_{T\rightarrow \infty} \frac{1}{T} \int_{0}^{\infty} e^{-i (E_m-E_k) t} \,dt}_{\delta_{E_k-E_m}}\\
	&=& \sum_{E_k = E_m} \braket{y}{\phi_k} \braket{\phi_k}{x}\braket{x}{\phi_m} \braket{\phi_m}{y}.  \label{continuouslimiting}
\ee
How fast does the time-averaged distribution converge to the limiting one? When computing 
\be
	\Delta_T(x\rightarrow y) = \bar{p}_T (x\rightarrow y) - \pi(x\rightarrow y),
\ee
the terms that do not subtract out come from the finite-$T$ integral in \eqref{continuouslimiting}, and only from those terms where we sum over nonequal pairs of energies, i.e.
\be
\bar{p}_T (x\rightarrow y) - \pi(x\rightarrow y)
 = \sum_{E_k \neq E_m} \braket{y}{\phi_k} \braket{\phi_k}{x}\braket{x}{\phi_m} \braket{\phi_m}{y} \frac{e^{-i (E_m-E_k) T}-1}{-i(E_m-E_k)T}. \label{inverseEs}
\ee
Thus for the continuous-time quantum walk starting at vertex $x$, the total probability distribution distance 
\be
	\norm{\bar{p}_T(x) - \pi(x)} = \sum_{y} \left| \bar{p}_T (x\rightarrow y) - \pi(x\rightarrow y) \right|
\ee
from the limiting distribution will converge towards 0 with growing $T$ as a function of the gaps in the energy spectrum.  The {\em mixing time}\footnote{This is analogous to \eqref{eq:quantum mixing}, where we defined the mixing time for discrete-time quantum walks.} $\mathcal M^\mathrm{q}_\epsilon$ is then a time beyond which the total probability distribution distance from the limiting distribution is smaller than a precision parameter $\ep$, i.e. 
\begin{align}
	\forall T\geq \mathcal M^\mathrm{q}_\epsilon: \quad \norm{\bar{p}_T(x) - \pi(x)} \leq \ep.
\end{align}
Note that it is not governed only by the energy gap between the ground state and in the first excited state, as for classical Markov chains, but grows because of small gaps anywhere in the spectrum.

Continuing with our example --- the CTQW on a cycle of length $N$, we can compute the limiting distribution plugging \eqref{cyclevecs} and \eqref{cycleeigs} into \eqref{continuouslimiting}. 

\begin{exercise}
For a cycle with odd length $N$, with an initial state concentrated on a single vertex $x$, show that the limiting distribution is
\begin{align}
	\pi(x\rightarrow y) = \frac{1+\delta_{y,x}}{N}-\frac{1}{N^2},
\end{align}
while for even-length cycles, we get
\begin{align}
	\pi(x\rightarrow y) = \frac{1+\delta_{y,x}+ \delta_{y,x+N/2}}{N}-\frac{2}{N^2}.
\end{align}
\end{exercise}
Both of these are not uniform, having small corrections of the order $N^{-2}$ at each vertex, which make up for the different amplitude at the initial vertex (or the vertex directly opposite $x$ for the even-cycle case). Analyzing the inverses of differences of the non-equal eigenvalues in \eqref{inverseEs}, it can be shown that the mixing time for the CTQW on a cycle scales as 
\begin{align}
	\mathcal M^\mathrm{q,cycle}_\epsilon \leq \ep^{-1} N \log N.
\end{align}
This (superlinear scaling in $N$) can be connected to the linear spreading of the maximum of the wavefront as seen in Fig.\ref{besselfigure}. This behavior is interesting for uses in the computational models related to the Feynman computer, e.g. in \cite{NagajThesis}.

The scaling of the mixing times for CTQW has been also investigated for necklace graphs, which are cyclic graphs made by connecting many copies of a certain subgraph (pearl) \cite{necklacewalk}, and also utilized in a quantum-walk based  model of quantum computation \cite{universal2local}.


\subsubsection{Symmetries and Continuous-time Quantum Walks}
\label{sec:gluedtrees}

When the graphs we walk on have symmetries, analyzing the dynamics simplifies a lot. The coherent evolution of superpositions also often gives interesting results, when constructive interference at specific times can significantly raise the probability of being at specific vertices. 

For example, let us look at the continuous-time quantum walk on the {\em hypercube} (see also Section \ref{sec:graphsearching} and Figure \ref{fig:hypercubefigure}). The adjacency matrix of an $n$-dimensional hypercube is conveniently written in terms of the Pauli matrices as
\be
	A = \sum_{j=1}^{n} \sigma_x^{(j)}.
\ee 
Schr\"{o}dinger time evolution of the initial state $\ket{00\cdots0}$ with the Hamiltonian $H=-A$ is surprisingly simple, as all the $\sigma_x^{(j)}$ matrices commute. The evolution is thus an independent rotation on each qubit: 
\be
	e^{-iHt} \ket{00\cdots0} 
	= \bigotimes_{j=1}^{k} \left( e^{it \sigma_{x}^{(j)}} \ket{0}_{j} \right)
	= \bigotimes_{j=1}^{k} \left( \cos t \ket{0}_j + i \sin t \ket{1}_1 \right).
\ee
At times $t=\frac{(2k+1)\pi}{2}$, all the qubits are rotated into the state $\ket{1}$, so the continuous-time quantum walk starting in the state $\ket{00\cdots 0}$ traverses the hypercube completely in constant time. However, it is important to remember that it is not only the time that counts as `cost' of a continuous-time quantum walk algorithm. Rather, it is the dimensionless parameter $\norm{H}t$.
\begin{exercise}
Show that the norm (largest eigenvalue) of the Hamiltonian $A$ is linear in $n$.  
\end{exercise}
When we rescale $A \rightarrow \frac{1}{\norm{A}}A$ to give it norm $1$, the time required for traversal becomes $\frac{\pi n}{2}$, linear in $n$, as we have seen in Section \ref{sec:graphsearching} for the discrete quantum walk on a hypercube. Compare this to the classical random walk on a hypercube, where the walk rapidly mixes towards the uniform distribution, where the probability of finding the string $\mathtt{11\cdots 1}$ is $2^{-n}$.

What was the symmetry that comes into play in this example? Due to the permutational symmetry of the hypercube (and a Hamiltonian that preserves it), one can also view the evolution as a quantum walk on a line of states (the $n+1$ symmetric superpositions with Hamming weight $k=0, \dots, n$), with transition coefficients $\sqrt{(n-k)(k+1)}$.

\begin{figure}
\begin{center}
\includegraphics[width=4.5in]{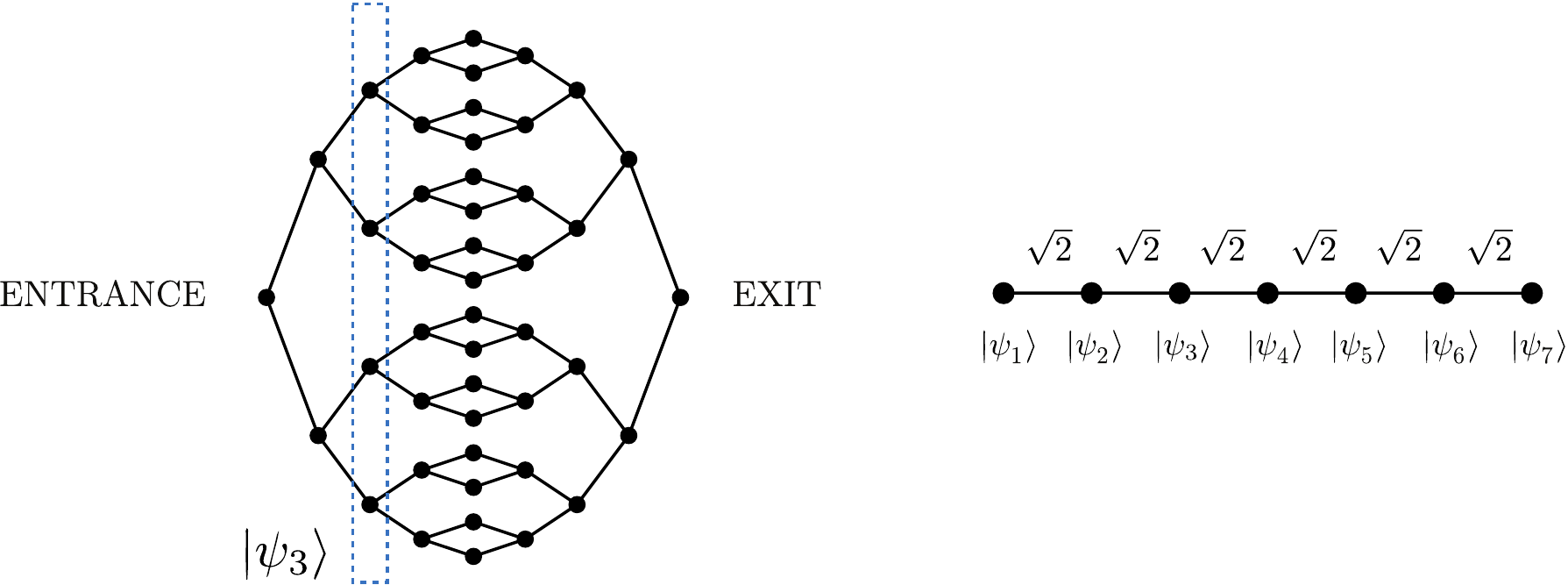}
\end{center}
\caption{Two binary trees with 4 layers make up the graph $G_4$ when glued together. The line on the right depicts the equivalent quantum walk on a line of the column states $\ket{\psi_n}$, with weight $\sqrt{2}$ on each link.
}\label{gluedtreesfigure}
\end{figure}

The next example of symmetries simplifying the evolution comes from Childs et al. \cite{ChildsWalkDifference}. Imagine we walk on a graph $G_n$ that is made from two glued binary threes of depth $n$ (with $2^{n-1}$ leaves each) as in Figure \ref{gluedtreesfigure}. This is reminiscent of (but much simpler than) the graph\footnote{The actual graph there is much more complicated, and we don't yet know how to naturally quantize the WALKSAT algorithm besides the general ``Groverizing'' approach.} encountered by the WALKSAT algorithm for 3-SAT (Section \ref{sec:SATsolving}), with each node having 2 ways to go towards the center, but only 1 way to go towards the endpoints. However, what does a continuous-time quantum walk see here? When we choose the Hamiltonian to be the negative of the adjacency matrix
\be
	H_{tt} = - A_{tree_1} - A_{tree_2}, \label{Htwotrees1}
\ee
its action turns out to be very simple if we view it in terms of the ``column'' states
\be
	\ket{\psi_k} = \frac{1}{\sqrt{n_k}}\sum_{j=1}^{n_k} \kets{x^{(k)}_j}. \label{columnstates}
\ee
Each $\ket{\psi_k}$ is a uniform superposition of the $n_k$ states located at the vertices of the $k$-th column of the graph.
In this basis $\{ \ket{\psi_k} \}_{k=1}^{2n-1}$, the Hamiltonian is the adjacency matrix of a quantum walk on a line of length $2n-1$ with weights $\sqrt{2}$, as
\be
	H_{tt} \ket{\psi_k} &=&  - \sqrt{2}\ket{\psi_{k+1}} - \sqrt{2}\ket{\psi_{k-1}}, \qquad 2\leq k \leq 2n-2, \\
	H_{tt} \ket{\psi_1} &=&  - \sqrt{2}\ket{\psi_{2}}, \label{twotreeaction} \\
	H_{tt} \ket{\psi_{2n-1}} &=&  - \sqrt{2}\ket{\psi_{2n-2}}. 
\ee
We have already seen how this walk behaves\footnote{Recall that we started the walk from the center of a finite line (or any point on a cycle) in Section \ref{sec:continuouswalk1D}. Nevertheless, if one starts the walk from an endpoint, the probability to be in another vertex again starts to spread linearly with time.} in Section \ref{sec:continuouswalk1D}, because the extra weight $\sqrt{2}$ only multiplies the whole Hamiltonian.
Thus the time it takes for the ``walker'' to move from the left endpoint to the right endpoint (with constant probability) scales as $O\left(\frac{2n-1}{2\sqrt{2}}\right)$, only linearly with the tree depth $n$. 

Compare this to a classical random walk without memory, which will (highly probably) get stuck in the center region of the graph. Note though, that there exists a classical recursive $O(n^2)$ algorithm for traversing this type of graph (see Section \ref{sec:graphsearching}). It crucially depends on the possibility of identifying the center column vertices by their degree.

\begin{figure}
\begin{center}
\includegraphics[width=4.5in]{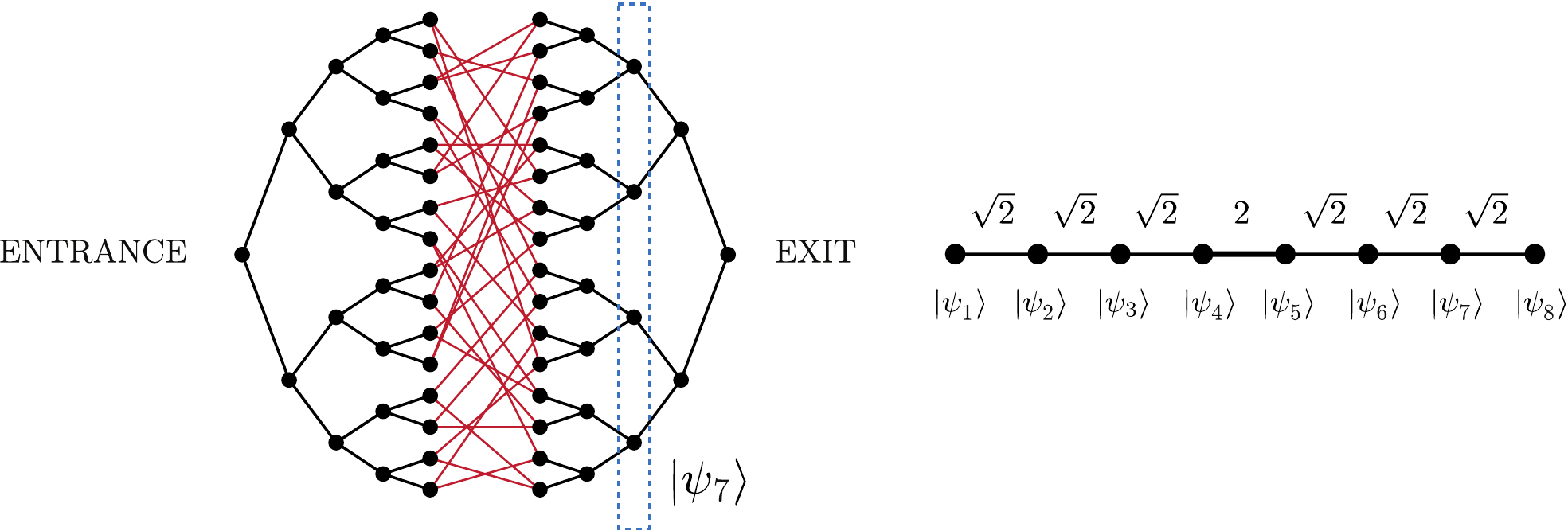}
\end{center}
\caption{Two binary trees with 4 layers, glued by a randomly chosen cycle of length $2^{n+1}$ with vertices alternating between the two trees. Every vertex (besides the endpoints) in the graph now has three neighbors. The line on the right depicts the equivalent quantum walk on a line of the column states $\ket{\psi_n}$ with modified weights.
}\label{gluedtreesfigure2}
\end{figure}

Soon afterwards, Childs et al. \cite{ChildsTreesExp} modified the previous construction, gluing the two trees by a cycle of length $2^{n+1}$ alternating between the leaves of the two trees as in Figure \ref{gluedtreesfigure2}, ensuring the degree of the central vertices is also 3.
This gave them one of the few examples with a provable exponential separation in query complexity between the best possible classical algorithm and a simple quantum-walk algorithm for an oracle problem. Let us thus investigate the unitary evolution governed by the Hamiltonian
\be
	H_{tt2} = - A_{tree_1} - A_{cycle} - A_{tree_2}, \label{Htwotrees}
\ee
where $A_{tree}$ is the adjacency matrix of a tree and $A_{cycle}$ the adjacency matrix of the central cycle. Because of symmetry, we can again choose a basis of $2n$ ``column'' states $\ket{\psi_n}$ \eqref{columnstates}, with
\be
	n_k = \begin{cases}
					2^{k-1} 	& 1 \leq k \leq n, \\
					2^{2n-k} 	& n+1 \leq k \leq 2n.
	\end{cases}
\ee
The action of \eqref{Htwotrees} on $\ket{\psi_n}$ differs from \eqref{twotreeaction} only in the middle, where accounting for the cycle connections we obtain
\be
	H_{tt2} \ket{\psi_{n}} &=& - \sqrt{2} \ket{\psi_{n-1}} - 2 \ket{\psi_{n+1}}, \\
	H_{tt2} \ket{\psi_{n+1}} &=& - 2 \ket{\psi_{n}} - \sqrt{2} \ket{\psi_{n+2}}.
\ee
Therefore, the evolution with $H_{tt2}$ (in the symmetric subspace) is the same as the continuous-time quantum walk on a line depicted in Figure \ref{gluedtreesfigure2}, with weights $\sqrt{2}, \dots, \sqrt{2}, 2, \sqrt{2}, \dots, \sqrt{2}$. When we let the state $\ket{\psi_1}$ evolve, it now starts to be reasonable to expect that after time $O(n)$, there will be a significant probability of measuring the state located at the root of the tree on the right, i.e. for $e^{-iH_{tt}t}\ket{\psi_1}$ to have a large overlap with $\ket{\psi_{2n}}$.
As shown in \cite{ChildsTreesExp}, we can solve for eigenvectors and eigenvalues of this walk similarly to what we did on a line, and then show a mixing result to prove the probability of reaching \textsc{exit} in $O(n)$ time is substantial. 

The quantum walk on two-trees glued by a cycle solves a Hamiltonian oracle problem -- to find the vertex named \textsc{exit}. We can turn it into an oracle problem in the usual quantum circuit model. For this, we consider a black-box oracle returning the names of neighbors of a given vertex. We can then efficiently simulate \cite{HamiltonianSimulation} the quantum walk with the Hamiltonian \eqref{Htwotrees}, because the Hamiltonian is sparse. 

Unlike the two simply glued trees traversal problem in Figure \eqref{gluedtreesfigure}, the oracle problem with two trees glued by a cycle as in Figure \eqref{gluedtreesfigure2} is difficult classically \cite{ChildsTreesExp}. Here is a sketch of the proof. First, we can think of the vertices having random names of length $3n$. Only $2^{2n}$ out of those $2^{3n}$ strings thus correspond to graph vertices, so that it is hard to guess a name of a vertex actually belonging to the graph. This restricts any classical algorithm to explore only a connected part of the graph around the \textsc{entrance} (as the oracle can supply the names of the neighbors of a given vertex). However, whichever way we explore the graph, we can embed the part we have seen into the glued trees + cycle at random. For a number of queries that is exponential in $n$ but much less than the number of vertices of one of the trees, the probability of finding the \textsc{exit} (or even reaching the same vertex we already have been in by a different path) will remain exponentially small.

Recently, another quantum algorithm for traversing randomly glued trees has been found \cite{AQCtrees}
in the adiabatic quantum algorithm (AQC) model. It is inspired by the quantum walk-based one presented above. However, the AQC algorithm differs from the usual approaches \cite{FarhiAdiabatic} in that it utilizes two lowest energy levels (instead of only the ground state), and does not care that the eigenvalue gap between these two levels is exponentially small.


\subsection{Spatial Search}
\label{ContinuousSearchSection}

We have seen how to perform searches for marked vertices on graphs using coined quantum walks in Section \ref{sec:graphsearching}. The goal was to concentrate the amplitude of the evolved state on a specially marked vertex (which had different coin-flipping or scattering properties). We can perform the same feat with continuous-time quantum walks. In fact, the development of the corresponding algorithms was often almost parallel, with continuous quantum walks leading the way \cite{ChildsSearch} (finite-dimensional lattices) and later catching up \cite{ChildsDirac} to the competing discrete-time method \cite{AmKeRi05}.

Search algorithms based on continuous-time quantum walks are based on the following common idea. We start with a uniform superposition $\ket{s}$ over all vertices and let it evolve with the Hamiltonian
\be
	H_{search} = - \gamma A_{graph} - \sum_{w\in W} \ket{w}\bra{w}, \label{Hsearch}
\ee
where $A$ is the adjacency matrix (or sometimes, the Laplacian) of the graph, $W$ is a set of marked vertices and $\gamma$ is a tunable parameter. Often, the ground state of \eqref{Hsearch} and its first excited state will both have large overlap with the uniform superposition $\ket{s}$ and some marked state $\ket{w}$. When we let the state $\ket{s}$ evolve, due to quantum tunneling, the state $\ket{s}$ will then transform into a state with a large overlap with the state $\ket{w}$, on the timescale of $\frac{1}{E_1-E_0}$. Let us look at a few examples of how this happens, starting with the unstructured search problem, and a quantum walk algorithm that is an analog analogue of Grover's algorithm \cite{AnalogAnalogue}.


\subsubsection{Complete Graph}

An unstructured search of $N$ items is equivalent to a spatial search on a complete graph with $N$ vertices (where we can freely jump from any vertex to any other vertex) as in Figure \ref{fig:CG}. For simplicity, let us now think only of a single marked vertex.
In Section \ref{sec:cg}, we have seen how this problem is solved by a scattering quantum walk. Let us now look at it in 
a Hamiltonian formulation, with the marked state specified by an oracle Hamiltonian
\be
	H_w = - \ket{w}\bra{w},
\ee
where the special vertex $\ket{w}$ (the ``winner'') gets a decrease in energy compared to the other vertices. Our goal is to prepare the state $\ket{w}$ (if there is such a state).
We will follow Farhi et al. \cite{AnalogAnalogue}, who worked this out soon after Grover's algorithm appeared.
The approach is to add an instance-independent Hamiltonian -- the rescaled adjacency matrix of the full graph.
We will rescale the adjacency matrix $A$, so that its norm is $O(1)$ (corresponding to $\gamma=\frac{1}{N}$ in \eqref{Hsearch}). The reason for this is the time-energy tradeoff, as increasing the norm of the Hamiltonian obviously decreases the required runtime.
Together with the oracle Hamiltonian, we will thus use 
\be
	H = - \frac{1}{N} A + H_w 
	= - \ket{s}\bra{s} - \ket{w}\bra{w},  \label{Hgrover}
\ee
where $\ket{s}=\frac{1}{\sqrt{N}}\sum_{j=1}^{N} \ket{j}$ is the uniform superposition of computational basis states. 

The algorithm is rather simple. We start in the uniform superposition $\ket{s}$ over all vertices, and let the system evolve (according to the Schr\"{o}dinger equation) for time $O\left(\sqrt{N}\right)$. Let us look at why it works. The Hamiltonian \eqref{Hgrover} never takes the evolution of $\ket{s}$ outside of the two-dimensional Hilbert space spanned by $\ket{w}$ and $\ket{s}$. We can thus rewrite the Hamiltonian in (and restricted to) the basis $\{\ket{w},\kets{w^\perp}\}$, where
\be
	\kets{w^\perp} = \frac{\ket{s}-\delta \ket{w}}{\sqrt{1-\delta^2}},
\ee
with $\delta = \frac{1}{\sqrt{N}}$ (this generalizes to $\sqrt{M/N}$ if there are $M$ marked states). In the new basis, the Hamiltonian reads
\be
	H &=& - \left(\ket{w}+\sqrt{1-\delta^2}\kets{w^\perp}\right)
	\left(\bra{w}+\sqrt{1-\delta^2}\bras{w^\perp}\right) - \ket{w}\bra{w} 
	\nonumber\\
	&=& - \left[\begin{array}{cc}
		\delta^2 + 1 & \delta \sqrt{1-\delta^2} \\
		\delta \sqrt{1-\delta^2} & 1-\delta^2
	\end{array}\right] 
	= -\iii + \delta\left(
	\delta \sigma_z
	+ \sqrt{1-\delta^2} \sigma_x\right).
\ee
Using $e^{i \alpha \vec{\hat{n}}\cdot\vec{\sigma}} = (\cos\alpha)\iii + i (\sin \alpha)  \vec{\hat{n}}\cdot \vec{\sigma}$, the time evolution of $\ket{s} = \delta \ket{w} + \sqrt{1-\delta^2} \kets{w^\perp}$, up to an insignificant phase is
\be
	e^{-iHt}\ket{s} &=& 
	\left[
	\begin{array}{cc}
		\cos (\delta t) -i \delta \sin(\delta t) & -i \sqrt{1-\delta^2} \sin(\delta t) \\
		-i \sqrt{1-\delta^2} \sin(\delta t) & \cos (\delta t) +i \delta \sin(\delta t)
		\end{array}
	\right]
	\left[
		\begin{array}{c}
			\delta\\
			\sqrt{1-\delta^2}
		\end{array}
	\right] \nonumber\\
	&=& 
	\left[
		\begin{array}{c}
			\delta \cos(\delta t) - i \sin(\delta t)\\
			\sqrt{1-\delta^2}\cos(\delta t)
		\end{array}
	\right]
	= \cos(\delta t) \ket{s} - i \sin(\delta t) \ket{w}.
\ee
Therefore, it is enough to wait time $T = \frac{\pi}{2\delta} = \frac{\pi}{2}\sqrt{N}$
to obtain the marked state $\ket{w}$ with probability 1.

In fact, this algorithm is optimal, and the reader can find the instructive proof -- an alternative to the BBBV proof \cite{BBBV} in the Hamiltonian oracle model -- in the paper\footnote{An Analog Analogue of a Digital quantum computation} by Farhi et al. \cite{AnalogAnalogue}, who were quick to realize they got the Grover search algorithm in continuous time soon after Grover published his original paper \cite{Grover97}. The proof is based on analyzing the evolution of a known state towards a marked state $\ket{w}$ with the oracle Hamiltonian $-\ket{w}\bra{w}$ plus an arbitrary driver Hamiltonian (without knowledge of the solution). For this algorithm (which doesn't know $w$) to work for all $N$ possible choices of $w$, the evolution of an initial state must substantially differ for all of those. That in turn lower bounds the required evolution time $T$ by $T\geq O\left(\frac{\sqrt{N}}{\norm{H}}\right)$.


\subsubsection{Searching on the Hypercube and on Finite-dimensional Lattices}
The next example where continuous-time quantum search provides a fast algorithm is search on the hypercube (cf. Section \ref{sec:discretehitting} for the discrete-time approach). Following Childs et al., \cite{ChildsSearch, ChildsDirac}, we will investigate the Hamiltonian  
\be
	H = - \gamma A_{hc}  - \ket{w}\bra{w},  \label{Hhyper}
\ee
where $A_{hc}=\sum_{j=1}^{n} \sigma_x^{(j)}$ is the adjacency matrix of a $n$-dimensional hypercube (described by $n$-qubit states). 
The spectrum of this Hamiltonian was analyzed by Farhi et al. in \cite{FarhiAdiabatic}. It turns out that for a specific choice of $\gamma = \frac{2}{n} + O\left(n^{-2}\right)$, the energy gap is 
$\frac{1}{\sqrt{2^n}} \left[1+O\left(n^{-1}\right)\right]$, and the ground 
and the first excited state of \eqref{Hhyper} are $\frac{1}{\sqrt{2}}\left(\ket{w}\pm\ket{s}\right)$ up to terms of order $O\left(n^{-1}\right)$. Thus, evolving the state $\ket{s}$ for time of order $\sqrt{2^n}$, the probability of finding $\ket{w}$ is of order 1. This is again a quadratic speedup over a classical algorithm, just as the one using discrete-time quantum walk search \cite{Kempe05}.

Let us now move to a $d$-dimensional lattice. Childs et al. \cite{ChildsSearch} investigate the Hamiltonian \eqref{Hsearch} and find a critical point in $\gamma$, where the gap is small, but the ground and the excited states again have large overlap with $\ket{w}$ and $\ket{s}$. Altogether, their algorithm has running time $O\left(\sqrt{2^n \log 2^n } \right)$ in dimensions $d\geq 4$. There it gives a quadratic speedup (up to a log-factor for $d=4$) over the best classical algorithm. However, the continuous quantum walk search does not work for $d<4$. 


The discrete-time lattice search algorithm \cite{ShKeWh03} has runtime $\sqrt{2^n}$ all the way down to $d=2$. Can continuous quantum walks work there too? It turns out yes, but it requires augmentation. Childs et al. \cite{ChildsDirac} showed that adding an extra degree of freedom (spin) of the walking particle and replacing the Laplacian (or the adjacency matrix) with the massless Dirac Hamiltonian helps. The dynamics is now governed by the Dirac equation (for a massless particle). 
The energy spectrum near $E=0$ now has linear (instead of quadratic) dispersion (dependence on the momentum). This is the underlying reason for the gap at the critical point being large enough to give a speedup for the continuous-time quantum walk algorithm. The result is then an algorithm with running time $O\left(\sqrt{2^n \log 2^n } \right)$ in dimensions $d=2$ and $d=3$, matching the discrete-time walk, at the cost of a larger Hilbert space.



\subsection{NAND Trees and Games}
\label{sec:NAND}

Let us have a look at another application of continuous-time quantum walks. In the early days of quantum computing, the paper {\em Quantum Computation and Decision Trees} \cite{FarhiWalk} investigated a system whose Hamiltonian is the adjacency matrix of a binary tree, and looked at the transmission and spreading of wavepackets (and their possible concentration on searched-for vertices). Here, we will see how the transmission-reflection properties of a graph (under time-evolution with an adjacency-matrix Hamiltonian) are affected by a certain graph property. The result of this analysis is a continuous quantum walk-based algorithm \cite{FarhiNAND} for evaluating NAND trees of depth $n=\log_2 N$, whose query complexity $O\left(N^{0.5}\right)$ beats the best possible classical algorithm which requires $O\left(N^{0.753}\right)$ oracle queries. This quantum algorithm developed in the Hamiltonian oracle model has been quickly translated into the more common discrete-query-model \cite{ChildsNAND}, and then generalized to arbitrary trees and optimized by Ambainis et al. \cite{AmbainisNAND}.

\begin{figure}
\begin{center}
\includegraphics[width=4in]{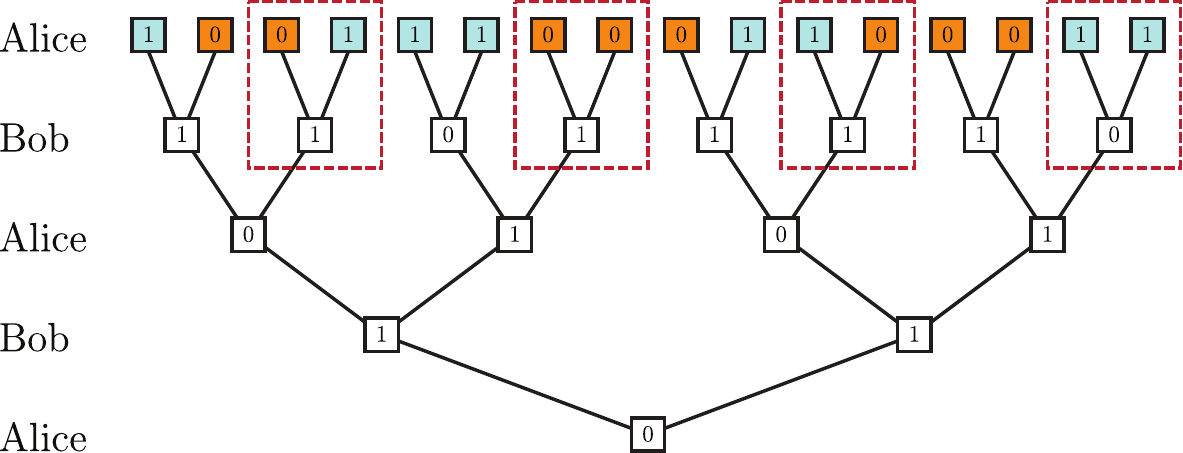}
\end{center}
\caption{A game tree for a game with two players alternating moves, starting at the root of the tree. If Alice (who chooses her move first) ends up at a leaf labeled 1, she wins. Given a leaf labeling, we fill the vertices using the NAND operation (as in the marked rectangles). For this particular game, Alice does not have a winning strategy, assuming best possible play from Bob.
}\label{nandgamefigure}
\end{figure}

Alice (who starts) and Bob (going second) are playing a simple game for two alternating players, making one of two possible moves. This game can be described by a binary tree of depth $n$, with the $N=2^n$ leaves  labeled by the results of the game as in Figure \ref{nandgamefigure}. Does Alice (the first player to move) have a winning strategy, ensuring she ends up at a leaf labeled 1? We can figure it out by working our way down the tree from the leaves to the root. If Alice reaches a leaf labeled 1, she wins, if she gets to a leaf with a 0, she loses. When Bob chooses his move on the layer $n-1$, he will surely choose a win for himself by sending Alice to a 0-labeled leaf (if one like that exists). Thus, if there is a 0-leaf above his position, he has a winning strategy, and we can label that graph vertex 1. On the other hand, if both leaves above Bob's vertex are 1, he can not win (with best play from the opponent), and so we will label it 0 as in the marked rectangles in Figure \ref{nandgamefigure}. This works at any level of the tree, so we can work our way down to the root using the NAND function on the bits on the children of each node $x = \neg (c_1 \wedge c_2)$. The problem of determining the existence of a winning strategy for Alice in this game is thus equivalent to evaluating a binary NAND tree with a set of game results fixed on the leaves. 

The provably optimal \cite{NANDoptimality} classical algorithm for this problem uses randomized recursive branch evaluation. To evaluate a given vertex, we randomly pick one of its children and continue with the recursive evaluation. If we conclude with a 0, the value at the original vertex is certainly 1 (because of the property of the NAND). Only if the evaluation produces a 1, we need to do more work and evaluate the other branch as well. For a game with a balanced\footnote{In a balanced tree, the lengths of the subtrees from a node do not differ by more than 1.} binary tree, this recursive procedure requires $O(N^{0.753})$ calls to the oracle encoding the game results at the leaves. 

\begin{figure}
\begin{center}
\includegraphics[width=2in]{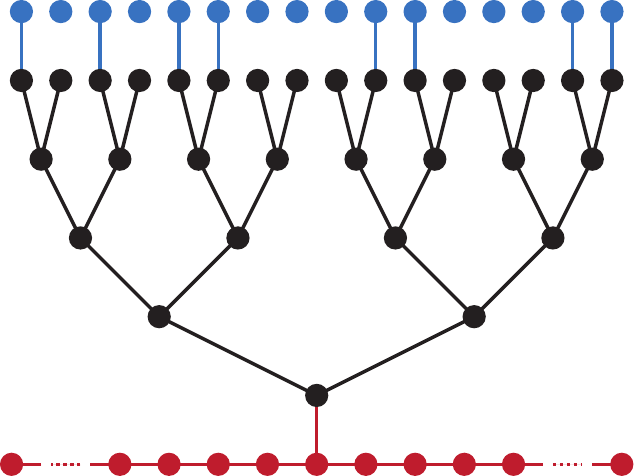}
\end{center}
\caption{A graph for the quantum walk of the NAND-tree algorithm, corresponding to the game in Figure \ref{nandgamefigure}. Its centerpart is a balanced binary tree. Above it we have an extra row of vertices, connected to the tree by edges encoding a particular game (there is an extra edge only for those tree vertices that originally carried a ``1'' label). Finally, the tree is connected to a $\sqrt{N}$-long line (runway).
}\label{nandtreerunway}
\end{figure}

The basis of the quantum algorithm of Farhi et al. \cite{FarhiNAND} is a continuous-time quantum walk on a graph, whose main part is a balanced tree with extra leaves at the vertices with game results equal to 1. This tree is rooted above the middle of a long line of vertices\footnote{As shown in \cite{AmbainisNAND}, this long line can be shortened to just one vertex on each side, with additional weight $\sqrt{N}$.} as in Figure \ref{nandtreerunway}. On the runway (the long line), the eigenvectors of an adjacency-matrix Hamiltonian have to be combinations of plane waves. 
When we look at eigenvectors with energies close to 0 (and momenta close to $\pi/2$), we find that the graph possesses interesting scattering properties. We look for eigenvectors 
\be
	\ket{\psi} \propto \sum_{x\leq0} e^{-i p x} \ket{x} + R \sum_{x\leq0} e^{i p x} \ket{x} + T \sum_{x>0} e^{-i p x} \ket{x} + \ket{\varphi_{tree}},
\ee
which are a combination of a right-moving wave with a reflected and transmitted part, plus something inside the tree.
For NAND-trees evaluating to 1 (with a winning strategy for Alice, the starting player), Farhi et al. claim that the close-to-zero-energy eigenvectors have little support on the root of the tree. The eigenvalue equation at the vertex $x=0$ right below the tree root then reads
\be
	\bra{x=0} H \ket{\psi} = e^{ip} + R e^{-ip} + T e^{ip} 
	= \left(e^{-ip} + e^{ip}\right) + T \left( e^{ip} - e^{-ip} \right).
\ee
Continuity dictates $1+R=T$. Assuming $E \ket{0} \approx 0$ then implies $T=1$, perfect transmission for close to E=0 wavepackets.
Small overlap on the root of the tree thus implies support on the exit runway of the graph. On the other hand, if the NAND-tree evaluates to 0, the close-to-zero-energy eigenvectors do have significant support on the root of the tree, which in turn implies no support on the exit line for these states. Viewed as a scattering problem, we can think of putting a close-to-zero-energy wavepacket on the input line of the graph, let it evolve, and then measure the system to determine whether the wavepacket has reflected from the tree (NAND=0), or passed through it (NAND=1). Let us sketch why this is the case, utilizing a the recursive evaluation of the tree. 

\begin{figure}
\begin{center}
\includegraphics[width=4.5in]{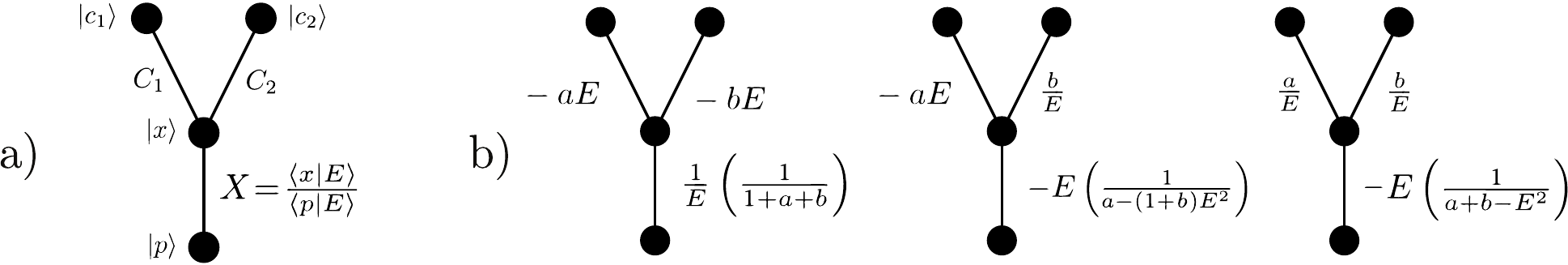}
\end{center}
\caption{a) Defining the ratios of amplitudes at the vertex $x$ and its children $c_i$ and parent $p$ for a NAND tree. b) Evaluating the ratio $X$ from known ratios $C_i$ using \eqref{recursion2E}. When at least one of the $C_i$'s is large and positive, we get a small $X$, just as (at least) a single 0 for a child implies a 1 on a NAND tree.
}\label{nandtreerecursionEfigure}
\end{figure}

When we look for the eigenstates of the adjacency matrix of the tree, we would like to have $H\ket{E}=E\ket{E}$. At vertex $x$, it means that
\be
	\bra{x}H\ket{E} &=& E\braket{x}{E} \\
	\bra{x}H\ket{E} &=& \braket{p}{E} + \sum_{i}\braket{c_i}{E},
\ee 
where $c_i$ are the children of $x$ and $p$ is the parent vertex of $x$. The values of $\braket{c_i}{E}$ and $\braket{x}{E}$ thus determine $\braket{p}{E}$ as 
\be
	\braket{p}{E} &=& E\braket{x}{E} - \sum_{c}\braket{c}{E}, \label{recursion1}
\ee 
which is especially simple for $E=0$. 
Denoting the ratios of amplitudes $C_i = \frac{\braket{c_i}{E}}{\braket{x}{E}}$ and $X = \frac{\braket{x}{E}}{\braket{p}{E}}$ as in Figure \ref{nandtreerecursionEfigure}a) translates \eqref{recursion1} to
\be
	X = \frac{1}{E-C_1-C_2}. \label{recursion2E}
\ee
We will now think about eigenstates with small $E>0$. We now want to show the following correspondence between the ratios $X$, the overlaps $\braket{x}{E}$ and the NAND value of the subtree rooted at vertex $x$:
\be
	\textrm{small } X  \quad \longleftrightarrow \quad \textrm{negligible overlap } \braket{x}{E} 
		\quad &\longleftrightarrow& \quad \textrm{NAND=1 at vertex $x$} \nonumber\\
	\textrm{large } X  \quad \longleftrightarrow \quad \textrm{reasonable overlap } \braket{x}{E} 
		\quad &\longleftrightarrow& \quad \textrm{NAND=0 at $x$, NAND=1 at $p$.} \nonumber
\ee
Let us start at a vertex $x$ which is a leaf. It has no children (whose $C_i$'s are thus zero), making its $X$ large and positive. Using \eqref{recursion2E}, we can work out the ratios down the tree as in Figure \ref{nandtreerecursionEfigure}b). When both $C_i$'s are small and not positive, $X$ is going to be large and positive. On the other hand, if at least one of the $C_i$'s is large, $X$ is going to be small and negative. In this sense, \eqref{recursion2E} evaluates the NAND operation, as described by the mapping above. What does it mean for the case when $x$ is the root of the tree, and $p$, its parent is already part of the runway? 
Small $X$ implies that the overlap $\braket{x}{E}$ is tiny, while the whole tree has NAND=1. On the other hand, a large ratio $X$ would imply NAND=0 for the whole tree, and $\braket{x}{E}$ not entirely small. 


This was a sketch of the actual proof of Farhi et al. \cite{FarhiNAND}, who show that for eigenstates of $H$ with $|E| < \frac{\ep}{16\sqrt{N}}$, the overlap at the root in the NAND=1 case is no larger than $\ep$. This happens because a ``small'' overlap on some far-out vertices can not grow much more than by doubling when we go down the tree from the leaves two levels at a time using \eqref{recursion2E}. 

The peculiar scattering properties of the NAND tree (arising from the small overlap of small $E$ states on the root of the NAND=1 trees) shown by Farhi et al. are retained for non-regular NAND trees, as shown by Ambainis et al. \cite{AmbainisNAND}. One extra ingredient in their approach is that they shorten the runway to just 3 vertices, and add extra weight to the connections. Furthermore, evaluating whether the NAND=0 or 1 for the tree is done by phase-estimation of the quantum walk on this graph.


\subsection{Quantum Walks and Universal Quantum Computation}
\label{sec:universality}

We have investigated continuous-time quantum walks with Hamiltonians that are adjacency matrices of graphs, i.e. they contain only 0's and 1's (or possibly, with some weights at the edges). The time-evolution of simple initial states can be used for searching (on regular lattices), to traverse graphs in interesting (e.g. for glued trees) ways, or to investigate graph properties (such as evaluating NAND trees). Thanks to Childs et al. \cite{ContinuousUniversal, ContinuousUniversal2}, there is one more surprising application -- {\em universal quantum computation}. It comes in two varieties.

First, there is a way to translate an arbitrary unitary transformation on $n$-qubits into the evolution of a continuous-time quantum walk on a non-weighted graph with maximum vertex degree 3 \cite{ContinuousUniversal}. This idea has also been later translated to the {\em discrete} quantum walk model by Lovett et al. \cite{DiscreteUniversal}.  Although this first model is universal for quantum computation, we can not hope to construct the graph physically, as the number of vertices is exponential in $n$. However, this implies that simulating  a quantum walk (the evolution of a single excitation) on a sparse, easily computable and describable, unweighted graph is as difficult as general quantum computation. Thus, it is one more example of a problem that can be asked purely classically, in terms of $0/1$ matrices, but which has a deep relationship to quantum computation (cf. for example \cite{stringrewriting}). 

Second, a novel idea recently appeared in \cite{ContinuousUniversal2}. Instead of using a single walker (a wavepacket with a certain momentum) and a ``wire'' for each of the $2^n$ computational basis states, Childs et al. decided to use multiple walkers -- one walker per qubit. The state of a qubit is encoded by an excitation moving in real time on a dual-rail\footnote{A dual rail construction has two wires per qubit. An excitation on the first line corresponds to the state $\ket{0}$, while an excitation on the second wire corresponds to $\ket{1}$. The state on a dual-rail graph can also be in a coherent superposition of $\ket{0}$ and $\ket{1}$ -- encoding a qubit.}. Single-qubit gates are performed just as in the first model, while they found a way to implement a 2-qubit gate (C-PHASE) by scattering of excitations on two neighboring wires via a gadget graph connecting the two wires (utilizing an extra ancillary wavepacket with a different momentum). 
The overall size of the graph is no longer exponential, now it is only $poly(n,L)$ (a rather large polynomial at the moment), with $n$ the number of qubits used and $L$ the number of gates in the circuit. The analysis of the underlying scattering process is beyond the scope of this article, so we choose to explain only the first, simpler (but exponential in space) construction.

\begin{figure}
\begin{center}
\includegraphics[width=4.8in]{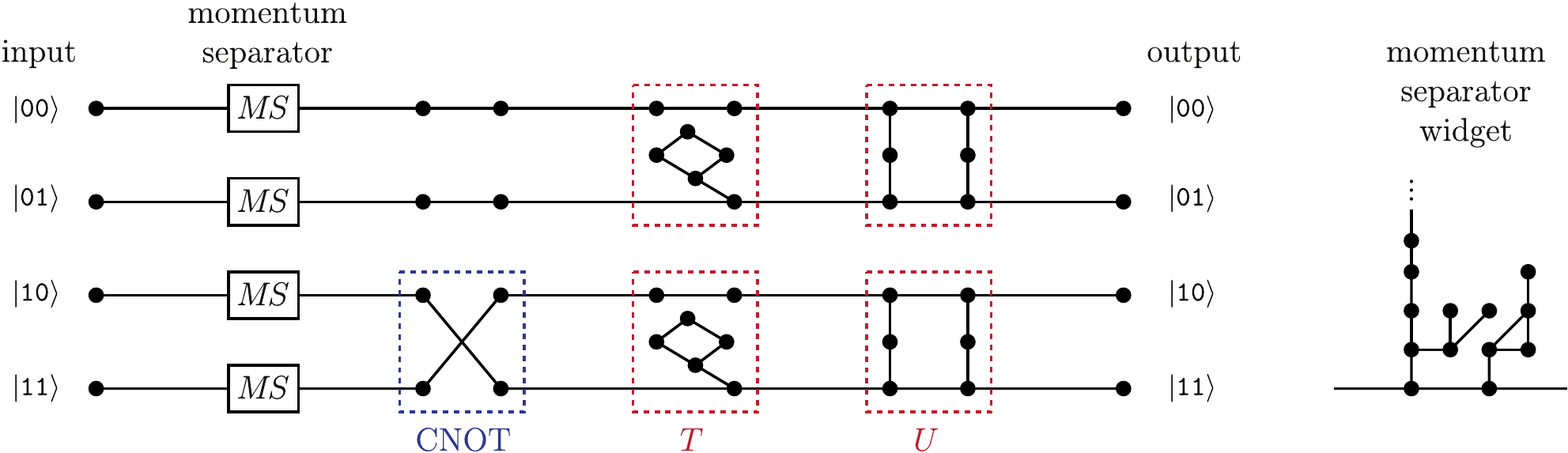}
\end{center}
\caption{A graph corresponding to a 2-qubit circuit has 4 wires. It starts with momentum separator widgets (depicted in detail on the right), and then 3 gate widgets. First, the CNOT gate between the two qubits, then the $T$ gate \eqref{Tgate} on the second qubit, and finally a $U$ gate \eqref{Ugate} on the second qubit.
}\label{universalfigure}
\end{figure}

Let us now sketch the first idea. How can we achieve arbitrary unitary transformations just from letting a continuous quantum walk run on a non-weighted graph? For an $n$-qubit unitary, we will need $2^n$ quantum wires (corresponding to the $2^n$ basis states) of $poly(n)$ length, as in Figure \ref{universalfigure}. These wires will be connected by gate widgets, according to what quantum circuit we want to apply. If we worked with infinitely long wires, the eigenvectors of this Hamiltonian would be plane waves $\kets{\tilde{k}}$ characterized by their momenta $k$, with
\be
	\braket{x}{\tilde{k}} =e^{ikx},
\ee
at location $x$ and normalized to $\braket{\tilde{k}'}{\tilde{k}}=2\pi \delta(k-k')$, corresponding to eigenvalues $2 \cos k$, as we have seen for the walk on a line in Section \ref{sec:continuouswalk1D}. The graph here consists of incoming and outgoing lines (not infinite), plus the the graph widgets on which the plane waves can scatter. On the lines, the eigenvectors of $H$ can only be linear combinations of the plane waves traveling right or left (with momenta $\pm k$), possibly also with imaginary momenta, giving bound states. Given an incoming plane wave with particular momentum, scattering theory allows us to compute the transition (and reflection) coefficients for the wave on the incoming/outgoing lines. The graph widgets in \cite{ContinuousUniversal} are constructed in such a way that the reflection
coefficients for a plane wave with momentum $k=-\frac{\pi}{4}$ are exactly zero, while the transmission amplitudes correspond to desired unitary transformations. 

Universal computation requires interaction of qubits. This part is simple, as a \textsc{CNOT} gate between qubits $l$ and $m$ will be implemented by a crossing of the wires $\mathtt{\cdots 1}_l \, \mathtt{\cdots 0}_m\mathtt{\cdots}$ and $\mathtt{\cdots 1}_l \,\mathtt{\cdots 1}_m\mathtt{\cdots}$ as in Figure \ref{universalfigure}.
However, besides the \textsc{CNOT} operation, we also need a universal set of single-qubit gates. A single qubit gate on the $m$-th qubit can be implemented by adding a graph widget to the $\mathtt{\cdots 0}_m \mathtt{\cdots}$ and $\mathtt{\cdots 1}_m \mathtt{\cdots}$ wires. Having two types of gates at hand is sufficient for single-qubit universality. First, we need the $\frac{\pi}{8}$ gate\footnote{Up to an overall phase, it is equivalent to multiplying the $0$-amplitude by $e^{i\frac{\pi}{8}}$ and the $1$-amplitude by $e^{-i\frac{\pi}{8}}$.}
\be
	T = \left[
		\begin{array}{cc}
			1 & 0 \\
			0 & e^{i\frac{\pi}{4}}
		\end{array}
	  \right], \label{Tgate}
\ee
which can be implemented by the widget in Figure \ref{universalfigure} (repeated for all values of qubits not  involved in the $T$ gate). Second, we can utilize a basis-changing gate
\be
	U = -\frac{1}{\sqrt{2}} \left[
		\begin{array}{cc}
			i & 1 \\
			1 & i
		\end{array}
	  \right], \label{Ugate}
\ee
implemented as in Figure \ref{universalfigure}. 
While $(T,U)$ is is not the usual universal gate set ($x$ and $z$ rotations, or phase gate and Hadamard), observe that we can implement  the Hadamard gate\footnote{The Hadamard gate switches between the $x$ and $z$-bases as $H \ket{x\pm} =  \ket{z\pm}$.} $H$ using $T$ and $U$ as $H=i T^2 U T^2$, and the phase gate simply as $T^2$. 

Finally, because the single qubit gate widgets require a plane wave with specific momentum to work, we need to use a momentum separator/filter. This is a graph widget with transmission coefficients close to zero for momenta away from $k=-\frac{\pi}{4}$ and $\frac{3\pi}{4}$, while it lets the selected ones through completely. Its second part separates wavepackets with these two momenta in time, as the effective path length of the widget is different for them. The computation then consists of initializing the system with wave packets on corresponding input lines (e.g. only on the $\mathtt{00\cdots 0}$ line, and letting the system evolve with the graph adjacency matrix Hamiltonian. After time linear in the number of gate applications, we measure the amplitude of the desired output line\footnote{Note that for BQP universality it is enough to be able to initialize the system in the state $\ket{0}^{\otimes n}$, apply a quantum circuit and make a projective measurement onto the state $\ket{0}^{\otimes n}$ afterwards.}.


\subsection{Connecting Continuous Time and Discrete Time Quantum Walks}
\label{sec:connectthewalks}

It is now time to look at the similarities and differences for the two approaches to quantum walks. 
Coined or two-register discrete-time quantum walks are directly implementable using quantum circuits. On the other hand, continuos-time quantum walks are more natural in their interpretation as the dynamics of an excitation in a physical system, are generally easier to analyze, but require small degree of the vertices. Both methods have brought forth successful algorithms (discrete: graph search, element distinctness, continuous: glued trees traversal, NAND tree evaluation), many of which have been soon translated from one model to the other. We can see a clear analogy for the search algorithm on a 2D lattice, where the continuous-time quantum walk \cite{ChildsDirac} needs a ``coin'' -- an extra spin degree of freedom -- to work as fast as the discrete quantum walk algorithm. Meanwhile, a successful translation of the glued-trees traversal walk to a discrete-time quantum walk utilizing a 3-headed coin \cite{TrFlMaKe03} should be possible, but we don't have a rigorous analysis showing it. Worse than that, until recently we didn't know how to formulate a CTQW algorithm for the element-distinctness algorithm discussed in Section \ref{subsetfinding}. 

\begin{figure}
\begin{center}
\includegraphics[width=13.45cm]{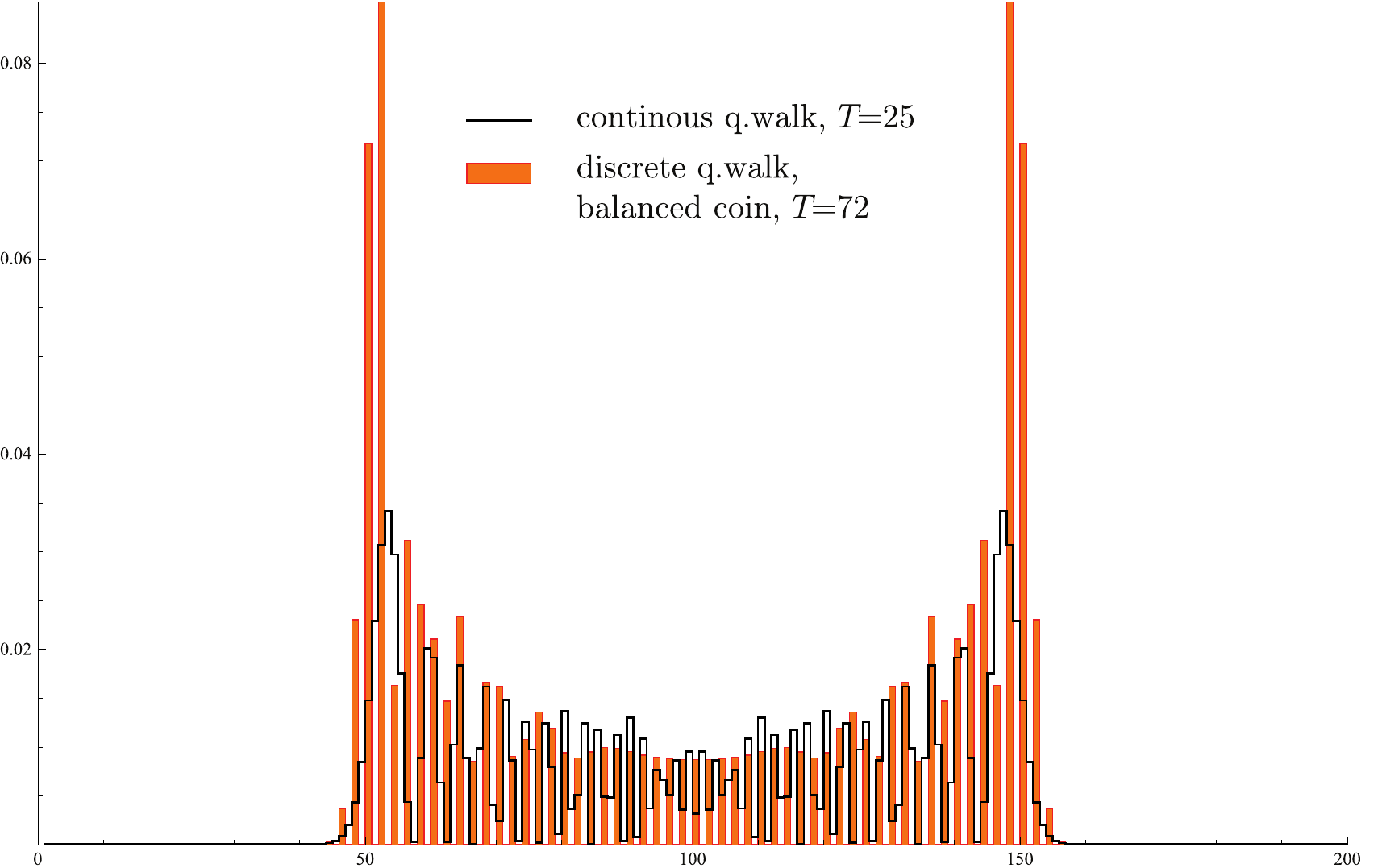}
\end{center}
\caption{Comparison of probabilities of finding the walker at a given spot for continuous and discrete quantum walks (balanced coin \eqref{eq:balancedcoin}, initial state $\ket{x=100}\otimes\left(\ket{\leftarrow}+\ket{\rightarrow}\right)/\sqrt{2}$) on a line, starting at the center ($x=100$). Note that the discrete quantum walk has nonzero probabilities only at odd sites. The speed of spreading is linear in both cases: $2$ for the continuous and $1/\sqrt{2}$ for the discrete quantum walk.
}\label{walkcomparisonfigure}
\end{figure}

Does there exist a simple correspondence between the two models, arriving at CTQW's as a short time-step limit of some DTQW? This can't be possible directly, as DTQW's are implemented in a larger Hilbert space -- including a coin register besides a position register (or on two position registers in scattering quantum walks or in Szegedy's formulation of quantized Markov chains, see Chapter \ref{sec:qmc}). However, the behavior of DTQW's and CTQW's is remarkably similar in several cases, e.g. in 1D (see Figure \ref{walkcomparisonfigure}). In \cite{ChildsDiscreteContinuous}, Andrew Childs has found a way of reproducing the dynamics of any Hamiltonian (thus including a CTQW) as a limit of discrete-time quantum walks. This in turn has implications for Hamiltonian simulation. Let us shortly sketch this construction.

Our goal is to take a CTQW with Hamiltonian $H$, and find a corresponding DTQW (in a limiting sense). The Hilbert space for the CTQW is spanned by the states $\ket{j}$. First, we will find a particular DTQW in Szegedy's two-register formulation, implemented in the space spanned by the vectors $\ket{j,k}$. To go between the state space of the CTQW and the DTQW, we will use an isometry
\begin{align}
	T:= \sum_{j=1}^N \ket{\psi_j}\bra{j}, \label{DCisometry}
\end{align}
defined in the following way. 
Let $\ket{d}=\sum_{j=1}^N d_j \ket{j}$ be the largest-eigenvalue eigenvector of the matrix
$\mathrm{abs}(H) = \sum_{j,k} |H_{jk}| \ket{j}\bra{k}$ (whose elements are the absolute values of the elements of $H$).
From the Perron-Frobenius theorem, we know that all $d_j$ must be nonnegative. We denote the largest eigenvalue of $\mathrm{abs}(H)$ as $\norm{\mathrm{abs}(H)}$.
The orthonormal (easily checkable) states $\ket{\psi_j}$ in the isometry \eqref{DCisometry} are then defined as 
\begin{align}
	\ket{\psi_j}:= \frac{1}{\sqrt{\norm{\mathrm{abs}(H)}}} \sum_{k=1}^N \sqrt{H^*_{jk}\frac{d_k}{d_j}} \ket{j,k}.
\end{align}
Using the states $\ket{\psi_j}$, we can define this discrete-time quantum walk: 
\begin{enumerate}
	\item Map an initial state $\ket{j}$ to a two-register state $\ket{\psi_j}$ using the isometry $T$. 
	\item Apply a reflection about the span of the states $\ket{\psi_j}$, swap the registers and add a $\frac{\pi}{2}$ phase, i.e. apply
	\begin{align}
		U = iS \Big( 2\sum_{j} \ket{\psi_j}\bra{\psi_j} -\id  \Big)= iS (2TT^\dagger - \id)
	\end{align}
	\item Repeat step 2 (a given number of times). After that, use the inverse isometry $T^\dagger$ to get back to the original state space.  
\end{enumerate}
This DTQW is interesting, because if the rescaled eigenvalues $\lambda$ of $\frac{H}{\norm{\mathrm{abs}(H)}}$ are small, the eigenvalues of the DTQW can be shown to be approximately $\mu_{\pm} \approx \pm e^{\pm i \lambda}$. Furthermore, with additional rotations at the beginning and end, we obtain an approximation (valid up to $O(\lambda)$) of the evolution according to a CTQW for time $\tau$
\begin{align}
	e^{-i\frac{H}{\mathrm{abs}(H)}\tau}=\sum_{\lambda} e^{-i\lambda \tau} \ket{\lambda}\bra{\lambda}
	\approx 
	T^\dagger \frac{(1-iS)}{\sqrt{2}} \,(iU)^\tau\, \frac{(1+iS)}{\sqrt{2}} T, \label{DCapprox}
\end{align} 
in the form of a DTQW with $\tau$ reflections and register-swaps. 

Still, this is working only when all of the $\lambda$'s are small, i.e. when $h=\frac{\norm{H}}{\norm{\mathrm{abs}(H)}}$ is small. We can make $h$ small by taking a lazy walk instead of the original one. A lazy walk makes a step only with a very small probability $\epsilon$. The trick is to enlarge the Hilbert space, and instead of $T$ \eqref{DCisometry}, use the isometry $T_\epsilon$ with modified states 
\begin{align}
	\kets{\psi_j^\epsilon} :=  \sqrt{\epsilon} \ket{\psi_j} + \sqrt{1-\epsilon} \ket{\perp_j},
\end{align}
where $\ket{\perp_j}$'s is a set of orthogonal states, orthogonal also to all of the $\ket{\psi_j}$'s and the states that come from $\ket{\psi_j}$'s by swapping their registers. The reflection about the span of $\kets{\psi_j^\epsilon}$ will thus have only a small part that corresponds to the reflection about the span of $\ket{\psi_j}$.
It can be shown that the modified simulation procedure acts the same way as the previous one, but as if we evolved with $\epsilon H$ instead of $H$. 

Altogether, when we want to simulate $H$ for time $t$ up to precision $\delta$, we choose a large enough number of simulation steps $\tau = O\left((\norm{H}t)^{\frac{3}{2}}\delta^{-\frac{1}{2}}, \norm{\mathrm{abs}(H)}t\right)$, and use a ``lazy'' walk with $\epsilon = \norm{\mathrm{abs}(H)}\frac{t}{\tau}$. For this all to work, we need to be able to analyze the 
principal eigenvector of $\mathrm{abs}(A)$. However, it is also possible to use different states $\ket{\psi_j}$ without the knowledge of the principal eigenvector, but that might in turn result in some increase of the required simulation resources \cite{ChildsDiscreteContinuous}. Using this translation from CTQW to DTQW (and vice versa!), Childs has obtained a CTQW version of the element distinctness algorithm with $O(N^{\frac{2}{3}})$ queries (in the Hamiltonian query model) and a DTQW version of the randomly glued-tree traversal algorithm, with a polynomial runtime.


\subsection{Summary}
Continuous-time quantum walks (CTQW) are a powerful tool for describing the dynamics of an excitation in a system. The Hamiltonian of the system can be as simple as the adjacency matrix of a graph. We have seen that the spreading of a wavepacket in this model can be similar to DTQW and that it can be used in lattice-search algorithms (see Section \ref{ContinuousSearchSection}). Furthermore, the dynamics have interesting properties and can be analyzed with usual quantum-mechanical tools -- utilizing symmetries allowed us to see the simple movement of superpositions in the randomly-glued tree traversal problem (see Section \ref{sec:gluedtrees}), while investigating the scattering properties of graphs resulted in the NAND-tree evaluation algorithm (see Section \ref{sec:NAND}). In fact, scattering properties of simple graphs can be used as a universal model of quantum computation (see Section \ref{sec:universality}). Finally, we have looked at how CTQW can be connected to discrete-time quantum walks in Section \ref{sec:connectthewalks}.



\section*{Acknowledgements}
\fancyhead[LO]{Acknowledgements}
\addcontentsline{toc}{section}{Acknowledgements}

We acknowledge support from the projects VEGA QWAEN, LPP-0430-09, APVV-0646-10
COQI, FP7 COQUIT and APVV SK-PT-0008-10.
We would also like to thank Yasser Omar for valuable discussions.

\newpage
\section*{Appendices}
\appendix
\fancyhead[LO]{Appendices}
\addcontentsline{toc}{section}{Appendices}


\setcounter{equation}{0} \setcounter{figure}{0} \setcounter{table}{0}
\section{Limiting Distribution of Classical Random Walks}
\label{sec:classicallimit}

To show the results of Sec.~\ref{sec:RWproperties}, let us consider random walk on graph $\mathcal G=(V,E)$, $|V|=N$, given by eq.~(\ref{eq:unbiasedRW}) where at each position the walker decides unbiasedly, where to go next. Further suppose, that the graph is connected (for unconnected graphs we would consider its connected subgraphs) and non-bipartite. The transition matrix $M$ of this random walk can be expressed as $M=AD^{-1}$ where
\[
D=\diag(d(1),d(2),\ldots,d(N))
\]
is the diagonal matrix having degrees of vertices on its diagonal and $A$ is the adjacency matrix
\[
A_{ij}=\begin{cases}
1\text{, if $ij\in E$,}\\
0\text{, otherwise.}
\end{cases}
\]
For undirected graphs $M$ is not symmetric in general and may not have spectral decomposition. On the other hand $A$ is symmetric. Applying a similarity transformation on $M$,
\[
D^{-1/2}MD^{1/2}=D^{-1/2}AD^{-1/2}\equiv Q,
\]
we see, that $M$ can be spectrally decomposed as well as it has the same eigenvalues as $Q$. Let $\lambda_1,\lambda2,\ldots,\lambda_N$ be the eigenvalues of $Q$ and $v_j,j=1,2,\ldots,N$ its corresponding eigenvectors. Then eigenvectors $w_j,j=1,2,\ldots,N$ of $M$ are connected to corresponding eigenvectors of $Q$ by equality $w_j=D^{1/2}v_j$. We can further suppose that the eigenvalues are ordered,
\[
\lambda_1\geq\lambda_2\geq\ldots\geq\lambda_N.
\]
Now we can write
\[
Q=\sum_j\lambda_j v_jv_j^T
\]
with vectors $v_j$ fulfilling normalization $v_j^Tv_i=\delta_{ij}$ and for the purpose of the evaluation of evolution $p(m)=M^mp(0)$ we write
\begin{eqnarray}
\label{eq:RWexpansion}
p(m) &=& M^mp(0)=\left(D^{1/2}QD^{-1/2}\right)^mp(0)=D^{1/2}Q^mD^{-1/2}p(0)\notag\\
&=& \sum_j\lambda_j^mD^{1/2}v_jv_j^TD^{-1/2}p(0).
\end{eqnarray}

Now that we know, that $M$ can be spectrally decomposed, we can concetrate on its eigenvalues, which are the same also for $Q$. We will show, that all the eigenvalues lie in the range $-1\leq\lambda_j\leq 1$ and that $\pi$ is the unique eigenvector with eigenvalue $1$.

\begin{lemma}
Eigenvalues of matrix $M$ lie in the interval $[-1;1]$ and the vector $\pi$ is the unique eigenvector of $M$ with eigenvalue $1$.
\end{lemma}

To prove this, let us consider some eigenvector $w$ with eigenvalue $\lambda$ and choose such $j$, that $|w(j)|/d(j)$ is maximal, where $w(j)$ is the $j$-th component of vector $w$. In other words we consider inequality
\[
\frac{|w(j)|}{d(j)}\geq\frac{|w(i)|}{D(i)}
\]
for all indices $i$. We also know, that
\[
\lambda w(j)=\sum_{i\st ij\in E}M_{ji}w(i)=\sum_{i\st ij\in E}\frac{w(i)}{d(i)}.
\]
Now we write
\begin{eqnarray}
\label{eq:eigenvalueRW}
|\lambda||w(j)| &=& \left|\sum_{i\st ij\in E}\frac{w(i)}{d(i)}\right|\leq \sum_{i\st ij\in E}\frac{|w(i)|}{d(i)}\notag\\
&\leq& \sum_{i\st ij\in E}\frac{|w(j)|}{d(j)}=|w(j)|.
\end{eqnarray}
This shows, that $|\lambda_j|\leq 1$ for all eigenvectors $v_j$. Moreover, equality can be obtained only when
\begin{equation}
\label{eq:ratios}
\frac{|w(j)|}{d(j)}=\frac{|w(i)|}{D(i)}
\end{equation}
for all neighbors $i$ of $j$ in the graph. The equality Eq.~(\ref{eq:ratios}) can be easily extended to all vertices, i.e.~to all indices $i$, as the graph is connected and finite, meaning, that whenever the equality would not hold, there would be at least one edge in the graph connecting vertices with different ratios which would be in contradiction with the proof. We see, that eigenvalue $\lambda_1=1$ uniquely determines the eigenstate, as the one fulfilling condition Eq.~(\ref{eq:ratios}): we can always consider $w_1(1)$ to be positive, in which case we can proceed in a similar manner as in Eq.~(\ref{eq:eigenvalueRW}), but without the absoulute values to find, that
\[
\frac{w_1(j)}{d(j)}=\frac{w_1(i)}{D(i)}.
\]
Taking this ratio to be equal to $1/2|E|$ we recover $\pi$:
\[
\pi=\frac{1}{2|E|}(d(1),d(2),\ldots,d(N))^T.
\]
This concludes the proof.

As $\pi$ is a multiple of $w_1=D^{1/2}v_1$ we can also find, that
\[
v_1=\sqrt{2|E|}D^{-1/2}\pi.
\]
We can conclude now, that
\[
1=\lambda_1>\lambda_2\geq\lambda_3\geq\ldots\geq\lambda_N\geq -1.
\]
Furthermore, from Eq.~(\ref{eq:RWexpansion}) we can observe, that in the limiting case of $m\to\infty$ all the terms with eigenvalues $\lambda_j$, $j=2,3,\ldots,N$ tend to zero and so
\[
\lim_{m\to\infty}p(m)=D^{1/2}v_1v_1^TD^{-1/2}p(0)=\pi\left(2|E|\pi^TD^{-1}p(0)\right)=\pi.
\]
This shows, that the limiting distribution is indeed $\pi$.

\begin{exercise}
Show that $2|E|\pi^TD^{-1}p=1$ for any probability distribution $p$.
\end{exercise}

\begin{exercise}
In previous proofs we have disregarded the possibility for eigenvalue of $M$ to be $-1$. Show, that we could do it, as only bipartite graphs lead to this case. [Hint: Take the eigenvector with eigenvalue $-1$ and apply $M^2$ to it]
\end{exercise}


\setcounter{equation}{0} \setcounter{figure}{0} \setcounter{table}{0}\newpage
\section{Evolution of Hadamard Walk in Detail}
\label{sec:stationary_point}

\subsection{Method of Stationary Phase}

The method of mathematical analysis called \emph{the method of stationary phase}, taken e.g.~from Ref.~\cite{AmBaNaViWa01} and from advanced mathematical textbooks, allows us to estimate our integrals, Eqs.~(\ref{eq:integrale}). Based on a fact, that in integrals of type
\[
I(m)=\int_a^b g(k) \e^{\ii m\phi(k)\dd k}
\]
the largest contribution comes from areas, where the oscillations in phase are small and that is near stationary points of function $\phi(k)$. Basically, under reasonable suppositions of smooth $g(k)$ that does not vanish in the only stationary point $a$ of the order $p$ of interval $[a;b]$ and with $t\to\infty$ we can make approximation
\[
I(m)\sim g(a)\exp\left\{\ii m\phi(a)+\sgn[\phi^{(p)}(a)]\frac{\ii\pi}{2p}\right\}\left[\frac{p!}{m\no{\phi^{(p)}(a)}}\right]^{\frac{1}{p}}\frac{\Gamma\left(\frac{1}{p}\right)}{p},
\]
where $\phi^{(p)}(a)\neq 0$ from assumption and $\sgn$ is a sign function.

Especially for the case of $p=2$ we get
\begin{equation}
\label{eq:approx_p2}
I(m)\sim\sqrt{\frac{\pi}{2m\no{\phi''(a)}}}g(a)\exp\left\{\ii m\phi(a)+\sgn[\phi''(a)]\frac{\ii\pi}{4}\right\}.
\end{equation}

\subsection{Hadamard Walk Evolution Approximation}
\label{sec:hadamard_approximation}

Integrals from Eqs.~(\ref{eq:integrale}) can be transformed to suit Eq.~(\ref{eq:approx_p2}) first by setting $x=\lambda m$, when these integrals obtain form
\begin{equation}
\label{eq:approx_gen_hadamard}
I(m;\lambda)=\frac{1}{2\pi}\int_{-\pi}^\pi g(k) \e^{\ii\phi(k;\lambda)}m \dd k,
\end{equation}
where $\phi(k;\lambda)=k\lambda-\omega_k$ and $g(k)$ is either an even or an odd function. In this slightly generalised case, there are no stationary points for $\lambda>1/\sqrt{2}$ or $\lambda<-1/\sqrt{2}$ and $I(m;\lambda)$ by getting with $\lambda$ further from zero decreases exponentially fast. For\footnote{We do not care about boundary points much, as these do not play an important role.} $\lambda\in(-1/\sqrt{2};1/\sqrt{2})$ we find stationary points $\pm k_\lambda\in[0;\pi]$ of order $p=2$, where
\[
\cos k_\lambda=\frac{\lambda}{\sqrt{1-\lambda^2}}
\]
and
\[
\frac{\partial^2\phi}{\partial k^2}(\pm k_\lambda;\lambda)=\pm(1-\lambda^2)\sqrt{1-2\lambda^2}\qquad(=-\omega''_{k_\lambda}).
\]
Under these conditions, and by dividing the integration range $[-\pi;\pi]$ in Eq.~(\ref{eq:approx_gen_hadamard}) into four subintervals by points $0$ and $\pm k_\lambda$ we find
\[
I(m;\lambda)=\frac{2g(k_\lambda)}{\sqrt{2\pi m\no{\omega''_{k_\lambda}}}}\begin{cases}
\cos\displaystyle{\left[m\phi(k_\lambda;\lambda)+\frac{\pi}{4}\right]}\text{, for $g$ even,}\\
\ii \sin\displaystyle{\left[m\phi(k_\lambda;\lambda)+\frac{\pi}{4}\right]}\text{, for $g$ odd.}
\end{cases}
\]
Finally, for Eqs.~(\ref{eq:integrale}), we obtain
\begin{subequations}
\label{eq:approximations}
\begin{eqnarray}
\alpha^m(\lambda m) &\sim& \frac{2}{\sqrt{2\pi m\no{\omega''_{k_\lambda}}}}\cos\left[m\phi(k_\lambda;\lambda)+\frac{\pi}{4}\right],\\
\beta^m(\lambda m) &\sim& \frac{2\lambda}{\sqrt{2\pi m\no{\omega''_{k_\lambda}}}}\cos\left[m\phi(k_\lambda;\lambda)+\frac{\pi}{4}\right],\\
\gamma^m(\lambda m) &\sim& -\frac{2\sqrt{1-2\lambda^2}}{\sqrt{2\pi m\no{\omega''_{k_\lambda}}}}\sin\left[m\phi(k_\lambda;\lambda)+\frac{\pi}{4}\right].
\end{eqnarray}
\end{subequations}
Now the probability of being at position $\lambda t$ after $m$ steps is
\begin{equation}
\label{eq:approx_hadamard_app}
P^m(\lambda m)\sim\frac{2(1+\lambda)}{\pi m(1-\lambda^2)\sqrt{1-2\lambda^2}}[1+\lambda\sqrt{2}\cos\theta],
\end{equation}
where
\[
\theta=2m\phi(k_\lambda;\lambda)+\frac{\pi}{2}+\mu
\]
and
\[
\tan\mu=\frac{\sqrt{1-2\lambda^2}}{1+2\lambda}.
\]
Note also following equality for further reference:
\begin{equation}
\label{eq:alphabeta}
\sgn\left[\alpha^m(x)\beta^m(x)\right]=\sgn x.
\end{equation}
\begin{exercise}
Prove Eq.~(\ref{eq:alphabeta}) [Hint: use approximated formulas].
\end{exercise}


\setcounter{equation}{0} \setcounter{figure}{0} \setcounter{table}{0}\newpage
\section{Catalan Numbers}
\label{sec:catalan_numbers}

Catalan numbers $C_n$ \cite{CatalanWiki} comprise a sequence of numbers that occur in combinatorics in many problems based on recurrence relations. For example, they enumerate the paths of a particle on a semi-infinite line starting from position 0 and returning to this position after $2n$ steps, i.e.~when only the non-negative positions are available. In Section \ref{sec:RWproperties}, we utilize Catalan numbers for computing a hitting time for a classical random walk, and in Section \ref{sec:discreteboundary}, they help us analyze a quantum walk on a line with an absorbing boundary. 

Considering all the paths returning to origin after $2k$ steps ($k=1,2,\ldots,n$) we can arrive at recurrence relation
\[
C_n=\sum_{k=1}^nC_{k-1}C_{n-k}=\sum_{k=0}^{n-1}C_kC_{(n-1)-k}
\]
with $C_0=1$. Writing the generating function for Catalan numbers, $c(x)=\sum_{n=0}^\infty C_nx^n$ we can find, that
\begin{equation}
\label{eq:catalan_genfun_recurrence}
c(x)=1+xc^2(x).
\end{equation}
Solving the quadratic equation we get
\begin{equation}
\label{eq:catalan_genfun}
c(x)=\frac{1-\sqrt{1-4x}}{2x}=\frac{2}{1+\sqrt{1-4x}}.
\end{equation}
The other solution of the quadratic equation is not acceptable, as $c(x)$ is a power series at $x=0$ and so it cannot have a pole.

On the other hand we know, that power series for $\sqrt{1+y}$ at $y=0$ is
\[
\sqrt{1+y}=1-2\sum_{n=1}^\infty\binom{2n-2}{n-1}\frac{(-1)^n}{4^n}\frac{y^n}{n}.
\]
Setting $y=-4x$ we can obtain also power series for Catalan generating function from Eq.~(\ref{eq:catalan_genfun_recurrence}),
\[
c(x)=\frac{1-\sqrt{1-4x}}{2x}=\sum_{n=0}^\infty\frac{1}{n+1}\binom{2n}{n}x^n.
\]
This shows, that Catalan numbers have form
\begin{equation}
\label{eq:catalan_number}
C_{n}=\frac{1}{n+1}\binom{2n}{n}.
\end{equation}
For large $n$ one can use approximation
\begin{equation}
\label{eq:catalan_limit}
C_n\simeq\frac{4^n}{n^{3/2}\sqrt{\pi}}.
\end{equation}

\begin{exercise}
By employing
\[
\sqrt{2\pi}n^{n+1/2}\e^{-n}\leq n!\leq\e n^{n+1/2}\e^{-n}
\]
show that
\begin{equation}
\label{eq:catalan_ineq}
C_{n+1}\geq\frac{2\sqrt{\pi}}{\e^2}\frac{4^k}{(k+1)^3/2}.
\end{equation}
\end{exercise}

\begin{exercise}
Show that
\begin{equation}
\label{eq:catalan_connection}
C_{n+1}=\frac{2(2n+1)}{n+2}C_n.
\end{equation}
\end{exercise}


\setcounter{equation}{0} \setcounter{figure}{0} \setcounter{table}{0}\newpage
\section{Grover's Fixed-point Search}
\label{sec:fixedpoint}

Grover's algorithm \cite{Grover97} for searching for one of $M$ marked items among $N$ elements 
in its original form is not a fixed-point search. If we run it for longer than its optimal time, 
the probability of finding a marked element begins to decrease, and after some time, we even get back to our original initial state. In \cite{BrassardScience}, Gil Brassard wrote ``The quantum search algorithm is like baking a souffle\dots you have to stop at just the right time or else it gets burnt.'' When we view the Grover algorithm as a quantum walk, the same is true --- the walker concentrates on a marked vertex at some special time, but then moves away from it. Could we modify the quantum algorithm (and the walk) to stay near a solution?

This algorithm can be modified \cite{GroverFixPoint,GroverFixPoint2} to work as a fixed-point one, producing a solution with high probability even if we ``overshoot'' the runtime. However, it loses the square root speedup (in number of oracle calls). Nevertheless, it can be useful when the number of marked items is not small\footnote{When this algorithm is used as a subroutine in Section \ref{sec:quantizeMCMC}, the ratio of marked/unmarked items is a constant.}. Let us sketch this modification, based on iterative phase-$\pi/3$ search.

First, let us assume we have an algorithm (a unitary transformation $U$ plus a final measurement) which fails to produce a marked state with probability $f$, when starting in the uniform superposition state $\ket{s}$. 
We will iteratively intersperse its uses with other (oracle-calling) transformations to increase the probability of finding a solution.
The additional ingredients are the selective phase shifts
\begin{align}
	R_s^{\theta} &= e^{i\theta} \Pi_{s} + (\iii-\Pi_s), \\
	R_M^{\phi} &= e^{i\phi} \Pi_{M} + (\iii-\Pi_M),
\end{align}
for the uniform superposition $\ket{s}$ and the marked subspace $M$. Note that for $\theta=\phi=\pi$ these are the original reflection operators (up to a sign) from Grover's algorithm. However, let us now choose $\theta=\phi=\frac{\pi}{3}$ and look at the operation
\begin{align}
	U_1 = U R_{s}^{\frac{\pi}{3}} U^\dagger R_{M}^{\frac{\pi}{3}} U. \label{fpsearch}
\end{align}
The original algorithm $U$ fails with probability $f$ (which can be even very close to 1). The unitary $U$ thus transforms $\ket{s}$ as
\begin{align}
	U \ket{s} = \sqrt{f} \ket{w'} + \sqrt{1-f} \ket{w},
\end{align}
with $\ket{w}$ a vector from the marked subspace, and $\ket{w'}$ some unmarked vector. Because of unitarity of $U$, there exists a vector $\ket{s'}$ orthogonal to $\ket{s}$, which is transformed by $U$ as
\begin{align}
U \ket{s'} =  \sqrt{1-f} \ket{w'} - \sqrt{f} \ket{w},
\end{align}
which allows us to express
\begin{align}
\ket{w}  = U \left( \sqrt{1-f}\ket{s} - \sqrt{f} \ket{s'} \right). 
\end{align}
To compute the failure probability of the new algorithm, let us apply $U_1$ to the initial state $\ket{s}$.
\begin{align}
	U_1 \ket{s} &= U R_{s}^{\frac{\pi}{3}} U^\dagger R_{M}^{\frac{\pi}{3}} U \ket{s} 
	= U R_{s}^{\frac{\pi}{3}} U^\dagger R_{M}^{\frac{\pi}{3}} \left(
		\sqrt{f} \ket{w'} + \sqrt{1-f} \ket{w}
	\right) \\
	&= U R_{s}^{\frac{\pi}{3}} U^\dagger \left(
		\sqrt{f} \ket{w'} + e^{i\pi/3}\sqrt{1-f} \ket{w}
	\right) \\
	&= U R_{s}^{\frac{\pi}{3}} U^\dagger \left(
		\sqrt{f} \ket{w'} + \sqrt{1-f}\ket{w} + \left(e^{i\pi/3}-1\right)\sqrt{1-f} \ket{w}
	\right) \\
	&= U R_{s}^{\frac{\pi}{3}} U^\dagger \left(
		U\ket{s} + e^{2i\pi/3}\sqrt{1-f} \, U \left( \sqrt{1-f}\ket{s} - \sqrt{f} \ket{s'} \right)
	\right) \\
	&= U R_{s}^{\frac{\pi}{3}}  \left(
		\left(1+ e^{2i\pi/3}(1-f)\right)
		\ket{s} - e^{2i\pi/3}\sqrt{(1-f)f} \ket{s'}
	\right) \\
	&= U  \left(
		e^{i\pi/3} \left(1+ e^{2i\pi/3}(1-f) \right)
		\ket{s} - e^{2i\pi/3} \sqrt{(1-f)f} \ket{s'}
	\right) \\
	&=  \left(
		e^{i\pi/3} - 1 + f \right) 
		U \ket{s}
		- e^{2i\pi/3} \sqrt{(1-f)f} \, U\ket{s'}
\end{align}
After the final substitution for $U\ket{s}$ and $U\ket{s'}$, we collect the coefficients in front of $\ket{w'}$ and obtain
the probability of failure for the new algorithm
\begin{align}
	\left| \bra{w'} U_1 \ket{s} \right|^2 &= \left| \sqrt{f}\left(e^{i\pi/3}-1+f\right) - e^{2i\pi/3} (1-f) \sqrt{f} \right|^2 \\
	 &= f \left| \left( e^{i\pi/3}-1 - e^{2i\pi /3}\right) + f \left(1+e^{2i \pi /3}\right) \right|^2 \\
	 &= f^3  \left|1+e^{2i \pi /3}\right|^2 = f^3.
\end{align}
The original failure probability was $f<1$, so the new failure probability $f^3$ is smaller. 
This composite algorithm \eqref{fpsearch} reminds us of an error-correcting scheme.

Let us compute how the failure probability decreases with the number of queries.
At each level of composition, it decreases as $f \rightarrow f^3$.
Assuming the original algorithm $U$ takes $c_0=n$ oracle calls,
the cost of the new approach $U_1$ is $c_1=3n+1$. At the next level, composing $U_2 = U_1 R_s^{\pi/3} U_1^\dagger R_M^{\pi/3}U_1$ uses
$c_2 = 9n + 3 +1$ oracle calls.
\vspace*{-0.1cm}
\begin{exercise}
Show that the number of oracle calls for $k$-levels of composition becomes $q_k=3^k n + \half\left(3^k-1\right)$.
\end{exercise}
It turns out that we can even use $U_0=\iii$ (without any oracle calls) for our starting algorithm. Alternatively, we could use a single iteration from Grover's algorithm $U_G = R^{\pi}_s R_M^{\pi}$. 
\vspace*{-0.1cm}
\begin{exercise}
For the strategy $U_0$ with $c_0=0$, show that the failure probability at the $k$-th level of composition $f_k = f_0^{3^k}$ decreases with the number of oracle calls as
$f(q_k) = f_0^{2 q_k + 1} = \left(1-\frac{M}{N}\right)^{2 q_k + 1}$.
\end{exercise}
Thus, the fixed-point quantum algorithm needs need $O\left(\frac{N}{M}\right)$ queries to find a marked item with reasonable probability. This is obviously inferior to the $O\left(\sqrt{\frac{N}{M}}\right)$ scaling of Grover's original algorithm, but can still be useful for ratios $\frac{M}{N}$ that are not too small. Finally, let us compare this scaling to the classical case, where the failure probability after querying $q$ elements decreases with the number of oracle calls as $f(c_k) = {N-M \choose c_k}/ {N \choose c_k} \approx \left(1-\frac{M}{N}\right)^{c_k}$ for $M, c_k\ll N$.

\newpage
\fancyhead[LO]{References}
\section*{}
\vspace*{-1.5cm}
\addcontentsline{toc}{section}{References}


\newpage
\medskip
\thispagestyle{empty}
\noindent
\begin{wrapfigure}{l}{3.1cm}
  \begin{center}
    \vspace*{-0.33cm}
    \includegraphics[width=3.1cm,clip]{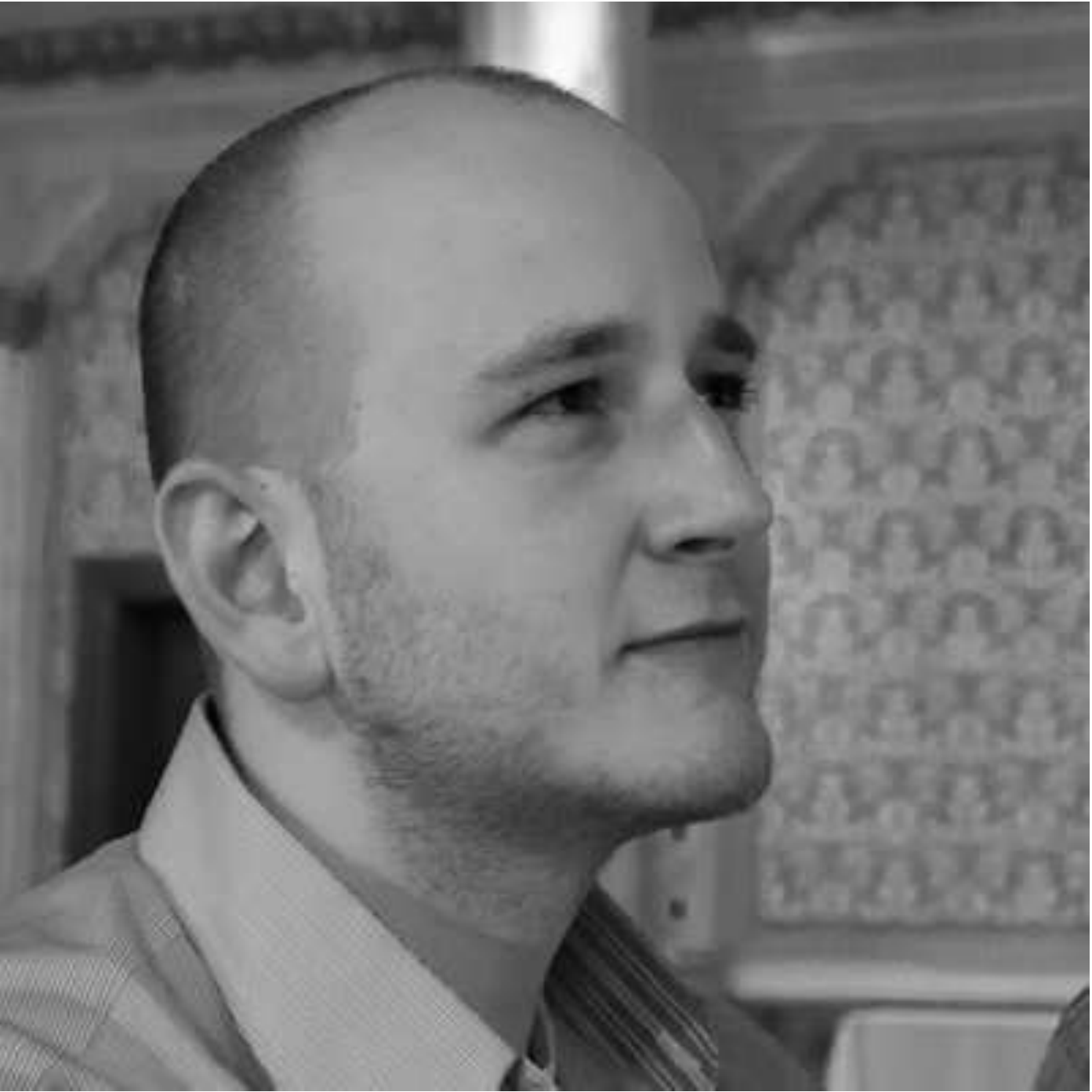}
  \end{center}
\vspace*{-0.7cm}
\end{wrapfigure}
{\bf Dr. Daniel Reitzner} received his PhD in General and mathematical physics
from Slovak Academy of Sciences and Comenius University in 2010. He
was employed at the Research Center for Quantum Information between
years 2010 and 2011. Since 2011 he is employed at the Technical
University in Munich. He specializes in quantum walks and their
algorithmic applications as well as in some foundational aspects of
quantum mechanics connected to joint measurability and compatibility.\\

\noindent
\begin{wrapfigure}{l}{3.1cm}
  \begin{center}
    \vspace*{-0.77cm}
    \includegraphics[width=3.1cm,clip]{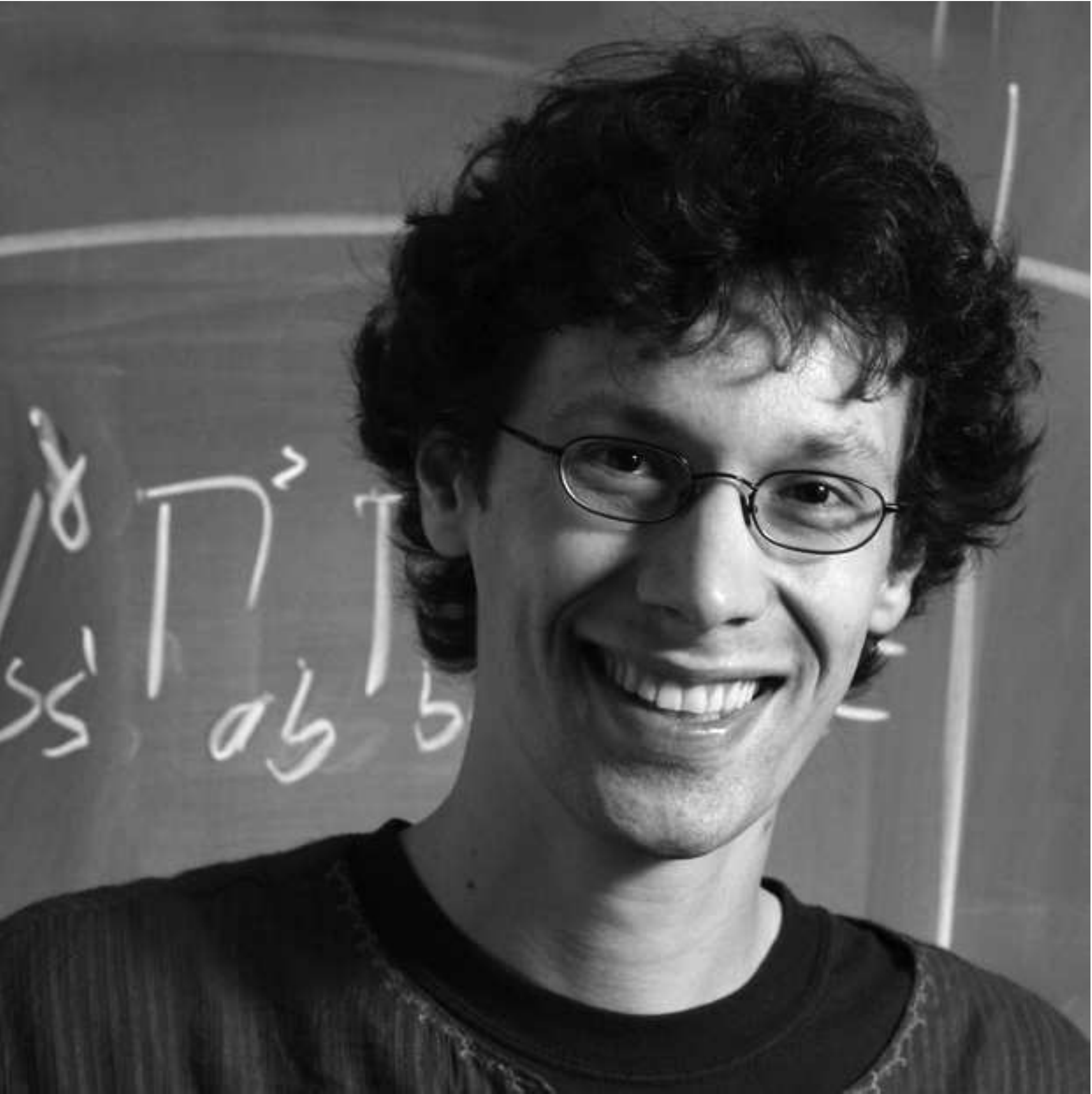}
  \end{center}
\vspace*{-0.7cm}
\end{wrapfigure}
{\bf Dr. Daniel Nagaj} received his PhD in theoretical physics from MIT in 2008. \hspace*{-0.0cm}Between 2008 and 2012, he was employed at the Research Center for Quantum Information at the Institute of Physics of the Slovak Academy of Sciences. He mainly studies the computational capabilities of nature (quantum mechanical) - adiai\-ba\-tic quantum computation, quantum walks, and Hamiltonian complexity. He is also interested in numerical methods for condensed-matter physics based on DMRG and tensor product states.\\

\noindent
\begin{wrapfigure}{l}{3.1cm}
  \begin{center}
    \vspace*{-0.8cm}
    \includegraphics[width=3.1cm,clip]{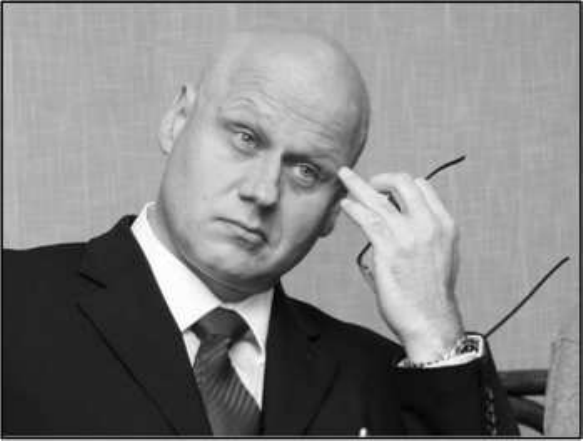}
  \end{center}
\vspace*{-0.7cm}
\end{wrapfigure}
{\bf Prof. Vladim\'{\i}r Bu\v{z}ek}
has graduated at the Moscow State University (both MSc and PhD). His research interests include quantum optics, quantum information sciences, foundations of quantum theory, quantum thermodynamics, and quantum measurement theory. He is an author and co-author of more than 220 research papers and 15 chapters in monographs and books. These papers have been cited more than 7000 times ($H=42$).  During his academic career he was a visiting professor at a number of academic institutions including the Imperial College, London (UK), the National University of Ireland, Maynooth (Ireland), and the University of Ulm (Germany). He has been serving on various national and international advisory, scientific, and editorial boards.
Prof. Bu\v{z}ek is the president of the Learned Society of the Slovak Academy of Sciences and a foreign corresponding member of the Austrian Academy of Sciences. He is a fellow of the Institute of Physics (UK) and of the Optical Society of America (USA). For his research achievements he was awarded the Ernst Abbe Medal and the International Commission for Optics Prize, the Humboldt Research Award (Germany) and the E.T.S Walton Award (Ireland).\\

\end{document}